\RequirePackage{etoolbox}
\pretocmd\PackageWarning{%
  \edef\pkgname{#1}\edef\hyperrefname{hyperref}%
  \ifx\pkgname\hyperrefname
    \expandafter\gobblethree
  \fi
}{}{\undefined}
\newcommand*{\gobblethree}[3]{}

\documentclass{fpfslut}

\usepackage[T1]{fontenc}
\usepackage[utf8]{inputenc}
\usepackage{graphicx}
\usepackage{amsthm}
\usepackage{amssymb}
\usepackage{epigraph}
\usepackage{epstopdf}
\usepackage{url}
\usepackage{amsthm}
\usepackage{amsmath}
\usepackage{wasysym}
\usepackage{anysize}
\usepackage{academicons}
\usepackage{scalerel}
\usepackage{tikz}
\usetikzlibrary{svg.path}

\usepackage{pdfpages} 

\usepackage{times}
\usepackage{newtxtext,newtxmath}

\marginsize{3.5cm}{2.5cm}{1cm}{1cm}

\usepackage[acronym,section=subsection,nomain]{glossaries} 
\usepackage{glossary-mcols}
\usepackage{glossary-long}

\setglossarystyle{mcolalttree}
\definecolor{orcidlogocol}{HTML}{A6CE39}
\tikzset{
  orcidlogo/.pic={
    \fill[orcidlogocol] svg{M256,128c0,70.7-57.3,128-128,128C57.3,256,0,198.7,0,128C0,57.3,57.3,0,128,0C198.7,0,256,57.3,256,128z};
    \fill[white] svg{M86.3,186.2H70.9V79.1h15.4v48.4V186.2z}
                 svg{M108.9,79.1h41.6c39.6,0,57,28.3,57,53.6c0,27.5-21.5,53.6-56.8,53.6h-41.8V79.1z M124.3,172.4h24.5c34.9,0,42.9-26.5,42.9-39.7c0-21.5-13.7-39.7-43.7-39.7h-23.7V172.4z}
                 svg{M88.7,56.8c0,5.5-4.5,10.1-10.1,10.1c-5.6,0-10.1-4.6-10.1-10.1c0-5.6,4.5-10.1,10.1-10.1C84.2,46.7,88.7,51.3,88.7,56.8z};
  }
}

\newcommand\orcidicon[1]{\href{https://orcid.org/#1}{\mbox{\scalerel*{
\begin{tikzpicture}[yscale=-1,transform shape]
\pic{orcidlogo};
\end{tikzpicture}
}{|}}}}

\usepackage{hyperref}
\newglossary[fc-glg]{functions}{fc-gls}{fc-glo}{Glossary and various symbols}
\makeglossaries

\input{glossary-agn.gls}

\title{{\em Selected Chapters on}\\[12pt]Active Galactic Nuclei as Relativistic Systems}
\supervisor{ }
\yearofsubmission{2021}

\author{V. Karas, J. Svoboda, \& M. Zaja\v{c}ek}

\newcommand{\xmm}{{XMM-Newton}}

\def\keywordsname{\bf Keywords} 
\def\keywords#1{\unskip\par\smallskip\noindent{\emph{\keywordsname:} \rm #1}}

\newcommand{\cov}{\mathrm{cov}}

\begin{document}

\maketitle

\pagenumbering{arabic}

\mainmatter
\setcounter{tocdepth}{2}

\begin{annotation}
Astrophysical black holes  in nuclei of galaxies are of indisputable relevance to the current research. These extreme objects are of immense interest to  pure relativists as well as observational astronomers. This diversification is reflected in varied approaches that people adopt to the problem. Mathematically minded theorists take more interest in analytical solutions to Einstein’s equations and idealized systems where the role of strong gravity can be exposed. Observers want to actually \textit{see\/} the gas flows swirling in the form of an accretion disk or a torus. Astronomers often invoke disk-type accretion to explain spectral features observed in galactic nuclei, although the detailed physics of accretion flows still remains incomplete.\hfill~\\[6pt]
Despite that the processes in strong gravity govern only the innermost regions of accretion disks, their role is quickly diminished with the distance from the source. Non-gravitational physics plays a crucial role and it often dominates the acceleration and emission processes. In order to see where the limitations of our models are, it is useful to discuss simplified and general analytical models along with particular astronomical objects.\hfill~\\[6pt]
These lecture notes present an overview of selected aspects of physical processes occurring in the inner regions of Active Galactic Nuclei (AGN). Observational evidence strongly suggests that strong gravitational fields play a significant role in governing the energy output of AGN and their influence on the surrounding medium due to the presence of a supermassive black hole (SMBH) or a system of SMBHs undergoing a merger. We begin by an account of observational properties of AGN, their basic structural components and unification scenarios, and the arguments for the presence of SMBHs in their cores. Subsequently, in more detail, we discuss selected phenomena that are related to black-hole accretion and relevant for the emerging radiation signal and the acceleration of matter in AGN cores. \hfill~\\[6pt]
In order to reduce an unnecessary overlap with numerous reviews and textbook chapters on the similar subject, we focus on just a few topics, such as the launching scenarios of nuclear outflows and jets that start and pre-collimate in the immediate vicinity of black holes, at distance of only a few gravitational radii, and then can reach spectacular length-scales of $\sim$Mpc, extending far beyond the host galaxy. We also discuss the interaction between magnetic and gravitational fields in the strong-gravity regime of General Relativity. Some more recent aspects of the AGN unification scheme are included and quite a large number of references are provided to relevant papers and textbooks at the end of the volume. The interested reader is invited to consult the original works, which can provide an exciting context to the lively research on Active Galactic Nuclei.\hfill~\\[6pt]

 \begin{minipage}[b]{.48\linewidth}
    {\bf Cover photo:} Cygnus A active galaxy is shown on a collage across the electromagnetic spectrum: the X-ray signal (blue) from the Chandra Observatory have been superimposed over the radio emission (red), which extends to nearly 300,000 light-years. Jets of relativistic particles emanate to either side from the galaxy's central supermassive black hole and mark the hot spots at both ends at the impact onto the surrounding material. Optical data (yellow) from Hubble Space Telescope and the Digital Sky Survey complete the multi-wavelength view.
 \end{minipage}
 \hfill
 \begin{minipage}[b]{.48\linewidth}
    \includegraphics[angle=-90,width=\linewidth]{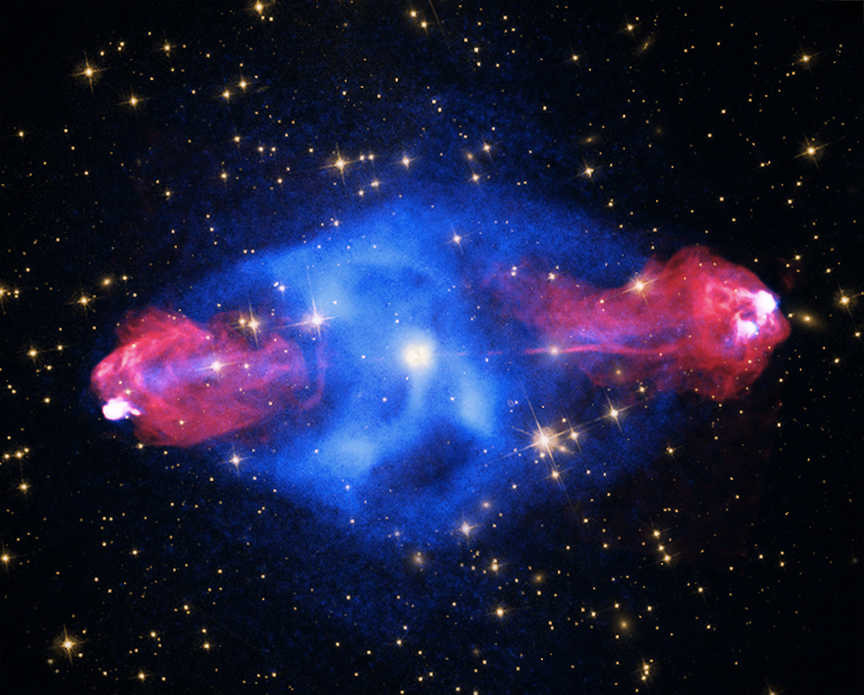}
    
    \medskip
    
    {\small Cygnus A composite image credits: X-ray NASA/CXC/SAO; Optical NASA/STScI; Radio NSF/NRAO/AUI/VLA}
 \end{minipage}
\vspace*{15cm}
\end{annotation}

\tableofcontents
\thispagestyle{uheadings}

\addtocontents{toc}{\hspace*{3.5ex}Preface}

\newpage
\section{Defining properties of Active Galactic Nuclei}
\epigraph{\textit{``Do not undertake a scientific career in quest of fame or money. There are easier and better ways to reach them. Undertake it only if nothing else will satisfy you; for nothing else is probably what you will receive. Your reward will be the widening of the horizon as you climb. And if you achieve that reward you will ask no other."}}{--- \textup{Cecilia Payne-Gaposchkin}}
   Active galactic nuclei (AGN) are compact cores of a special category of galaxies that are characterized by high luminosity of non-stellar origin. They are distinguished also by the spectral energy distribution (SED) of the emerging radiation which spans a wide range of electromagnetic spectrum -- from low-frequency radio waves across millimeter, infrared, and optical light, and further up to X-rays and high-energy gamma-rays. Most AGN are characterized by a strong emission in UV, known as the Big Blue Bump, and emission in X-rays much exceeding the typical X-ray luminosity of normal, non-active, galaxies. Some AGN exhibit powerful jets,  collimated to a very narrow cone and moving by strongly relativistic velocities up  to the intergalactic space. There is a plenty of different observational evidence of the nuclear activity in various sources that lead to classification of several different kinds of AGN (occasionally referred to as an AGN zoology). This various appearance strongly depends on the orientation of the AGN with respect to the observer (as we will more discuss in sect.~\ref{AGN_structure}). Our aim is not to fully describe all observed kinds of AGN, but rather introduce the main characteristic features.

Based on a broad spectrum of observational evidence it has been generally accepted that the common denominator for AGN activity is the presence of at least one supermassive black hole (SMBH) that accretes the surrounding matter. A variety of physical processes must operate and a wide range of physical conditions must occur, namely, density, pressure and temperature of the diluted environment, and intensity of the magnetic field and of the radiation field pervading the SMBH neighborhood. Accretion rate and the spin of the black hole appear to be the main parameters influencing the intrinsic properties of active nuclei. Whereas numerical simulations are necessary to model the evolution of the system, it turns out that idealized scenarios can reflect underlying physical mechanisms in their mutual interplay. We will attempt to supplement the vast literature that deals with the subject of AGN by the context of simplified toy-model approaches, where the number of degrees of freedom are reduced and only selected ingredients are taken into account.

Let us mention a subtle mystery that governs seemingly different and unrelated objects: the mass scaling and the corresponding length scales. An appropriate change of the scale by many orders of magnitude can bring us from one type of object to another, from SMBH in the cores of galaxies to their bulges and even beyond. This fact has been extensively investigated in the context of feedback mechanisms and even the impact on the large-scale structure of the Universe.

We will summarize arguments for the presence of such a giant black hole in AGN. The multiple convincing evidence has also been found in our own Galaxy. Unlike the cores of active galaxies, the $10^6$ solar mass black hole in the center of the Milky Way (Sagittarius A*) is, however, currently inactive; we will discuss this in section~\ref{SgrA}. In the following chapters, we will review the main radiation processes relevant for black-hole accretion (Chapt.~\ref{Radiation_processes}) and jet-launching mechanisms (Chapt.~\ref{Jets}).
Finally, in Chapter~\ref{AGN_across_mass}, we will compare accretion on SMBH in AGN with the case of stellar-mass black holes in X-ray binaries. Readers are referred to numerous monographs and review articles \citep[e.g.,][]{1999agnc.book.....K,1991RPPh...54..579O,2006eac..book.....S, 2013peag.book.....N} for further information about physics, observational properties and rich phenomenology of different kinds of AGN. A very recent exposition can be found in \citep{2017A&ARv..25....2P}. The importance of coevolution of SMBHs and their host galaxies, processes of SMBH growth and the resulting relation between the black hole mass and the bulge mass in AGN have been reviewed by \citet{2013ARA&A..51..511K}.
   \subsection{AGN across wavelengths}
   \label{basics}
   The subject of AGN has grown enormously over recent three decades of intense investigation. Nowadays, it is an arena of a far too vast and interconnected observational and theoretical research to be covered in a single review. Even if we concentrate our attention on a restricted volume of the inner regions near SMBH, we have to omit many topics, including those ones that are of immense importance. We will attempt to formulate a compromise and provide a very limited introductory overview supplemented by several less-discussed aspects of current interest.


\begin{figure}[tbh!]
 \centering
 \begin{tabular}{cc}
 \includegraphics[width=0.35\textwidth]{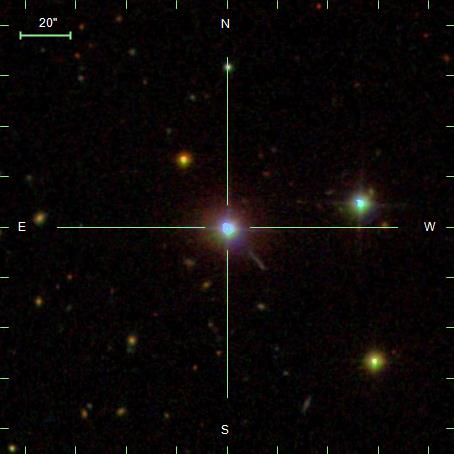} & \includegraphics[width=0.5\textwidth]{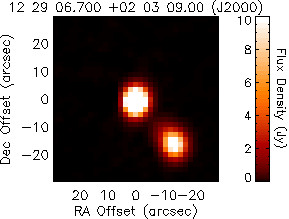}
 \end{tabular}
 \caption{Left panel: Optical image of quasar 3C 273 adopted from SDSS optical survey \citep{2002AJ....123..485S}. Right panel: The radio counterpart of 3C 273 at the frequency of $1.4\,{\rm GHz}$. Figure from the FIRST survey \citep{1995ApJ...450..559B}.}
 \label{3C273}
\end{figure}

\begin{figure}[tbh!]
 \centering
 \includegraphics[width=0.8\textwidth]{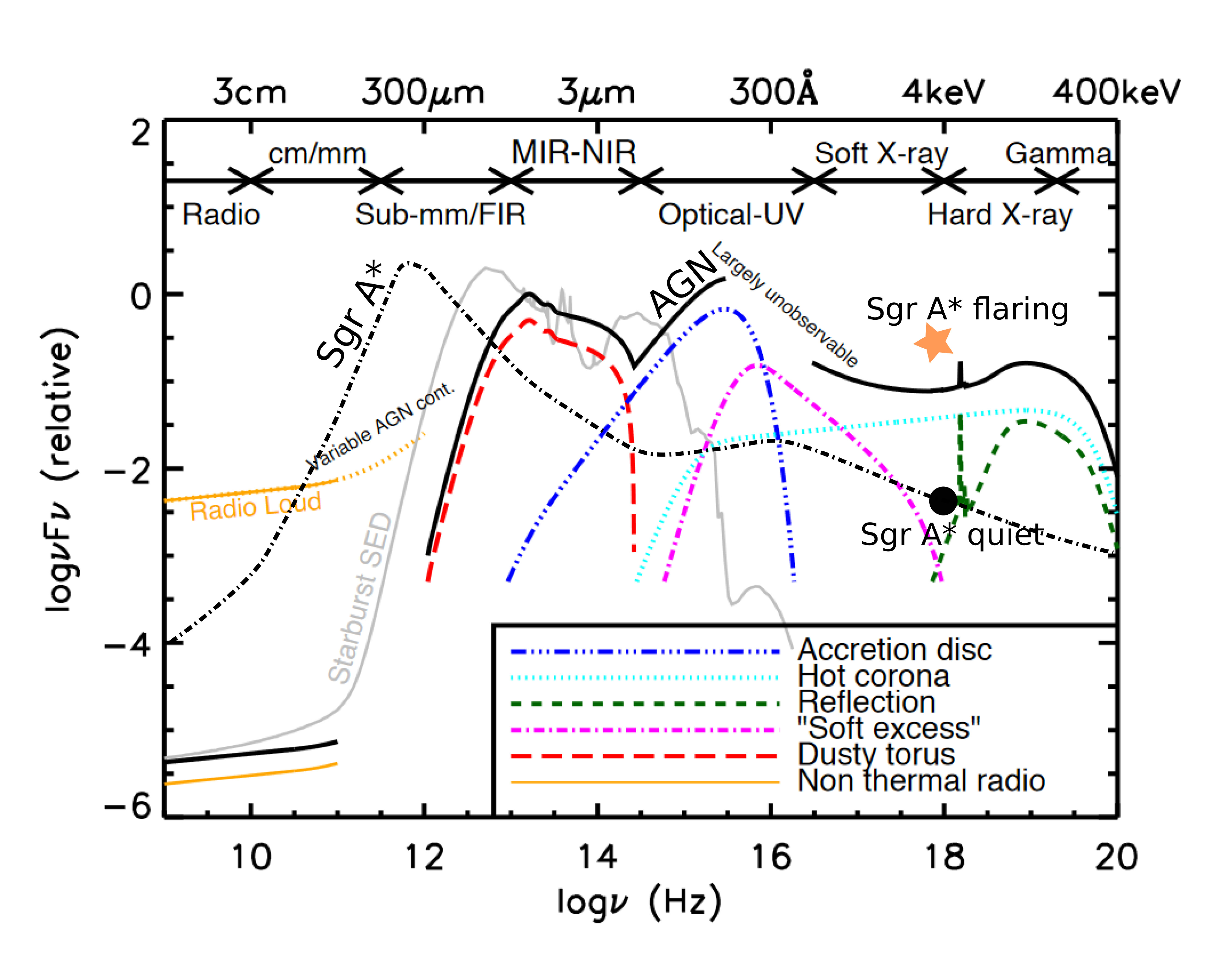}
 \caption{An illustrative view of the AGN spectral energy distribution (SED) based on the actual observational data of radio-quiet AGN \citep[see references][ for more details]{Elvis1994, Richards2006}. The total AGN SED is depicted with a solid black line, while individual components (accretion disc, hot corona, reflection component, soft excess, dusty torus, and non-thermal radio component) are coloured. Towards sub-mm/FIR and radio domains, the non-thermal radio component depends on the jet activity (radio-loud vs. radio-quiet). For comparison, the SED of an extremely low-luminous AGN, represented by Sgr~A*, is shown using a dot-dashed line. Sgr~A* exhibits flares in the NIR/X-ray domains, which can increase the X-ray flux density by as much as several orders of magnitude. The solid grey line depicts the SED of a starburst galaxy, which peaks in the sub-mm/FIR domain \citep[based on the SED of M82 adopted from the GRASIL library; ][]{1998ApJ...509..103S}. The image was adopted from \citet{2014PhDT.......357H} and modified to add the SED of Sgr~A* according to \citet{2014ARA&A..52..529Y}. 
 }
\label{agn_sed}
\end{figure}

Active galaxies differ from normal, inactive galaxies, such as our Milky Way, by significantly broader wavelength range of their electromagnetic spectrum, which can extend over the entire observable interval from radio wavelengths, through infrared, optical, UV, X-ray, up to gamma-ray wavelengths. Each wavelength puts a different piece into the mosaic.  An example of an AGN is quasar 3C 273, whose image in optical and radio is shown in Fig.~\ref{3C273}. While the optical image reveals bright emission from a point-like source (nucleus of the galaxy), the radio image reveals two spatially separated knots of the relativistic jet. However, the jet as a thin faint line can also be identified in the optical image, which is due to the broad-band synchrotron emission that can also extend from the radio to the optical domain. Multi-frequency observations are therefore necessary to fully understand the AGN structure and physics.

The SED of normal galaxies can be understood as the superposition of thermal spectra produced by stars of different surface temperature. Since the range of the stellar temperature is rather narrow, $T_{\rm eff} \gtrsim 3000\,{\rm K} $  and $T_{\rm eff} \lesssim 40\,000\,{\rm K}$, most of the bolometric luminosity of normal galaxies is emitted in the infrared, optical, and ultraviolet parts of the electromagnetic spectrum.
AGN therefore require a different mechanism from the stellar thermal emission to explain their prominent radio, X-ray, and even gamma-ray emission. Furthermore, the enormous radiation output of AGN cannot often be detected directly; instead it is obscured by a lot of gas and dust that absorbs, modifies and reemits much of the characteristic observational signatures \citep{2018arXiv180604680H}. This reprocessing challenges observational methods to uncover the complete population and understand the evolution of SMBHs.

Within a limited energy range, the AGN SED can often be described as a power-law function,
\begin{equation}
\label{powerlaw}
L_{\nu} \propto \nu^{-\alpha},
\end{equation}
where $\alpha$ is the spectral index. A schematic view of an AGN SED is shown in Fig.~\ref{agn_sed}. Different physical components and mechanisms contribute to the SED and produce a more complicated spectrum, namely, a broken power-law with multiple values of the spectral index, or even a superposition with additional components and cut-off at a certain maximum energy. Thermal emission from the accretion disk, non-thermal emission from the hot corona, X-ray reflection, and reprocessed emission by the dusty torus, they all contribute to the SED. When the jet is present, the AGN SED extends to radio and gamma-rays and the jet contribution can even dominate the signal when pointing along the line of sight towards us (the case of so-called blazars). The AGN SED often peaks in the ultraviolet domain due to the multi-temperature thermal emission of an accretion disc, which is referred to as the big blue bump. In contrast, the spectrum of starburst galaxies (shown by a grey curve in Fig.~\ref{agn_sed}) is limited mainly to the infrared and optical frequency bands, and peaks in the sub-mm/FIR domain. For comparison, in Fig.~\ref{agn_sed}, we also show the SED of Sgr~A*, the compact radio source in the center of the Milky Way. Sgr~A* is a typical representative of extremely low-luminous galactic nuclei that accrete several orders of magnitude below the Eddington limit, see Eq.~\eqref{eq_Eddington_luminosity} for the definition. Low-luminous, quiescent galaxies are typically powered by optically thin and geometrically thick hot accretion flows \citep{2014ARA&A..52..529Y} that are in contrast to optically thick and geometrically thin accretion discs that are present for highly-accreting AGN. The result is that for Sgr~A*, the peak of its SED shifts towards longer wavelengths to the sub-mm/mm domain. In comparison with the thermal big blue bump, the sub-mm/mm peak arises due to the synchrotron emission of thermalized electrons of the magnetized hot flow.

\subsubsection*{Radio band}
The radio band was the first band to discover AGN and quasars. The normal stars are very weak in radio band. Thus, the discoveries of optical point sources that have also strong radio emission led to the first identification of quasars. The quasar discovery was made by \citet{1963Natur.197.1040S} who noticed a very large redshift ($z=0.158$) of emission lines in a star-like source from the 3C (The Cambridge 178 MHz radio survey) sample  \citep{1959MmRAS..68...37E}. 

The AGN radio luminosity can enormously span the range over several orders of magnitude, and therefore, AGN are distinguished into radio-loud vs. radio-quiet sources. The majority of AGN are radio-quiet, while only about 10$\%$ are radio-loud
\citep{Kellermann1989, Balokovic2012}. The reason for this dichotomy is the presence or absence of the strong relativistic jet launched by the SMBH. However, the radio-loud/quiet distinction and the reason behind it is not yet a settled problem. Different ideas were proposed to explain this dichotomy, including the spin paradigm \citep{Moderski1998, Sikora2007}, observational selection effects \citep{2000ApJS..126..133W, Cirasuolo2003} or evolution of the spectral states \citep{Koerding2006, Svoboda2017}. In the study of radio dichotomy, it is also important whether the total radio luminosity or only the core radio luminosity is taken into account \citep{2011MNRAS.417..184B}. This distinction is obviously possible only for local AGN but not for unresolved distant quasars.

\begin{figure}[tbh!]
    \centering
    \includegraphics[width=\textwidth]{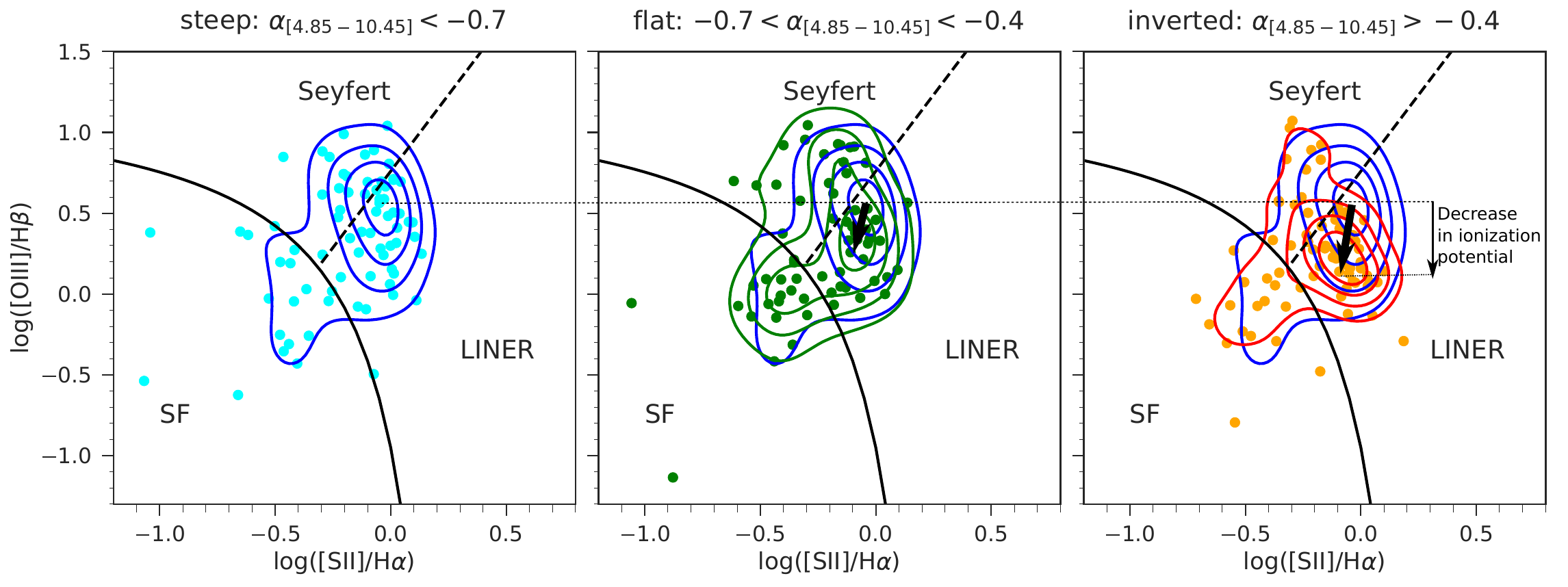}
    \caption{Distribution of sources with different radio spectral index in the narrow emission-line diagnostic diagram, also known as the Baldwin--Phillips--Terlevich (BPT) plot \citep{1981PASP...93....5B}, where three spectral classes of galaxies are denoted: star-forming (SF), Seyfert, and LINER. From the left to the right, we see that steep sources have a tendency to group towards the Seyfert sources, i.e. a higher ionization potential, while inverted sources peak in the LINER region, i.e. towards a lower ionization potential. Figure adopted from \citet{2020past.conf..248Z}; see also \citet{2019A&A...630A..83Z} for a detailed analysis and a discussion.}
    \label{fig_specindex}
\end{figure}

The radio luminosity, spectrum and morphological classification depends on how bright the jet core is versus extended radio lobes created by the jet dissipation into the intergalactic matter. The original morphological classification by \citet{Fanaroff1974} distinguishes between the radio-core (FR type I) and radio-lobes (FR type II) dominated sources. Later on, radio point sources were categorized as FR type 0.
The radio spectral slope differs accordingly. The core-dominated spectra are typically radio-flat with $\alpha=0\pm 0.5$, while the radio lobe-dominated spectra are steep with $\alpha<-0.5$. A special class of inverted radio spectra was established for indices $\alpha>0.5$. Many more classes were introduced based on the radio spectral slope, e.g., Compact Steep Sources \citep{1990A&A...231..333F}, Giga-Hertz Peak Sources \citep{1991ApJ...380...66O}, etc. The spectral index distribution is also correlated with the optical properties of host galaxies, specifically their ionizing potential traced by narrow emission-line ratios. In particular low-ionization narrow emission region (LINER) sources have a tendency for a flatter spectral index and a lower ionization potential expressed by the narrow emission-line ratio [OIII]/H$\beta$, while Seyfert sources have a steeper spectral index and a larger ionization potential, see Fig.~\ref{fig_specindex} for the graphical representation in the optical diagnostic diagram and \citet{2019A&A...630A..83Z} for a detailed analysis. This distinction could be related to the duty cycle of the sources \citep{2009ApJ...698..840C,2016A&ARv..24...10T}, with Seyfert galaxies being active long enough to develop radio lobes, while LINER sources lack large radio lobes due to their old age or because of just recently restarted activity. In both cases, a compact self-absorbed radio core with a compact jet would dominate the total radio output, which would result in a flat or a potentially inverted radio spectral index due to the self absorption of the synchrotron emission.  

\begin{figure}[tbh!]
    \centering
    \includegraphics[width=0.99\textwidth]{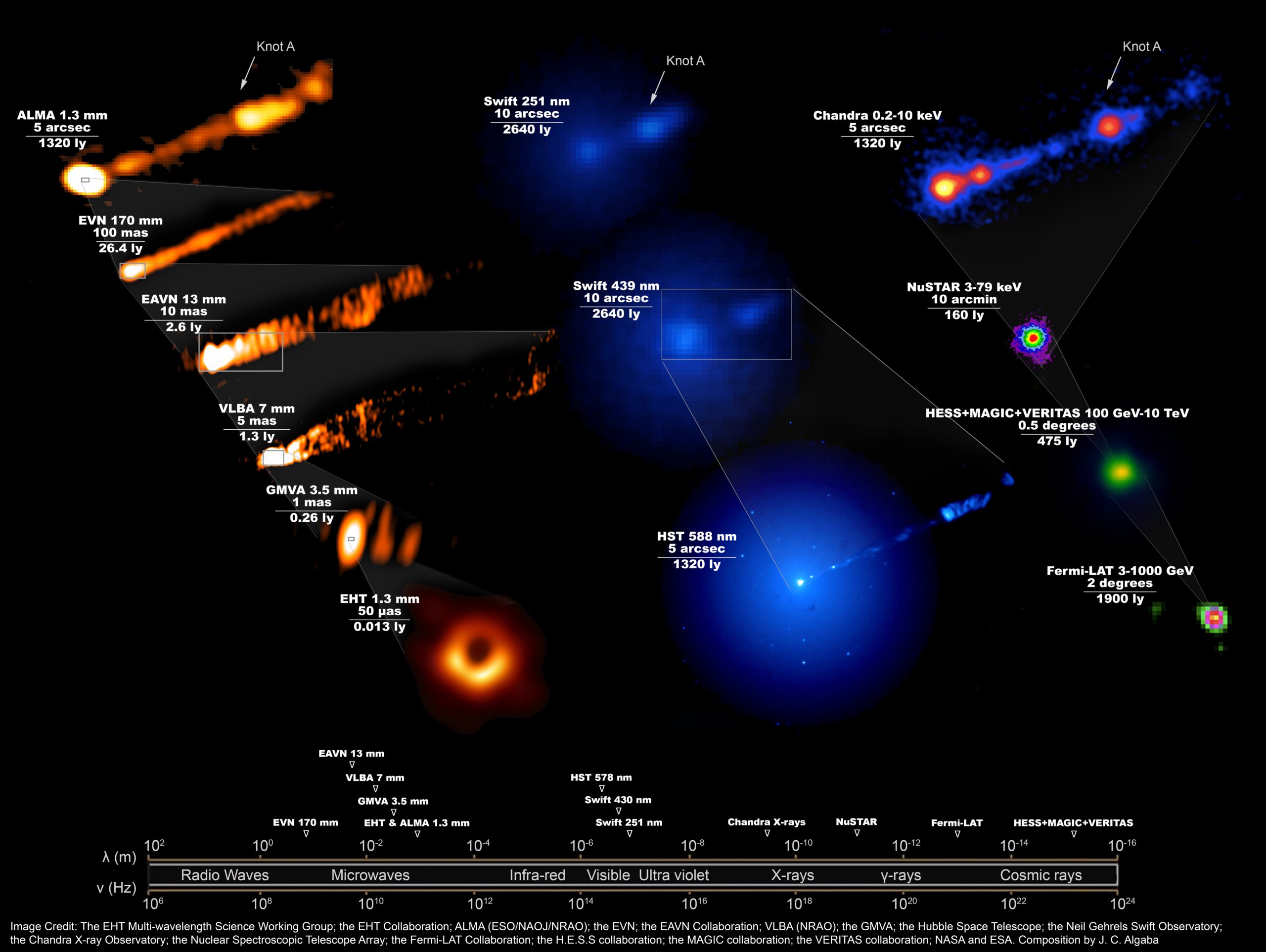}
    \caption{A collage of multi-wavelength images during the 2017 observing campaign of M87. 19 different instruments on the ground and in space had to be used to obtain the presented images. In the upper left corner, the ALMA image at 1.3 mm (orange) shows a large-scale jet structure, with a clear core region and knot A. This is comparable in projected scale to the optical HST image in the center as well as to the X-ray Chandra image in the upper right corner. Image credits: The EHT Multi-wavelength science working group; the EHT Collaboration; ALMA (ESO/NAOJ/NRAO); the EVN; the EAVN Collaboration; VLBA (NRAO); the GMVA; the Hubble Space Telescope; the Neil Gehrels Swift Observatory; the Chandra X-ray Observatory; the Nuclear Spectroscopic Telescope Array; the Fermi-LAT collaboration; the H.E.S.S. collaboration; the MAGIC collaboration; the VERITAS collaboration; NASA and ESA. The multi-wavelength composition  by J. C. Algaba. (Based on the M87 multi-wavelength analysis by \citet{2021ApJ...911L..11E}.)}
    \label{fig_M87_multiwavelength}
\end{figure}

The high-resolution radio observations using very-long-baseline interferometry (VLBI) have played a major role in understanding the jet kinematics, collimation, and the launching mechanism \citep[see][ for a review]{2017A&ARv..25....4B}. Starting initially at centimeter wavelengths, the VLBI technology made it possible to resolve out the jet kinematics on parsec scales. The launching mechanism and initial plasma flow acceleration remained largely unresolved, since at longer wavelengths (smaller frequencies), the plasma is optically thick for non-thermal synchrotron radiation due to self-absorption. The onset of mm-VLBI observations mitigated this problem and the high-resolution VLBI studies at mm wavelengths has brought unprecedented view into the inner portions of jets, including the black hole shadow of M87* \citep{2019ApJ...875L...1E}, which is at the center of the giant elliptical galaxy M87 (Virgo A, NGC4486) at the center of the Virgo cluster of galaxies, having a distance of $\sim 16.4$ million parsecs. Combining eight radio telescopes all around the world (Arizona, Hawai’i, Mexico, Chile, Spain, and the South Pole) that observed M87 simultaneously in 2017 at the wavelength of 1.3 mm as a part of the Event Horizon Telescope (EHT), the angular resolution of a few 10\,$\mu{\rm as}$ was reached as well as a sufficient coverage of the observer's \textit{uv} plane in the Fourier space \citep{2019ApJ...875L...1E}. This enabled the astronomers to make use of \textit{Van Cittert--Zernike theorem} to produce the intensity distribution in the source/sky plane out of the visibility distribution, or rather a mutual coherence function, in the observer's plane, or the \textit{uv} plane, which is defined by the telescope baselines. Mathematically, \textit{van Cittert--Zernike} theorem can be formulated as,
\begin{equation}
    \Gamma_{12}(u,v,0)=\iint I(l,m)\,\exp{\big[- 2\pi\Im (ul+vm)\big]}\mathrm{d}l\,\mathrm{d}m\,,
    \label{eq_vancittert_zernike}
\end{equation}
where $I(l,m)$ is the intensity distribution as a function of direction cosines, $l$ and $m$, in the source plane, and $\Gamma_{12}(u,v,0)$ is a mutual coherence function between two observing stations as a function of $u$ and $v$ coordinates (in units of the observing wavelength), or $x$- and $y$-direction, in the observer's plane.
At these scales, the dominantly elongated morphology of a collimated jet changes into the ring-like feature with an asymmetric surface-brightness distribution due to an emitting plasma orbiting at the fraction of a light speed around the supermassive black hole of $(6.5 \pm 0.7)\times 10^9\,M_{\odot}$, which results in a mild Doppler beaming in the observer's frame \citep{2019ApJ...875L...1E}. Combining the cm- and mm-VLBI, it is thus possible for M87 to connect the small-scale acceleration region with the large-scale jet structures. In Fig.~\ref{fig_M87_multiwavelength}, we show the multi-wavelength image of M87 during the observational campaign in 2017, combining images at mm wavelengths, UV/Optical, X-ray as well as $\gamma$-ray domains. In this sense, M87 serves as a unique example of an active galactic nucleus, which plays a crucial role in terms of comparing the results of general relativistic  magnetohydrodynamics and general relativistic ray-tracing models with observational data.   

\subsubsection*{Infrared band}
Most AGN emission in the infrared band is not a direct emission of the accreting black hole, but a reprocessed emission in the dusty environment. Because of the strong photoionizing effects of AGN emission, the dust can hardly survive in the innermost region. Instead, a large reservoir of dust occurs at a distance $\gtrsim 1\,{\rm pc}$ from the black hole. This dust structure is not spherically symmetric, but in overall shape it concentrates around the equatorial plane and creates a toroidal structure with an opening angle of $\approx 30-60$ degrees, the so-called dusty molecular torus.

The inner radius of the dusty torus is set by the dust sublimation temperature $T_{\rm sub}$. According to \citet{1987ApJ...320..537B}, the dust sublimation radius depends primarily on the dust sublimation temperature (here scaled to $1\,500\,K$), the AGN UV luminosity (scaled to $10^{45}\,{\rm erg\,s^{-1}}$), and the UV optical depth, which is generally set to zero in the first-order estimates,
\begin{equation}
    R_{\rm sub}=0.37\,\left(\frac{T_{\rm sub}}{1\,500\,{\rm K}}\right)^{-2.8}\left(\frac{L_{\rm AGN}}{10^{45}\,{\rm erg\,s^{-1}}} \right)^{1/2} \left(\frac{a_{\rm dust}}{0.05\,{\rm \mu m}} \right)^{-1/2}\exp{(-\tau_{\rm UV}/2)}\,{\rm pc}.
    \label{eq_sublimation_radius}
\end{equation}
The sublimation temperature depends on the composition, we mainly distinguish silicate and graphite grains, with graphite grains having a larger sublimation temperature in the range $\sim 1500-1900\,K$, while the silicate grains sublime at $\sim 1000-1400\,{\rm K}$. \citet{2007A&A...476..713K} also explicitly add the grain size in Eq.~\eqref{eq_sublimation_radius}, which can be of relevance. The dependency on the grain size arises due to the grain absorption efficiency, which is proportional to the dust grain size in the near-infrared domain. The outer radius of the torus is expected to be located at $R_{\rm torus}\sim YR_{\rm sub}$, where $Y\sim 5-10$ \citep{2006ApJ...648L.101E,2008ApJ...685..160N}.

The contribution of the infrared emission is particularly large for the type-II AGN,
a class of AGN that are heavily obscured in the optical emission by the dusty torus. The absorbed radiation heats the dust whose reprocessed emission appears in the mid-IR band. There are some AGN classified solely based on their excess in this mid-IR band, while their optical and X-ray emission would not distinguish them from normal galaxies. 

The infrared emission is essential to understand the nature and geometry of the torus. Based on the infrared observations, it is now getting more and more evident that the torus is composed of many individual clouds and is clumpy rather than having a homogeneous structure
\citep{2002ApJ...570L...9N, 2006A&A...452..459H, 2009ApJ...702.1127R},
as originally proposed by \citet{1988ApJ...329..702K}.
It was also realized that the inner radius of the torus as well as its opening angle depends on the AGN luminosity \citep[see, e.g.,][]{2008ApJ...679..140T, 2009A&A...502...67T}. For a recent review on AGN torus physics and structure, see, e.g., \citet{2015ARA&A..53..365N}, or an overview by \citet{2016SPIE.9907E..0RB} of the results achieved by recent most-advanced infrared-interferometry measurements.

\subsubsection*{Optical/UV}
The optical spectrum of AGN is distinguishable from normal galaxies mainly due to their strong broad-band continuum and strong emission lines often with large line widths that can be explained by the Doppler broadening due to the motion of the gas around the centre with a supermassive black hole. Not all AGN have similar properties of their emission lines, and according to the width of the lines, AGN have been classified into two main types: type-I with very broad lines and type-II with only narrow lines. Several intermediate types (such as type-I.2, type-I.5, I.9) were also identified \citep[see, e.g.,][]{1989agna.book.....O}.

The explanation of the absence of the broad lines in type-II galaxies is due to the obscurer along the line-of-sight. Most of the AGN emission is absorbed in these galaxies, and the observer only sees very dimmed continuum, the reprocessed emission and the narrow emission lines from the polar regions, i.e. a narrow-line region on the scale of $\sim 100\,{\rm pc}$. The broad line wings appear only in the polarized emission. The first type II source where the broad lines represented by the Balmer lines and the permitted FeII line are revealed in the polarized light was NGC 1068 using spectropolarimetric measurements, see \citet{1985ApJ...297..621A} for details. This discovery led to the so-called AGN unification scheme, i.e. that AGN have the similar emission components and the morphology with a various detected strength because of being viewed at different viewing angles \citep{Antonucci1993,1995PASP..107..803U}.

\begin{figure}[tbh!]
 \centering
 \begin{tabular}{cc}
 \includegraphics[angle=0,width=0.7\textwidth]{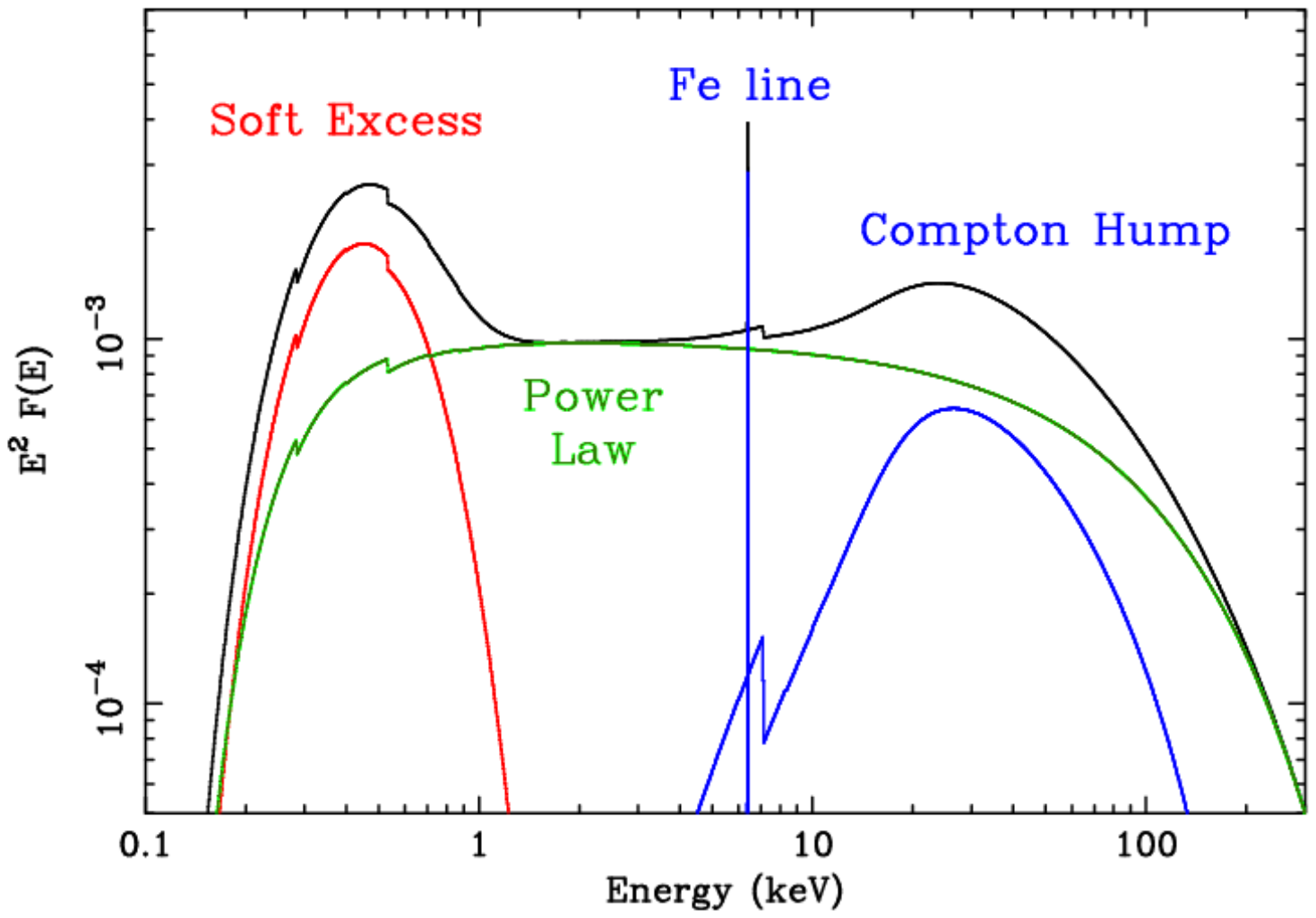} 
 \end{tabular}
 \caption{Schematic view of AGN X-ray spectrum in the energy range 0.1 keV $<E<$ 300 keV \citep{2006AN....327..943F,2000PASP..112.1145F}. Customarily, the ordinate shows the radiation flux scaled by $E^2$, so that the central part of the power-law component (see the green line) becomes approximately horizontal.}
\label{xrayspec}
\end{figure}

The AGN SED peaks in the UV band (near 10\,eV or 1240\,\AA). This peak, known as the big blue bump, is traditionally associated with the expected peak of the accretion-disk thermal emission \citep{Czerny1987}. In some quasar spectra, the big blue bump was indeed well modelled by the Shakura--Sunyaev accretion disk model \citep{2011MNRAS.415.2942C}. However, for most AGN and quasars, the peak in the SED is not as sharp as would be predicted by an accretion-disk thermal emission,
and other model components are needed to explain SED towards infrared and X-ray band. The observational obstacle is the fact that at the energy range of 13.6\,eV -- 0.1\,keV, the emission is so obscured by the Galactic interstellar gas.

\subsubsection*{X-rays and $\gamma$-rays}
The AGN X-ray emission comes from the closest neighborhood of the supermassive black hole and possesses the most relevant evidence about the central source. The primary X-ray emission has a power-law like shape and it is dominated by the inverse Compton scattering of accretion-disk thermal photons in radio-quiet AGN, while the boosted jet emission prevails the spectrum in radio-loud sources (see Sec.~\ref{observed_radiation}). The X-ray power-law radiation is partly reprocessed at the accretion disk surface. The main imprints of such X-ray reflection are the iron line and the Compton hump, for a review, see e.g., \citet{2003PhR...377..389R, 2009A&ARv..17...47T}. A schematic figure of the typical AGN X-ray spectrum is shown in Figure~\ref{xrayspec}. 

The soft X-ray part of the AGN spectrum often exhibits an excess that can be due to ionized reflection \citep{2006MNRAS.365.1067C} or due to Comptonization of thermal photons in a warm corona in contrast to the hot corona that creates power-law going to harder X-rays \citep{Done2012}. The soft X-ray emission is very often subjected to a mildly ionized absorption due to the so-called warm absorber. The absorption lines are often observed blueshifted, indicating an outflow of the warm absorber. In the most extreme cases, the outflow reaches $v_{\rm outflow} \approx 0.3$\,c \citep{Tombesi2010}. Such outflows are referred to as UFOs (ultra-fast outflows).

The very hard X-ray emission of radio-quiet AGN is very weak. The primary X-ray power-law is diminished by an exponential decrease of the flux around the so-called cutoff energy, as predicted by the Comptonization models. The first measurements of the cutoff energy suggest it to differ from source to source, usually it is over 100\,keV \citep{Fabian2015}.

AGN with significant $\gamma$-ray emission are typically radio-loud objects, indicating the origin of the $\gamma$-rays in relativistic jets. The mechanism producing $\gamma$-rays is most likely the inverse Compton upscattering of either low-frequency synchrotron photons of the jet emission or thermal photons of an accretion disc. The most luminous $\gamma$-ray sources are blazars. These sources have their collimated jets pointed towards us and the emission is strongly boosted by the Doppler beaming effect. Depending on the strength of broad emission lines, blazars can further be distinguished into BL Lac sources that lack optical emission lines (or they are only very weak) and flat-spectrum radio quasars (FSRQ) that exhibit prominent broad emission lines in their optical spectra. The $\gamma$-ray emission of blazars is variable on short timescales of the order of days and even hours, which suggests the compact size of $\gamma$-ray emitting zones, see e.g. \citet{2021MNRAS.503.3145B}. In addition, several sources exhibit a significant correlation between the radio and the $\gamma$-ray emission, with the radio emission lagging behind the high-energy emission, suggesting a spatial separation of $\gamma$-ray and radio-emission unit opacity zones as disturbances propagate downflow the jet \citep{2014MNRAS.445..428M}, but also vice versa in some cases \citep{2019Galax...7...72B,2021MNRAS.502.5245P}, which implies the existence of two separate emission zones for the low- and high-energy radiation production.


\subsection{AGN variability}

AGN were found variable across all wavelengths with different amplitudes \citep{2002MNRAS.332..231U,2003MNRAS.345.1271V,2016MNRAS.461.3145V}. In general, when the source is observed in a certain waveband over time, the simplest way to represent its variability is to show its light curve, i.e. display $N$ flux densities $f_i$ at specific times $t_i$.

The variability is mainly related to \textit{primary sources} of the radiation, in particular, the thermal emission of the accretion disk or the non-thermal emission of the jet. In addition, there is also variability of \textit{secondary emission sources} that are powered by the emission of primary sources, for instance, the free-free emission of photo- or collisionally ionized broad-line region clouds (see Subsection~\ref{circumnuclear}). In that case, the light curve of the secondary emission is significantly correlated with the primary emission light curve,
and can be used to derive the scales of the emission regions, such as the extent of broad line region. The method is known as the \textit{reverberation mapping} ans is discussed in more detail in Subsection~\ref{circumnuclear}.

The AGN variability can be studied using different formalisms, each of which has its strengths as well as weaknesses. One of them is the \textit{structure function} (SF), which transforms the light curve to the variability amplitude-timescale space and SF can be constructed in the following way. For each flux density $f_i$ at time $t_i$, we calculate the difference of logarithms for each flux $f_j$ at later times $t_j$, $\delta \log{F}(\delta t)=\log{(F_j/F_i)}$ where $\delta t=t_j-t_i$. The structure function is then defined as the mean absolute value of $2.5\delta \log{F(\delta t)}$, in other words the mean absolute value of the magnitude difference over different time intervals. A more precise definition of the structure function is based on the decomposition of the flux into the signal and noise components, $f_i=s_i+n_i$, with the corresponding variances $\sigma_{s}^2$ and $\sigma_{n}^2$. The definition stems from the covariance $\cov(s_i,s_j)$ of the light curve with its time-shifted copy by time $\Delta t=t_i-t_j$ (in the rest frame of the source). In particular \citep{2016ApJ...826..118K},
\begin{equation}
    \mathrm{SF_1}(\Delta t)^2=2[\sigma_{ s}^2-\cov(s_i,s_j)]+2\sigma_n^2\,.
    \label{eq_sf1}
\end{equation}
Using the measured light curves, the structure function is commonly estimated as,
\begin{equation}
    \mathrm{SF_2}(\Delta t)^2=\frac{1}{N_{\Delta t \text{pairs}}}\sum_{i=1}^{N_{\Delta t \text{pairs}}}(f_i-f_j)^2\,,
    \label{eq_sf2}
\end{equation}
where $ \mathrm{SF_1}(\Delta t)^2=\mathrm{SF_2}(\Delta t)^2-2\sigma_{n}^2$, i.e. one needs to subtract correctly the noise variance $\sigma_n^2$. If this is not done properly, the structure-function slope may be flatter than it should be from basic theoretical and modelling principles, see \citet{2016ApJ...826..118K} for details.

The variability level can also be characterized using the \textit{mean fractional variance}, which is defined as,
\begin{equation}
    F_{\rm var}=\frac{\sqrt{\sigma^2-\delta^2}}{\langle f\rangle}\,,
    \label{eq_mean_variance}
\end{equation}
where $\langle f\rangle$ is the mean flux density of $N$ measurements,
\begin{equation}
    \langle f\rangle=\frac{1}{N}\sum_{i=1}^{N}f_{i}\,.
    \label{eq_mean_flux}
\end{equation}
The flux density variance $\sigma^2$ can then be calculated as,
\begin{equation}
    \sigma^2=\frac{1}{N}\sum_{i=1}^{N}(f_i-\langle f\rangle)^2\,,
\end{equation}
while the mean square flux uncertainty is simply expressed as,
\begin{equation}
    \delta^2=\frac{1}{N}\sum_{i=1}^{N}\delta_i^2\,.
\end{equation}
The mean fractional variance expressed by Eq.~\eqref{eq_mean_variance} corrects for the flux measurement uncertainties, and in this sense, it estimates the true intrinsic AGN variability. It is also referred to as the \textit{excess variance}.

The AGN variability is aperiodic and driven by a \textit{stochastic} process in most cases \citep{2016MNRAS.461.3145V}. However, there are a few cases when on top of the stochastic variability, one can also detect a quasi-periodic flux density changes. Such a periodic process can be driven by the accretion-flow perturbations due to a bound perturber, in particular a secondary supermassive or an intermediate-mass black hole, a neutron star, or even a wind-blowing main-sequence star. \citet{2021arXiv210208135S} present a general relativistic magnetohydrodynamic model for the accretion-disc perturbations driven by a bound companion and discuss the connection to several observed sources, both active galactic nuclei and quiescent SMBH, such as Sgr~A*, where stars with pericenter distances of a few 100 gravitational are already detected \citep{2020ApJ...889...61P,2020ApJ...899...50P}.  

\begin{figure}[tbh!]
    \centering
    \includegraphics[width=0.6\textwidth]{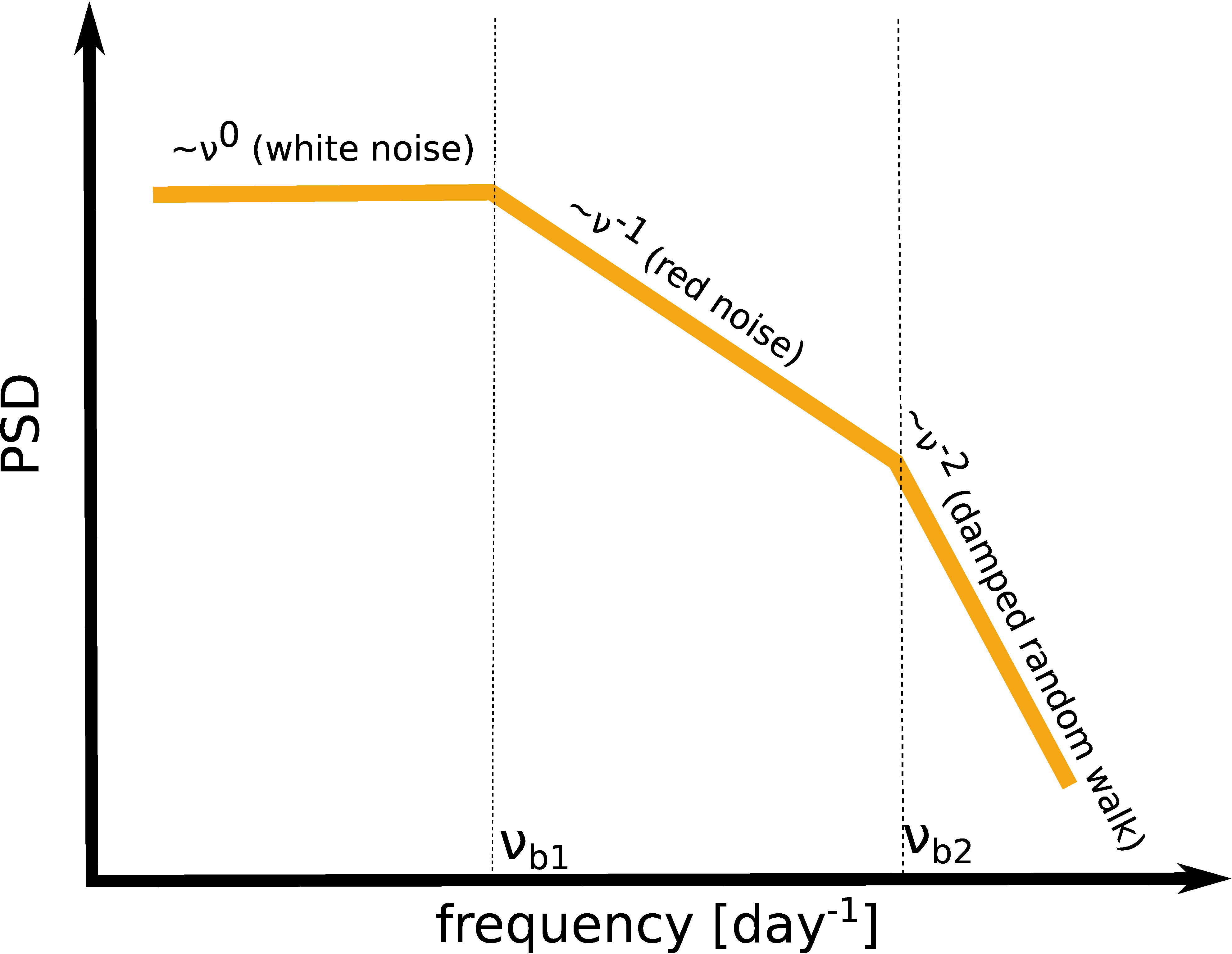}
    \caption{An illustrative power spectral density (PSD) shape described as a broken power-law function with two spectral breaks, $\nu_{\rm b1}$ and $\nu_{\rm b2}$. At low frequencies, the PSD is dominated by the flat ``white''-noise process, while at the highest frequencies, the slope steepens to $-2$ and the PSD is modelled well by the damped random walk (DRW) process. At intermediate frequencies, between $\nu_{\rm b1}$ and $\nu_{\rm b2}$, the PSD is dominated by the ``red'' noise process with the slope close to $-1$.}
    \label{fig_psd_sketch}
\end{figure}

The \textit{power spectral density} (PSD) allows one to study the nature of the stochastic mechanism as well as on which timescales most of the variability occurs. To construct the PSD, we need to obtain the Fourier transform of the light curve $C(t)$, which is by default in the time domain. $\Tilde{C}(\nu)$ in the frequency domain is:
\begin{equation}
    \Tilde{C}(\nu)=\int_{-\infty}^{+\infty} C(t)\exp{(-i2\pi \nu t)}\mathrm{d}t\,.
    \label{eq_fourier_transform}
\end{equation}
The PSD is then obtained by the multiplication of $\Tilde{C}(\nu)$ by its complex conjugate $\Tilde{C}^{*}(\nu)$. Over a certain temporal frequency range ($\nu$ is expressed in Hz or day$^{-1}$), the PSD can be well fitted by a single power-law function, i.e.,
\begin{equation}
    \text{PSD}\propto \nu^{\alpha}\,,
    \label{eq_PSD_powerlaw}
\end{equation}
where $\alpha\leq 0$. Over a broader frequency range, from the smallest frequencies (longest timescales) to the highest frequencies (shortest timescales), PSD is generally described as a broken power-law function with one or two break frequencies. This stems from the fact that the total power is obtained by integrating over all frequencies and it needs to be finite. In particular, by integrating from a fixed frequency to the highest frequencies, it is required to have $\alpha<-1$ for convergence, while integrating towards the smallest frequencies, the power-law slope needs to increase, i.e. $\alpha>-1$. Hence, at least two power-law frequency breaks are expected. The general PSD shape is sketched in Figure~\ref{fig_psd_sketch}. From the general shape, $\text{PSD}\propto 1/f^{1-2}$, it is evident that most of the variability occurs on longer timescales (small temporal frequencies). At intermediate frequencies, between the PSD frequency breaks at $\nu_{\rm b1}$ and $\nu_{\rm b2}$, the PSD shape is mostly driven by the ``red-noise'' process with the PSD slope of $-1$, while at higher frequencies, it gets steeper with the slope of $-2$, which is described well by the damped random walk process \citep[DRW; ][]{2009ApJ...698..895K,2010ApJ...721.1014M}. At low frequencies, the PSD is nearly independent of the frequency, which is typical of the ``white noise". Mock light curves can be generated from the assumed PSD broken power-law shape using the \textit{Timmer--Koenig} algorithm (see \citet{1995A&A...300..707T} for details).

The two PSD breaks are difficult to capture observationally, as at low frequencies, the timescales are too long, which requires a long monitoring program. On the other hand, at high frequencies, the short timescales are difficult to capture because of the constraints on the signal-to-noise ratio, i.e. the shorter the timescale is, the larger the observational noise. The break frequency between $\sim \nu^{-1}$ and $\sim \nu^{-2}$ is generally observed for both AGN and stellar-mass black holes, with the typical characteristic timescale of $T_{\rm B}=1/\nu_{\rm B}$ of days for SMBHs and seconds for stellar-mass black holes. The second break indicated in Figure~\ref{fig_psd_sketch} between $\sim \nu^{-1}$ and $\sim \nu^0$ has been mainly confirmed for stellar-mass black holes at low frequencies (long timescales). 

The characteristic variability timescale $T_{\rm B}$ has empirically been determined to be proportional to the black hole mass and indirectly proportional to the accretion rate \citep{2005MNRAS.363..586U,2006Natur.444..730M},
\begin{equation}
    T_{\rm B}\propto \frac{M_{\rm BH}^{1.12}}{\lambda_{\rm Edd}^{0.98}}\,,
    \label{eq_TB_0}
\end{equation}
where $\lambda_{\rm Edd}=L_{\rm bol}/L_{\rm Edd}$ is an Eddington ratio, i.e. the ratio between the bolometric luminosity $L_{\rm bol}=\eta \dot{M}c^2$ and the Eddington accretion rate $L_{\rm Edd}=4\pi GM_{\rm BH}m_{\rm p}c/\sigma_{\rm T}$, see also Eq.~\eqref{eq_Eddington_luminosity} and the related text for the derivation. The Eddington ratio can then be expressed $\lambda_{\rm Edd}\propto \dot{M}/M_{\rm BH}$, which leads to the relation between the characteristic variability timescale $T_{\rm B}$, the black hole mass $M_{\rm BH}$, and the accretion rate $\dot{M}$,
\begin{equation}
    T_{\rm B}\propto \frac{M_{\rm BH}^{2.1}}{\dot{M}^{0.98}}\sim \frac{M_{\rm BH}^2}{\dot{M}}\,,
    \label{eq_TB_1}
\end{equation}
which is valid for accreting SMBHs as well as quiescent stellar binaries \citep{2005MNRAS.363..586U,2006Natur.444..730M} and can be referred to as ``variability fundamental plane''.

Since $T_{\rm B}\propto M_{\rm BH}$ the variability timescale (break frequency) has been associated with the standard timescales of an accretion disc. However, the orbital dynamical timescale $t_{\rm dyn}$ is too short for AGN, of the order of hours, while the viscous timescale $t_{\rm vis}$ for an efficient accretion tends to be too long, of the order of years given $H/R\ll 1$. The association of the break frequency with the thermal instability timescale is more promising \citep{2006Natur.444..730M}. In addition, radial accretion flow fluctuations propagating from outer radii inwards have also been successful in interpreting the PSD frequency break \citep{1997MNRAS.292..679L,2001MNRAS.321..759C}. Another interpretation is the association of the characteristic variability timescale with the viscous timescale at the truncation radius, $t_{\rm vis}\sim t_{\rm dyn}\alpha_{\rm ss}^{-1}(H/R)^{-2}$, where $\alpha_{\rm ss}$ is the $\alpha$-viscosity parameter of the geometrically thin, optically thick Shakura--Sunyaev disc \citep{1973A&A....24..337S}. At the truncation radius, the optically thick disc transitions into geometrically thick and optically thin adiabatically dominated accretion flow \citep{2004PThPS.155...99Z}. The viscous timescale depends on the accretion rate, which determines the position of the truncation radius and in this sense the dynamical timescale. The height-to-thickness ratio $(H/R)$ also depends on the accretion rate, in particular, for accretion rates close to and exceeding the Eddington rate, the flow is geometrically puffed up and transitions into the slim disc solution with $H/R\sim 1$ \citep{2010LNP...794..203M,2019ApJ...884L..37L}. In this way, the viscous timescale can be altered by the accretion state. Finally, the characteristic variability timescale could be determined by the characteristic timescale of the production of hard X-rays, specifically, by the cooling timescale of electrons in the Comptonization radiation process that is responsible for the production of the hard X-ray power-law spectral distribution. According to the derivation by \citet{2012A&A...540L...2I} based on the first principles, the cooling timescale of hot-corona electrons that are responsible for the upscattering of soft seed photons follows the same dependency as indicated by Eq.~\eqref{eq_TB_1}.

\subsection{AGN structure and the unification scenarios}
A large and stable luminosity that is comparable and often outshines the whole galaxy is best explained by an infall of matter -- gas and dust -- through the potential well of a very compact concentration of matter. For an object with mass $M$ and radius $R$, on which a matter of mass $m$ falls from infinity onto its surface, the released gravitational potential energy is,

\begin{equation}
  \Delta E_{\rm grav}=G\,\frac{Mm}{R}\,.
  \label{eq_grav_energy}
\end{equation}

This energy source can be compared to other relevant source of energy -- the thermonuclear fusion. For the conversion of hydrogen of mass $m$ into helium, the released energy would be
\begin{equation}
 \Delta E_{\rm nuc}\simeq0.007 mc^2\,.
 \label{eq_nuclear_energy}
\end{equation}

The ratio of the gravitational and nuclear energies is essentially the function of the dimensionless \textit{mass-to-radius ratio} $GM/Rc^2$, which is also denoted as \textit{compactness parameter}, defined by
\begin{equation}
\epsilon \,\equiv\,\frac{GM}{Rc^2},
\end{equation}
where $R$ is a typical length-scale of the system. This compactness parameter can be expressed also as
\begin{equation}
 \frac{\Delta E_{\rm grav}}{\Delta E_{\rm nuc}} \simeq 143\,\epsilon.
 \label{eq_grav_nuclear_ratio}
\end{equation}
For black holes, the typical length-scale is given by the gravitational radius,
\begin{equation}
\label{rg}
R_{\rm g}=\frac{GM}{c^2}
\simeq1.5\times10^{13}\,M_8\,\rm{cm}
\end{equation}
($M_8{\equiv}{M}/10^8M_{\odot}$). For high compactness,
the release of gravitational potential energy can be even two orders of magnitude larger than the energy available via thermonuclear fusion. For black holes, the compactness ratio is naturally the largest. For neutron stars with the mass of $M\approx 1\,M_{\odot}$ and radius of $R\approx 10\,{\rm km}$, the compactness parameter is $\epsilon=GM_{\rm NS}/(R_{\rm NS}c^2)\approx 0.1$, which yields $\Delta E_{\rm grav}/\Delta E_{\rm nuc}$ of the order of 10. In case we fix the mass of the gravitating object to one solar mass, $M=1\,M_{\odot}$, the ratio $\Delta E_{\rm grav}/\Delta E_{\rm nuc}$ is equal to unity for the radius of $R=200\,{\rm km}$, i.e. for all other stable astrophysical objects the thermonuclear fusion is more efficient source than the release of the gravitational potential energy  \footnote{For white dwarfs, we get $\epsilon\approx 10^{-4}$ and $\epsilon\approx 10^{-6}$ for the Sun.}. For sun-like stars, the ratio $\Delta E_{\rm grav}/\Delta E_{\rm nuc} \simeq 10^{-4}$, i.e. the accretion yield is several thousand times smaller than the thermonuclear yield. Nevertheless, even for main-sequence stars the accretion can be of importance in symbiotic systems where the matter is transferred from one star to the other. 

On the other hand, in binary stellar systems that contain a white dwarf, a fast thermonuclear fusion on the surface can be visible as a bright \textit{nova} and generally outshine the luminosity due to the accretion. The important quantity for all energy-generation processes is the timescale $\tau$, on which the energy $\Delta E$ is released. The total bolometric luminosity across the whole electromagnetic spectrum can then be defined as,
\begin{equation}
L=\frac{\Delta E}{\tau}
 \label{eq_luminosity}
\end{equation}
Let us note that another dimensionless quantity can also be denoted as the compactness parameter in the theory of accretion onto compact objects \citep{1980ApJ...238L..63C},
\begin{equation}
 \tilde{\varepsilon}\equiv\frac{L\,\sigma_{\rm T}}{R m_{\rm e}c^3}
 \label{cp}
\end{equation}
with Thomson cross-section for electrons is $\sigma_{\rm T}=6.65\,246\times10^{-25}\;{\rm cm}^2$;
$\tilde{\varepsilon}$ has an advantage of taking the effects of radiation pressure into account.

Considering the symmetric stationary accretion on a compact object -- assuming a supermassive black hole (hereafter SMBH) of mass $M_{\bullet}$ -- it is possible to derive a useful analytic expression for the maximum accretion luminosity, for which the stationary symmetric accretion stops to proceed. Let us assume that the black hole is surrounded by the ionized hydrogen gas. In that case the radiation pressure force mainly acts on electrons via the Thomson scattering, since for protons the scattering is smaller by a factor of $(m_{\rm e}/m_{\rm p})^2$. Since the electron and the proton are bound by a Coulomb force, the radial outward radiation force acting on the whole pair can be calculated as $F_{\rm rad}=\sigma_{\rm T} S/c$ (rate of momentum absorption), where $S$ is the bolometric radiation flux, $S=L/(4 \pi r^2)$, and $c$ is the speed of light. The Coulomb pair is attracted inwards mainly due to the proton-black hole gravitational interaction $F_{\rm grav}=GM_{\bullet}(m_{\rm p}+m_{\rm e})/r^2\simeq GM_{\bullet}m_{\rm p}/r^2$. Finally, the total inward force acting on the electron-proton pair can be calculated as,

\begin{equation}
F_{\rm tot}\simeq \frac{1}{r^2} \left(GM_{\bullet}m_{\rm p}-\frac{\sigma_{\rm T}L}{4\pi c}\right)\,.
\label{eq_total_force_Eddington}
\end{equation}

The total force is zero when the term in parentheses disappears, which is for the luminosity,

\begin{equation}
 L_{\rm Edd}=\frac{4\pi G M_{\bullet} m_{\rm p} c}{\sigma_{\rm T}}=1.3 \times 10^{46} \left(\frac{M_{\bullet}}{10^8\,M_{\odot}} \right)\,{\rm erg\,s^{-1}}\,.
  \label{eq_Eddington_luminosity}
\end{equation}

Expression given by Eq.~\eqref{eq_Eddington_luminosity} is referred to as the \textit{Eddington luminosity}, a.k.a.\ the \textit{Eddington limit}. This limit was derived under the assumption of the steady spherical accretion, which is in real systems quite unrealistic since the accretion proceeds in the axially symmetric disks in the first approximation as will be discussed in Section~\ref{Radiation_processes} in more detail.

\begin{figure}[tbh!]
 \centering
 \begin{tabular}{cc}
 \includegraphics[width=0.8\textwidth]{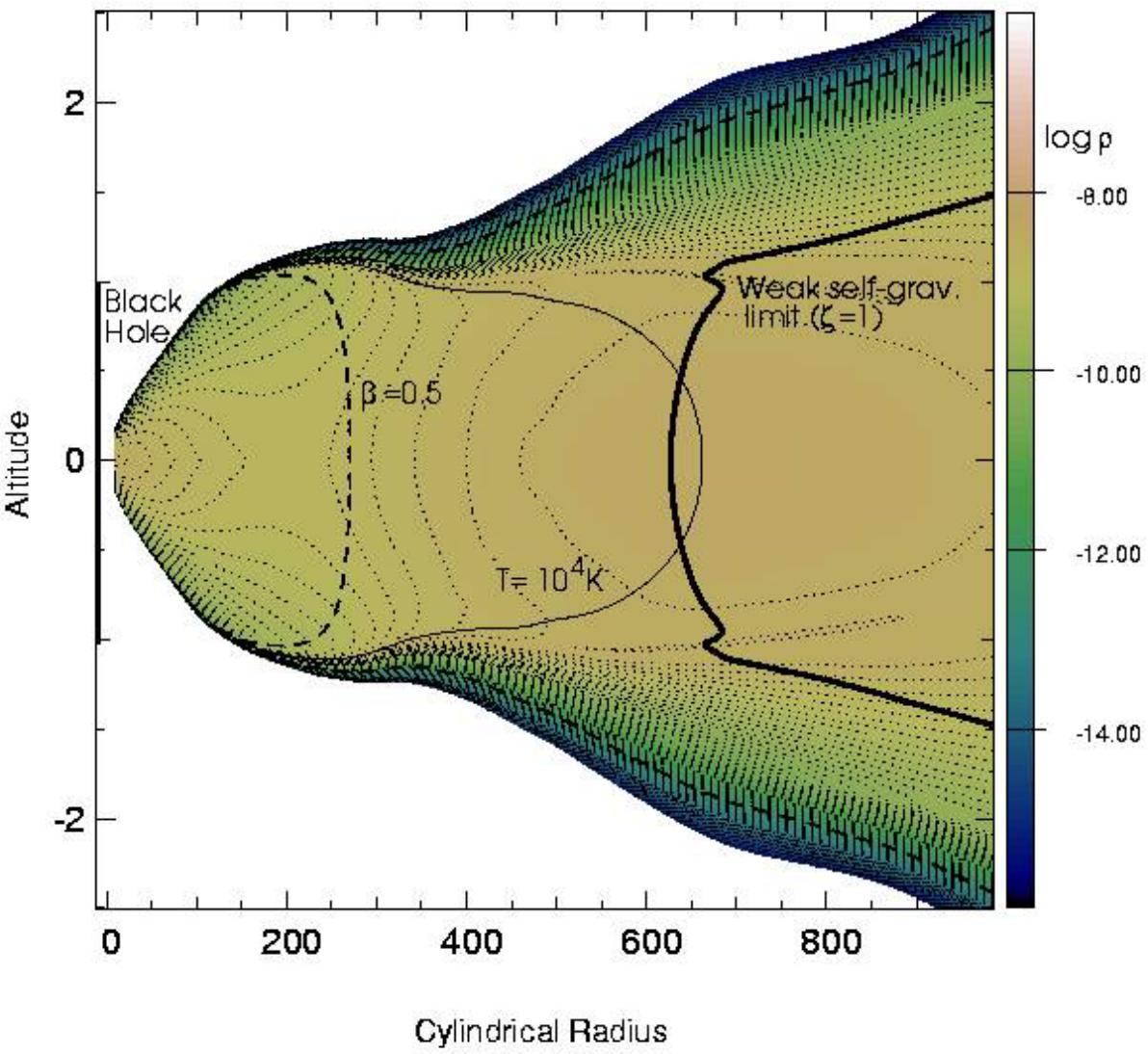}
 \end{tabular}
 \caption{The geometrical profile in the azimuthal ($\phi={\rm const}$) section and the distribution o density (color coded in arbitrary units) within the accretion torus feeding a $M_\bullet=10^8M_\odot$ SMBH \citep[figure adapted from][]{2000A&A...358..378H}. The black hole is located in the center of the matter distribution. The gaseous environment becomes weakly self-gravitating beyond the radius $\approx6\times10^2\,GM_\bullet/c^2$, as characterized by the Toomre criterion and indicated by the numerically constructed $\zeta=1$ contour \citep[for further details, see ref.][]{2004CQGra..21R...1K}.}
\label{fig-hure-disk}
\end{figure}

However, the Eddington limit still serves as a useful value with which the observed bolometric luminosity $L_{\rm bol}$ is compared. The ratio of the observed luminosity to the Eddington luminosity is called the Eddington ratio, $\lambda_{\rm Edd}\equiv L_{\rm bol}/L_{\rm Edd}$. In order to calculate the Eddington ratio, one first needs to convert the luminosity in a given wavelength (frequency band) to the bolometric luminosity, using the SED of a given AGN spectral class, see also Fig.~\ref{agn_sed}. The calculation of the Eddington ratio also requires the determination of the supermassive black hole mass $M_{\bullet}$, for which more methods can be employed, depending on the AGN host type and the redshift of the source. These methods will be summarized later in this Subsection.

In general, the observations show that the AGN luminosity in most galaxies is \textit{sub-Eddington}, $\lambda_{\rm Edd}< 1$. For low Eddington ratios, $\lambda_{\rm Edd} \leq 0.3$, the optically thick and geometrically thin accretion disks provide good fits to the SED of the AGN \citep{1999PASP..111....1K,2001ApJ...563..560B,2011ApJ...728...98D}. For higher accretion rates, the disk turns into the geometrically thick or ``slim'' mode, with the effects of photon trapping and advection becoming relevant \citep[see Fig.~\ref{fig-hure-disk};][]{1988ApJ...332..646A,2014MNRAS.439..503S,2014ApJ...782...45D,2014ApJ...793..108W}. The slim accretion disks with \textit{super-Eddington} luminosities, $\lambda_{\rm Edd} \geq 1$, have different SED properties than standard thin disk, mainly the high-energy cut-off and the strong anisotropy in the emitted radiation, which is expected to arise due to the self-shadowing of the inner parts of slim disks \citep{2014ApJ...797...65W}.

   \label{AGN_structure}

\begin{figure}[tbh!]
\begin{center}
\includegraphics[width=0.9\textwidth]{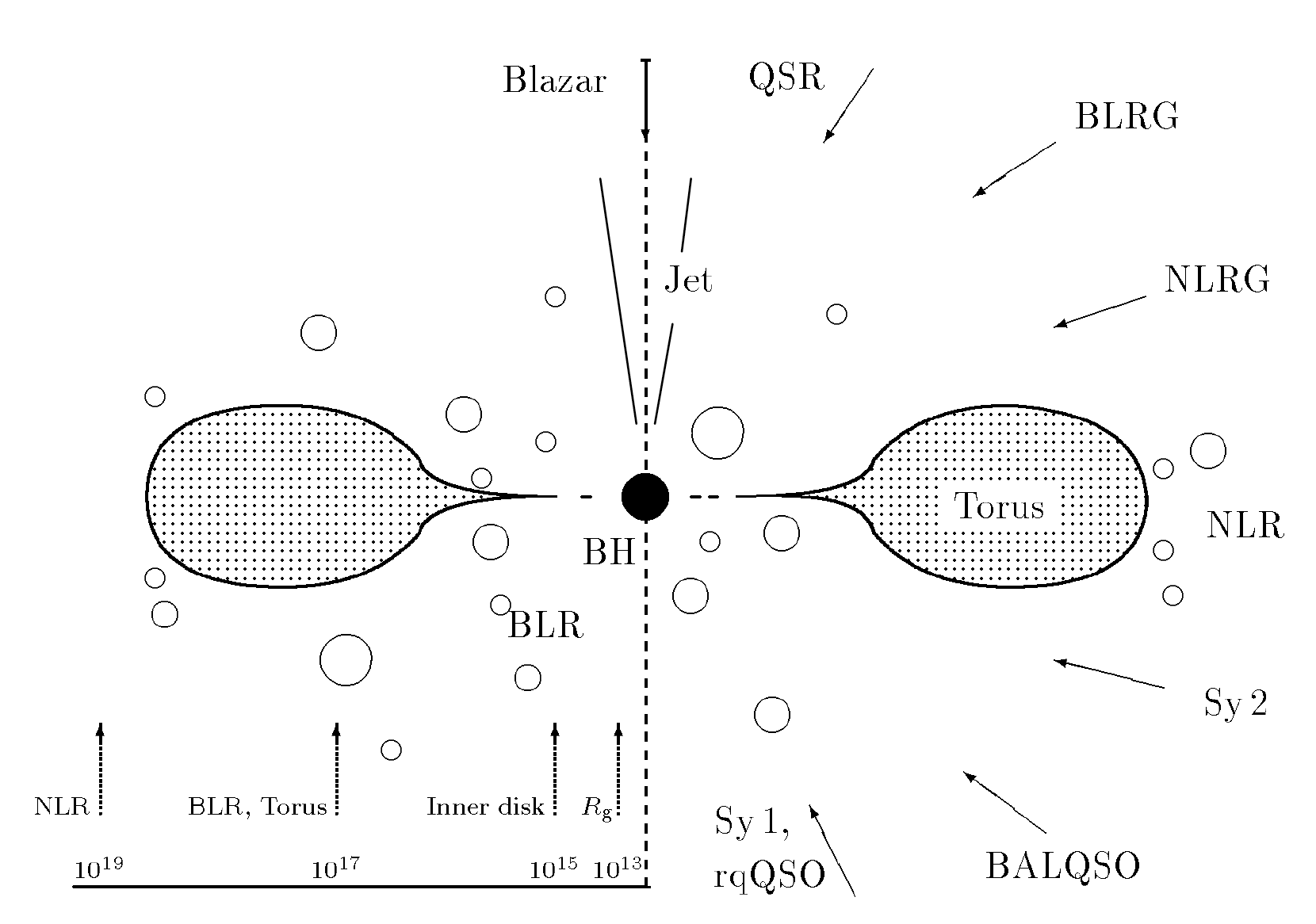}
\caption{A schematic picture of AGN unification. 
Radio-quiet sources are divided into two groups of Seyfert 1 (a.k.a. Sy1) and Seyfert 2 (Sy2) types, as shown at the bottom, while radio-loud galaxies (shown in the top) are distinguished also according to the kinetic power of the jet and its orientation with respect to the observer. Notes: FR I and II are radio galaxies with the morphology according to the Fanaroff--Riley classification. NLRG are Narrow-Line Radio Galaxies, BLRG are Broad-Line Radio Galaxies, and FSRQ are Flat-Spectrum Radio Quasars.}
\label{AGN_unification}
\end{center}
\end{figure}


AGN have various appearances of their electromagnetic spectrum. The most prominent distinction among different AGN kinds is at the radio band, where the flux of radio-loud and radio-quiet AGN differs by several orders of magnitude. The radio-loud AGN are associated with sources launching a powerful and collimated relativistic jet from the close vicinity to the supermassive black hole (see Section~\ref{Jets}).
The other major distinction of different AGN kinds is the presence/absence of the broad emission lines and suppressed continuum of the obscured sources. The suppressed continuum and the absence of broad emission lines are related and caused by a large obscuration of the central region by a dusty circumnuclear structure, called torus. 

The optical polarization measurements reveal that the broad wings of the lines are not completely absent in type-II objects. The broad lines that are otherwise hidden in the total spectra have appeared in the polarized spectra because the polarized emission originates by the scattering of the central-region photons at the polar region. The polarization thus served as a tool to view the central region of type-II AGN like via a periscope. These measurements proved that the obscured type-II sources have also the same central region as unobscured type-I AGN, which led to the first AGN unification explaining the difference between different kinds of AGN based on the various viewing angle \citep{Antonucci1993, 1995PASP..107..803U}. Type-I AGN are viewed from ``above'', while type-II AGN are viewed from the edge and thus with the line-of-sight intercepting the torus. 

Figure~\ref{AGN_unification}, adopted from \citet{Beckmann2012}, schematically shows the unification scenario for radio-quiet (bottom) as well as radio-loud AGN (top).
The figure displays the main constituents of AGN. There is a supermassive black hole in the centre surrounded by an accretion disk and electron plasma. About 10\% of the sources also have a strong relativistic jet and are radio-loud sources. The emission lines originate in the broad-line and narrow-line regions. The narrow-line region is more distant and is always visible. However, the broad-line region can be hidden at some high-inclined orientations.
The direct emission from the central region is visible in type-I objects, while this radiation is absorbed and only partly transmitted through a torus in type-II objects. The dominant emission in type-II sources might be the reflected emission at the torus and  the scattered emission in the narrow-line region. The broad-line region is thus well visible only for low-inclined sources. 

A special class of AGN are blazars that are pointing their jets directly towards the observer. Due to the strong relativistic beaming, the whole SED of blazars is dominated by the emission from the jet. The blazars can be divided into two groups -- BL Lac and FSRQ (Flat-Spectrum Radio Quasars). The difference between these two classes is in radio properties (FSRQ having a flat spectral index), but mainly in the presence of the prominent broad optical emission lines in FSRQs, while they are missing in BL Lac sources. The most accepted explanation for the difference between these two classes is the difference between their accretion rates. While BL Lac sources are supposed to be in an inefficient accretion mode with not enough radiation to photoionize and create the broad-line region clouds, FSRQs are highly-accreting luminous sources with a significant thermal emission from accretion disks sufficient to create and photoionize the broad-line region \citep{Fossati1998, Maraschi2003}.

The accretion rate is clearly another factor playing a significant role in the AGN classification. Similarly to the absence of the broad emission lines in BL Lac sources, also some Seyferts show the lack of the broad emission lines in their spectra, while they do not show any evidence of obscuration due to the torus blocking the view to the central region. Such sources are known as true-type-II Seyferts \citep{2002A&A...394..435P, Bianchi2012}, and may represent low-accreting AGN with absent or not sufficiently illuminated broad-line region clouds.

Let us note that the radio-loud AGN can be divided according to their jet power. Their morphological classification was proposed to correspond to FR-I type according to Fanaroff--Riley when the radio emission arises from a compact region. Oppositely, the high-power sources were proposed to correspond to FR-II type, where the most radio emission origintace from extended radio lobes. Alternatively, it hes been proposed that the FR classification might be related to the properties of the intergalactic medium rather than the intrinsic properties of the jet.

The circumnuclear obscurer is not a homogeneous body (as was already discussed in previous Section reviewing the infrared observations of AGN), but is rather formed by individual clumps. Therefore, it can happen that a clear view of the central region is possible in some sources also from a relatively high inclination, depending on the actual distribution of the obscuring clouds. And because the whole structure is dynamic (the clouds are in a motion), AGN types can vary from one type to the other. The change of the spectral type of an AGN was indeed observed several times, most prominently by sources called changing-look AGN \citep{LaMassa2015,MacLeod2016}.
However, it was argued that variable absorption due to passing clouds can hardly explain variability seen in the large extended optical region of AGN emission and the intrinsic changes of the accretion flow were suggested to be more credible interpretation
\citep{LaMassa2015, 2018arXiv180507873N}.

The basic AGN classification on type-I and type-II sources is not ubiquitous at any wavelength. Although the obscuration in the optical and X-ray energy domain was found to be related \citep{1994ApJ...422..521G, 2016A&A...586A..28B},
it was shown by \citet{Merloni2014} on a large AGN sample that only about 70\% AGN in fact agree on the optical and X-ray classification. The rest 30\% can be divided in two sub-groups: (i) to low-luminosity objects that are X-ray unobscured but lack the broad lines, and (ii) luminous AGN with the broad lines but with prominent X-ray absorption.


    \label{Supermassive black holes in AGN cores}
    \label{SMBH}
The presence of the supermassive black hole in AGN cores was historically proposed to explain large megaparsec scales of jets in quasi-stellar objects (QSOs) as well as the measured emitted large energies and the variability on very short-timescales. Let us briefly summarize here the main arguments:

\begin{itemize}
  \item For several radio sources, the measured jet sizes reach several megaparsecs, $L\approx 10^6\,{\rm pc}$. Taking the upper limit of the expansion speed as the speed of light $c$, we obtain the lower limit for the age of the central engine, $\tau > L/c\approx 10^6\,{\rm pc}/3\times 10^8\,{\rm m\,s^{-1}}\approx 3\times 10^6\,{\rm yr}$.
 \item Luminous quasars can reach bolometric luminosities of $L_{\rm QSO} \approx 10^{47}\,{\rm erg\,s^{-1}}$. Then the minimum emitted energy over the engine lifetime is, $E_{\rm QSO}\approx L_{\rm QSO}\tau \approx 10^{61}\,{\rm erg}$.
 \item Typical AGN are variable sources on all timescales, even as short as one day and less; $\tau_{\rm{var}}\approx 1\,{\rm day}$. This timescale sets an upper limit on the emitting region since the maximum velocity with which the changes can take place is the speed of light $c$ in order for the whole emitting region to be causally connected. Then the maximum length-scale can be inferred from a light travel time of the order of a few days, $R_{\rm emit}\lesssim \tau_{\rm var} c\approx 1\,{\rm day} \times 3\times 10^8\,{\rm m\,s^{-1}}\approx 2.6 \times 10^{13}\,{\rm m}\approx 1 {\rm mpc}= 170\,{\rm AU}$. 
\end{itemize}

Hence, the central engine has the length-scale of the Solar System and needs to be stable enough for at least millions of years and be able to produce luminosities of the order of $10^{13}\,L_{\odot}$. However, the mechanism of transferring the gas from the outer parts of the galaxy towards the nucleus and then feeding the central black hole remains the subject of many debates \citep{Storchi-Bergmann2019}. It appears that the estimated rate of mass inflow, about 0.01 to a few solar masses per year, is often higher by three orders of magnitude than the mass accretion rate onto the central SMBH. A question arises what the actual mass of such a compact and energetic system is. The mass estimates may be derived from the assumption of the \textit{virial theorem} that should hold for the self-gravitating systems of gas, dust and stars in hydrostatic equilibrium. i.e. those that neither contract nor expand. 

The virial theorem may simply be derived for bound orbits of gas and stars that orbit black holes in the centre of galaxies. The characteristic Keplerian velocity is given from the Newtonian theory by setting the centripetal force 
($mv^2/r$) acting on an object equal to the gravitational force ($GM_{\bullet}m/r^2$):
\begin{equation}
  v_{\rm K}=\sqrt{GM_{\bullet}/r}\,.
  \label{eq_kepler_vel}
\end{equation}

Accreting matter forms a disk-like structure, which consists essentially of many individual elements; we can
imagine them as discrete blobs of plasma, each one moving along an (approximately) Keplerian orbit with the characteristic velocity $v(R)=v_{\rm K}$. Due to the viscosity of the disk, blobs gradually lose the angular momentum and the gravitational potential energy while gaining the kinetic energy:
\begin{equation}
  E_{\rm K}=\textstyle{\frac{1}{2}} m v_{\rm K}^2=\textstyle{\frac{1}{2}} GM_{\bullet}m/r\,.
\end{equation}
The gravitational potential energy of the blob is $E_{\rm P}=-GM_{\bullet}m/r$ and the corresponding 
kinetic energy during the accretion process is equal to
\begin{equation}
  E_{\rm K}=\textstyle{\frac{1}{2}} \frac{GM_{\bullet}m}{r}=-\textstyle{\frac{1}{2}}E_{\rm P}\,.
  \label{eq_ep_loss_one_half_EK}
\end{equation}
The total mechanical energy reads
\begin{equation}
  E_{\rm tot}=E_{\rm K}+E_{\rm P}=\textstyle{\frac{1}{2}} E_{\rm P}\,.
\end{equation}
Finally, the virial theorem for the self-gravitating system in hydrostatic equilibrium (without taking time-dependency into account),

\begin{equation}
  2\langle E_{\rm K}\rangle_{\rm R}+\langle E_{\rm P}\rangle_{R}=0\,,
  \label{eq_virial}
\end{equation}
where the brackets $\langle...\rangle_{\rm R}$ denote the averaging over the whole dynamical system of the length-scale $R$.  

The importance of the virial-theorem for the accretion onto black holes is that one-half of the loss of the potential energy is converted into the gas kinetic energy while the other half is available for the gas heating, which leads to the emission of electromagnetic radiation in the X-ray, ultraviolet, and visible domains.

The Keplerian velocity profile of the emitting material bound to the central supermassive black hole directly implies that the emission lines of the hot gas should be significantly Doppler-broadened. The observed line profile is given by the summing of the flux of all orbiting blobs or clouds around the central black hole. For some AGN, not all, very broad lines with the line widths corresponding to $<v>=10^3$--$10^4\,{\rm km\,s^{-1}}$ have been detected -- these are AGN of type 1. Given the compactness of the emitting region obtained from the variability timescale, we can infer the total mass in this region from Eq.~\eqref{eq_kepler_vel},

\begin{equation}
  M_{\rm center}=\frac{\langle v\rangle\;^2R}{G}\approx 10^5\mbox{--}10^7\,M_{\odot}\,.
  \label{eq_mass_center}
\end{equation}

Based on the virial theorem, Eq.~\eqref{eq_virial}, the accretion luminosity can be expressed as,

\begin{equation}
   L_{\rm acc}=\frac{1}{2}\frac{GM_{\bullet}\dot{M}_{\rm acc}}{R_0}\,
   \label{eq_accretion_luminosity}
\end{equation}
where $R_0$ is the initial radius, from which the matter falls onto the compact object. In case of black holes, naturally infalling matter does not hit any surface but passes through the event horizon. In accretion disks around black holes, hot gas spirals in due to the extraction of the angular momentum, until it reaches the innermost stable circular orbit (ISCO). Subsequently, the matter falls into the black hole since no stable orbit exists inside the ISCO. Therefore, we can set $R_0=R_{\rm ISCO}$. For a non-rotating black hole, we have $R_0=6GM_{\bullet}/c^2$ and the accretion luminosity becomes,

\begin{equation}
  L_{\rm acc}=\eta \dot{M}_{\rm acc} c^2\,,
  \label{eq_accretion_efficiency}
\end{equation}
where $\eta=1/12$, i.e. the efficiency for a non-rotating black hole is $8.3\%$ and the correct relativistic calculation would give $\sim 5.7\%$ for an accretion from a thin accretion disk. For a maximally rotating black hole, $R_{\rm ISCO}=GM_{\bullet}/c^2$, which yields $\eta=1/2$, however the general relativistic calculation again gives a smaller value of $\sim 29\%$. We see that the accretion efficiency $\eta$ can reach values as large as $\sim 10\%$, which is one order of magnitude larger than the efficiency of the energy release via the thermonuclear fusion, $\epsilon \lesssim 0.8\%$, where the uppermost limit corresponds to the the conversion of hydrogen into iron. In other words, the energy released through the thermonuclear fusion of hydrogen is $\sim 6 \times 10^6\,{\rm eV}$ per hydrogen atom, while the accretion from a thin disk on a non-rotating black hole yields the energy rate larger by one order of magnitude.  

If we take $\eta=0.1$, then the accretion rate for typical quasar luminosities of $L_{\rm acc}\approx 10^{46}\,{\rm erg\,s^{-1}}$  is,

\begin{equation}
  \dot{M}_{\rm acc}=1.8 \left(\frac{L_{\rm acc}}{10^{46}\,{\rm erg\,s^{-1}}}\right)\left(\frac{\eta}{0.1}\right)^{-1}\,M_{\odot}{\rm yr^{-1}}\,.
  \label{eq_macc}
\end{equation}

This means that quasars must accrete close to the Eddington limit $L_{\rm Edd}$, see Eq.~\eqref{eq_Eddington_luminosity}, when considering black holes of mass $M_{\bullet}=10^8\,M_{\odot}$. The Eddington accretion rate is derived by setting the accretion luminosity, Eq.~\eqref{eq_accretion_efficiency}, equal to the Eddington luminosity, Eq.~\eqref{eq_Eddington_luminosity},

\begin{align}
   \dot{M}_{\rm Edd} & = \frac{4\pi GM_{\bullet}m_{\rm p}}{\eta \sigma_{\rm T} c}\,\notag\\
                   & \approx 2.6 \left(\frac{\eta}{0.1} \right)^{-1} \left(\frac{M_{\bullet}}{10^8\,M_{\odot}} \right)\,M_{\odot}\,{\rm yr^{-1}}\,.
   \label{eq_eddington_rate}
\end{align}

The stable and very compact configuration of the order of million Solar masses can be explained by the presence of supermassive black hole. Other compact objects, white dwarfs and neutron stars, are too light -- white dwarfs, which are prevented from further gravitational collapse by the pressure of degenerate electrons, have the upper mass limit given by the Chandrasekhar limit,

\begin{equation}
M_{\rm Ch}\approx m_{\rm B}[\hbar c/(Gm_{\rm B}^2)]^{3/2} \approx 1.4\,M_{\odot}\,
\end{equation}

and neutron stars have a slightly larger, but still too small upper limit given by the Tolman--Oppenheimer--Volkoff limit, $M_{\rm TOV}\leq 3\,M_{\odot}$. 

Main-sequence stars can theoretically reach the masses of the order of $100\,M_{\odot}$ when they consist purely of hydrogen and helium plasma, but they are too short-lived, less than million years. Hence, classical relativistic theories of stellar structure and evolution allow only one stable configuration to explain the characteristics of the AGN -- supermassive black holes. However, quantum theories of gravity could explain the compact large concentrations using models of quantum nature (boson stars, macroquantumness). In general, the presence of the event horizon is difficult to prove by observations of the emerging electromagnetic waves \citep{2002A&A...396L..31A}. 


The reader may consider it as strange and mysterious that black holes of $10^{6}$--$10^{10}\,M_{\odot}$ reside in the cores of luminous quasars and following the Soltan argument \citep{1982MNRAS.200..115S}, in the nuclei of most nearby galaxies. The classical stellar theories and numerical studies show that stellar black holes are the final end-states of the stellar collapse with the remnant mass of $>3\,M_{\odot}$. The initial stellar masses have to be in excess of $45\,M_{\odot}$\, depending on the stellar metallicity. Numerical studies that employ Cold Dark Matter (CDM) models show that the first stars in the early Universe were very massive, $\gtrsim 100\,M_{\odot}$, with very low metallicities \citep{2001ARA&A..39...19L}. The first baryonic structures started to form at the redshift of $z\sim 30$ based on the principle of the Jeans gravitational instability and they started to reionize the neutral hydrogen in their surroundings. This so-called reionization era finished at $z\sim 6$, i.e. the Universe was fully ionized at that redshift based on the quasar observations and the detection of the so-called \textit{Ly$\alpha$ forest} in their spectra. The formation of first baryonic structures, such as stars and black holes, provides a transition from a rather simple, smooth state of the Universe with small, primordial density fluctuations to the current, coarse state that is full of different structures and large differences in the mass density. Necessary tools to study this transition include computationally demanding cosmological hydrodynamic simulations \citep{1999ApJ...527L...5B,2000ApJ...540...39A} as well as the observations in the mid-infrared bands, in particular by the James Webb Space Telescope \citep{2001ARA&A..39...19L}.

The Ly$\alpha$ forest is a set of absorption lines in the spectra of distant quasars short of the Ly$\alpha$ line at the corresponding redshift. Due to the presence of a small amount of neutral hydrogen along the line of sight, Ly$\alpha$ photons are effectively scattered and suppressed, which yields a set of absorption lines at the blueshifted part of the actual Ly$\alpha$ line. In case the ratio of neutral to ionized hydrogen in the intergalactic medium reaches larger values than $10^{-3}$, the absorption lines become largely blended and a so-called \textit{Gunn--Peterson} through is formed. The Gunn--Peterson through was detected for a quasar at $z\approx 6.28$, but not for quasars at smaller redshifts \citep{2001AJ....122.2850B}. This indicates that at redshift $z\sim 6$ the reionization of the Universe by the radiation of first stars and the activity of first black holes is largely completed.     

\begin{figure}[tbh!]
    \centering
    \includegraphics[width=0.8\textwidth]{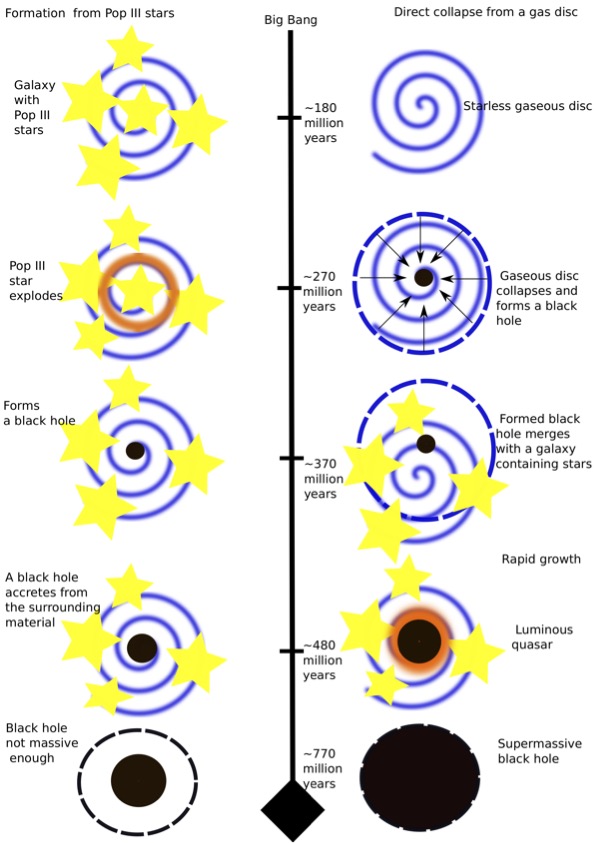}
    \caption{Two formation channels of the first massive black holes residing at the cores of the first quasars at $\sim 770$ Myr after the Big Bang: (i) the first scenario, depicted on the left, shows the formation of seed stellar black holes from Population III (Pop III) stars; (ii) the second scenario shown on the right involves a neutral, irradiated disk of hydrogen that collapses directly into a more massive black hole seed. This larger seed then merges with a cluster of Population III stars and gets quickly more massive by a rapid accretion. The second scenario is more consistent with the occurrence of the first quasars $\sim$ 770 Myr after the Big Bang. This visualisation was adapted from an inspiring article on the feedback processes by \citet{2018SciAm.318b..24N}.}
    \label{fig_first_bh}
\end{figure}

So far two basic channels of the formation of black hole seeds have been investigated. One includes very massive hydrogen-helium stars (Population III) that started to form after the \textit{recombination epoch} ($\sim 380\,000$ years after the Big Bang, which corresponds to the redshift $z=1100$) during the first hundred million years after the Big Bang. These very massive stars of $\sim 100$ Solar masses produced seed black holes of $10-100$ Solar masses that accreted from the local environment. The first stars formed in gaseous molecular disks that cooled mostly via $H_2$ cooling and fragmented subsequently and formed stars. The second scenario starts with a gaseous disk again, but instead of cooling through molecular hydrogen, it contains mainly atomic hydrogen, which is not such an efficient coolant. This scenario also includes external irradiation, which prevents molecular hydrogen from forming. The gaseous disk then becomes gravitationally unstable as a whole and the matter is channeled into the centre where the black hole is formed directly -- a so-called direct collapse black hole \citep[see also][ for an overview and a popular account]{2018SciAm.318b..24N}. 
\citet{2018SciAm.318b..24N} describes the two scenarios, (i) on one side the formation of seed black holes from Population III stars, and (ii) the direct collapse of an externally illuminated gaseous disk on the other side. The problem of the scenario that involves Population III stars as the origin of black-hole seeds is that the black holes are not massive enough at $\sim 770$ million years after the big bang when the first quasars are expected to appear. This ``under-weight'' problem is solved by the model of the direct collapse of a gaseous disk, which can lead to the much faster growth of the first black holes. We illustrate both scenarios in Figure~\ref{fig_first_bh}.

The basic model of the black hole growth by accretion can be constructed from the first principles. Let us assume that the black hole accretes with the efficiency of $\epsilon$ and the rate of $\dot{M}_{\rm acc}$, which yields the accretion luminosity of
\begin{equation}
  L_{\rm acc}=\epsilon \dot{M}_{\rm acc} c^2\,.
  \label{eq_accretion_luminosity2}
\end{equation}  
The complementary fraction of the accretion rate contributes to the black hole mass increase,
\begin{equation}
  \dot{M}_{\bullet}=(1-\epsilon)\dot{M}_{\rm acc}\,.
  \label{eq_mass_growth}
\end{equation}
which can be rewritten in terms of the Eddington luminosity and the Eddington ratio using Eqs.~\eqref{eq_accretion_luminosity}, \eqref{eq_accretion_luminosity2}
\begin{equation}
  \dot{M}_{\bullet}=\frac{1-\epsilon}{\epsilon}\frac{L_{\rm Edd}}{c^2}\lambda_{\rm Edd}\,.
  \label{eq_mass_increase_ratio}
\end{equation}
Using the Eddington luminosity, the relation~\eqref{eq_mass_increase_ratio} can be rewritten in such a way so that the mass is on the left side and the time on the other in a differential form. Subsequently, one can integrate the left side with the limits $(M_{\bullet}(0),M_{\bullet})$ and the right side with the limits $(0,t)$,
\begin{equation}
  \int_{M_{\bullet}(0)}^{M_{\bullet}} \frac{\mathrm{d}M_{\bullet}'}{M_{\bullet}'}=\int_0^t \frac{1-\epsilon}{\epsilon} \lambda_{\rm Edd} \frac{4\pi Gm_{\rm p}c}{\sigma_{\rm T}c^2} \mathrm{d}t'\,,
  \label{eq_growth_integral}
\end{equation}
which after the integration yields an exponential growth model,
\begin{equation}
  M_{\bullet}(t)=M_{\bullet}(0)\exp{\left(\frac{1-\epsilon}{\epsilon}\lambda_{\rm Edd}\frac{t}{\tau_{\rm Sal}}\right)}\,.
\end{equation}
Hence the black hole mass grows exponentially with time, $M_{\bullet} \propto \exp{(t/\tau_{\rm Sal})}$, where the $e$-folding timescale $\tau_{\rm Sal}$ is the \textit{Salpeter time},
\begin{equation}
\tau_{\rm Sal}=\frac{M_{\bullet}c^2}{L_{\rm Edd}}=\frac{\sigma_{\rm T}c}{4\pi G m_{\rm p}}\simeq 4.5 \times 10^8\,{\rm yr}\,. 
 \label{eq_Salpeter_time}
\end{equation}
Apparently, $\tau_{\rm Sal}$ depends on the fundamental constants only.

\begin{figure}[tbh!]
    \centering
    \includegraphics[width=0.8\textwidth]{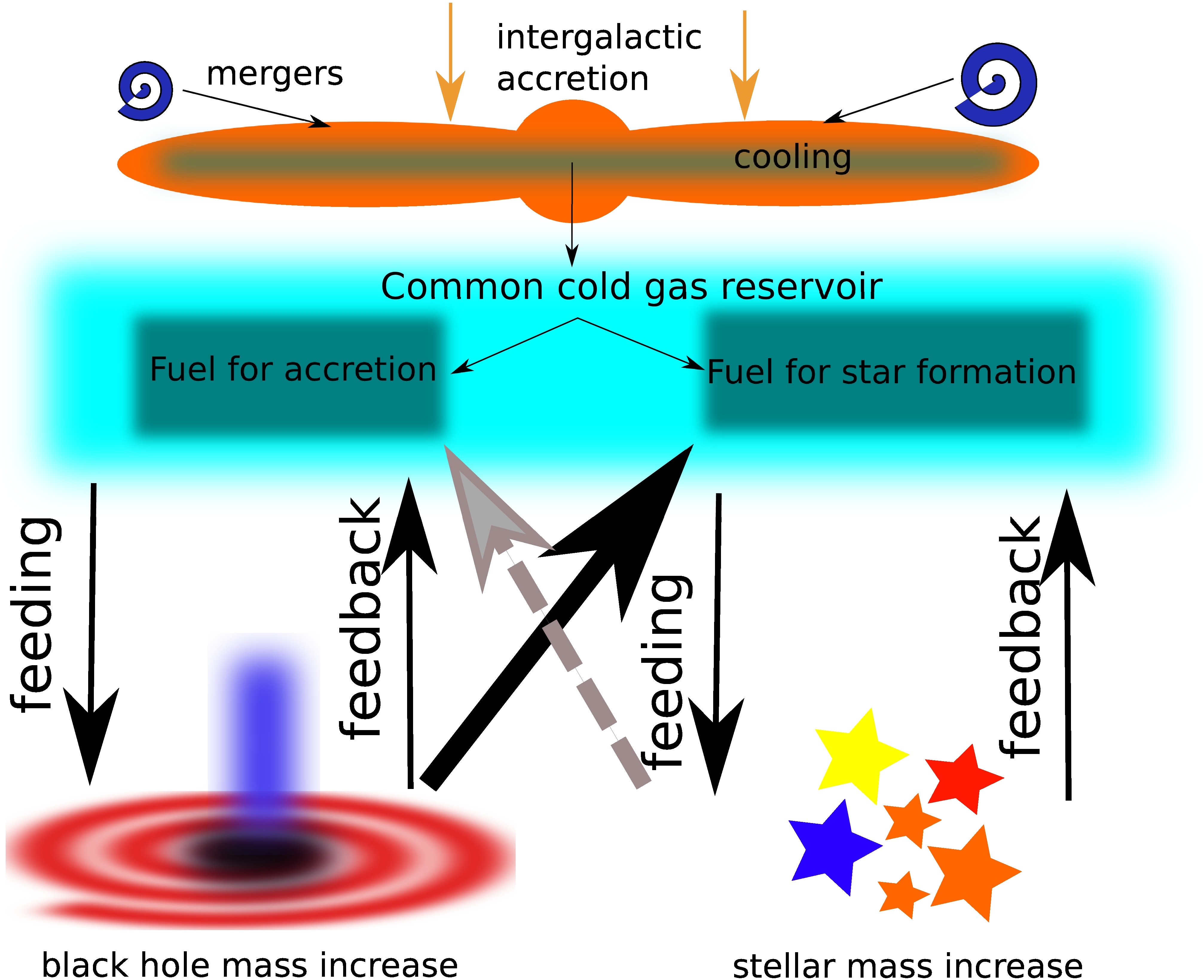}
    \caption{Illustration of the AGN feeding and feedback processes. From the top, we show the main sources of gas in an AGN host -- intergalactic accretion, gas-rich mergers, and galactic sources (supernovae, stellar winds). The cold gas is formed via the radiative and the adiabatic cooling. Subsequently, it fuels the black hole accretion as well as star-formation. Both processes lead to the mechanical and radiation feedback, which both remove the cold gas and cause the drop in the accretion and the star-formation rate. This illustration was inspired by \citet{2017NatAs...1E.165H} and adapted from \citet{2020past.conf..248Z}. .}
    \label{fig_feedback}
\end{figure}

The black hole growth via the accretion close to the Eddington limit, i.e. the quasar stage, is not a continuous process. This is mainly due to the feedback process, which we understand here as depositing the momentum onto the surrounding cold gas as well as its heating and ionization by intense radiation. This leads to the decrease of the amount of dense cold gas, which is needed for both the accretion as well as star-formation. This is a so-called \textit{negative feedback}. Occasionally, the propagating jet leads to the shocks and the compression of the ambient gas, which can trigger star-formation (\textit{positive feedback}). This complicated interplay among the cold gas reservoir, accretion/jet activity, and star-formation is illustrated in Figure~\ref{fig_feedback}. The negative feedback on the cold-gas reservoir can be in the form of \textit{radiative mode} (wind or quasar), where the gas is expelled due to quasar winds, jet and radiation and this mode is typical of luminous AGN, or in the form of \textit{kinetic mode} (radio or jet), where the cold gas is heated up and this mode occurs for low-excitation radio-loud AGN \citep{2012NewAR..56...93A}.

The star-formation rate is globally related to the density of gas in galaxies via the \textit{Kennicutt--Schmidt law}, which expresses the star-formation rate surface density, expressed in ${\rm M_{\odot}\,yr^{-1}\,pc^{-2}}$, as a power-law function of the gas surface-density, expressed in ${\rm g\,pc^{-2}}$ \citep{1959ApJ...129..243S},

\begin{equation}
    \Sigma_{\rm SFR} =A \Sigma_{\rm gas}^{n}\,,
    \label{eq_kennicutt_schmidt}
\end{equation}
where $A$ is an absolute star-formation efficiency.
Initially, according to the analysis of \citet{1959ApJ...129..243S}, $n$ was found to be $\approx 2$ close to the Solar neighbourhood based on the abundance of helium and the occurrence of white dwarfs. Using the sample of HI, CO, $H\alpha$ spiral galaxies and CO and far-infrared starburst galaxies, \citet{1998ApJ...498..541K} demonstrated that the Schmidt law expressed by Eq.~\eqref{eq_kennicutt_schmidt} applies globally for one-zone models of galaxies, i.e. for a disk-averaged star-formation rate, and the power-law index was constrained to be $n=1.4 \pm 0.15$.

    \subsection{AGN circumnuclear medium: the interplay of gas and stars}
    \label{circumnuclear}

In this subsection, we focus on the interstellar medium in the central parts of active galactic nuclei in the context of the mutual interaction among gas, stars, and the supermassive black hole. The motion of the gaseous-dusty circumnuclear medium in the potential dominated by the central supermassive black hole (SMBH) is necessary to account for broad emission lines with line widths of the order of several $1000\,{\rm km\,s^{-1}}$. Such large line widths arise due to the Doppler-broadening of the ionized gas emission due to the dominantly orbital motion of cloudlets around the SMBH.  The gas can flow in towards the SMBH inside the Bondi radius, where its gravitational potential prevails over that of the thermal gas pressure of temperature $T_{\rm g}$,

\begin{equation}
    r_{\rm Bondi}\approx \frac{GM_{\bullet}}{c_{\rm s}^2}=1\,\left(\frac{M_{\bullet}}{2\times 10^8\,M_{\odot}} \right)\left(\frac{T_{\rm g}}{10^8\,{\rm K}} \right)^{-1}\,{\rm pc}\,,
    \label{eq_bondi_radius}
\end{equation}
which eventually sets the basic length-scale of the region of our interest. Whether in free-fall as in the spherical Bondi accretion, for which the matter has lost its angular momentum, or already circularized gas with the initial angular momentum, the velocities of the order of $\sim 1000\,{\rm km\,s^{-1}}$ are reached already well before reaching the innermost stable circular orbit (ISCO) of the central black hole,

\begin{equation}
    r_{\rm BL,kin}=f \frac{GM_{\bullet}}{v_{\rm K}^2}=0.86f\left(\frac{M_{\bullet}}{2\times 10^8\,M_{\odot}}\right)\left(\frac{v_{\rm K}}{1000\,{\rm km\,s^{-1}}}\right)^{-2}\,{\rm pc}\sim 45\,000f\,R_{\rm s}\,,
    \label{eq_blregion}
\end{equation}
where $f$ is the kinematical/geometrical factor of the accreting gas and $R_{\rm s}$ represents the Schwarzschild radius of the SMBH, $R_{\rm s}=2GM_{\bullet}/c^2=1.9\times 10^{-5}(M_{\bullet}/2\times 10^8\,M_{\odot})\,{\rm pc}$. This gives a basic estimate of the \textit{mean kinematic radius of the broad-line region}, which is comparable to the Bondi radius in Eq.~\eqref{eq_bondi_radius}.

We will refer in the following analysis to both the \textit{interstellar} and the \textit{circumnuclear} medium of active galactic nuclei (AGN) interchangeably. This stems from the fact that stars as members of dense nuclear star clusters (NSCs) typically do not appear as one of the components in the unified AGN models \citep{1993ARA&A..31..473A,1995PASP..107..803U}. However, they are expected to be present and can actually provide a fraction to both the accreted matter and the outflow \citep{1980A&A....82...99B,2018MNRAS.479.4778Y} or at least they are expected to influence hydrodynamics in the central regions of galaxies. In fact, nuclear star clusters have been detected at the photometric as well as dynamical centers of a significant fraction (60$\%$-75$\%$) of both early-type and late-type galaxies in the local Universe \citep{1998AJ....116...68C,2002AJ....123.1389B,2006ApJS..165...57C,2011MNRAS.413.1875N}. In general, they are thought to co-evolve with massive black holes at galactic centers, where they mutually influence their growth \citep{2012AdAst2012E..15N}. Having typical effective radii of a few parsecs and the total masses of a few $10^5$--$10^8\,M_{\odot}$, NSCs are one of the densest stellar systems in galaxies \citep{2004AJ....127..105B,2005ApJ...618..237W,2014CQGra..31x4007S} with the mean surface densities up to $\sim 10^5\,M_{\odot}\,{\rm pc^{-2}}$, similar to globular clusters and other compact stellar systems, but brighter. In addition, there is an evidence of co-existence with active massive black holes at their centers \citep{2008ApJ...678..116S,2009MNRAS.397.2148G,2012AdAst2012E..15N}. Also, kinematically resolved NSCs show a Keplerian rise in velocities in their inner portions, $\sigma_{\star}\propto r^{-1/2}$. The influence radius of SMBHs $R_{\rm inf}$ is often estimated by the radius at which the Keplerian circular velocity of stars around the SMBH is equal to their one-dimensional (line-of-sight) stellar velocity dispersion,

\begin{equation}
    R_{\rm inf}=\frac{GM_{\bullet}}{\sigma_{\star}^2}\approx 86 \left(\frac{M_{\bullet}}{2\times 10^8\,M_{\odot}}\right) \left(\frac{\sigma_{\star}}{100\,{\rm km\,s^{-1}}}\right)^{-2}\,{\rm pc}\,
    \label{eq_influence_radius1}
\end{equation}
which can be larger than the NSC effective radius for massive black holes, but only as small as $\sim 2\,{\rm pc}$ for the Galactic centre black hole of $M_{SgrA*}\simeq 4\times 10^6\,M_{\odot}$ \citep{2014CQGra..31x4007S}.

To what extent the NSC as a whole influences the standard parameters of AGN (black hole mass, accretion rate, accretion disc structure, broad- and narrow-line region clouds) is still not known. A self-consistent treatment of the influence of stars on the accretion disc was done by Wang et al. \cite{2011ApJ...739....3W,2012ApJ...746..137W}, who build on the earlier models of \citep{1985MNRAS.213..665P,2002A&A...387..804V}. They show that stars that form beyond the self-gravitation radius in the accretion disc drive outflows of hot gas beyond the disc plane through supernova explosions. This hot gas leads to the cyclic production of broad-line region clouds as well as the metallicity gradient in the accretion disc. The general picture is more complex as the NSCs themselves experience recurrent \textit{in-situ} star-formation \citep{2006AJ....132.1074R}, possibly depending on the cycles of cold gas inflows from larger scales. Although NSCs are spheroidal, Seth et al. \citep{2006AJ....132.2539S} show that for some galaxies there are younger flattened subsystems embedded within the older spheroidal structure. Subsequent dynamical mechanisms \citep[resonant and non-resonant relaxation processes;][]{2013degn.book.....M} are required to turn young stellar disks/rings into spheroidal stellar systems. Seth et al. \citep{2006AJ....132.2539S} estimate the timescale of $\sim 0.5\,{\rm Gyr}$ between individual star-formation episodes.        

\begin{figure}[tbh!]
    \centering
    \includegraphics[width=0.8\textwidth]{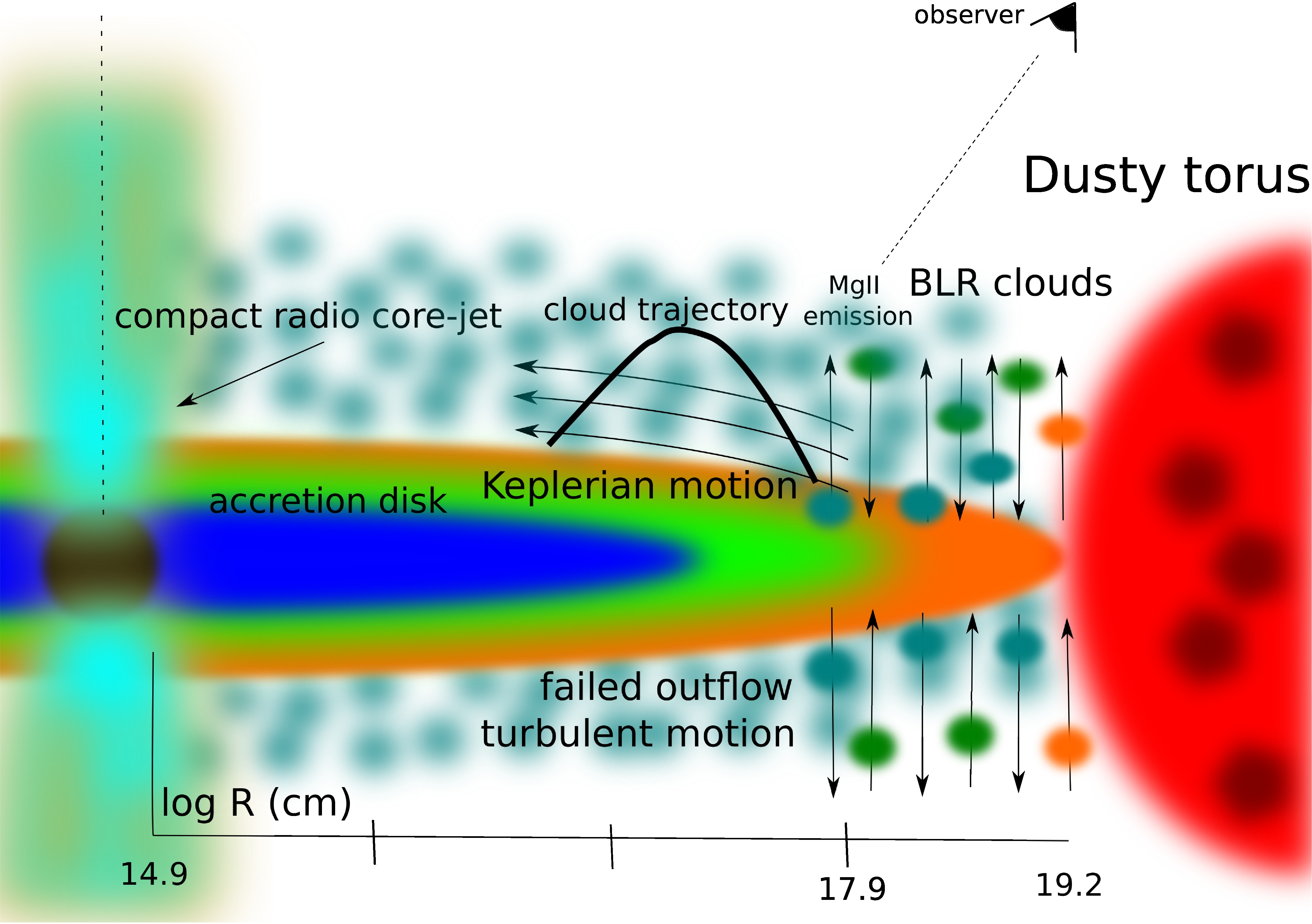}
    \caption{Illustration of the flattened distribution of the broad-line region clouds based on the MgII reverberation-mapping of the luminous intermediate-redshift quasar HE 0413-4031 ($z=1.38$). Adopted from \citet{2020ApJ...896..146Z}.}
    \label{fig_BLR_HE0413}
\end{figure}

The prominent broad emission lines associated with the broad line region (hereafter BLR) are generally observed in AGN \citep{2000ApJ...533..631K,2004ApJ...613..682P,2009ApJ...705..199B,2013ApJ...764...47G}, i.e. about $\sim 10\%$ of galactic nuclei with the high accretion rate of the order of $1\,M_{\odot}\,{\rm yr^{-1}}$ in comparison with the majority of quiescent sources, such as our Galactic centre. The strength of BLR lines implies the high covering fraction, up to $\sim 30-50\%$ \citep{2004ApJ...606..749K}, of the AGN disc continuum emission. On the other hand the BLR absorption lines are rare, which points towards the generally flattened structure, potentially linked to the disc, but not entirely in the disc plane as this would lead to the lower covering fraction, see general BLR reviews by \citet{1999agnc.book.....K,2013peag.book.....N}. We illustrate the flattened BLR geometry in Fig.~\ref{fig_BLR_HE0413}, in which the spatial scale at the bottom is based on the reverberation-mapping program that monitored the broad MgII line for the luminous quasar HE 0413-4031 \citep{2020ApJ...896..146Z}

The basic length-scale of the BLR region is given by Eq.~\eqref{eq_blregion}, and it was only recently kinematically resolved out on sub-parsec scales using the Very Large Telescope Interferometer (VLTI) GRAVITY in the quasar 3C273 \cite{2018Natur.563..657G}. They reported a spatial offset of $0.03\,{\rm pc}$ between the red-shifted and blue-shifted photocenters of the broad Pa$\alpha$ emission line along the direction perpendicular to the jet. The detected velocity gradient implies the Keplerian rotation of the emitting gaseous material. The data are well-fitted by a thick disc of BLR cloudlets, which is in a Keplerian rotation around the black hole of $M_{\rm 3C273}\approx 3\times 10^8\,M_{\odot}$. The inferred BLR radius is 150 light days $\sim 0.126\,{\rm pc}$, which is consistent with the general Keplerian estimate for the comparable black hole mass in  Eq.~\eqref{eq_blregion}.

It was shown that the size of the BLR region is linked to the monochromatic luminosity and the BLR flux from a given emission line is mainly produced at a region with a specific combination of the gas density and the density of ionizing photons \citep{1995ApJ...455L.119B}. This locally optimally emitting cloud model (LOC) explains the observed variation of the BLR size with the monochromatic luminosity, which is also known as the BLR ``breathing". The BLR ``breathing" has been apparent from early reverberation mapping studies of broad Balmer emission lines \citep[mainly H$\beta$ line centered at 4861\AA\,; ][; see the details below]{2003ApJ...587..123G,2004ApJ...606..749K,2006MNRAS.365.1180C,2009ApJ...692..246D,2012ApJ...747...30P,2015ApJS..217...26B,2016ApJ...821...33R} that belong to recombination lines. The basic feature of the BLR breathing can observationally be described as follows. As the central ionizing luminosity increases, so does the BLR radius approximately as $R_{\rm BLR}\propto L^{1/2}$, since more distant BLR clouds are photoionized and produce a given BLR line. In relation to that, the width of the broad line decreases approximately as $\Delta v_{\rm BLR}\propto R_{\rm BLR}^{-1/2}\propto L^{-1/4}$, which holds under the assumption that the corresponding BLR emitting material is virialized. The LOC model is physically motivated in a sense that the line emission is given by the sum of the emission from individual clouds with different densities and distances from the central source of ionizing continuum. The whole system has an axisymmetric geometry \citep{1995ApJ...455L.119B}. The clouds are characterized by the distance distribution $f(r)\propto r^{\Gamma_{\rm r}}$ and the density distribution $g(n)\propto n^{\gamma_{\rm n}}$, where $\Gamma_{\rm r}\approx -1$ and $\gamma_{\rm n}\approx -1$ \citep{1995ApJ...455L.119B}. Then the line luminosity is given by,
\begin{equation}
    L_{\rm line}\propto \iint_{R_{\rm in}}^{R_{\rm out}}\, r^2\, F(r) f(r)\, g(n)\, \mathrm{d}r\,\mathrm{d}n\,,
    \label{eq_line_luminosity}
\end{equation}
where $F(r)$ is an emission-line flux of a single cloud at the distance $r$. Using the integration scheme according to Eq.~\eqref{eq_line_luminosity}, the largest contribution comes from the clouds with the most efficient response to the continuum, i.e. clouds of a certain density range and at a distance where the ionizing photon flux reaches an optimum value.  
The LOC model implies that the overall BLR region is in fact much larger than the most emitting region that is observed.

The standard, powerful method to infer the length-scale of the BLR for a given source is the measurement of the time-delay between the observed variable continuum emission and the emission of one of the broad lines, typically Balmer lines (H$\alpha$, H$\beta$) or lines of some ions (MgII, CIV, CIII], FeII). This so-called \textit{reverberation mapping} is possible since the variable continuum and line emission are significantly correlated, which in other words means that the ionizing continuum powered by the accretion disc is the main driver of the BLR variability and the BLR emission revealed in the form of emission lines is just the reprocessed UV/optical emission of the disc \citep{1982ApJ...255..419B}. Mathematically, the reprocessing of the continuum flux by the BLR material located further away can be expressed via the so-called \textit{transfer function} $\psi(\tau)$ \citep{1982ApJ...255..419B,2004ApJ...613..682P}, which is a function of the rest-frame time-delay that depends on the distance of reprocessing clouds from the source $r$ as well as on the polar angle $\theta$ with respect to the line of sight,
\begin{equation}
    \tau=\frac{r}{c}(1+\cos{\theta})\,.
    \label{eq_time_delay_theta}
\end{equation}
The observer then sees the delayed line emission originating in all the clouds that are intersected by the iso-delay surface given by Eq.~\eqref{eq_time_delay_theta}.

The transfer function $\psi(\tau)$ can then be determined as follows. Let $j_{\rm l}(t)$ be the emissivity of a broad line. In general, it responds in a non-linear way to the continuum flux changes $F_{\rm c}-\langle F_{\rm c}\rangle$ with respect to the mean value. If we approximate the line emissivity as a function of the continuum flux via the Taylor expansion around the mean value $\langle F_{\rm c}\rangle$, assuming the small changes of $F_{\rm c}$ above the mean value, we obtain,
\begin{equation}
    j_{\rm l}(F_{\rm c})=j_{\rm l}(\langle F_{\rm c}\rangle)+\frac{\partial j_{\rm l}}{\partial F_{\rm c}}(F_{\rm c}-\langle F_{\rm c}\rangle)+\ldots\,.
\end{equation}
The change in the line luminosity $\delta L_{\rm l}(t)$ can then be expressed as the convolution of the transfer function with the continuum line curve,
\begin{equation}
    \delta L_{\rm l}(t)=\int_{0}^{\infty} \psi(\tau)\big[L_{\rm c}(t-\tau)-\langle L_{\rm c}\rangle\big]\mathrm{d}\tau\,,
    \label{eq_line_luminosity_conv}
\end{equation}
where the transfer (or response) function is defined via the relation,
\begin{equation}
    \psi(\tau)=c\int_{S}\frac{f(\mathbf{r})}{4\pi r^2}\frac{\partial j_{\rm l}}{\partial F_{\rm c}} \mathrm{d}A\,,
    \label{eq_transfer_function}
\end{equation}
where the surface of integration is the isodelay surface for the time-delay $\tau$, as given by Eq.~\eqref{eq_time_delay_theta} and $f(\mathbf{r})$ is the local volume filling factor at the position $\mathbf{r}$ from the source. The determination of the transfer function $\psi(\tau)$, which is centered at the mean time-delay $\tau$, is generally time-consuming. This is related to the fact that to obtain $\psi(\tau)$ one needs to make an inversion of the convolution in Eq.~\eqref{eq_line_luminosity_conv}, i.e. to perform the following Fourier transform
\begin{equation}
    \psi(\tau)=\int e^{-i2\pi \nu \tau}\frac{\hat{L}_{\rm l}(\nu)}{\hat{L}_{\rm c}(\nu)} \mathrm{d}\nu\,,
\end{equation}
where $\hat{L}_{\rm l}(\nu)$ and $\hat{L}_{\rm c}(\nu)$ are Fourier transforms of the line-emission and continuum light curves. To map out $\psi(\tau)$ with a sufficient precision, the Fourier-transforms of the light curves that are functions of $\propto \tau^{-1}$ need to be determined using the time-step smaller than $\Delta t<\tau/2$. In case the broad line is divided into several wavelength bins, in particular the line center and at least two line-wing regions, which correspond to different line-of-sight velocities $v_{\rm LOS}=v_{\rm orb}\sin{\theta}$ of the emitting gas, one can determine the transfer function $\psi(v_{\rm LOS},\tau)$ as a function of the line-of-sight velocity and the time delay, which is also referred to as the \textit{wavelength-resolved reverberation mapping}.

The observed time-delay can be determined using the interpolated cross-correlation function (ICCF), its discrete version (DCF), $z$-transformed DCF (zDCF), damped random-walk modelling of the continuum emission using the JAVELIN module, the $\chi^2$ method, data regularity/randomness estimators (von Neuman, Bartels), see e.g. \citet{2019AN....340..577Z,2020ApJ...896..146Z,2020A&A...642A..59R,2021ApJ...912...10Z} for the application of several methods and references therein. Once the time-delay was determined in the observer's frame, it is straightforward to obtain the estimate of the rest-frame time-delay using the source redshift, $\tau_{\rm rest}=\tau_{\rm obs}/(1+z)$. For the mean distance of the BLR line-emitting material, we can simply use $R_{\rm BLR}=c\tau_{\rm rest}$.

From the series of spectra, one can obtain the length-scale of the BLR $R_{\rm BLR}$ as well as the line width, which serves as a proxy for the gas orbital velocity in type I AGN. Using the full width at half maximum (FWHM) of the broad line expressed in km/s, one can obtain the gas orbital velocity as $v_{\rm BLR}=f_{\rm c}v_{\rm FWHM}$, where the factor $f_{\rm c}$ depends on the BLR viewing angle, geometry, and the kinematics. Assuming that the BLR cloud motion is virialized, we can use the \textit{virial theorem} given in Eq.~\eqref{eq_virial} to obtain the \textit{virial black hole mass}, 

\begin{align}
    M_{\rm vir} &=\frac{R_{\rm BLR}v_{\rm BLR}^2}{G}\,,
    \,\notag\\
    &=f_{\rm c}^2\frac{c\tau_{\rm rest}v_{\rm FWHM}^2}{G}\,,\notag\\
    &=f_{\rm vir}\frac{c\tau_{\rm rest}v_{\rm FWHM}^2}{G}\,,
    \label{eq_virial_BHmass}
\end{align}
where we have denoted $f_{\rm vir}=f_{\rm c}^2$, which is a so-called \textit{virial factor}. It depends on the BLR geometry, kinematics as well as the viewing angle approximately in the following form \citep{2006A&A...456...75C,2018NatAs...2...63M},
\begin{equation}
    f_{\rm vir}=[4(\sin^2{\iota}+(H_{\rm BLR}/R_{\rm BLR})^2)]^{-1}\,,
    \label{eq_virial_factor}
\end{equation}
where $\iota$ is the viewing angle of the BLR plane with respect to the line of sight.
In case $f_{\rm vir}$ depends only on the viewing angle, $f_{\rm vir}\approx (4\sin^2{\iota})^{-1}$, $f_{\rm vir}\approx 1$ for $\iota\approx 30^{\circ}$. However, fixing the virial factor to a single value of the order of unity, which is usually done in single-epoch spectroscopic mass determination \citep{2015ApJ...801...38W}, can introduce an error by a factor of $\sim 2-3$ in the virial black hole mass determination measurements \citep{2018NatAs...2...63M}. By comparing the SMBH mass inferred from the spectral energy distribution fitting and the reverberation mapping \citet{2018NatAs...2...63M} found out that the virial factor is inversely proportional to the FWHM of broad lines, $f_{\rm vir}\propto \mathrm{FWHM}^{-1}$, which reflects the dependence on the BLR viewing angle or the effect of the radiation pressure on the BLR geometrical distribution. 

Furthermore, it was found out that the size of the BLR or the rest-frame time delay are correlated with the monochromatic optical/UV continuum luminosity of the AGN \citep{2000ApJ...533..631K,2005ApJ...629...61K,2013ApJ...767..149B,2019ApJ...880...46C,2021ApJ...912...10Z,2021arXiv210301973Y}. For instance, the H$\beta$ time delay is proportional to the $\sim 5000$\,\AA\, luminosity, while MgII time lag is larger for larger $3000\,$\AA\, luminosity. This is a so-called radius-luminosity relation (RL), which has a simple power-law form, $\tau_{\rm rest}\propto L^{\alpha}$, where $\alpha\approx 0.5$ based on the simple photoionization theory. The power-law slope of $0.5$ stems from the definition of the ionization parameter of BLR clouds,
\begin{equation}
    U=\frac{Q(H)}{4\pi R_{\rm BLR}^2 c n(H)}\,,
    \label{eq_ionization_parameter}
\end{equation}
where $R_{\rm BLR}$ is the cloud distance from the source of ionizing continuum, $n(H)$ is the cloud hydrogen density, and $Q(H)=\int_{\nu_{\rm i}}^{\infty} L_{\nu}/(h\nu) \mathrm{d}\nu$ is the hydrogen-ionizing photon flux in ${\rm cm^{-2} s^{-1}}$, where $\nu_{\rm i}$ is the lowest frequency of photons that can ionize the hydrogen atom. Under the assumption that $Un(H)$ is approximately constant for BLR clouds across different sources, we obtain $R_{\rm BLR}\propto Q^{1/2}(H)\propto L^{1/2}$.

\begin{figure}[tbh!]
    \centering
    \includegraphics[width=\textwidth]{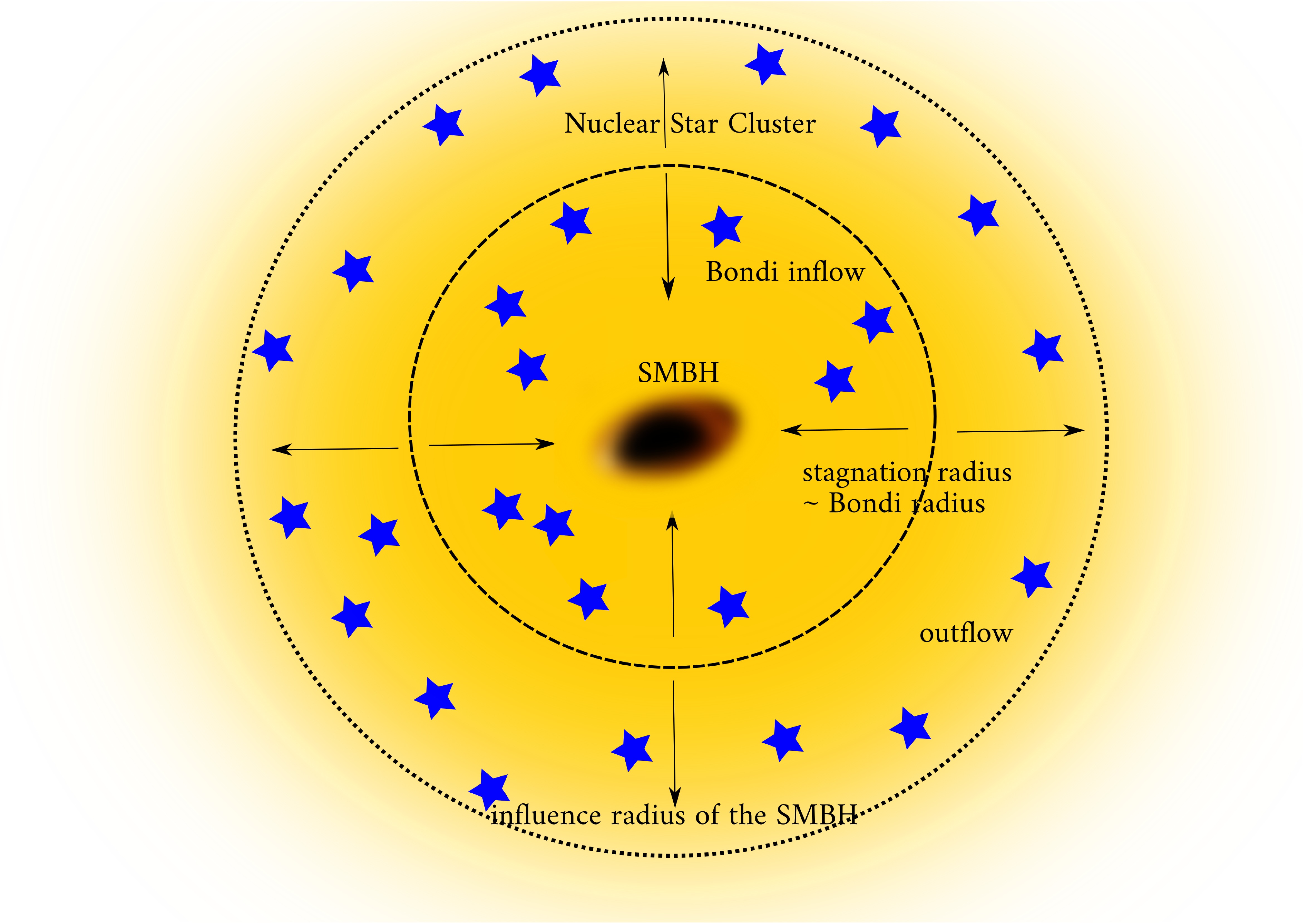}
    \caption{Illustration of the basic scales in the quasispherical Nuclear Star Cluster (NSC), inside the sphere of influence of the SMBH. Stellar outflows influence the hydrodynamics in such a way so that the hot flow develops a stagnation radius where its radial velocity reaches zero. Inside the stagnation radius, nearly radial Bondi flow can develop and power the accretion on the low-luminous SMBH, such as Sgr~A*. Drawn not to the scale.}
    \label{fig_stag_radius_NSC}
\end{figure}

In low-luminosity AGN, specifically the extremely low-luminous Galactic centre, $\sim $ 200 massive OB stars \citep{2006ApJ...643.1011P,2010ApJ...708..834B} can provide essentially all the material that is partially accreted at the corresponding Bondi radius of $M_{\bullet}\simeq 4\times 10^6\,M_{\odot}$, but $>99\%$ forms an outflow that flattens the density profile of the accretion flow, $n\propto r^{-3/2+p}$, where $p\approx 1$ \citep{2013Sci...341..981W}. The mass-loading by stellar winds supplied by point sources or a spherical matter distribution was dealt in both semi-analytical and numerical approaches \citep{2004ApJ...613..322Q,2008MNRAS.383..458C,2015MNRAS.453..775G,2018MNRAS.479.4778Y,2019MNRAS.482L.123R}. Steady-state 1D inflow-outflow solutions of the mass-loaded accretion flows contain a stagnation radius where the bulk velocity of the flow goes through zero \citet[see e.g. ][]{2015MNRAS.453..775G}. Inside the stagnation radius, inflow takes place, while outside it, the gas escapes the system. Under certain circumstances, both the inflow and the outflow reach supersonic velocities \citep{2018MNRAS.479.4778Y}. The stagnation radius is proportional to the gravitational parameter of the SMBH as for the Bondi radius, but it is inversely  proportional to the square of the terminal wind velocity instead of the sound speed, under the assumption that stellar winds are faster than the stellar velocity dispersion, $v_{\rm w}>\sigma_{\star}$. Depending on the power-law slope $\Gamma$ of the stellar brightness profile, the stagnation radius can be expressed as \citep{2015MNRAS.453..775G}, 

\begin{align}
  R_{\rm stag} & \approx \left(\frac{13+8\Gamma}{4+2\Gamma}-\frac{3\nu}{2+\Gamma}\right)\frac{GM_{\bullet}}{\nu v_{\rm w}^2}\,\notag\\
            & \approx
 \begin{cases}
   15\,\left(\frac{M_{\bullet}}{2\times 10^8\,M_{\odot}}\right)\left(\frac{v_{\rm w}}{500\,{\rm km\,s^{-1}}}\right)^{-2}\,{\rm pc} &,  \text{core\, ($\Gamma=0.1$)}\,,\\
   8\,\left(\frac{M_{\bullet}}{2\times 10^8\,M_{\odot}}\right)\left(\frac{v_{\rm w}}{500\,{\rm km\,s^{-1}}}\right)^{-2}\,{\rm pc} &,  \text{cusp\, ($\Gamma=0.8$)}\,,
  \end{cases}
  \label{eq_stagnation_radius}
\end{align} 
where we considered the core and the cusp stellar brightness distributions for the power-law slope $\Gamma$. The gas density slope $\nu\equiv -\mathrm{d}\rho/\mathrm{d}r|_{R_{\rm stag}}$ at the stagnation radius can be approximated using the stellar brightness slope $\nu\approx 1/6(4\Gamma+3)$. In general, the ratio of the stagnation radius to the Bondi radius is of the order of unity,

\begin{equation}
  \frac{R_{\rm Stag}}{R_{\rm B}} \approx \frac{13+8\Gamma}{(2+\Gamma)(3+4\Gamma)}\,.
  \label{eq_stag_Bondi}
\end{equation}
The basic length-scales inside the sphere of influence of the SMBH are depicted in Figure~\ref{fig_stag_radius_NSC}, where we assumed quasi-spherical distribution of stellar sources sources as well as the spherical distribution of hot plasma.

Extending the analysis above for AGN would require certain modifications, namely the extra radiation field (non-thermal continuum from the very centre and the thermal disk emission) and the gravitational influence of the disk and the dusty torus. In the AGN, there needs to be extra material available for the accretion apart from the stellar wind supply. For quasars, the accretion rate is of the order of $\dot{M}_{\rm acc}\approx 1\,M_{\odot}\,{\rm yr^{-1}}$, for Seyfert galaxies it is about two orders of magnitude smaller. In comparison, in the Galactic centre, based on the polarization measurements of sub-mm emission, the accretion rate is much smaller, $\dot{M}_{\rm acc}\approx 10^{-7}-10^{-9}\,M_{\odot}{\rm yr^{-1}}$ \cite{2007ApJ...654L..57M}. In general, the structure of the central parts of galaxies differs quite significantly between quasars and low-luminosity nuclei, depending thus on the accretion rate and thus the overall supply of the material.  

Nevertheless, stars, stellar clusters and supernovae are expected to influence the circumnuclear medium in AGN and stars have been employed to explain some of the observables in the broad line region \citep{1985MNRAS.213..665P,2002A&A...387..804V}.

\subsubsection*{Observational signatures of the broad line region}

AGN show prominent emission lines in the UV as well as the optical part of the electromagnetic spectrum on the top of UV and optical non-stellar continuum. According to the full width at half maximum (FWHM), these lines are generally classified as broad (FWHM$\approx 1000-25000\,{\rm km\,s^{-1}}$) and narrow (FWHM$\approx 200-1000\,{\rm km\,s^{-1}}$), but there are AGN sources that do not exhibit any emission lines or only during a low-state of the central variable continuum source (in particular blazars, \citep{2000ARA&A..38..521S}). The unified model of AGN can generally explain the presence and the absence of broad emission lines in their spectra for type I and II AGN, respectively, depending on the inclination of the system with respect to the observer and hence the blocking of UV/optical photons by an optically and geometrically thick dusty torus that is clumpy, which leads to the disappearance of broad lines for type II AGN when observed in the total, unpolarized intensity (see e.g. \citep{1985ApJ...297..621A,1993ARA&A..31..473A,1995PASP..107..803U}). Narrow-line components are mostly due to forbidden transitions, they are emitted by a low-density gas ($n\sim10^3\, \rm cm ^{-3}$), which is spatially extended and partially resolved for nearby AGN  \citep{2006A&A...459...55B}. While the broad lines are dominantly caused by permitted transitions in the denser medium with the number density of $n>10^8\,{\rm cm^{-3}}$. Furthermore, narrow lines are often superposed on the broader component, which resulted in the subclassification of Seyfert galaxies into intermediate classes depending on the relative strength of the broad and narrow components (hereafter BC and NC, respectively; \citep{1976MNRAS.176P..61O,1977ApJ...215..733O,1981ApJ...249..462O,1992MNRAS.257..677W}). BC emission lines respond to the continuum variation, but with a shorter time delay, which indicates that broad-line emitting region is photoionized and optically thick to ionizing continuum. On the other hand, NC emission lines do not show a correlated variation since they are emitted by a spatially extended region where the ionizing continuum is geometrically diluted and the recombination time scale is very long.

Broad lines led to the identification of quasars \citep{1963Natur.197.1040S} and remain essential in studying accretion processes, outflows from the nuclear region as well as the effect of the central variable continuum source on the circumnuclear medium as well as the host galaxy as a whole. The AGN cover a large redshift range from the local sources such as M81 at $d=3.2\,{\rm Mpc}$ or the brightest AGN in the X-ray domain Cen-A \citep{1993nag..conf...47M} at $d=4.4\,{\rm Mpc}$ up to $z\approx 7.1$ (ULAS J1120+0641; \citep{2011Natur.474..616M}), at which the first quasars started to contribute to the reionization of the intergalactic medium. Therefore different broad lines from the rest-frame optical and UV part of the spectrum are suited for their monitoring in the observer-frame optical domain depending on the source redshift: HI H$\alpha$, $z=0.0-0.5$; H$\beta$, $z=0.0-1.0$; MgII$\lambda2800$, $z=0.3-2.6$; CIII] $\lambda1909$, $z=0.8-4.2$ (semi-forbidden line); CIV$\lambda1549$, $z=1.2-5.4$; HI Ly$\alpha$, $z=1.9-7.2$ \citep{2000ARA&A..38..521S}.

\begin{figure}[tbh!]
    \centering
    \includegraphics[width=\textwidth]{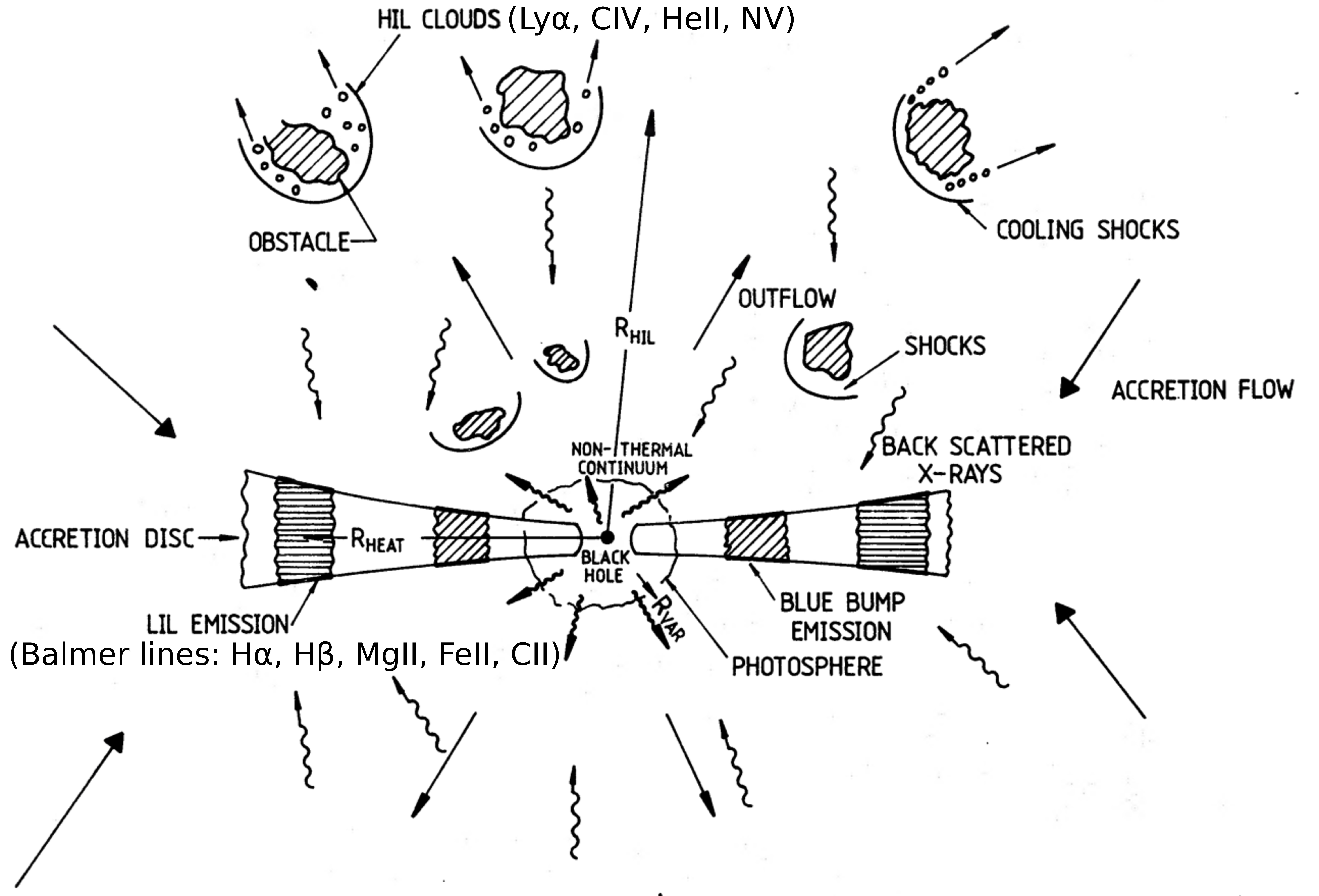}
    \caption{An illustration of the low- and high-ionization components of the Broad Line Region (BLR) in AGN (denoted as LIL and HIL). While the LIL component is mostly virialized and positioned close to the disc plane, the HIL part is associated with the nuclear outflows. The original illustration is from \citet{1988MNRAS.232..539C}, where we added examples of both low-ionization lines used for the reverberation-mapping and high-ionization lines.}
    \label{fig_lil_hil}
\end{figure}

A two-component model was proposed for the BLR region \citep{1988MNRAS.232..539C} that consists of low- and high-ionization lines, see Figure~\ref{fig_lil_hil} for the illustration. The emission lines that belong to the low-ionization lines (LIL, with an ionization potential $\lesssim$20 eV), such as most of Balmer lines as $H\beta$, CII $\lambda1336$, and MgII$\lambda2800$ and the pseudo-continuum of FeII, form in the mildly ionized, higher density regions $(n>10^{11}\,{\rm cm^{-3}})$ close to or directly in the extended parts of the accretion disc with no inflow/outflow signatures. However, when the accretion rate is close to the Eddington limit, low ionization lines show the presence of outflows (e.g. \citep{2018A&A...620A.118N}). The high-ionization lines (HIL, with an ionization potential $\gtrsim$40 eV), such as Ly$\alpha$, CIV $\lambda1549$, HeII $\lambda1640$, and NV $\lambda1240$, form in the highly ionized, lower density regions $(n < 10^{10}\,{\rm cm^{-3}})$ associated with an outflow beyond the disc, but still close to the central black hole, as inferred from the reverberation mapping studies. Blueward asymmetries and blueshifted peaks with respect to the rest-frame of the HIL indicate the existence of disk winds in AGN \citep{1995ApJ...451..498M, 2009NewAR..53..140G, 2017A&A...608A.122S}. Considering the general behavior of the LIL, observations indicate the coexistence of virial and radiation forces in the BLR. The disk wind model also supports the existence of the broad absorption outflows in the UV and X-ray domain of the AGN spectra, which are generally blueshifted \citep{2004ApJ...616..688P, 2010MNRAS.408.1396S}. 

Around $1\%$ of the type-1 AGN show a double-peak profile mainly in the Balmer lines. Due to the variation in the peak profiles of months to years, this effect is due to a rotating disk \citep{1997ApJ...490..216E}, where the line-emission of the disk has a small optical depth and/or the line-emitting
annulus of the disk is narrow \citep{2007ApJS..169..167G, 2008ApJ...686..138F}.

\subsubsection*{Formation and stability of BLR clouds}

The BLR is generally envisaged as a collection of clouds with a certain geometrical distribution and a velocity field, which is determined mainly by the potential of the SMBH since the BLR clouds are located well inside its sphere of gravitational influence, see Eq.~\eqref{eq_influence_radius1}, as well as by radiation pressure of an accretion disc and potentially other effects, such as gas pressure gradient and magnetic field. Already early models of the BLR propose that its dynamics and ionized emission correspond to that of the clumpy, cloudy medium (see e.g. \citet{1990agn..conf...57N}). The intercloud medium is of a lower density and higher ionization and therefore does not contribute significantly. The photoionization equilibrium is attained when the photoionization rate is balanced by the recombination rate, which can be expressed by an ionization parameter $U_{\rm ion}$,

\begin{equation}
    U_{\rm ion}=\frac{Q_{\rm ion}(H)}{4\pi r^2 c n_{\rm e}}\,,
\end{equation}
which reflects the ratio of the density of hydrogen-ionizing photons and the electron density that stands for the recombination rate. The number of ionizing photons $Q_{\rm ion}(H)=\int_{\nu_{\rm i}}^{+\infty}(L_{\nu}/h\nu) \mathrm{d}\nu$ is obtained by integrating the ratio $L_{\nu}/h\nu$ over the photon spectrum that can ionize hydrogen and $L_{\nu}$ is the ionizing luminosity of the AGN.  The BLR emission line strengths require the photoionization equilibrium at the temperature of $T\sim 10^4\,{\rm K}$ similar to plasma in HII regions and planetary nebulae \cite{1990agn..conf...57N,2000ARA&A..38..521S}.

It has been recognized early on that the BLR region consists of two distinct components with a likely different origin \citep{1988MNRAS.232..539C}:
\begin{itemize}
 \item[(i)] \textit{the low-ionization line (LIL) component}, revealed by lines with a low ionization potential (<20 eV), such as H$\beta$, MgII, and FeII, which does not reveal a significant inflow-outflow motion, and that likely forms in the denser region closer to the accretion disc,
 \item[(ii)] \textit{the high-ionization line (HIL) component}, whose lines (e.g. Ly$\alpha$, HeII, CIV) with a higher ionization potential (>40 eV) exhibit an outflow motion revealed via blueward asymmetries and blueshifted line peaks, that forms in the lower-density, outflowing gas, but still close to the SMBH as revealed via the measured time-lags of HIL lines. 
\end{itemize}
For an illustration of the LIL and HIL components of the BLR, see Fig.~\ref{fig_lil_hil} and the corresponding more detailed discussion. In the following text, we focus on the formation mechanisms and the stability of the LIL component.


Rare detection of broad absorption features in type-I AGN (AGN with broad emission lines) suggests a flattened structure of the BLR.
On the other hand, the clouds cannot be too close to the disc or within the disk since the covering factor of the irradiating continuum flux needs to be $\sim 30-50\%$ \citep{2004ApJ...606..749K} to account for the observed significant correlation between the continuum and the broad-line emission.
Therefore, the BLR clouds are more likely distributed in a torus-like or a ring-like geometry. 

The BLR inner radius is set up by a specific formation mechanism, such as the dust formation within the disk that can lead to a dust-driven wind \citep{2011A&A...525L...8C}. When such a wind gets high enough above the equatorial plane, the dust is exposed to intensive UV irradiation from the AGN and eventually destroyed. The dust-driven wind fails to further expand and the gas cloud drops back to the disk plane. This mechanism can well explain the appearance of the BLR clouds at the distance of roughly the same temperature around 1000\,K in the accretion disk, as well as the turbulent velocity field due to the mixture of inflowing and outflowing components. The accretion disk self-shielding or the inner accretion disk wind can provide more shielding from the central irradiation, allowing the gas clouds to get further above the equatorial plane. This can explain the observed X-ray absorption by BLR clouds observed in intermediate AGN types \citep{2015A&A...578A..96S}.

The BLR outer radius can be considered to coincide with the inner radius of the dusty molecular torus -- i.e. the dust sublimation radius, which can be estimated based on \citet{2008ApJ...685..160N},
\begin{equation}
    R_{\rm d}\simeq 0.4 \left(\frac{L}{10^{45}\,{\rm erg\,s^{-1}}} \right)^{1/2} \left(\frac{T_{\rm sub}}{1500\,{\rm K}}\right)^{-2.6}\,{\rm pc}\,,
\end{equation}
which is almost identical to Eq.~\eqref{eq_sublimation_radius} in terms of the normalization, but with a small difference of the power of the dust sublimation temperature $T_{\rm sub}$, i.e. $2.6$ instead of $2.8$, which stems from more detailed radiative transfer calculations of \citet{2008ApJ...685..160N}. The mean rest-frame distance of the BLR is given by the measured time-delay between the continuum and the specific line-emission light curves, $R_{\rm BLR}\sim c\tau_{\rm BLR}$, where $\tau_{\rm BLR}$ depends on the specific line emission. It was revealed that there is a power-law relation between $\tau_{\rm BLR}$ and the monochromatic luminosity of the accretion disc, $L_{\rm mon}$, with the expected dependence $\tau_{\rm BLR}\propto L_{\rm mon}^{1/2}$ based on the constant photoionization parameter and the cloud density as we discussed in Subsection~\ref{subsec_circumnuclear}, see Eq.~\eqref{eq_ionization_parameter}. This is a so-called radius-luminosity (RL) relation. For the broad H$\beta$ line, the RL relation is consistent with the square-root law with a scatter of $0.13$ dex for the initial sample, see \citet{2013ApJ...767..149B},
\begin{equation}
    \log{\left(\frac{\tau_{H\beta}}{1\,{\rm lt-day}}\right)}=1.527^{+0.031}_{-0.031}+0.533^{+0.035}_{-0.033}\log{\left(\frac{L_{5100}}{10^{44}\,{\rm erg\,s^{-1}}} \right)}\,.
    \label{eq_RL_Bentz}
\end{equation}
As the number of sources increased, especially including less variable and highly accreting sources, the scatter along the RL relation became more prominent. It was shown that the departure from the best-fit RL relation is correlated with the accretion rate, i.e. it tends to be larger for the high-accretion rate sources, see \citet{2019ApJ...883..170M} and references therein. One of possible interpretations is that at higher accretion rates, the disc becomes geometrically thicker, i.e. it is a slim disc instead of a thin Shakura--Sunyaev type, and shields the radiation from its inner parts \citep{2014ApJ...797...65W}. In this sense, the radiation field that the BLR clouds see is modified and generally anisotropic. In particular, the measured time-delay corresponds to a smaller luminosity and is thus effectively shorter \citep{2014ApJ...797...65W}.

The relevance of the RL relation lies in the fact that it is essential as a secondary black hole mass estimator using just a single spectrum, without the necessity to measure the BLR time delay for each source \citep{1998ApJ...505L..83L,1999ApJ...526..579W,2002MNRAS.337..109M,2006ApJ...641..689V}, which is challenging to determine for fainter AGN towards higher redshifts. From Eq.~\ref{eq_RL_Bentz}, it is possible to determine the rest-frame time delay from the monochromatic luminosity and the line FWHM can be obtained from the same spectrum. The virial black hole mass can then be calculated from Eq.~\eqref{eq_virial_BHmass} under the assumption of a virial factor. In this way, the RL relation is crucial for studying black-hole mass evolution across the cosmic history \citep{2002MNRAS.337..109M}. While the RL relation for H$\beta$ line has been verified for $\sim 100$ sources \citep{2019ApJ...883..170M}, it is essential to test the dependency $\tau_{\rm BLR}\propto L_{\rm mon}^{\alpha}$ for other lines for which the correlation and the time delay is detected. In particular, it is of interest to compare the best-fit slopes $\alpha$ between different broad emission lines. Towards higher redshifts, MgII line moves into the optical bands and the reverberation mapping using MgII can be performed with optical monitoring telescopes. \citet{2019ApJ...880...46C} showed the RL relation also exists for MgII line, with the further confirmation by \citet{2020ApJ...896..146Z} using eleven sources, who showed that the scatter and the time-delay shortening for high-accreting sources is also present in the MgII RL relation. Using in total 69 sources, including the Sloan Digital Sky Survey Reverberation Mapping data \citep[SDSS-RM;][]{2020ApJ...901...55H}, \citet{2021ApJ...912...10Z} determined the intercept of $1.67\pm 0.05$ and the slope of $0.30 \pm 0.05$ for the MgII RL relation, i.e. smaller than the canonical value, but with a large scatter of $\sim 0.30$ dex. In summary, the RL relation can be used for estimating the characteristic length-scale of the BLR region once the luminosity of the source is known.    

The direct confirmation of the flattened structure of the BLR was provided by the first images of the BLR obtained using broad infrared lines resolved by the infrared Very Large Telescope Interferometer GRAVITY \citep{2017A&A...602A..94G}. The flattened geometry and the velocity field is consistent with an orbiting gas bound to the central black hole, see in particular observations of 3C273 using broad Pa$\alpha$ line by \citet{2018Natur.563..657G}, NGC3783 using broad Br$\gamma$ line by \citet{2021A&A...648A.117G}, and IRAS 09149-6206 analyzed using broad Br$\gamma$ line by \citet{2020A&A...643A.154G}. For NGC3783, observations are consistent with the thick, rotating ring of clouds, whose radial distribution peaks close to the inner edge. The temperature of the BLR clouds is kept at $\sim 10\,000-20\,000\,{\rm K}$ due to the atomic-transition, radiative heating/cooling in the partially ionized plasma, independent of the ionizing flux across different sources. The very intense irradiation by a hard ionizing continuum of the AGN leads to the complete ionization and the heating/cooling balance is established at $\sim 10^7\,{\rm K}$ due to Compton heating/cooling. In the plasma irradiated by an AGN, thermal instability is expected to develop and the partially ionized clouds having $\sim 10^4\,{\rm K}$ are in an approximate pressure equilibrium with the hot and diluted phase at $\sim 10^7\,{\rm K}$ \citep{1981ApJ...249..422K}, i.e. the clouds are confined by a hot intercloud medium. At this two-phase balance, the evaporation timescale of the colder plasma due to thermal conduction is comparable to the timescale, during which the hot phase cools down and condenses onto clouds, see also the discussion and the estimates of the evaporation/condensation timescales below. Hence, the lifetime and the stability of BLR clouds is difficult to determine, but supposedly it is of the order of the orbital timescale or its fraction as we will demonstrate based on the evaporation/condensation as well as tidal timescales. 

For the estimate of the basic timescales, we assume the column density of the BLR gas of $N_{\rm cl}\sim 10^{24}\,{\rm cm^{-2}}$ and the volume number density of $n_{\rm cl}\sim 10^{12}\,{\rm cm^{-3}}$, from which the basic length-scale or a cloud radius can be estimated as $R_{\rm cl}\sim N_{\rm cl}/n_{\rm cl}\sim 10^{12}\,{\rm cm}$ \citep[see e.g.][]{2019OAst...28..200C,2016ApJ...831...68A}. The average mass of BLR clouds is thus of the order of $M_{\rm cl}\sim 4/3 \pi R_{\rm cl}^3 m_{\rm p}n_{\rm cl}\sim 7.0 \times 10^{24}\,{\rm g}\sim 7.5\,M_{\rm Ceres}$, which is comparable to the mass of a large asteroid (a few times of the mass of Ceres). The basic intrinsic dynamical timescale of BLR clouds is given by the perturbation propagation timescale, which can be estimated as the ratio of the cloud radius to the sound speed inside the cloud, where we assume the isothermal sound speed for $T_{\rm cl}\sim 10^4\,{\rm K}$,
\begin{align}
    \tau_{\rm dyn}&=\frac{R_{\rm cl}}{c_{\rm cl}}\,\notag\\
    &\simeq 13\,\left(\frac{R_{\rm cl}}{10^{12}\,{\rm cm}}\right)\left(\frac{T_{\rm cl}}{10^4\,{\rm K}}\right)^{-1/2}\text{days}\,.\label{eq_sound_timescale}
\end{align}
Due to the coexistence of two fluids with different densities in the BLR region (hot diluted intercloud medium and colder, denser BLR clouds) as well as the likely existence of the velocity shear $v_{\rm shear}$ between the two phases, the BLR cloud will be significantly perturbed or disrupting on a \textit{crushing timescale} that is equal to a shock-propagating timescale \citep[see e.g.][]{2015MNRAS.449....2M},
\begin{align}
    \tau_{\rm crush}&=\frac{R_{\rm cl}}{v_{\rm shear}}\sqrt{r_{\rho}}\,\notag\\
    &=3.7\,\left(\frac{R_{\rm cl}}{10^{12}\,{\rm cm}} \right)\left(\frac{v_{\rm shear}}{1000\,{\rm km\,s^{-1}}} \right)^{-1}\left(\frac{r_{\rho}}{10^3} \right)^{1/2}\text{days}\,,\label{eq_crushing_timescale}
\end{align}
where $r_{\rho}=n_{\rm cl}/n_{\rm h}$ denotes the density ratio between the cloud and the hot intercloud medium. Assuming the pressure equilibrium and the confinement of the BLR clouds by hot diluted intercloud medium, we obtain $r_{\rho}\sim n_{\rm cl}/n_{\rm h}=T_{\rm h}/T_{\rm cl}\sim 10^3$. The development of the Kelvin-Helmholtz instability due to the velocity shear as well as the Rayleigh-Taylor instability due to the centrifugal acceleration in the shock propagating through the cloud occurs on timescales comparable to the crushing timescale expressed by Eq.~\eqref{eq_crushing_timescale}. This timescale related to the hydrodynamical instabilities would imply the BLR cloud lifetime of at least a few days. 

However, the lifetime of clouds is plausibly longer than indicated by Eq.~\eqref{eq_crushing_timescale} due to stabilisation mechanisms that prevent the quick development of large-scale hydrodynamical instabilities. In the BLR region, these mechanisms contribute:
\begin{itemize}
    \item[(i)] slow cooling or warm circumnuclear medium, see e.g. discussion in \citet{2012A&A...548A.113D},
    \item[(ii)] ionizing radiation from external sources, in particular X-ray/UV emission of the inner accretion disc \citep{2014MNRAS.437..843G}, which heats up the surrounding gas content, which is linked to point (i),
    \item[(iii)] circumnuclear magnetic field \citep{2014A&A...561A.152V}.
\end{itemize}
In the BLR region, all the conditions (i), (ii), and (iii) are present to a smaller or a larger extent. The ionizing radiation, which keeps the circumnuclear medium warm, increases the thermal gas pressure that suppresses the growth of Rayleigh-Taylor fingers and other instabilities in BLR clouds. This is confirmed by numerical hydrodynamical simulations. In \citet{2007A&A...472..141V}, authors present dynamical models of cold clouds moving through the hot interstellar medium. Such a set-up is common in several astrophysical settings, including cooling flows in galactic halos, star-forming regions where cold molecular clouds are surrounded by young massive stars, and it is also applicable to the BLR region in the gravitational influence of the SMBH. They demonstrate that heat conduction smooths steep temperature as well as density gradients at the cloud boundary since the heat conduction timescale is shorter than the cooling timescale. As a result, there is a decrease in the shearing velocity between the cloud surface and the hot medium. According to Eq.~\eqref{eq_crushing_timescale}, this results in the prolongation of the crushing timescale as well as the timescale for the development of the Kelvin-Helmholtz instability and colder BLR clouds can survive the passage through the hot intercloud medium, at least for a significant fraction of their orbital period.

The BLR region, as described earlier, can be envisaged as a multiphase medium similar to other astrophysical set-ups \citep{1977ApJ...218..148M}. It consists of at least two phases -- cold, dense BLR clouds and hot, diluted intercloud medium -- that are approximately in pressure equilibrium since the circumnuclear medium is externally heated by an AGN. The BLR clouds cool down radiatively via atomic transitions in the partially ionized plasma, while the hot, diluted intercloud medium that is fully ionized cools down via Compton cooling. This way the two phases can stay in pressure equilibrium with BLR clouds having $\sim (10-20)\times 10^3\,{\rm K}$ and the intercloud medium with $\sim 10^7\,K$ \citep{1981ApJ...249..422K}. The BLR region is certainly not stationary but dynamic. There is a certain velocity shear between the clouds and the hot intercloud medium. The basic estimate of the velocity difference is given by the broad line FWHM of a few $1000\,{\rm km\,s^{-1}}$. If we consider the value of $v_{\rm FWHM}=5000\,{\rm km\,s^{-1}}$ for a circular velocity, this is associated with the characteristic BLR radius of
\begin{align}
    r_{\rm BLR}&=\frac{GM_{\bullet}}{f_{\rm vir}v_{\rm FWHM}^2}\,\notag\\
    &=0.017\left(\frac{M_{\bullet}}{10^8\,M_{\odot}}\right)\left(\frac{v_{\rm FWHM}}{5000\,{\rm km\,s^{-1}}} \right)^{-2}\left(\frac{f_{\rm vir}}{1} \right)^{-1}\,{\rm pc}\,. \label{eq_BLR_radius}
\end{align}
The BLR distance can also be expressed in terms of gravitational radii,
\begin{align}
    \frac{r_{\rm BLR}}{r_{\rm g}}&=f_{\rm vir}^{-1}\left(\frac{c}{v_{\rm FWHM}} \right)^2\,\notag\\
    &=3600 f_{\rm vir}^{-1}\left(\frac{v_{\rm FWHM}}{5000\,{\rm km\,s^{-1}}} \right)^{-2}\,.\label{eq_radius_BLR_rg}
\end{align}
Since according to \citet{2018NatAs...2...63M} the virial factor $f_{\rm vir}$ is inversely proportional to the line FWHM, the dependence of $r_{\rm BLR}/r_{\rm g}$ on $v_{\rm FWHM}$ is expected to be formally weaker. However, for the broad component of H$\beta$ line, the virial factor dependency on the line FWHM is \citep{2018NatAs...2...63M},
\begin{equation}
    f_{\rm vir}^{H\beta}=\left(\frac{v_{\rm FWHM}}{4550\pm 1000\,{\rm km\,s^{-1}}} \right)^{-1.17\pm 0.11}\,,
    \label{eq_fvir_FWHM_restrepo}
\end{equation}
which for FWHM close to $5000\,{\rm km\,s^{-1}}$ yields $f_{\rm vir}\approx 1$. We note that the typical BLR region length-scale given by Eqs.~\eqref{eq_BLR_radius} and \eqref{eq_radius_BLR_rg} is comparable to the S cluster size in the Galactic center \citep{1996Natur.383..415E,1997MNRAS.284..576E}, see also Subsection~\ref{galactic_center} for further details.

The orbital timescale can be expressed as, under the assumption of a circular motion,
\begin{align}
    \tau_{\rm orb}&=\frac{2\pi GM_{\bullet}}{v_{\rm FWHM}^3}\,\notag\\
    &=21.1 \left(\frac{M_{\bullet}}{10^8\,M_{\odot}} \right)\left(\frac{v_{\rm FWHM}}{5000\,{\rm km\,s^{-1}}} \right)^{-3}\,{\rm yr}\,.\label{eq_orbital_time_BLR}
\end{align}
The relative velocity of BLR clouds with respect to the intercloud medium $v_{\rm rel}$ is essentially between zero (clouds comoving with the dynamical hot medium) and $v_{\rm FWHM}$ (the hot medium is stationary). In case we assume the stationary hot medium, the BLR clouds move supersonically, i.e. $M=v_{\rm rel}/c_{\rm s}>1$, within the radius,
\begin{align}
    r_{M>1}&<\frac{GM_{\bullet}\mu m_{\rm H}}{k_{\rm B}T_{\rm h}}\,\notag\\
    &<2.6\,\left(\frac{M_{\bullet}}{10^8\,M_{\odot}} \right)\left(\frac{T_{\rm h}}{10^7\,{\rm K}} \right)^{-1}{\rm pc}\,,\label{eq_supersonic_region}
\end{align}
which is essentially the whole BLR region, since the light-travel time in the rest frame for the maximum $r_{M>1}$ is $\tau_{c}=r_{M>1}/c\sim 3100$ days, which according to the radius-luminosity relation for the broad H$\beta$ line \citep{2013ApJ...767..149B} would correspond to the very large monochromatic luminosity at 5100\AA,.
\begin{align}
    \log\left(\frac{\tau}{1\,\text{lt-day}}\right) &= 1.527^{+0.031}_{-0.031}+0.533^{+0.035}_{-0.033}\log{\left(\frac{L_{5100}}{10^{44}\,{\rm erg\,s^{-1}}}\right)}\,\notag\\
    L_{5100}&=10^{44+(\log{\tau_{\rm BLR}}-1.527)/0.533}\,{\rm erg\,s^{-1}}=10^{47.69\,{\rm erg\,s^{-1}}}\,.\label{eq_RL_luminosity5100}
\end{align}
The bolometric luminosity can be estimated from $L_{5100}$ using the bolometric correction, $L_{\rm bol}=\text{BC}\times L_{5100}$. According to \citet{2019MNRAS.488.5185N}, the bolometric correction factor is a power-law function of the observed luminosity based on thin-disk calculations and X-ray properties of AGN,
\begin{equation}
    \text{BC}=40\times \left(\frac{L_{5100}}{10^{42}\,{\rm erg\,s^{-1}}} \right)^{-0.2}\approx 2.9\,,
    \label{eq_netzer_bc_corrections}
\end{equation}
which is estimated for $L_{5100}=10^{47.69}\,{\rm erg\,s^{-1}}$. The bolometric luminosity then is $L_{\rm bol}=2.9\times 10^{47.69}\,{\rm erg\,s^{-1}}=10^{48.15}\,{\rm erg\,s^{-1}}$. The Eddington ratio can be estimated using Eq.~\eqref{eq_Eddington_luminosity},
\begin{equation}
    \lambda_{\rm Edd}=\frac{L_{\rm bol}}{L_{\rm Edd}}\approx 112\,,
\end{equation}
which is two orders of magnitude above the Eddington limit. In this sense, the limitation for the supersonic motion of BLR clouds, expressed by the relation~\eqref{eq_supersonic_region}, is met by a majority of AGN unless the hot intercloud is dynamic and partially comoving with the BLR clouds.

The shock propagates through the BLR cloud on the dynamical timescale, see Eq.~\eqref{eq_sound_timescale}, which is much shorter than the orbital timescale. Therefore the BLR cloud is shaped by stripping off the material that is located at the cloud-medium border, which results in the formation of the tail behind the denser head. This is also consistent with the detection of comet-shaped absorbers of the X-ray emission along the line of sight \citep{2010A&A...517A..47M}. 

Since BLR clouds are confined by the hot intercloud medium, their evolution is affected strongly by the heat conduction. One can adopt in the first approximation the results of \citet{1977ApJ...211..135C}, who studied the effect of heat conduction on spherical clouds embedded in hot plasma. In the diffusion approximation, the heat flux due to the heat conduction can be expressed as \citep{2007A&A...472..141V},
\begin{equation}
    \mathbf{q_{diff}}=-\kappa_{\rm h} \mathbf{\nabla T}\,,
    \label{eq_heat_flux}
\end{equation}
where the conductivity of the hot phase can be calculated as
\begin{equation}
    \kappa_{\rm h}=\frac{1.84\times 10^{-5} T_{\rm h}^{5/2}}{\ln{\Psi}}\,{\rm erg\,s^{-1}K^{-1}cm^{-1}}\,,
\end{equation}
where the Coulomb logarithm can be estimated as follows,
\begin{equation}
    \ln{\Psi}=29.7+\ln{\left(\frac{T_{\rm e,6}}{\sqrt{n_{\rm e}}} \right)}\,,
\end{equation}
where the electron density $n_{\rm e}$ is expressed in ${\rm cm^{-3}}$ and the electron temperature $T_{\rm e,6}$ is scaled to $10^6\,{\rm K}$.

\citet{1977ApJ...211..135C} derive a classical mass-loss rate due to heat conduction, in other words the evaporation rate,
\begin{equation}
    \dot{m}_{\rm evap}=-\frac{16 \pi \mu m_{\rm p} \kappa_{\rm h} R_{\rm cl}}{25k_{\rm B}}\,,
    \label{eq_evaporation_rate}
\end{equation}
from which we can derive the evaporation timescale under the assumption of a constant cloud density, $m_{\rm cl}=4/3\pi R_{\rm cl}^3\rho_{\rm cl}$. Then the evaporation rate can also be expressed as $\dot{m}_{\rm evap}=4\pi R_{\rm cl}^2 \dot{R}_{\rm cl} \mu n_{\rm cl}m_{\rm p}$. After plugging in Eq.~\eqref{eq_evaporation_rate}, specifically on the left side, we obtain the expression for the evaporation timescale,
\begin{equation}
    \tau_{\rm evap}=\frac{25 k_{\rm B}n_{\rm cl} R_{\rm cl}^2}{8\kappa_{\rm h}}\,,
    \label{eq_evaporation_timescale}
\end{equation}
which depends on the cloud density, a square of its radius, and the temperature as well as the density of the surrounding hot medium.
The conductivity for the hot medium is $\kappa_{\rm h}\approx 2.69 \times 10^{11}\,{\rm erg\,s^{-1}\,K^{-1}\,cm^{-1}}$ under the assumption of the pressure equilibrium. Then the evaporation timescale of the BLR clouds can be estimated as follows,
\begin{equation}
  \tau_{\rm evap}\simeq 51\,\left(\frac{n_{\rm cl}}{10^{12}\,{\rm cm^{-3}}}\right)\left(\frac{R_{\rm cl}}{10^{12}\,{\rm cm}}\right)^2\left(\frac{\kappa_{\rm h}}{2.69\times 10^{11}\,{\rm erg\,s^{-1}\,K^{-1}\,cm^{-1}}} \right)^{-1}\,{\rm yr}\,,
  \label{eq_evap_timescale}
\end{equation}
which is significantly larger than the hydrodynamical crushing timescale that is of the order of days, see Eq.~\eqref{eq_crushing_timescale}.

The quasi-Keplerian motion of clouds in the potential of the SMBH can lead to the formation of broad emission lines with the FWHM of a few $1000\,{\rm km\,s^{-1}}$. The smooth line profile implies the high abundance of such dense clouds. The physical origin of the BLR line-emitting gas has remained largely unknown despite a large number of theoretical, observational, and computational studies. The BLR clumps may generally be formed in the inflow, in-situ in the disc via the gravitational instability, as a part of the irradiated accretion disk surface, or in the outflow. Reverberation mapping measurements imply that the Keplerian motion dominates but inflow or outflow signatures are also seen \citep{grier2012}. A more complicated picture involving all of these processes is also plausible \citep{2011ApJ...739....3W,2012ApJ...746..137W}. These scenarios are separately discussed below:
\begin{itemize}
\item \textit{Inflow models.} Accretion onto the central black hole can, perhaps partially, proceed in the form of the BLR infall. Circumnuclear material/interstellar medium contains colder clouds embedded in a hotter intercloud medium. Colder material is more likely to accrete since its thermal velocity of gas particles is much lower that the escape velocity, and the infall will happen in the absence of the angular momentum barrier. Numerical simulations of the multi-phase medium show the infall of the cold clouds and the outflow of the hot X-ray emitting gas in the gravitational field \citep{barai2012,elvis2017}. Convincing observational arguments for the inflow has been found by \citet{Hu2008}, and the consequences of this discovery to the BLR structure were discussed by \citet{2009ApJ...707L..82F}. Since the force multiplier in the Fe II emitting region is very large, the infall is possible only if most of the clouds are actually shielded from the central source. We thus observe predominantly non-irradiated faces of the optically thick clouds, and this solves the long-standing problem of modelling Fe II to H$\beta$ ratio. On the other hand, observational determination of Fe II shifts is very complex and the shift claim might still be the artefact of the data analysis method \citep{bon2018}. The analysis of CIV line emission in NGC 5548 has indicated a large 2D or 3D random motion on the top of the radial infall, see \citep{1996ApJ...463..144D}. In the context of inflow models, \citet{2017ApJ...847...56E} presented the model of the ``quasar rain'', which involves a hot disk outflow, from which colder clumps can condense via the thermal instability \citep{1981ApJ...249..422K}. These heavier and cold clumps than fall back towards the disk since they are not supported enough by the radiation pressure.
\item \textit{In-situ cloud formation via gravitational instability -- affect of the nuclear star cluster.} AGN typically host massive and extended accretion disks that at larger distances are gravitational unstable due to the decreasing temperature and the increasing surface density. Hence, eventually the Toomre stability criterion drops below one and the disk fragments. At intermediate scales it can be supported against fragmentation by the turbulence, spiral waves or the magnetic field. The fragmenting self-gravitating disks have extensively been studied by several authors, see \citet{1999A&A...344..433C}, \citet{2008A&A...477..419C} and references therein. More recently, \citet{2011ApJ...739....3W} studied the complex processes taking place in the fragmenting star-forming AGN disks. In particular, supernova explosions lead to the ejection of gas above the disk plane. They can also contribute to the high, supersolar metallicity observed in quasars. There can generally be several cycles of BLR formation depending on starburst phases \citep{2012ApJ...746..137W}. In the broader text, once the nuclear star cluster has formed and a fraction of stars has evolved over the billions of years, the stars as such could give rise to the BLR clouds. \citet{1997MNRAS.284..967A} and \citet{1997MNRAS.285..891A} proposed the ``bloated-stars'' model where the broad lines could be produced by the collective emission of the extended envelopes of several $10^4$ stars located deep inside the sphere of influence of the SMBH. This model has, however, failed to explain some of the BLR characteristics, e.g. line profiles and the flattened geometry now detected by the GRAVITY instrument. For the evolved stellar population, we expect the orbital randomization by stellar resonant and non-resonant relaxation processes over the course of billions of years.      
\item \textit{Irradiated disk surface.} In the context of the HIL and LIL components of the BLR \citep{1988MNRAS.232..539C}, some LILs could be formed within the accretion disk on its surface since their velocity field is mainly given by the orbital Keplerian component. This is mainly the case of double-peak low-ionization lines \citep{2019MNRAS.488L...1B,2003ApJ...599..886E}. The velocity field of BLR clouds has the orbital Keplerian component as well as the turbulent component \citep{2013A&A...549A.100K}, which becomes more prominent for high-Eddington sources. The double-peak sources are generally associated with lower Eddington sources, where the orbital velocity component is dominant. From stationary models of the BLR associated with the accretion disk, \citet{2018MNRAS.474.1970B} showed the formation of the puffed-up stationary dusty atmosphere or rather ``dust-inflated accretion disk" within the range where the dust opacity dominates. \citet{2005MNRAS.359..545N} proposed that the accretion disc warping and precession due to an axisymmetric potential created by the previously formed stellar disk can be held responsible for lifting off some material above the original disk plane. This material can then obscure the AGN and contribute to the inclination-dependent emission of AGN.
\item \textit{Outflow models.} Outflow scenario is the most frequently adopted hypothesis, since a strong blueshift seen in high ionization UV lines, like for example CIV, is observed in numerous quasars \citep[e.g.][]{brotherton1994}. This outflow has been interpreted as an accretion disk wind \citet{1995ApJ...451..498M}, or bi-conical ejection along the symmetry axis \citep{zheng1990}. The bi-conical structure was motivated by observations of the elongated Narrow Line Region (NLR), well seen for example in [OIII] emission, and the NLR has been hypothesized to be the extension of the outflowing BLR, see e.g. \citep{2021MNRAS.503.3145B}. 

The common prediction of the outflow models of the BLR is that the broad lines are present only as long as the classical cold accretion disk extends all the way to the BLR launching radius or rather the range of radii. In other words, the BLR lines are present only as long as the Big Blue Bump is present. For sources with the low Eddington ratio, such as also true Seyfert~2 galaxies (no broad lines, no obscuration), the broad lines disappear. This is also the case of low-luminosity Sgr~A* and M87 where the accretion flow is represented by hot Advection-Dominated Accretion Flow or ADAF, see \citet{2014ARA&A..52..529Y} for a review. Hot flows are accompanied by radially dependent outflows but these are fully ionized and not capable of producing broad lines. However, for at least one candidate true Seyfert~2 (NGC3147) with a low accretion rate of $\lambda_{\rm Edd}\sim 10^{-4}$, a double peak broad H$\alpha$ with the base width of as much as $\sim 27\,000\,{\rm km\,s^{-1}}$ was detected, see \citet{2019MNRAS.488L...1B}. This finding to some extent questions the existence of true Seyfert 2 sources and the presence of the classical cold disc below $\sim 100$ gravitation radii for such low accretion rates is also beyond theoretical expectations. Below we list basic outflow models that are distinguished based on the driving force of the outflow: magnetically driven, thermally driven, and radiation-pressure driven winds. For a general overview of the disc-outflow mechanisms, see also \citet{2007ASPC..373..267P} and \citet{2019OAst...28..200C}. The formation, structure, and the evolution of the outflow is generally given by the equation of radiation magnetohydrodynamics \citep{2007ASPC..373..267P}
\begin{equation}
    \rho \frac{\mathrm{D\mathbf{v}}}{\mathrm{D}t}+\rho\Phi=-\nabla P+\frac{1}{4\pi}(\nabla \times \mathbf{B})\times \mathbf{B}+\rho\mathbf{F}_{\rm rad}\,,
    \label{eq_magnetohyd}
\end{equation}
where $\rho$, $\mathbf{v}$, and $P$ are the fluid density, velocity, and pressure. Furthermore, $\Phi$ denotes the gravitational potential, $\mathbf{B}$ stands for the magnetic field, and $\mathbf{F}_{\rm rad}$ is the radiation force per unit mass. To generate outflow, at least some of the terms on the right need to overcome the gravitational force, which can lead to the true outflow (escaping gas) or also a failed wind, for which some of the material returns back to the accretion disk. The mechanisms to generate the outflow are represented on the right-hand side: (a) the gas pressure gradient, (b) Lorentz force, and (c) the radiation-pressure force due to the radiation-pressure gradient. We discuss these mechanisms in more detail below.
\begin{enumerate}
    \item \textit{Thermally driven winds.} Spontaneous thermal outflows can be initiated when the outer parts of the disk are heated to approximately the virial temperature. The virial temperature can be derived assuming the virial equilibrium, see Eq.~\eqref{eq_virial}, for the rotating gas with the thermal kinetic energy of $K_{\rm therm}=3/2 N k_{\rm B} T_{\rm vir}$, where we assumed the ideal monoatomic gas with $N=M_{\rm gas}/(\mu m_{\rm p})$ particles. Then the virial temperature can be approximated as,
    \begin{align}
        T_{\rm vir}\approx \frac{GM_{\bullet}\mu m_{\rm p}}{3 k_{\rm B} r}=\frac{\mu m_{\rm p}}{3k_{\rm B}}v_{\rm gas}^2=\frac{\mu m_{\rm p}c^2}{3k_{\rm B}\hat{r}}\,,
        \label{eq_virial_temp}
    \end{align}
    where $\hat{r}$ is expressed in gravitational radii. For $\hat{r}=10^4$, we obtain $T_{\rm vir}\sim 1.8 \times 10^8\,{\rm K}$ or $k_{\rm B}T_{\rm vir}\sim 16\,{\rm keV}$, which corresponds to X-ray emitting plasma. The colder outer parts are heated by the warmer inner parts of the disc, mainly by X-ray irradiation. Specifically the Compton heating can bring the diluted cooler gas in the upper parts of the disc to the inverse Compton limit of $\sim 10^8\,{\rm K}$, see \citet{1983ApJ...271...70B} and \citet{1983ApJ...271...89B} for details. This balance is achieved by the Compton heating by hard photons and inverse Compton cooling by softer photons. In addition to the inverse Compton heating/cooling balance, a viscous dissipation close to the disk surface could be another source of heating \citep{1989MNRAS.236..843C}. Depending on whether the gas thermal speed exceeds the local escape speed from the disc, which is a function of radius, two solutions are possible -- in case the escape speed is not exceeded, which is more likely in the inner disk parts with large escape velocities, a stationary puffed-up hot atmosphere forms above the disc (corona). In the opposite case, a thermally-driven wind is launched when the thermal gas speed exceeds the escape velocity, which is more likely in the outer disk parts. This wind is almost fully ionized and hence it is not efficient in producing lines. In some cases, if launched from the outer BLR or the dusty torus, it could be responsible for the observed low-velocity intervening warm absorbers, see \citet{2005ApJ...625...95C}. Thermally driven winds operate at high temperatures of $\sim 10^7-10^8\,{\rm K}$; for lower temperatures, the other terms in Eq.~\eqref{eq_magnetohyd} dominate.
    \item \textit{Magnetically driven winds.} In case the accretion disc is threaded by a large scale magnetic field with a significant poloidal component, a magnetically-driven wind can be launched via the magneto-centrifugal force \citep{1982MNRAS.199..883B}. For this model to work, the magnetic field needs to be transported inwards from the circumnuclear medium further out. Models of magnetically-driven winds predict the concave shape of flow lines and the gas escape speeds can achieve large velocities. The magnetically driven wind is also employed to explain the high-ionization UltraFast Outflows (UFOs) that are detected as absorption features in the X-ray spectra \citep{2015ApJ...805...17F}. \citet{2015ApJ...805...17F} constrain from the X-ray spectra the launching radius of UFOs at $\sim 200\,r_{\rm ISCO}$, where $r_{\rm ISCO}$ is the innermost stable circular orbit.
    \item \textit{Radiation-pressure driven winds.} There are two potential sources of the opacity that we discuss further below -- (i) absorption lines of various ionized species in the gas, for which the force multiplier $M$ that is introduced as the ratio of the total radiation pressure to the radiation pressure attributed to the Thomson scattering, can reach values of a few hundred, and (ii) the dust opacity, which is relevant for the distance where the disk temperature drops below the dust sublimation temperature and the dust can coexist with the gas. Case (i) -- line-driven wind-- is also responsible for driving fast and diluted stellar winds of hot OB stars, which is also known as the CAK theory or Castor--Abbott--Klein, see \citet{1975ApJ...195..157C}, while case (ii) -- dust-driven wind-- is relevant for slower and more massive outflows of late-type evolved stars, see \citet{1995SSRv...73..211S} for a review.
    \begin{figure}
    \centering
    \includegraphics[width=0.7\textwidth]{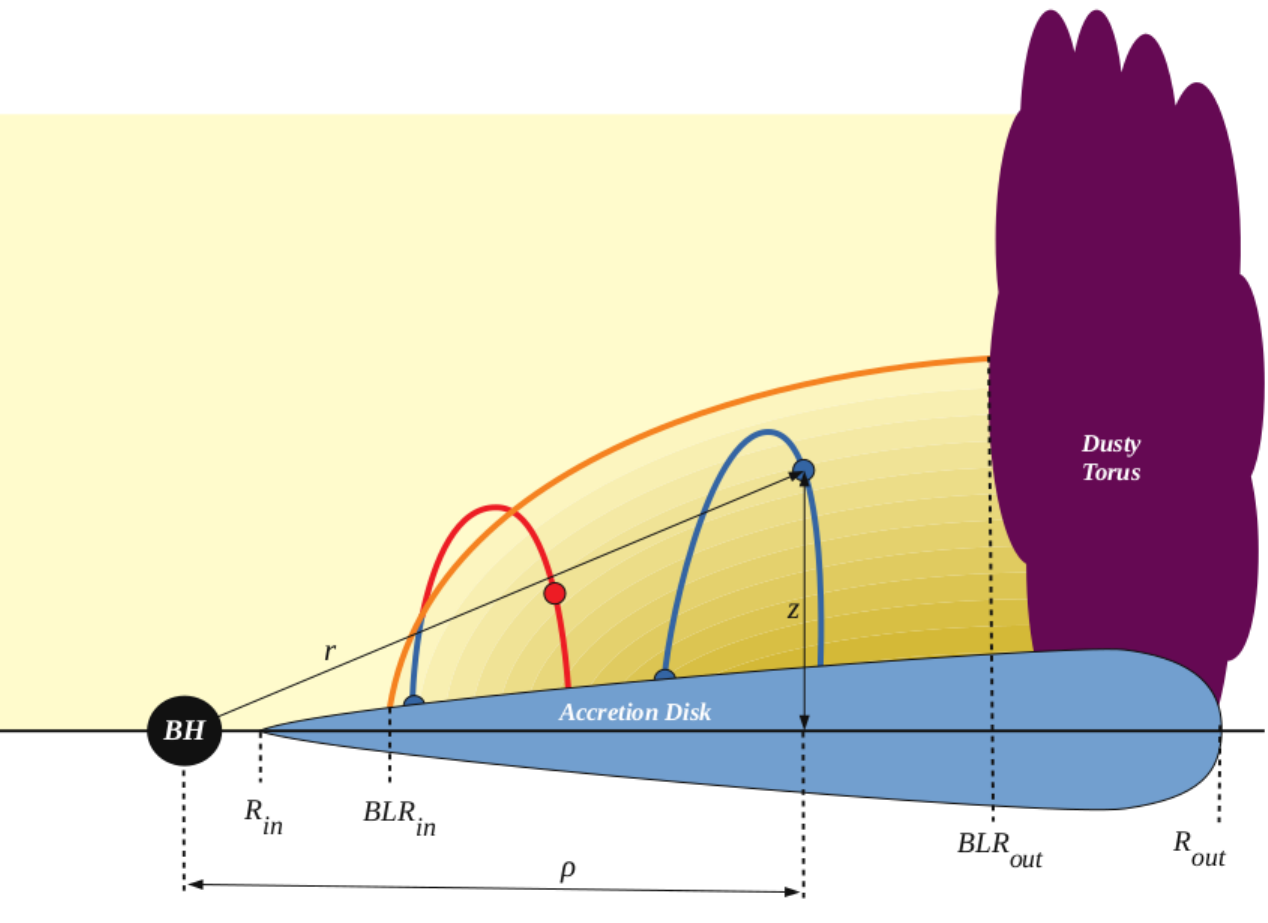}
    \caption{The illustration of the Failed Radiatively Driven Dusty Outflow or \textit{FRADO} model. In the disk zone where dust can form, i.e. where the effective disk temperature drops below the dust sublimation temperature, the dust opacity is high enough so that gaseous-dusty clumps are lifted off the disk plane due to the scattering of disk photons on dust grains. Once these BLR clouds are directly irradiated by external UV and X-ray photons, dust evaporates and clouds fall back to the disk since they are not supported by radiation pressure anymore. The dominant velocity component of the FRADO clouds is still orbital, i.e. Keplerian, but there is also an extra turbulent component due to the outflow-inflow of the BLR clouds. See also \citet{2011A&A...525L...8C} and \citet{2021arXiv210200336N} for details.}
    \label{fig_FRADO}
\end{figure}
    \begin{figure}
        \centering
        \includegraphics[width=0.9\textwidth]{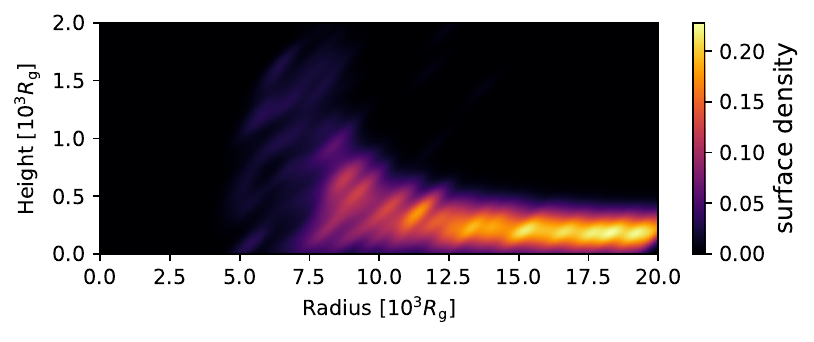}
        \caption{The instantaneous surface-density distribution of the BLR clouds based on the dynamical calculations of the FRADO model, see \citet{2021arXiv210200336N} for details.}
        \label{fig_FRADO_surface_plot}
    \end{figure}
    \vspace{\baselineskip}
    \textit{Line-driven winds.} The CAK-like mechanism was first applied to the AGN winds by \citet{1995ApJ...451..498M}. In the follow-up paper by \citet{1996ApJ...466..704C}, line profiles were calculated -- both double-peak and single-peak lines could be produced by line-driven disk outflows. The inclusion of the large radial shear in the line-driven wind was necessary to reproduce the single-peak line-profile as well as the response (transfer) function of lines consistent with observations, including their red- and blue-wing responses, see also \citet{1997ApJ...474...91M}. \citet{1995ApJ...454L.105M} also applied the disk wind driven by UV resonance lines to interpret broad absorption features in spectra. In recent years, there has been a large progress in hydrodynamical models of AGN winds. \citet{2016ApJ...827...53W} used these simulations to calculate echo images, line profiles and response functions of BLR lines for different viewing angles. The wind hydrodynamics and the ionization state were not calculated simultaneously, instead the radiative transfer was performed a posteriori. A general consistency is found with observations as well as previous BLR model calculations. Previously, \citet{2000ApJ...543..686P} and \citet{2004ApJ...616..688P} found from hydrodynamical simulations of AGN winds that the outflowing gas is virialized across a large range of radii, which allows to reliably measure black-hole masses using the reverberation mapping. In addition, the outflowing gas can shield itself from the central X-ray irradiation and it can be pushed up and accelerated above the disc plane just by the photons from the disk photosphere. Two wind modes are found -- a fast narrow-angle outflow closer to the SMBH and a slower wind close to the disk plane at larger radii.
    
\vspace{\baselineskip}
   \textit{Dust-driven winds.} A dust-driven outflow as an origin for the BLR was proposed later than line-driven winds, specifically in the letter by \citet{2011A&A...525L...8C}. The basic set-up is similar, i.e. the disk radiation field is capable of pushing off some gas above its plane. However, in case of dust-driven winds, the scattering of disk photons on dust grains is the main source of opacity. The gaseous-dusty clumps are accelerated upwards where they are exposed to harder UV and X-ray photons from the inner parts of the disk. The dust in the clumps evaporates and since the radiation force is diminished, the clouds fall back onto the disk surface. For this reason, the dust-driven wind as the origin for the BLR is also referred to as Failed Radiatively Accelerated Dusty Outflow or FRADO for short, see also \citet{2017ApJ...846..154C} and \citet{2021arXiv210200336N} for details. In this model, the inner radius of the BLR coincides with the thin disk radius, where the effective temperature drops below the dust sublimation temperature, which is in the range of $\sim 1000-1500\,{\rm K}$. For different sources, this radius is a function of monochromatic luminosity only, and does not depend on the SMBH mass or the accretion rate, which provides an elegant qualitative and quantitative explanation for the RL relation. The radius where the dust can form within the disk is still closer than the inner radius of the dusty molecular torus, which is set up by the direct illumination by UV photons, see Eq.~\eqref{eq_sublimation_radius}. The illustration of the FRADO model is in Fig.~\ref{fig_FRADO}. The dominant velocity component of the FRADO clouds is still orbital, i.e. Keplerian, following the disk rotation, but there is an extra turbulent component due to the outflow-inflow of clouds, see also Fig.~\ref{fig_BLR_HE0413} for a spatial 3D view of the model. Here we draw the attention to the predicted appearance of the BLR, which is strikingly different from the usually assumed appearance of a quasi-spherical structure of cloudlets or a flattened system of orbiting gas clouds. As calculated by \citet{2021arXiv210200336N}, the instantaneous structure of the BLR towards the inner radius is thicker due to a reached higher altitude of the clouds, with some of them forming an outflow. Towards a larger distance, the BLR vertical profile is progressively narrower since the clouds reach only a relatively small altitude above the disk surface before falling back onto the disk at roughly the same radius. This is captured by the Gaussian-smoothed surface distribution of the BLR clouds in Fig.~\ref{fig_FRADO_surface_plot}.  
\end{enumerate}

\end{itemize}

    \label{subsec_circumnuclear}
   \subsection{AGN vs.\ inactive galaxies: the context of Milky Way's center}
   \label{galactic_center}
    \label{SgrA}

About $2.5$ billions of years after the Big Bang (redshift $\sim 2-3$), the Universe went through the period of `quasar era', when most large galaxies were accreting at rates close to the Eddington limit, see Eq.~\eqref{eq_Eddington_luminosity}. Quasars were 10\,000 times more common than in the nearby Universe. In other words, only about 1 quasar in 500 galaxies can be detected today.  In the local Universe, quasars and even medium-luminosity Seyfert-like sources are rare and $\gtrsim 90\%$ of galaxies are inactive or show only little activity. Since massive black holes cannot be destroyed or evaporate during the Universe lifetime, the argument is that massive black holes of $\sim 10^6-10^{10}\,M_{\odot}$ powering quasars are inactive or dormant in nuclei of most current galaxies, mainly massive ones \citep[`Soltan argument',][]{1982MNRAS.200..115S}.  

The pioneering works of \citet{1969Natur.223..690L} and \citet{1971MNRAS.152..461L} attributed large quasar luminosities to the conversion of the potential energy of infalling material into radiation -- they pointed out the large efficiency of the accretion mechanism in producing thermal and radiative energy. The inclusion of the SMBHs into the centre of most large galaxies led to solving many questions related to the quasar activity, mainly they could explain large luminosities and the jet generation. However, a more direct dynamical evidence for the presence of massive black holes was still missing. For such an experiment, nearby nuclei are much better suited because of instrumental resolution limits. In other words, it was necessary to look for the evidence in `dead quasars' of the local Universe where the activity of black holes is none or little at best. Therefore the previous discussion concerning the accretion in AGN manifested by the accretion disk and the jet and their radiative signatures has little relevance for inactive galaxies. In addition, broad-line region clouds disappear for low-luminous galactic nuclei with the bolometric luminosity of $L_{\rm bol}\leq 5\times 10^{39}\,(M_{\bullet}/10^7\,M_{\odot})^{2/3}\,{\rm erg\,s^{-1}}$ \citep{2009ApJ...701L..91E} as expected based on the disc-wind scenario of the BLR formation. With the Eddington ratio $\lambda_{\rm Edd}\sim 4\times 10^{-6}$, the accretion flow is hot, advection-dominated and radiatively inefficient \citep{2014ARA&A..52..529Y}. Therefore the reverberation-mapping technique and the associated virial-mass determination is not feasible for quiescent galaxies. This is also the case of both Sgr~A* and M87.

Massive black holes in galaxies are not isolated. Like observed for the center of the Milky Way, SMBHs are embedded in the dense stellar cluster, see \citet{1998AJ....116.2263L,2014CQGra..31x4007S,2016ApJ...821...44F}. The cusp-like stellar cluster surrounding the SMBH is also expected from a pure theoretical treatment, see \citet{1976ApJ...209..214B,1977ApJ...216..883B} and \citet{1977ApJ...217..287Y}. Therefore, the best dynamical probe of the black hole presence are bound stellar orbits inside the sphere of gravitational influence of SMBHs. Stars have significantly larger ratios of their mass to their cross-section than orbiting gas clouds. Therefore they are not sensitive to non-gravitational forces, which makes the potential mass determination reliable. However, the length-scale on which the potential of the SMBH prevails over the stellar cluster potential is rather small in comparison with the typical galaxy size. One can derive the radius of the influence of the SMBH in a galactic bulge by considering a typical Keplerian velocity at the distance $r$ from the SMBH of mass $M_{\bullet}$, $v_{\rm K}=\sqrt{GM_{\bullet}/r}$. By comparing the Keplerian velocity to the average one-of-sight velocity dispersion of stars $\sigma_{\star}$, one arrives to the critical length-scale, below which the gravitational potential is dominated by the point mass of $M_{\bullet}$,

\begin{equation}
  r\lesssim R_{\rm inf}\approx \frac{GM_{\bullet}}{\sigma_{\star}^2}\,.
\end{equation}

If we consider for the numerical evaluation the black hole mass of $M_{\bullet}=10^7\,M_{\odot}$ and the average velocity dispersion of stars in the bulge (on large scales), $\sigma_{\star}=100\,{\rm km\,s^{-1}}$, the influence radius typically extends on the scales from a fraction of a parsec to several parsecs,

\begin{equation}
  r\lesssim R_{\rm inf}\approx 4.3\left(\frac{M_{\bullet}}{10^7\,M_{\odot}}\right)\left(\frac{\sigma_{\star}}{100\,{\rm km\,s^{-1}}}\right)^{-2}\,{\rm pc}\,.
\end{equation}

Therefore it depends on the angular resolution of instruments in order to resolve the sphere of influence in our Galaxy as well as in other galaxies. Considering the eight-meter class telescopes in the infrared bands, which are frequently used for the imaging of the Galactic centre, a typical angular resolution given by the diffraction limit is $\theta\approx \lambda/D\approx 57\,{\rm mas}$, which was evaluated for the wavelength of $\lambda=2.2\,{\rm \mu m}$ (infrared $K$ band) and the telescope aperture of $D=8\,{\rm m}$, such as the aperture of unit telescopes at the European Southern Observatory in Chile. 

The angular size of the influence radius, when scaled to the distance of $1\,{\rm Mpc}$, reads
\begin{equation}
   \theta_{\rm inf}=887.2\,\left(\frac{M_{\bullet}}{10^7\,M_{\odot}}\right)\left(\frac{\sigma_{\star}}{100\,{\rm km\,s^{-1}}}\right)^{-2}\left(\frac{D}{1\,{\rm Mpc}}\right)^{-1}\,{\rm mas}\,,
   \label{eq_theta_inf}
\end{equation}
which implies that the Keplerian rise inside nuclear star clusters can be resolved to the distance of $D<17\,{\rm Mpc}$ for the SMBH mass of $10^7\,M_{\odot}$. This corresponds to the redshift range of $z\lesssim 0.004$. This limits the studies of the sphere of influence of the SMBH to nearby galaxies depending also on the mass of the SMBH, which directly influences the length-scale of the sphere of influence. For the most massive black holes of $M_{\bullet}\sim 10^{10}\,M_{\odot}$, the maximum distance, for which the influence radius can be resolved out, is $D\lesssim 15.6\,{\rm Gpc}$ that corresponds to the redshift of $z\lesssim 2$.

For the Galactic centre SMBH and the averaged one-dimensional stellar velocity dispersion of $\sigma_{\star} \approx 100\,{\rm km\,s^{-1}}$, we get \citep[see e.g.][]{2013degn.book.....M,2015MNRAS.453..775G},
\begin{equation}
  R_{\rm inf}\simeq GM_{\bullet}/\sigma_{\star}^2=1.7 \left(\frac{M_{\bullet}}{4\times 10^6\,M_{\odot}} \right) \left(\frac{\sigma_{\star}}{100\,{\rm km\,s^{-1}}} \right)^{-2}\,{\rm pc}\,,
  \label{eq_influence_radius}
\end{equation}
which corresponds to the angular scale of $\theta_{\rm inf}\approx 44''$. With eight-meter class telescopes in the NIR-bands, one can thus map the potential up to as small scales as of $\sim 52\,{\rm mas}\approx 2.1\,{\rm mpc}$ or $\sim 11\,000$ gravitational radii. With the mass of $\sim 4\times 10^6\,M_{\odot}$, the compact radio source Sgr~A* is at the lower end of the observed black-hole mass spectrum \citep{2013ARA&A..51..511K}. This is in agreement with the $M_{\bullet}$--$\sigma_{\star}$ relation, where the SMBH mass is related to the bulge stellar velocity dispersion via the power-law relation, $M_{\bullet}\propto \sigma_{\star}^{\beta}$ where $\beta\sim 4-5$ \citep{2000ApJ...539L...9F}. In addition, the bulge mass of the Milky Way is lower, which is in line with the SMBH mass--bulge mass ``Magorrian'' relation $M_{\bullet}=\epsilon_{\bullet}M_{\rm bulge}$, where $\epsilon_{\bullet}=(1-2)\times 10^{-3}$ \citep{1998AJ....115.2285M}. For an overview of the SMBH mass -- the bulge characteristics correlations, see \citet{2016ASSL..418..263G} and \citet{2016arXiv161107872B} for reviews. The Galactic center SMBH is clearly the nearest SMBH, closer by a factor of $\sim 100$ than the SMBH in the Andromeda galaxy M31 and closer by a factor of $\sim 2000$ than SMBHs in the nearest galaxy cluster in the Virgo constellation. Therefore, Sgr~A* has played an irreplaceable role in mapping out the gravitational potential as well as in studying mutual interactions among different components (stars, molecular, neutral, ionized gas, dust, SMBH) in the vicinity of the SMBH.

It is only inside $R_{\rm inf}$ that one can expect to measure the Keplerian rise in orbital velocities of stars, $v_{\star} \propto r^{-1/2}$. The Galactic centre played a key role in confirming the presence of the black hole since the Keplerian rise was not only detected but for several closest stars their orbits were determined. The first proper motions of stars were detected by \citet{1996Natur.383..415E,1997MNRAS.284..576E,1998ApJ...509..678G}, who also introduced the notation of fastest stars closest to the compact radio source Sgr~A* beginning with ``S'' (representing the infrared point source). Subsequently, the measured proper motions and relative velocities have been used to derive accelerations and stellar orbits for several S stars \citep{2002MNRAS.331..917E,2002Natur.419..694S,2008ApJ...689.1044G,2009ApJ...692.1075G,2016ApJ...830...17B,2017ApJ...837...30G}. In the future, this will allow to test the General Theory of Relativity based on the determined orbits of S stars (mainly using the brightest S star S2) and potentially stars that will be found on even tighter orbits \citep{2017FoPh...47..553E,2017ApJ...845...22P,2018MNRAS.476.3600W,2018arXiv180411014Z}. In fact, the NIR-observations of the S2 star moving around Sgr~A* on a highly elliptical orbit with the period of $\sim 16\,{\rm yr}$ led to the successful measurements of two relativistic effects -- the combined gravitational redshift and special relativistic transverse Doppler shift of $\sim 200\,{\rm km\,s^{-1}}$ \citep{2018A&A...615L..15G} based on the measurement of stellar absorption lines as well as the prograde Schwarzschild precession consistent with the general relativistic prediction of $\sim 12'$ per period \citep{2020A&A...636L...5G}, which can simply be estimated using the relation,

\begin{equation}
    \Delta \phi=\frac{6\pi G M_{\bullet}}{a_{\rm S2}(1-e_{\rm S2}^2)c^2}+\mathcal{O}(v^4/c^4)\approx 11.6'\,\text{per orbital period}\,,
    \label{eq_Schwarzschild_precession}
\end{equation}
where the semi-major axis $a_{\rm S2}$ and the orbital eccentricity $e_{\rm S2}$ for the S2 star were adopted from \citet{2018A&A...615L..15G}.

An indication of the prograde general relativistic periastron shift for the S2 star was determined earlier \citep{2017ApJ...845...22P}, being consistent with the predictions. The detection of the prograde relativistic periastron shift of the S2 star implies the negligible extended mass inside its orbit in comparison with the point-source mass of Sgr~A*. This allows to place an upper limit on the extended mass, at most $\sim 0.1\%$ of the Sgr~A* mass inside the S2 orbit, as well as on the potential second massive source within the inner arcsecond, $m_{2}\lesssim 1000\,M_{\odot}$, such as an intermediate-mass black hole \citep{2020A&A...636L...5G}. 

The observational and technical effort that resulted in the confirmation of the compactness of the central massive object of $4$ million Solar masses based on the monitoring of orbiting stars was distinguished with the Nobel Prize in Physics in 2020, whose one half was awarded jointly to Reinhard Genzel and Andrea Ghez (the other half was awarded to Roger Penrose ``for the discovery that black hole formation is a robust prediction of the general theory of relativity"). The concise summary of the development of the Galactic center high-precision observations with the focus on the near-infrared domain can be found in \citet{2021arXiv210213000G}. 

\begin{figure}[tbh!]
  \centering
  \includegraphics[width=\textwidth]{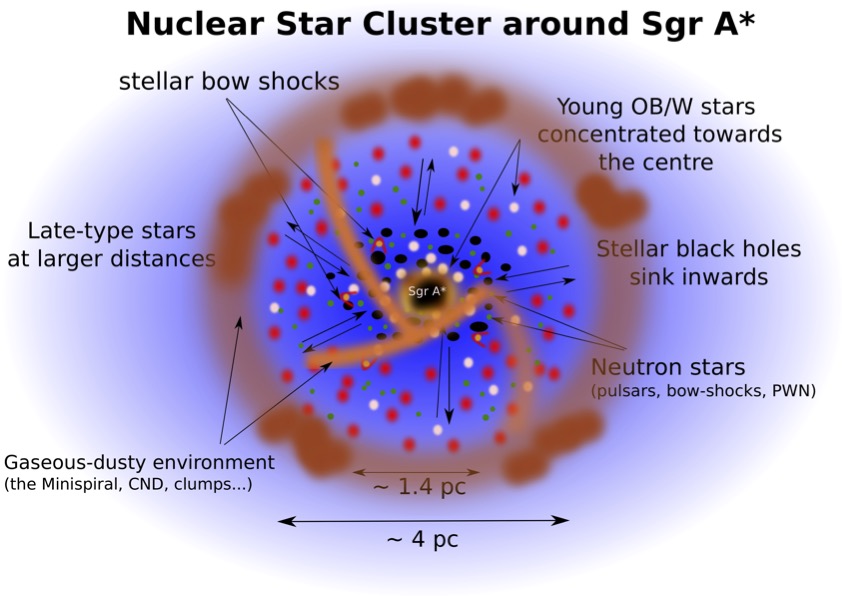}
\caption{An illustration of the multi-component stellar and gaseous-dusty environment around the compact radio source Sgr~A*. The stellar population consists of older, late-type stars whose density appears to decrease towards the centre, at least for bright red giants, forming a distinct flat core. The density of young, massive OB/Wolf-Rayet stars, on the other hand, increases towards Sgr~A*, creating a cusp. In the same region, where the Nuclear Star Cluster is located, the interstellar medium is filled with hot, ambient plasma. Denser and partially neutral minispiral arms orbit around Sgr~A*, while some of the wind-blowing stars move supersonically through them and form extended bow shocks that are more prominent at longer infrared wavelengths. The population of compact remnants is observationally unexplored in general. Its presence is, however, expected based on the history of massive star-formation in the Galactic centre \citep{1993ApJ...408..496M}. The heaviest members -- stellar black holes -- are expected to sink towards the very centre due to the dynamical friction and kinetic energy equipartition, where they might have formed a steep cusp-like dark cluster.}
  \label{img_nsc_sgra}
\end{figure}

Although the number density of stars towards the Galactic centre increases as $n_{\star}=n_0 (r/r_0)^{-\delta}$ for $\delta>0$, the total number of stars inside the radius $r$, $N_{\star}(<r)$, falls off with the deprojected radius for $0<\delta<3$ since the volume shrinks with $r^3$. The observationally determined value of the power-law slope of the stellar density is $\delta<3$, see e.g. \citet{2009A&A...502...91S} or \citet{2010RvMP...82.3121G}, with bright late-type stars forming a flat core and early-type stars forming a steeper cusp towards the centre, see Fig.~\ref{img_nsc_sgra} for an illustration. The number of stars can thus fall below one at a certain radius, which marks the volume which lacks stars or one can refer to it as a \textit{sparse region}, see also \citet{2018arXiv180411014Z} for a detailed discussion. As a consequence, it is quite implausible to detect a cluster of stars in the strong-gravity regime, i.e. below 100 Schwarzschild radii. To quantify the length-scale, on which the number of stars is expected to fall below one for a certain stellar type $T$, it is straightforward to calculate the radius $r_{1T}$ at which this is expected to happen. One can obtain the radius $r_{1T}$ from the integral $N_{\star}(<r)=\int_0^r 4\pi r^2 n_{\star}(r)\mathrm{d}r$ by setting $N_{\star}(<r)=1$  \citep[see also][]{2006ApJ...645.1152H,2018arXiv180411014Z},  

\begin{equation}
    r_{1T}=R_{\rm inf} (C_{\rm T}N_{\rm inf})^{-1/(3-\delta_{\rm T})}\,,
    \label{eq_radius_1T}
\end{equation}
where $R_{\rm inf}$ is the radius of gravitational influence, see Eqs.~\eqref{eq_influence_radius1}, \eqref{eq_influence_radius}, $N_{\rm inf}$ is the total number of main-sequence stars inside $R_{\rm inf}$, $C_{\rm T}N_{\rm inf}$ is the total number of stars of type $T$ inside  $R_{\rm inf}$ and $\delta_{\rm T}$ is the power-law slope of the stellar population of type $T$. \citet{2006ApJ...645.1152H} analyzed the inner radius of the stellar cluster for different types of objects, namely main-sequence stars (MSs), white dwarfs (WDs), neutron stars (NSs), and stellar black holes (BHs). They analyzed the properties of the nuclear star cluster and determined that there are in total $N_{\rm inf} \approx 3.4 \times 10^6$ main-sequence stars inside the influence radius. In Table~\ref{tab_inner_radii}, we list the inner radii of other stellar components of the cluster as calculated by \citet{2006ApJ...645.1152H}. According to Table~\ref{tab_inner_radii}, below a few 100 Schwarzschild radii, it is quite unlikely to detect a star that can be used for high-precision tests of general theory of relativity in the highly relativistic regime. This is in particular bad news for the chance of detecting a pulsar around the supermassive black hole in the Galactic center, since it could be used as a precise clock for probing the space-time metric around Sgr~A* as first put forward by \citet{1979IAUS...84..401P}. However, the inner radii listed in Table~\ref{tab_inner_radii} refer to the stellar cusp in a statistical sense -- there could still be a pulsar approaching Sgr~A* on the scale of $\sim 10-100$ gravitational radii that was deflected into a highly eccentric orbit due to e.g. dynamical scattering on a massive perturber \citep{2007ApJ...656..709P}, a binary-star disruption close to the SMBH \citep{2014A&A...565A..17Z}, or a natal velocity kick during the supernova type II explosion in the clockwise stellar disk containing massive OB stars \citep{2017A&A...602A.121Z,2017MNRAS.469.1510B}.  

\begin{table}[tbh!]
    \centering
    \caption{An overview of inner radii of the nuclear stellar cluster for different stellar types $T$ calculated according to Eq.~\eqref{eq_radius_1T} using the dynamical model of \citet{2006ApJ...645.1152H}.}
    \begin{tabular}{ccccc}
    \hline
    \hline
    Population $T$ & $C_{\rm T}$ & $\delta_{\rm T}$ & $r_{1T}\,[\rm pc]$ & $r_{1T}\,[r_{\rm S}]$\\
    \hline
     MS & $1$       & $1.4$ & $2 \times 10^{-4}$ & $523$ \\
     WD & $10^{-1}$ & $1.4$ & $7 \times 10^{-4}$ & $1831$ \\
     NS & $10^{-2}$ & $1.5$ & $2 \times 10^{-3}$ & $5231$ \\
     BH & $10^{-3}$ & $2$   & $6 \times 10^{-4}$ & $1569$ \\
     \hline
    \end{tabular}
    \label{tab_inner_radii}
\end{table}

\begin{figure}[tbh!]
    \centering
    \includegraphics[width=\textwidth]{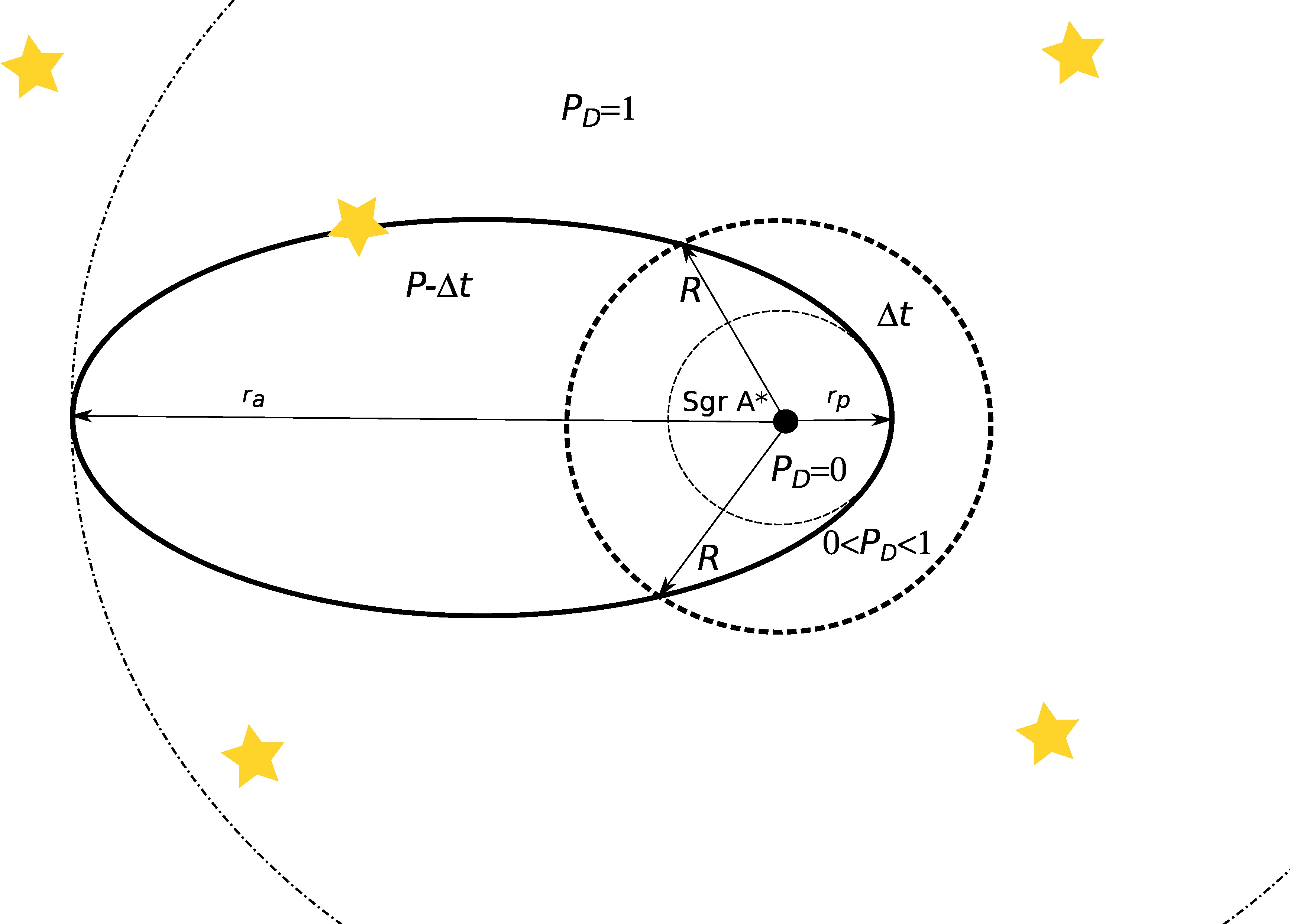}
    \caption{An illustration of the sparse region according to \citet{2018arXiv180411014Z}. The star on an eccentric orbit spends the time $\Delta t$ inside a certain region around Sgr~A*. The probability of the detection $P_{\rm D}$ can be estimated according to \citet{2018arXiv180411014Z}. Adopted from \citep{2018arXiv180411014Z}.}
    \label{fig_sparse_region}
\end{figure}

The set-up of the \textit{sparse region} is illustrated in Fig.~\ref{fig_sparse_region}. Since inside the S cluster with the radius of 1 arcsecond ($\sim 0.04\,{\rm pc}$) the gravitational potential is dominated by Sgr~A*, stellar orbits can be approximated with an ellipse as shown in Fig.~\ref{fig_sparse_region}. Although statistically stars are not expected in this region, they can be scattered into high-eccentric orbits by gravitational encounters with massive perturbers, such as molecular clouds or stellar clusters on the scale of a few parsecs from Sgr~A* \citep{2007ApJ...656..709P}. The filling of the inner arcsecond by stars via the dynamical scattering on massive perturbers is faster by a factor of $10$--$10^7$ in comparison with the classical two-body dynamical relaxation, which takes place on the timescales of a few billion years \citep{2013degn.book.....M}, 
\begin{align}
 t_{\rm r} &= \frac{0.34\sigma_{\star}^3}{G^2\rho_{\star}m_{\star}\log{\Lambda}}\,\notag\\
 &=0.95\times 10^{10}\left(\frac{\sigma_{\star}}{200\,{\rm km\,s^{-1}}}\right)^3 \left(\frac{\rho_{\star}}{10^6\,M_{\odot}{\rm pc^{-3}}} \right)^{-1}\left(\frac{m_{\star}}{1\,M_{\odot}}\right)^{-1}\left(\frac{\log{\Lambda}}{15} \right)^{-1}\,{\rm yr}\,
 \label{eq_relax_time}
\end{align}
where $\rho_{\star}$ is the stellar mass density, $m_{\star}$ is the mean mass of a single star, and $\log{\Lambda}$ is the Coulomb logarithm, which can be estimated as $\log{\Lambda}\approx \log{(M_{\bullet}/m_{\star})}\sim 15$ for $M_{\bullet}=4\times 10^6\,M_{\odot}$ and $m_{\star}=1\,M_{\odot}$.

The non-resonant two-body relaxation normally proceeds slowly and after $\sim 10$ Gyr, the relaxed, steady-state star cluster around a massive black hole is characterized by a cusp-like number density distribution with the slope of $n_{\rm BW}\propto r^{-7/4}$ for equal stellar masses, see \citet{1976ApJ...209..214B}, and $n_{\rm BW}\propto r^{-3/2}$ for unequal stellar masses \citet{1977ApJ...216..883B}. In the Galactic center, the cusp-like relaxed distribution appears to be the case for faint late-type stars, while bright late-type stars have a flatter core-like distribution \citep{2020A&A...641A.102S}, which was first recognized based on the drop in the CO bandhead absorption feature inside $\sim 0.5$ pc from Sgr~A*, see also \citet{1990ApJ...359..112S}. This can be attributed to various depletion mechanisms of red giant envelopes that will be described later on. 

However, there are faster dynamical processes which can induce plunging, highly-eccentric stellar orbits that refill the ``sparse'' or ``loss-cone'' region. First, the non-resonant two-body relaxation timescale expressed by Eq.~\eqref{eq_relax_time} can be shortened, for the fixed stellar density, by either (i) increasing the perturber mass $m_{\star}$ (molecular clouds, stellar cluster), see the scattering by massive perturbers analyzed by \citet{2007ApJ...656..709P}, or (ii) decreasing the 1D velocity dispersion $\sigma_{\star}$. The second scenario occurs in stellar disks where the velocity dispersion is small due to nearly corotating orbits in a certain radial bin, see \citet{2014ApJ...786..121S}. For instance, in the clockwise disk of young OB/Wolf-Rayet stars having the estimated age of $\sim 6$ Myr \citep{2006ApJ...643.1011P,2009ApJ...697.1741B,2009ApJ...690.1463L,2010ApJ...708..834B}, the two-body relaxation could have already affected the radial power-law density distribution, and hence the relaxation time is effectively shortened by two-three orders of magnitude in comparison with the spherical cluster of late-type stars.  

Once on a tight eccentric orbit, the star gradually descends down the potential well due to gravitational radiation \citep[extreme-mass ratio inspirals -- EMRIs; see][ for a review]{2018LRR....21....4A}. Due to the loss of the orbital energy, the semi-major axis decreases - its time derivative is negative. In addition, the gravitational radiation from the system decreases the eccentricity of an inspiralling star, making the orbit more circular \citep{1964PhRv..136.1224P}. The characteristic timescale for an EMRI inspiral, given the initial semi-major axis $a_0$ and the eccentricity $e_0$, is given by the Peters's relation \citep{1964PhRv..136.1224P},
\begin{equation}
    \tau_{\rm GW}(a_0,e_0)\simeq \frac{5}{256}\frac{c^5(1+q)^2}{G^3 q M_{\bullet}^3}\frac{a_0^4}{f(e_0)}\,,
    \label{eq_timescale_GW}
\end{equation}
where $q=m_{\star}/M_{\bullet}$ is the mass ratio that we set to $q=2.5\times 10^{-6}$ for $m_{\star}=10\,M_{\odot}$. For the initial distance of $a_0=1000\,r_{\rm g}$ and zero eccentricity, we get $\tau_{\rm GW}(1000\,r_{\rm g},0)\simeq 5$ Gyr. The non-zero eccentricity can significantly shorten the inspiral time, e.g. for $e_0\sim 0.9$, we obtain $\tau_{\rm GW}(1000\,r_{\rm g},0.9)\sim 4$ Myr, i.e. three orders of magnitude shorter inspiral time. The eccentricity evolution during the gravitational-wave inspiral can be quite complex in realistic scenarios. There are two competing effects as analyzed by \citet{2021PhRvD.103b3015C}: (i) due to the gravitational-wave radiation and other vector or scalar backreaction effects, the eccentricity decreases, i.e. the orbit becomes more circular; (ii) the environmental effects, in particular the accretion onto the star or the dynamical friction, lead to the eccentricity increase. The incorporation of these effects is necessary to correctly estimate the EMRI rate for different types of galaxies, in particular the AGN and the quiescent black holes are expected to differ in this regard, and the redshift-dependency of the EMRI rate is also anticipated.  

On the scale of a few 10 Schwarzschild radii, the presence of stars and compact objects can be revealed by a quasi-periodic variability due to the regular perturbations of the accretion flow \citep{2020ApJ...896...74L,2021arXiv210208135S}. Quasi-periodic eruptions (QPEs) in the X-ray domain that are present in both AGN as well as quiescent sources \citep{2021Natur.592..704A} are potentially caused by either a matter Roche-lobe overflow from a stellar companion \citep{2020MNRAS.493L.120K,2021arXiv210713015M} or by the regular passages of a perturber and the associated induced density waves \citep{2021arXiv210208135S}. The inspiral timescale due to the gravitational radiation given by Eq.~\eqref{eq_timescale_GW} indicates that two EMRIs can in principle approach each other since for e.g. a twice heavier star the inspiral timescale is shorter by a factor of approximately two. This can lead to gravitationally and even physically interacting pairs of EMRIs of uneven masses, which could be detected via the quasi-periodic pattern in the X-ray light curves, similar to the detected QPEs \citep{2017ApJ...844...75M,2021arXiv210713015M}. 

\begin{figure}[tbh!]
    \centering
    \includegraphics[width=\textwidth]{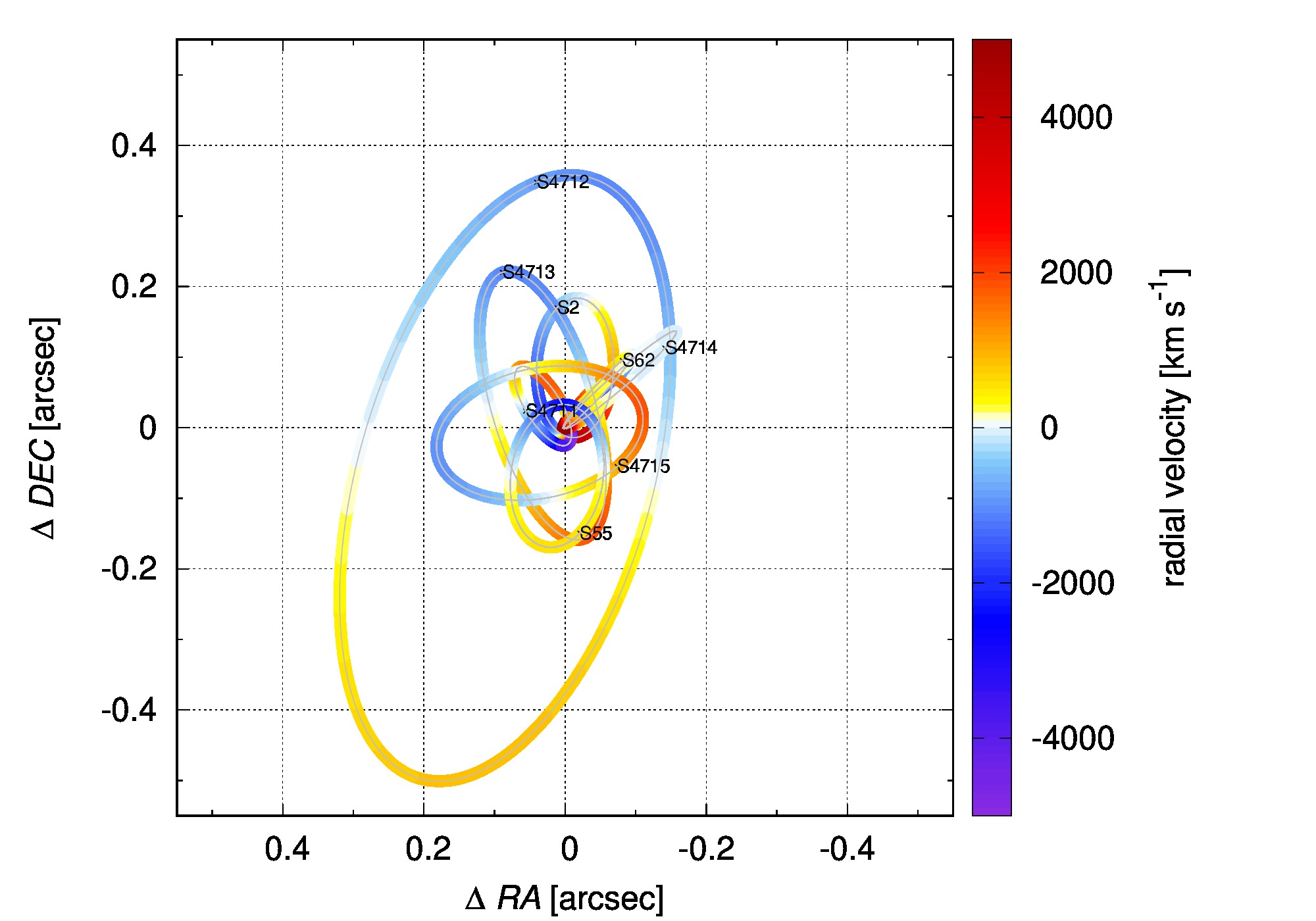}
    \caption{Best-fit orbital solutions of short-period S stars in the Galactic center, including S2 \citep{2018A&A...615L..15G}, S0-102/S55 \citep{2012Sci...338...84M}, and new faint S stars S62 and S4711-S4715 \citep{2020ApJ...889...61P,2020ApJ...899...50P}. The sky plane is represented by the right ascension and the declination offsets from Sgr~A* in arcseconds. The colour axis on the right represents radial velocities in ${\rm km\,s^{-1}}$. The positions next to text labels depicting individual fast-moving stars correspond to the epoch of $1990.0$.}
    \label{fig_faint_stars}
\end{figure}

Recently, fast-moving stars were also detected with either orbital periods or pericenter distances smaller than the archetypal S2 star. In terms of the orbital period, \citet{2012Sci...338...84M} identified S0-102/S55 star using W. M. Keck telescope ($K$-band magnitude of $17.1$) with the orbital period of $\sim 11.5$ years, see Fig.~\ref{fig_faint_stars} for the best-fit orbits, including radial velocities that are colour-coded according to the colour bar on the right. More recently, S62 ($K\sim 16.1$ mag) with the orbital period of $\sim 9.9$ years was identified by \citet{2020ApJ...889...61P}; see also \citet{2021A&A...645A.127G} and \citet{2021ApJ...918...25P} for updated analyses. \citet{2020ApJ...899...50P} discovered so far the shortest-period star S4711 ($K\sim 18.4$ mag) with the orbital period of only $\sim 7.6$ years. In addition to S62 and S4711\footnote{The nomenclature of S stars is not very organized. The ``S'' stands for the infrared source and the original numbering system is based on \citet{1996Natur.383..415E} and \citet{1997MNRAS.284..576E}; the UCLA numbering generally differs, see e.g. \citet{1998ApJ...509..678G}. The star S4711 and the other stars studied in \citet{2020ApJ...899...50P} are named after the house number 4711 in Cologne, Germany, (now Glockengasse 4) where the Original Eau de Cologne (perfume) was founded and produced. The numbering of houses in Cologne was introduced by French officers in 1794. In this sense, as an allegory, the naming of S4711 symbolizes the ``numbering'' of newly discovered stars in the Galactic center (Andreas Eckart, a private communication).}, several other faint S stars were identified, whose orbital fractions are at least partially inside the S2 orbit. We compare the orbital solutions of these fainter stars, including S62, S4711, and S4712-S4715, with the orbits of S2 and S55/S0-102 in Fig.~\ref{fig_faint_stars}. The most promising candidates for studying relativistic effects are S62 and S4714, which have the smallest pericenter distances of only $\sim 17.8\,{\rm AU}$ ($\sim 450.5$ gravitational radii) and $\sim 12.6\,{\rm AU}$ ($\sim 320.1$ gravitational radii), respectively. This results in the highest potential pericenter orbital velocities of $\sim 20\,000\,{\rm km\,s^{-1}}$ ($\sim 6.7\%$ of the light speed) and $\sim 24\,000\,{\rm km\,s^{-1}}$ ($\sim 8\%$ of the light speed) for S62 and S4714, respectively. These pericenter velocities are at least a factor of two larger than the pericenter velocity of the S2 star, which amounts to $\sim 7600\,{\rm km\,s^{-1}}$ or $\sim 2.5\%$ of the light speed.

The lack of stars on the scales of 10 Schwarzschild radii is, however, complemented by orbiting \textit{hot spots} that orbit Sgr~A* close to the innermost stable prograde circular orbit (ISCO) of a nearly Schwarzschild black hole or alternatively, at the retrograde ISCO of a highly-rotating Kerr black hole \citep{2018A&A...618L..10G}. These observations were performed by the near-infrared GRAVITY@ESO interferometer, which can resolve out the offset of an emission centroid from Sgr~A* with an angular resolution of $\sim 20$--$70\,{\rm \mu as}$. The instrument detected the continuous shift of the centroid position on the length-scales of $150\,{\rm \mu as}$ with the periodicity of $10-30$ minutes. The detected clockwise shift of the emission centroid is consistent with the nearly face-on orbital motion of a denser blob of relativistic electrons that emit non-thermal synchrotron NIR emission. According to the detected positions of the blob, the motion takes place on the scale of $6-10$ gravitational radii $(GM_{\bullet}/c^2)$. In general, due to uncertainties, the observations suffer from a spin-degeneracy, hence the hot spot orbits Sgr~A* close to the prograde ISCO of a nearly Schwarzschild black hole (low-rotating Kerr black hole) or close to the retrograde ISCO of a highly-rotating Kerr black hole. Based on the modulated light curves, flares of Sgr~A* modelled by an orbiting hot-spot model can be employed to probe the potential of Sgr~A* in a strong-gravity regime, which is currently inaccessible for stars. Light curves of X-ray flares were used to obtain an independent estimate of the mass of Sgr~A* \citep{2017MNRAS.472.4422K}, which is consistent with the mass determined from stellar astrometry.     

The connection between the multi-wavelength variability of Sgr~A* and blobs or \textit{hot spots} orbiting on ISCO scales was already proposed and studied more than a decade before the actual detection of an orbital motion by the GRAVITY experiment \citep{2005MNRAS.363..353B,2006MNRAS.367..905B,2006A&A...460...15M}. 
The orbiting hot-spot may result from shocks or magnetic reconnection events in the innermost parts of the hot, magnetized accretion flow \citep{2008A&A...492..337E,2008ApJ...682..361Y,2010A&A...510A...3Z,2015ApJ...810...19L}. These plasma instabilities can produce a blob of relativistic electrons with the Lorentz factor of $\gamma \sim 10^3-10^6$, which can explain the multi-wavelength variability of Sgr~A* that occurs on hourly timescales. The flares also exhibit the time lags of about one hour between the NIR/X-ray domain and flares at lower frequencies (submillimeter and radio domains), which can be explained by cooling via an adiabatic expansion \citep{2008A&A...492..337E,2008ApJ...682..361Y}. In addition, some NIR flares have simultaneous X-ray counterparts which implies the synchrotron-self-Compton (SSC) or inverse Compton process to explain the NIR/X-ray connection \citep{2004A&A...427....1E,2006A&A...450..535E,2006A&A...455....1E}. 

The NIR flares exhibit a correlation between flux modulations and changes in the polarization angle and degree \citep{2010A&A...510A...3Z}. \citet{2010A&A...510A...3Z} showed that the changes in the polarization angle during flares are consistent with the orbital motion of a compact hot spot rather than with the extended emission. This was also confirmed by the GRAVITY experiment \citep{2018A&A...618L..10G}, which detected a continuous swing of the polarization angle connected with the centroid shift with the same periodicity of $45\pm 15$ minutes as that of the orbital motion. Such a gradual change in the polarization angle may be explained by the hot spot motion in the poloidal magnetic field, i.e. which is perpendicular to the orbital plane. 

\begin{figure}[tbh!]
    \centering
    \includegraphics[width=0.8\textwidth]{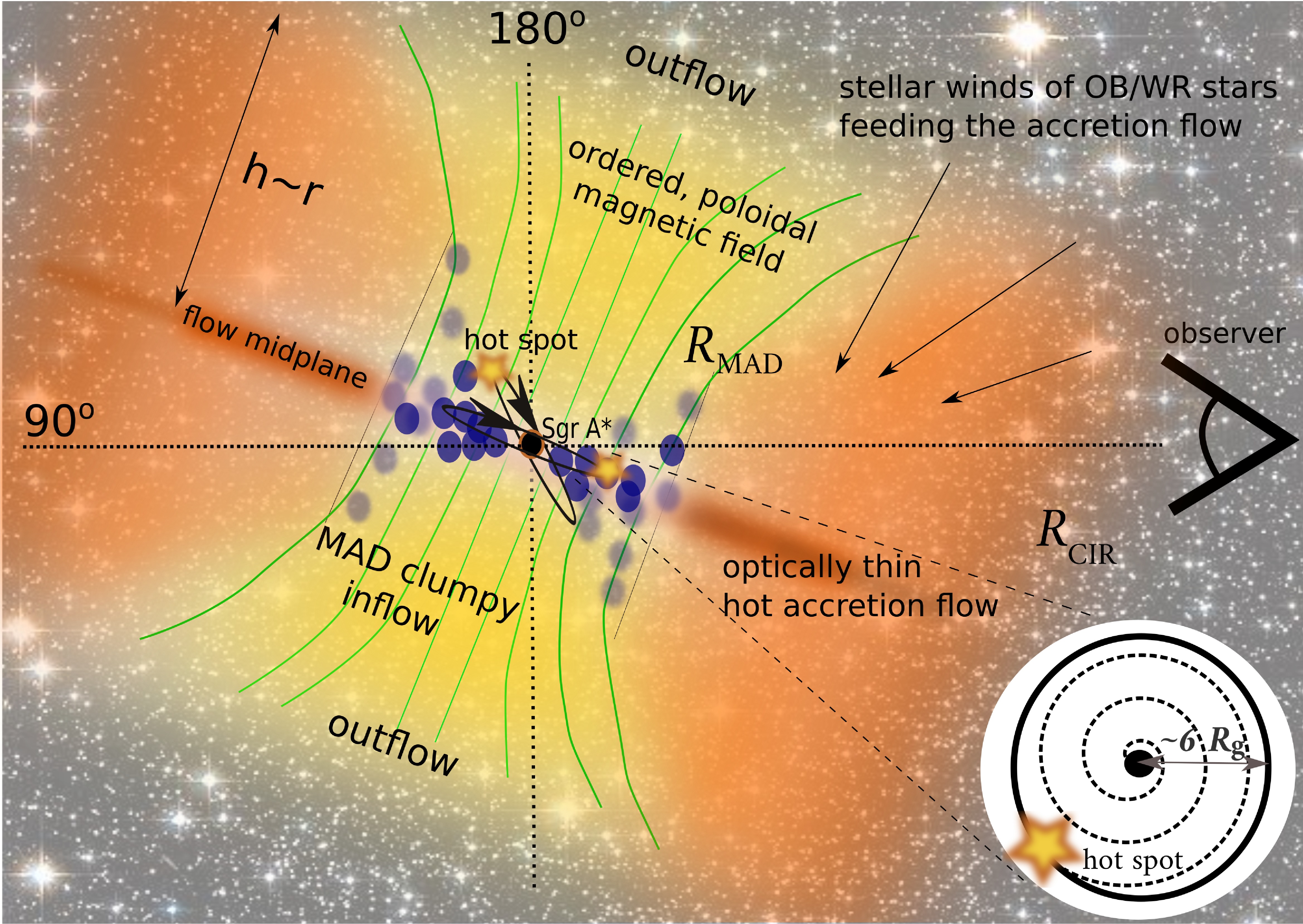}
    \caption{Illustration of the hot accretion flow feeding the variable radio, near-infrared, and X-ray source Sgr~A* in the Galactic center. On the larger scales, the cusp of $\sim 100$ mass-losing OB/Wolf-Rayet stars provides the material, a fraction of which is accreted at the Bondi radius \citep{1997ApJ...488L.149C,2006MNRAS.366..358C,2010ApJ...716..504S,2013Sci...341..981W,2020ApJ...888L...2C}, $r_{\rm Bondi}\sim 0.14(M_{\bullet}/4\times 10^6\,M_{\odot})(T_{\rm g}/1.5\times 10^7\,{\rm K})^{-1}\,{\rm pc}$, see Eq.~\eqref{eq_Bondi_radius}. The actual accretion rate of Sgr~A* within 100 gravitational radii, $\dot{M}_{\rm acc}\approx 10^{-9}-10^{-7}\,M_{\odot}\,{\rm yr^{-1}}$ \citep{2007ApJ...654L..57M}, is $\lesssim 1\%$ of the Bondi rate, $\dot{M}_{\rm Bondi}\approx 10^{-3}\,M_{\odot}{\rm yr^{-1}}$, which is supplied to a large extent by stellar winds \citep{2020ApJ...888L...2C}. The poloidal magnetic field frozen in the plasma is dragged inwards, which leads to the accumulation of the magnetic flux in the inner regions. This can result in the development of the magnetically arrested accretion flow \citep[MAD; ][]{1974Ap&SS..28...45B,1976Ap&SS..42..401B,2003PASJ...55L..69N}, which is characterized by a slow, diffusive accretion consisting of magnetically confined blobs accompanied by magnetic interchanges and reconnection events. Observationally, the heated expanding gas due to reconnection events could be detected as orbiting ``hot spots'' that are visible for about one orbital period or its fraction close to the innermost stable circular orbit (ISCO; positioned at 6 gravitational radii for a non-rotating, Schwarzschild black hole), see the figure inset in the right bottom corner. The background image portrays the globular cluster NGC 288 to represent the dense stellar field in the Galactic center; the stellar field of NGC 288 was adopted from the Hubble Space Telescope's Wide Field Channel of the Advanced Camera for Surveys (credit: ESA/Hubble \& NASA).}
    \label{fig_MAD_illustration}
\end{figure}

There are several indications that the magnetic field in the Galactic center is dynamically relevant and ordered on both large and small scales \citep{2015llg..book..391M,2018A&A...618L..10G}. Concerning the larger scales, the Faraday rotation measurements of the radio emission of the magnetar PSR J1745–2900 \citep{2013Natur.501..391E} put a lower limit on the line-of-sight component of the magnetic field, $B\gtrsim 8\,{\rm mG}$, which is at a deprojected distance of $\gtrsim 0.12\,{\rm pc}$. \citet{2013Natur.501..391E} also confirmed the ordered structure of the magnetic field at the scale of $0.1\,{\rm pc}$ that is at the intermediate distance between the ordered magnetized filaments in the Central Molecular Zone \citep{2015llg..book..391M} and the ordered, poloidal magnetic field close to the innermost stable circular orbit of Sgr~A* \citep{2015Sci...350.1242J,2018A&A...618L..10G}. Since the hot plasma is magnetized and its fraction accretes towards Sgr~A*, it is expected to lead to the accumulation of the magnetic flux as the magnetic field lines are dragged inwards by the flow. At a certain distance, at the so-called magnetospheric radius $R_{\rm MAD}$, the accretion flow becomes unstable and fragments since the gravitational potential energy of the flow is comparable to the electromagnetic pressure \citep{2003PASJ...55L..69N}. Following the energy equipartition, we can approximately derive $R_{\rm MAD}$ based on the equality for the axisymmetric flow \citep{2003PASJ...55L..69N,2020ApJ...897...99T},
\begin{align}
    \frac{GM_{\bullet}\Sigma}{R_{\rm MAD}^2} &\approx 2\frac{B_{R}B_z}{4\pi}\,\label{eq_equipartition}\,,\\
    R_{\rm MAD}&\sim \left(\frac{\sqrt{GM_{\bullet}}\dot{M}}{\sqrt{2}\epsilon B_{\rm pol}^2} \right)^{2/5}\,,\label{eq_mad_analytic}\\
    &\sim 68\,\left(\frac{M_{\bullet}}{4\times 10^6\,M_{\odot}}\right)^{1/5}\left(\frac{\dot{M}}{10^{-8}\,{\rm M_{\odot}\,yr^{-1}}}\right)^{2/5}\left(\frac{\epsilon}{10^{-2}}\right)^{-2/5}\left(\frac{B_{\rm pol}}{10\,{\rm G}}\right)^{-4/5}\,R_{\rm g}\,,\label{eq_mad_numerical}
\end{align}
where $\Sigma=\dot{M}/(2\pi R \epsilon v_{\rm ff})$ is the surface density of the flow, which depends on the accretion rate $\dot{M}$ and the radial velocity $v_{\rm R}=\epsilon v_{\rm ff}$ that is smaller than the free-fall velocity and their ratio is expressed by the parameter $\epsilon\sim 0.01-0.001$ \citep{2003PASJ...55L..69N}. While deriving Eq.~\eqref{eq_mad_analytic}, we assumed $B_{R}\sim B_z\approx B_{\rm pol}$. 
Below $R_{\rm MAD}$ the accretion proceeds discretely and diffusely through the poloidal magnetic field, which is refereed to as the magnetically arrested accretion disc \citep[MAD; ][]{1974Ap&SS..28...45B,1976Ap&SS..42..401B,2003PASJ...55L..69N}. For the Galactic center, we estimated $R_{\rm MAD}\sim 68\,R_{\rm g}$ using $\dot{M}\approx 10^{-8}\,M_{\odot}\,{\rm yr^{-1}}$, which is based on the Faraday-rotation measurements \citep{2007ApJ...654L..57M}, and $B_{\rm pol}\sim 10\,{\rm G}$ based on the near-infrared flare observations and modelling their timing and spectral properties \citep{2012A&A...537A..52E,2021ApJ...917...73W}. Inside the magnetospheric radius $R_{\rm MAD}$, the accretion flow fragments and proceeds at the radial velocity two to three orders of magnitude smaller than the free-fall velocity \citep{2003PASJ...55L..69N}. The accretion continues diffusively through the ordered, poloidal magnetic field via magnetic interchanges as well as magnetic reconnection events. The fragmented flow consists of magnetically confined islands. The released energy heats up the surrounding gas, which could observationally be detected as the hot spot orbiting Sgr~A* close to the innermost stable circular orbit or just below it, when the heated and outflowing gas spirals in towards the event horizon. This is illustrated in Fig.~\ref{fig_MAD_illustration}, which connects larger spatial scales, where the hot plasma is fed by stellar winds of $\sim 100$ mass-losing OB/Wolf-Rayet stars, and smaller scales, where the hot flow fragments and proceeds as the clumpy MAD flow below $R_{\rm MAD}$.

The MAD flow -- Magnetically Arrested Disk -- has a different dynamical set-up than the so-called SANE disk (Standard and Normal Evolution). For the SANE flow, which is only weakly magnetized, the angular momentum is transported outwards via the turbulence generated by the magnetorotational instability \citep{1992ApJ...400..610B}. The expected radial division between the SANE and the MAD flow for Sgr~A* is given by Eq.~\eqref{eq_mad_numerical}. Another important length-scale for the Sgr~A* accretion flow is the circularization radius $R_{\rm CIR}$, i.e. the distance where the hot flow that possesses an initial angular momentum circularizes. We assume that a circularized thick flow forms with the outer radius of $R_{\rm CIR}$ and that the initial stellar wind material was provided by a star orbiting at $r_{\star}$ with the wind total velocity of $v_{\rm g}$ (including the orbital and the terminal wind velocity). The circularized gas at the outer disk radius has the largest angular momentum, which initially corresponded to the wind material that had at most the escape velocity with respect to Sgr~A* at the distance $r_{\star}$, $v_{\rm g}\lesssim (2GM_{\bullet}/r_{\star})^{1/2}$,
\begin{align}
    R_{\rm CIR} \left(\frac{GM_{\bullet}}{R_{\rm CIR}} \right)^{1/2} &\approx \lambda r_{\star}v_{\rm g}\lesssim \lambda r_{\star}\left(\frac{2GM_{\bullet}}{r_{\star}}\right)^{1/2}\,\notag\\
    R_{\rm CIR}&\lesssim 2 \lambda^2 r_{\star}\,,
    \label{eq_circularization_radius}
\end{align}
where $\lambda$ is the retained fraction of the initial angular momentum of the stellar wind. The parameter $\lambda$ can be estimated from a few observational indications. Young massive OB/Wolf-Rayet stars with powerful winds are located beyond the S cluster at $r_{\star}\sim 0.1\,{\rm pc}$, i.e. the wind material is captured at the Bondi radius. The circularization radius as such is quite uncertain, however, there is an indication of the colder disk containing ionized $\sim 10^4\,{\rm K}$ gas orbiting around Sgr~A* as revealed by \citet{2019Natur.570...83M} using the hydrogen recombination line H30$\alpha$ at $1.3$ mm. For such a disk with $R_{\rm CIR}\sim 0.004\,{\rm pc}$, the retained fraction of the angular momentum is $\lambda\geq (R_{\rm CIR}/2r_{\star})^{1/2}\sim 0.14$, i.e. at least $\sim 14\%$. In the case of the stellar winds, the angular momentum is expected to be removed by shocks, wind-wind collisions, and the magnetohydrodynamical drag.    

\subsubsection*{Localization of Sgr~A and Sgr~A* -- the historical perspective}
The first radio source ever observed was the Galactic centre by Karl G. Jansky in 1932 at the frequency of $20.5\,{\rm MHz}$ and the angular resolution of $25^{\circ} \times 35^{\circ}$ while he was working at the Bell Telephone Laboratories. Grote Reber continued in his effort and developed his own radiotelescope and receivers between 1938 and 1948, partially during WWII. He detected the emission from the plane of the Milky Way at the frequency of $160\,{\rm MHz}$ ($1.9$ m), which was peaking towards the Galactic center. \citet{1951AuSRA...4..459P} detected a prominent source of radio emission at $1.2\,{\rm GHz}$ and $3\,{\rm GHz}$. The radio emission of the source appeared to be flat between $100\,{\rm MHz}$ and $1210\,{\rm MHz}$, similar to another radio source Taurus~A. They proposed that optically thin thermal gas is responsible for the emission and the source became to be known as Sagittarius A (Sgr~A). In the mid-1960s Barry Clark and Dave Hogg used the two-element Green Bank interferometer in West Virginia to analyze the small-scale structure of the radio emission with the angular resolution of $10''$. They detected a smaller structure inside the Sgr~A complex with the flux density of $0.3\,{\rm Jy}$ at $2.8\,{\rm GHz}$ $(11\,{\rm cm})$. However, the source was confused with the thermal emission of Sgr~A West \citep{1966ApJ...145...21C}.

These early observations of the Galactic centre region were essentially simultaneous with the discovery of quasars by \citet{1963Natur.197.1040S} in 1963 and the realization that they are extremely distant and luminous thanks to the identification of the redshifted Balmer series of hydrogen in the optical spectra of 3C273. The enormous energy in extragalactic sources -- which amounted to $4\times 10^{12}\,L_{\odot}$ for 3C273 in the optical bands -- was proposed to originate in large concentrations of mass in their centers and the proposal of `giant stars' powering Seyfert galaxies and quasars, see \citet{1963Natur.197..533H}. It was obvious from the first principles that the sudden collapse of mass into relativistic dimensions can yield enough energy to explain the power output of the AGN. Such a mechanism was proposed in the early 1960s by Vitaly L. Ginzburg in the context of the contraction of the central regions of galaxies and the subsequent formation of protostars \citep{1961AZh....38..380G}. Similar and other ideas and theories were presented at the First Texas Symposium on Relativistic Astrophysics in Dallas in December 1963. This event, which is often considered as the ``landmark'' in the rebirth of relativistic and gravitational astrophysics, brought together theorists and observers of the time. Although the Symposium did not provide final clues to the problem of quasar energetics, it triggered new enthusiasm and pointed astrophysicists in the right direction.

The first original theory of the accretion of gas and dust onto a collapsed object, which releases the thermal and radiative energy due to the conversion of the gravitational potential energy of the infalling matter, was proposed by Yakov Zel'dovich and Igor Novikov \citep{1972GReGr...3..119Z}. Independently, essentially the same process was described in 1964 by Edwin Salpeter at the Cornell University in the context of massive objects with the mass of $M\gtrsim 10^6\,M_{\odot}$ powering quasi-stellar objects \citep{1964ApJ...140..796S}. Donald Lynden-Bell generalized the quasar accretion mechanism to other galactic nuclei \citep{1969Natur.223..690L}. Subsequently, \citet{1971MNRAS.152..461L} discuss the Galactic centre specifically as a potential location of the ``old-quasar'' black hole and suggest to use the Very Long Baseline Interferometry (VLBI) to infer its flux density and the position. They also suggested to use objects in the surroundings of the black hole -- specifically line-emitting gas -- to determine its mass. Soon afterwards, Shakura \& Sunyaev presented the solution for an accretion disc where the macroscopic turbulence is responsible for the viscosity \citep{1973A&A....24..337S}. They introduced the $\alpha$ parameter, which relates the viscosity to the sound speed and the disc scale-height, $\nu=\alpha c_{\rm s} H$, and takes values between zero (no accretion) and one. The Shakura \& Sunyaev disc is in local thermal equilibrium and radiates away the accumulated viscous heat, cools down, and becomes geometrically thin and optically thick. The general relativistic solution for the thin disc was presented by \citet{1973blho.conf..343N}. In both works, it is assumed that the viscous torque vanishes at the innermost stable circular orbit (ISCO).

In the same year as the First Texas Symposium, Roy Kerr found the axisymmetric solution to the Einstein field equations \citep{1963PhRvL..11..237K}. The Kerr solution is applicable to any rotating uncharged mass, including compact objects. Two years later, Ezra Ted Newman generalized the Kerr solution to rotating charged bodies, which is nowadays referred to as Kerr-Newman solution \citep{1965JMP.....6..918N}, which is the most general asymptotically flat, stationary solution to Einstein-Maxwell field equations. These two solutions followed nearly 50 years after the first static solution of Karl Schwarzschild \citep{1916SPAW.......189S} and the non-rotating charged case of Reissner-Nordstr\"om \citep{1916AnP...355..106R,1918KNAB...20.1238N}. The overview of all fundamental solutions to Einstein field equations is in Table~\ref{tab_bh_solutions}.

\begin{table}[tbh!]
    \centering
    \caption{Overview of fundamental black-hole solutions including the year of their discovery.}
    \begin{tabular}{c|c|c}
    \hline
    \hline
     {Charge/Spin}    &  {Non-rotating $(J=0)$} & {Rotating $(J\neq 0)$} \\
     \hline
    {Uncharged $(Q=0)$}      &   Schwarzschild (1916)              &   Kerr (1963)            \\
    {Charged $(Q\neq 0)$}        & Reissner-Nordstr\"om (1916,1918)    & Kerr-Newman (1965)               \\
    \hline
    \end{tabular}
    \label{tab_bh_solutions}
\end{table}

Before the identification of the supermassive black hole in the Galactic center, other compact objects were envisaged and discovered, both theoretically and observationally, in parallel with the development of general relativity and quantum mechanics. First, these were the hot and small companions of some visible stars -- such as 40 Eridani A, Sirius A, and Procyon. In 1783, William Herschel discovered the two companions of 40 Eridani A -- 40 Eridani B and C, where 40 Eridani B was in 1910 spectroscopically identified to be of spectral type A by Williamina Fleming, despite being too faint for spectral type A, which indicated the size of the Earth. Sirius B, which is the nearest white dwarf to the Earth ($\sim 2.64$ pc) was discovered by Alvan Graham Clark on January 31, 1862 in Cambridgeport, Massachusetts. Sirius B as well as other similar faint objects (40 Eridani B, Van Maanen's star) were later identified as \textit{white dwarfs} (the name \textit{white dwarf} was introduced by Willem Jacob Luyten in 1922.), i.e. stellar remnants of low-mass stars supported by degenerate electron pressure with the mean mass density of $\sim 10^6\,{\rm g\,cm^{-3}}$. The internal structure of white dwarfs, where are effectively cooling off due to the lack of any thermonuclear fusion, was investigated by Ralph H. Fowler, Edmund Clifton Stoner, Wilhelm Anderson, and Subrahmanyan Chandrasekhar at the end of 1920s and early 1930s using the non-relativistic and relativistic equations of state of degenerate Fermi gas. Solving the hydrostatic equilibrium equation for the relativistic Fermi gas, Chandrasekhar also derived the limit on the mass of stable white dwarfs, which is a function of fundamental constants only,
\begin{equation}
    M_{\rm Ch}\approx \frac{\omega_3^0\sqrt{3\pi}}{2}\left(\frac{\hbar c}{G} \right)^{3/2}\frac{1}{(\mu_{e}m_{\rm H})^2}\sim 1.4\,M_{\odot}\,,
    \label{eq_chandra_limit}
\end{equation}
where $\omega_3^0\sim 2$ is the constant related to the solution of the Lane-Emden equation, $\mu_e$ is the average molecular weight per electron (here $\mu_e=2$), and $m_{\rm H}$ is the hydrogen atom mass. Since $M_{\rm Pl}=\sqrt{\hbar c/G}$ is the definition of the Planck mass, the Chandrasekhar limit can also be roughly estimated using the expression $M_{\rm Ch}\sim M_{\rm Pl}^3/m_{\rm H}^2\sim 1.9\,M_{\odot}$.
Then in 1963, Allan Sandage and Thomas A. Matthews found the first optical counterpart for the bright radio source 3C48 using the 5.1-m Hale telescope atop Mount Palomar, see \citet{1963ApJ...138...30M}. In the same year, also using the Palomar telescope, Maarten Schmidt identified the similar quasi-stellar source 3C273 \citep{1963Natur.197.1040S}, which is characterized by the radio-optical emission, with a distant extragalactic source with a relatively large redshift of $z=0.158$. As we mentioned, large luminosities of quasars exceeding the host luminosities by two orders of magnitude started to be interpreted by the accretion onto massive compact objects -- black holes. However, many experts still considered the concept of black holes purely mathematical. When the first pulsar (PSR B1919$+$21) with the periodicity of only 1.33730 seconds was discovered on August 6, 1967 by Jocelyn Bell \citep{1968Natur.217..709H}, it was clear that even more compact objects than white dwarfs do exist. The concept of fast-spinning neutron stars having the mean densities of $\sim 10^{14}\,{\rm g\,cm^{-3}}$ and with the radius of only $\sim 10$ km was already theoretically laid out by \citet{1934PNAS...20..259B} and \citet{1938Natur.141..333L}. Neutron stars provided a natural explanation for periods of the order of one second since any larger object than $R_{\rm limit}\sim [GM_{\rm Ch}P_{\rm rot}^2/(4\pi^2)]^{1/3}\sim 1700\,{\rm km}$ would be unstable and break up at the rotational periods of $P_{\rm rot}\sim 1\,{\rm s}$. Later discovered millisecond pulsars, the first of which with the period of $1.558\,{\rm ms}$ (PSR B1937$+$21) was discovered by \citet{1982Natur.300..615B}, constrained the radius to less than $\sim 170\,{\rm km}$. Hence, the concept of the continuing collapse, or rather of the inevitable formation of space-time singularities beyond certain threshold, later called the event horizon, see \citet{1965PhRvL..14...57P}, and hence the concept of black holes became rather realistic after the consecutive discoveries of white dwarfs, the first quasars, and pulsars. The mass limit for the formation of neutron stars during the stellar collapse is given by the Chandrasekhar limit, see Eq.~\eqref{eq_chandra_limit}, i.e. stellar end-products heavier than the Chandrasekhar mass will end up as neutron stars supported by the pressure of degenerate neutrons. The mass limit separating neutron stars from black holes is less clear but it has followed from the work of Richard C. Tolman, who developed the method for solving nonlinear Einstein's field equations for static spheres filled with fluid \citep{1939PhRv...55..364T}. Shortly before the WWII, in 1939, J. Robert Oppenheimer and George Volkoff derived the equation describing the static state of spherically symmetric spacetimes filled with fluid with pressure $P(r)$, now known as the Tolman-Oppenheimer-Volkoff (TOV) equation, which is essentially the general relativistic equation of hydrostatic equilibrium,
\begin{equation}
    \frac{\mathrm{d}P}{\mathrm{d}r}=-\frac{G}{r^2}\left[\rho+\frac{P(r)}{c^2}\right]\left[m(r)+\frac{4\pi r^3 P(r)}{c^2} \right]\left[1-\frac{2Gm(r)}{c^2}\right]^{-1}\,,
    \label{eq_tov_equation}
\end{equation}
where $m(r)$ is an enclosed mass within the distance $r$, $P(r)$ is the pressure, and $\rho(r)$ is density. When relativistic corrections are too small, the terms with $1/c^2$ can be neglected, which leads to the Newtonian equation of hydrostatic equilibrium describing static, spherically symmetric structure of matter, $\mathrm{d}P/\mathrm{d}r=-Gm(r)\rho(r)/r^2$. In the paper from 1939 titled ``On Massive Neutron Cores" \citep{PhysRev.55.374}, J. Robert Oppenheimer and George Volkoff derived the maximum mass of a neutron star, using the TOV equation and an equation of state of a degenerate Fermi gas of neutrons. They arrived at the value of $0.7\,M_{\odot}$, too low given the current values for the maximum or the TOV mass limit, which is in the range of $M_{\rm TOV}\approx 1.5-3\,M_{\odot}$ \citep{1983bhwd.book.....S,1996A&A...305..871B}. \citet{1996ApJ...470L..61K} found a slightly smaller range towards larger masses, $M_{\rm TOV}\approx 2.2-2.9\,M_{\odot}$. This discrepancy can be explained by the usage of an inadequate equation of state by Oppenheimer and Volkoff, which neglected the short-range nuclear repulsive force among neutrons. The value of the TOV limit was recently constrained based on the gravitational-wave detections of neutron star-neutron star mergers as well as based on the follow-up electromagnetic detections, mainly associated with the first such detection GW170817 \citep{2017ApJ...848L..12A}. The analysis of GW170817 data constrained the TOV limit to $M_{\rm TOV}\lesssim 2.3\,M_{\odot}$ \citep{2019PhRvD.100b3015S} for cold spherical neutron stars. Hence, the existence of the Chandrasekhar and TOV limits implies the three stable end-states of the stellar matter -- white dwarfs, neutron stars, and black holes. Black holes as such occupy the largest mass range from $M_{\rm TOV}$ up to essentially $\sim 10^{10}\,M_{\odot}$. In general, the division into stellar, intermediate-mass, and supermassive black holes is often introduced depending on the mass. These three classes of black holes likely have the common origin in stars or cosmologically in collapsing gas clouds and the further increment in mass depends on the dynamical and magnetohydrodynamic state of the surrounding medium where the black hole is located.  

Finally, a compact radio source later denoted as Sgr~A* was discovered by Bruce Balick and Robert Brown on February 13 and 15, 1974, which showed a brightness temperature in excess of $\sim 10^7\,{\rm K}$ and was unresolved at $<0.1''$. To resolve such a compact object among the extended radio emission of the angular size of $\lesssim 20''$ was made possible thanks to the newly commissioned interferometer Green Bank -- Huntersville (National Radio Astronomy Observatory -- NRAO) with the longest baseline of $35$\,km. It consisted of three $26$-meter telescopes in Green Bank, which were separated by $\lesssim 2.7\,{\rm km}$ and the newly-installed $14$-meter radio telescope on the mountaintop in Huntersville separated by $35\,{\rm km}$ from Green Bank. The Green Bank--Huntersville interferometer could observe at $11\,{\rm cm}$ and $3.7\,{\rm cm}$ simultaneously.

The compactness of the source, its high brightness temperature, and the association of the radio source with the Galactic center made Sgr~A* a suitable candidate for the Milky Way supermassive black hole as hypothesized by Lynden-Bell and Rees \citep{1971MNRAS.152..461L}. The basic properties were confirmed by higher-resolution Very Long Baseline Interferometry (VLBI) radio observations at $3.7$ cm by \citet{1975ApJ...202L..63L}, who resolved out Sgr~A* down to $\lesssim 0.02''$ using the $242\,{\rm km}$ baseline. The measured flux density of $F_{3.7}=0.6\pm 0.1\,{\rm Jy}$ implied the brightness temperature of
\begin{align}
    T_{\rm B}&=\frac{F_{\nu}c^2}{2\nu^2 k_{\rm B}\pi \theta^2}\,\notag\\
    &\sim 10^7 \left(\frac{F_{\nu}}{0.6\,{\rm Jy}}\right)\left(\frac{\nu}{8\,{\rm GHz}} \right)^{-2} \left(\frac{\theta}{0.02''}\right)^{-2} \,{\rm K}\,,
    \label{eq_brightness_temperature}
\end{align}
that is based on the Rayleigh-Jeans approximation, $h\nu \ll k_{\rm B}T_{\rm B}$, which is certainly valid in the radio domain ($h \nu\sim 5.4\times 10^{-17}\,{\rm erg}$ while $k_{\rm B} T_{\rm B}\sim 1.38 \times 10^{-9}\,{\rm erg}$, i.e. thermal energy per particle is 8 orders of magnitude larger).
In addition, based on the comparison with the previous radio data, they inferred that the source is variable.

\begin{figure}
    \centering
    \includegraphics[width=0.9\textwidth]{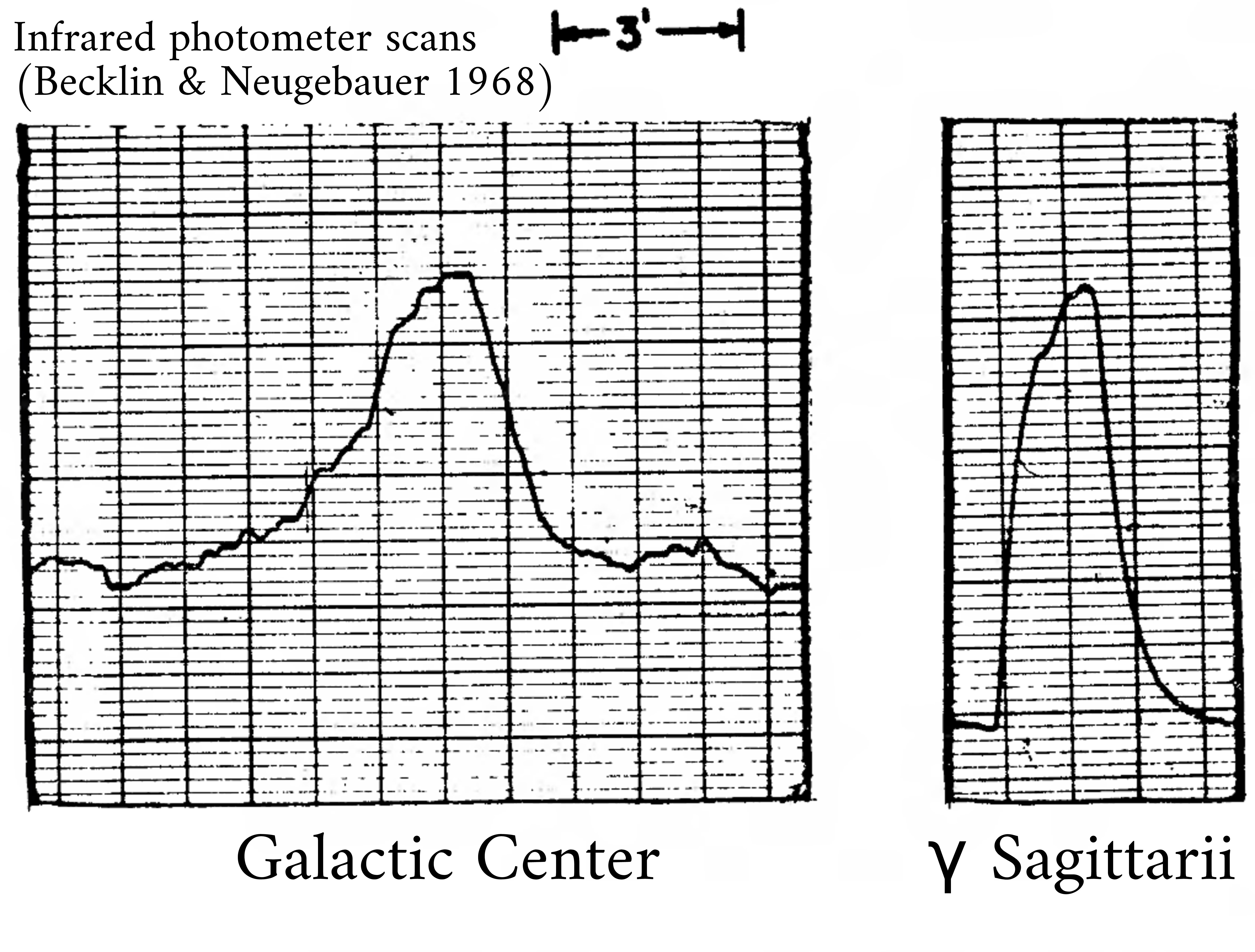}
    \caption{The infrared photometer scan along the right ascension at the wavelength of $2.2\,{\rm \mu m}$. To the left, the Galactic center brightness profile is shown, while to the right, the star $\gamma$ Sagitarii is shown for comparison. The strip-chart image was taken from \citet{1968ApJ...151..145B}.}
    \label{fig_bn_GC}
\end{figure}

In parallel to the first radio observations of the Galactic center, there had also been the first attempts to detect Sgr~A* and its surroundings in the infrared domain. Infrared light can penetrate through the dust along the line of sight much better than the optical or the UV radiation that is either absorbed or scattered. The first attempts to detect the Galactic center in the infrared domain were performed in 1945, but due to a lack of sensitivity as well as a coarse sampling , these observations conducted at 1 micrometer were not successful in detecting the Galactic center complex without confusion \citep{1947ApJ...106..235S}; see also \citet{1961SvA.....5..361M}. However, \citet{1947ApJ...106..235S} found an indication for the infrared source behind the dust clouds and suggest to perform observations at $2\,{\rm \mu m}$. This was followed up by Eric Becklin and Gerry Neugebauer \citep{1968ApJ...151..145B}, who used the military infrared photometer with the PbS detector mounted at the Mount Wilson and the Palomar observatories with the angular resolution between $1.8'$ and $0.08'$ that corresponds to the linear scales of $4.2$ and $0.2$ parsecs, respectively, at the distance of the Galactic center. They managed to resolve out the nuclear star cluster that was characterized by a sharp symmetric peak in the $2.2\,{\rm \mu m}$ flux density along the Galactic plane, see Fig.~\ref{fig_bn_GC}. In addition, they also resolved out individual stellar complexes within the cluster using individual bolometer scans. They interpret the central emission seen in the right-ascension scan as a stellar emission of the central cluster with the effective photosphere temperature exceeding $\sim 4000\,{\rm K}$. However, a non-thermal contribution to the thermal stellar emission cannot be excluded, especially in the central part.

The infrared observations using the fine-structure NeII line at $12.8\,{\rm \mu m}$ detected in emission revealed the supersonic and ordered motion of the ionized gas, with the centroid of the NeII line of $+75\pm 20\,{\rm km\,s^{-1}}$ and the velocity dispersion of $200\,{\rm km\,s^{-1}}$ \citep{1976ApJ...205L...5W}. In particular, \citet{1977ApJ...218L.103W} detected the preferentially redshifted neon-emitting gas in the eastern part of Sgr~A West that could have been separated from the blueshifted gas in the western part. Based on the Doppler velocities in the range from $+250\,{\rm km\,s^{-1}}$ to $-350\,{\rm km\,s^{-1}}$, the total enclosed mass within $1\,{\rm pc}$ was estimated to $\sim 4\times 10^6\,M_{\odot}$. A more detailed spatial distribution as well as the dynamics of Sgr~A West inferred from the NeII line were presented by \citet{1979ApJ...227L..17L} and \citet{1980ApJ...241..132L}. They revealed that the gas is kept ionized by stars with $T_{\rm eff}\lesssim 35\,000\,{\rm K}$ and that Sgr~A~West or minispiral clouds are transient in nature with the lifetime of $1\,000-10\,000$ years. With the inauguration of the Very Large Array in 1980, it was possible to produce the 5 GHz radio map of Sgr~A West with the comparable angular resolution of $2''\times 8''$ as that of the infrared images at $10\,{\rm \mu m}$ \citep{1981ApJ...250..155B}. Both the radio and the infrared images exhibited the similar structures, in particular the infrared thermal emission peaks corresponded to the peaks in the radio continuum. The ratio of the flux densities of the discrete sources between the infrared and the radio domains was high, $\sim 100-1000$, which indicated the ionization by the soft thermal spectral energy distribution corresponding to $\sim 25\,000\,{\rm K}$, i.e. by the OB stars in the central stellar cluster.

 The designation Sgr~A* was adopted for the first time by Robert Brown \citep{1982ApJ...262..110B} to distinguish the variable non-thermal compact source from the more extended thermal region Sgr~A West. 

\subsubsection*{Sgr~A* and the environment of its SMBH}
The detection of stellar proper motions, the measurement of radial velocities and mainly the fitted quasi-Keplerian orbits of stars in the innermost arcsecond from the compact radio source Sgr~A* led to the very reliable determination of the mass of the central dark object, $M_{\bullet}\simeq 4 \times 10^6\,{M_{\odot}}$ \citep{1996Natur.383..415E,1997MNRAS.284..576E,1998ApJ...509..678G,2002Natur.419..694S,2009ApJ...692.1075G,2016ApJ...830...17B,2017ApJ...845...22P,2017ApJ...837...30G}. 

Under the assumption that Sgr~A* is a black hole, the corresponding Schwarzschild radius is $R_{\rm Schw}=1.2 \times 10^{12}\,{\rm cm}(M_{\bullet}/4\times 10^6\,M_{\odot})$ and the expected mean density is,

\begin{equation}
  \rho_{\bullet}=1.7 \times 10^{25} \left(\frac{M_{\bullet}}{4\times 10^6 \, M_{\odot}}\right) \left(\frac{R_{\rm Schw}}{3.9 \times 10^{-7}\,{\rm pc}} \right)^{-3}\,M_{\odot}{\rm pc^{-3}}\,.
  \label{eq_density_blackhole}
\end{equation}

If the measurements of the orbiting matter (stars and hot gas) approach this value, we can consider Sgr~A* to be a black hole in a classical relativistic sense for simplicity, excluding extended configurations such as a cluster of old compact remnants. On the other hand, it is experimentally very difficult to distinguish a black hole from other very compact configurations of matter that lack an event horizon, such as boson stars, fermion balls, and other options \citep[for a general discussion and the notion of ``macroquantumness'', read more in][]{2017FoPh...47..553E}.

Among the determined stellar orbits in the innermost stellar cluster \citep[so-called S cluster; see the pioneering works by][]{1996Natur.383..415E,1997MNRAS.284..576E,1998ApJ...509..678G}, the tightest constraint for the density of the dark mass comes from the monitoring of B-type star S2 (or S0-2) with the pericentre distance of $r_{\rm P} \simeq 5.8 \times 10^{-4}\,{\rm pc}$ \citep{2017ApJ...845...22P,2002Natur.419..694S,2009ApJ...692.1075G,2017ApJ...837...30G}

\begin{equation}
  \rho_{\rm S2}=5.2 \times 10^{15} \left(\frac{M_{\bullet}}{4.3 \times 10^6\,M_{\odot}}\right)\left( \frac{r_{\rm P}}{5.8 \times 10^{-4}\,{\rm pc}} \right)^{-3}\,M_{\odot}{\rm pc^{-3}}\,.
  \label{eq_density_blackholeS2}
\end{equation}
The largest density constraint is given by 3$\sigma$ Very Long Baseline Interferometry (VLBI)\-measured source size of $\sim 37\mu {\rm as}$ \citep{2008Natur.455...78D}, which in combination with the lower mass limit of $M_{\rm SgrA*} \gtrsim 4 \times 10^5\,M_{\odot}$ based on the proper motion measurements \citep{2004ApJ...616..872R}, yields the density lower limit of $\rho_{\rm SgrA*} \geq 9.3\times 10^{22}\,M_{\odot}\;{\rm pc^{-3}}$. This is only about two orders of magnitude less than the density expected for a black hole of $\sim 4\times 10^6\,M_{\odot}$, see Eq.~\eqref{eq_density_blackhole}. When taking into account the overall stability of this dark concentration of mass over the Galaxy lifetime of $\sim 10^{10}\,{\rm yr}$, the most plausible stable configuration that can explain such a large concentration of mass within a small volume emerges within the framework of general relativity: a singularity surrounded by an event horizon -- a black hole, ruling out most of the unstable alternatives \citep{2017FoPh...47..553E}, such as dense stellar clusters and boson stars. 

From multi-wavelength observations over the past fifty years, ranging from radio, millimeter, sub-millimeter, through infrared, X-ray, up to gamma-ray wavelengths, see \citet{2010RvMP...82.3121G} for a review, it has become clear that the supermassive black hole associated with the compact radio source Sgr~A* is not surrounded by vacuum. Instead, it is embedded in the Nuclear Star Cluster (NSC), which is one of the densest stellar clusters in the Galaxy \citep{2014CQGra..31x4007S}, as well as in the multi-phase gas and dust environment \citep{2017A&A...603A..68M}. 

\begin{figure}[tbh!]
  \centering
  \includegraphics[width=0.6\textwidth]{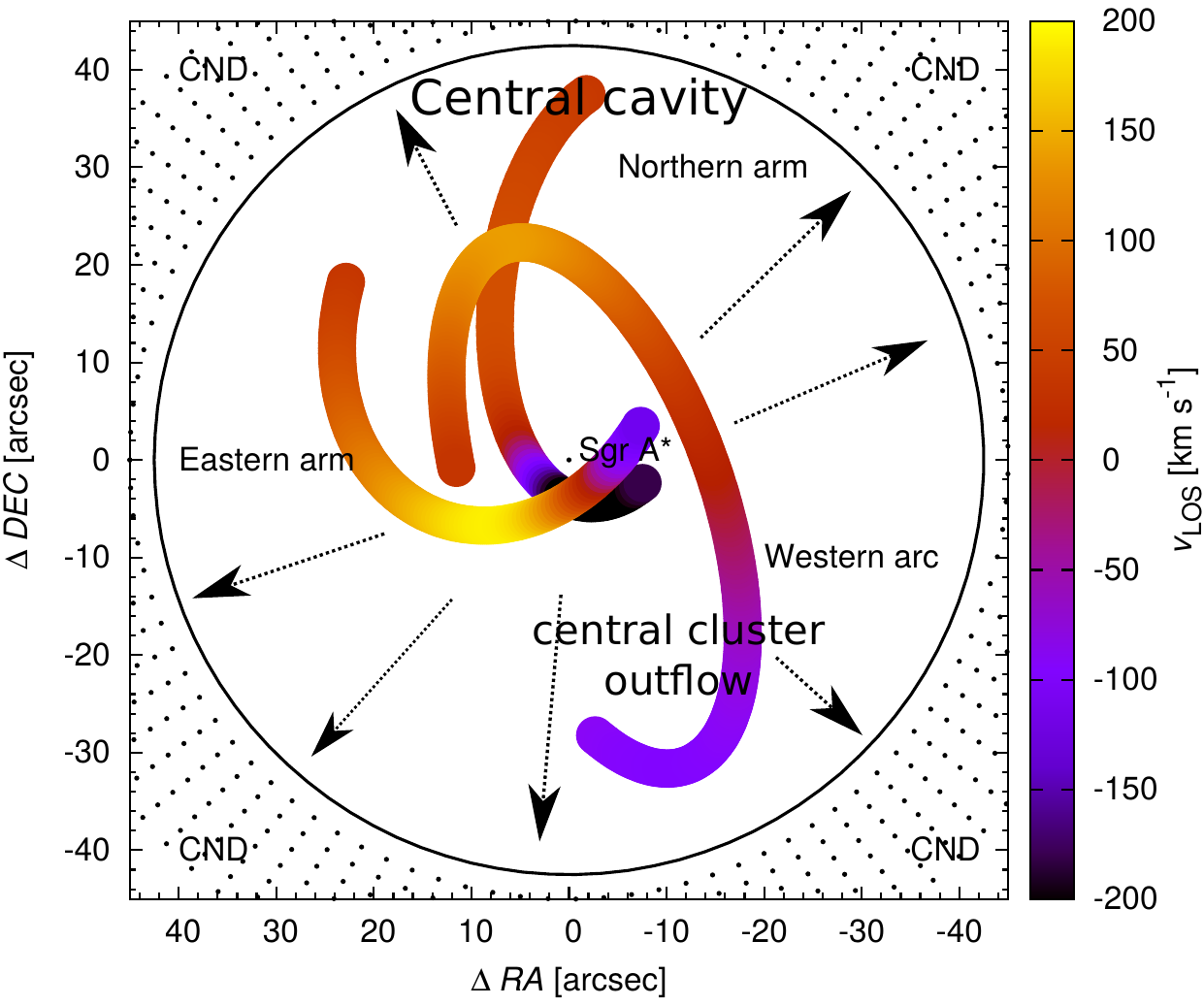}
  \caption{Projected positions of the Circum-Nuclear Disk (CND) and the Minispiral arms. The Minispiral is color-coded to express radial-velocities of the emitting ionized gas along the three arms according to the quasi-Keplerian model of \citet{2010ApJ...723.1097Z}. The inner rim of the clumpy CND is located at $\sim 1.5\,{\rm pc}$, where the region of rarefied ionized gas -- the so-called {\em Central Cavity} -- is located. The central outflow from the central stellar cluster centered at Sgr~A* plausibly interacts with the inner rim of the CND, whose material fills the cavity. This region is, however, not quite homogeneous and spherically symmetric. Several gaseous phases -- ionized, neutral, molecular -- as well as dust grains are located in the central region.}
  \label{img_cnd_cavity}
\end{figure}

In Fig.~\ref{img_nsc_sgra}, we illustrate different components of the inner portions of the Galaxy. The figure captures the clumpy Circum-Nuclear Disk (CND), which is mostly composed of warm $(\sim 100\,{\rm K})$ neutral and molecular gas of the total mass of $M_{\rm CND} \approx 10^5\,M_{\odot}$ and the number densities of $n_{\rm CND}\approx 10^4-10^7\,{\rm cm^{-3}}$ \citep{1985ApJ...297..766G,1986A&A...161..334S,1987ApJ...318..124G,1989A&A...209..337M,2005ApJ...622..346C,2012A&A...542L..21R,2017ApJ...850..192M}. The peak of the molecular emission is at $\sim 1.6\,{\rm pc}$ from Sgr~A*. However, the outer edge of the CND at $\sim 4 - 7\,{\rm pc}$ approximately coincides with the outer edge of the Nuclear Star Cluster. The inner edge of the CND is rather sharp at $\sim 1.5\,{\rm pc}$, where the gas density drops below $\sim 10^{4}\,{\rm cm^{-3}}$ \citep{2016MNRAS.459.1721B}. This inner part is also denoted as the \textit{central cavity}, which is composed mostly of rarefied gas ionized by hot massive OB stars in this region, see \citet{2016MNRAS.459.1721B} and references therein. The central cavity is considered to be refilled with the gas from the CND due to the interaction between its inner rim and an almost spherical outflow from the central stellar cluster, see Fig.~\ref{img_cnd_cavity} for an illustration.

\begin{figure}[tbh!]
  \centering
  \includegraphics[width=0.9\textwidth]{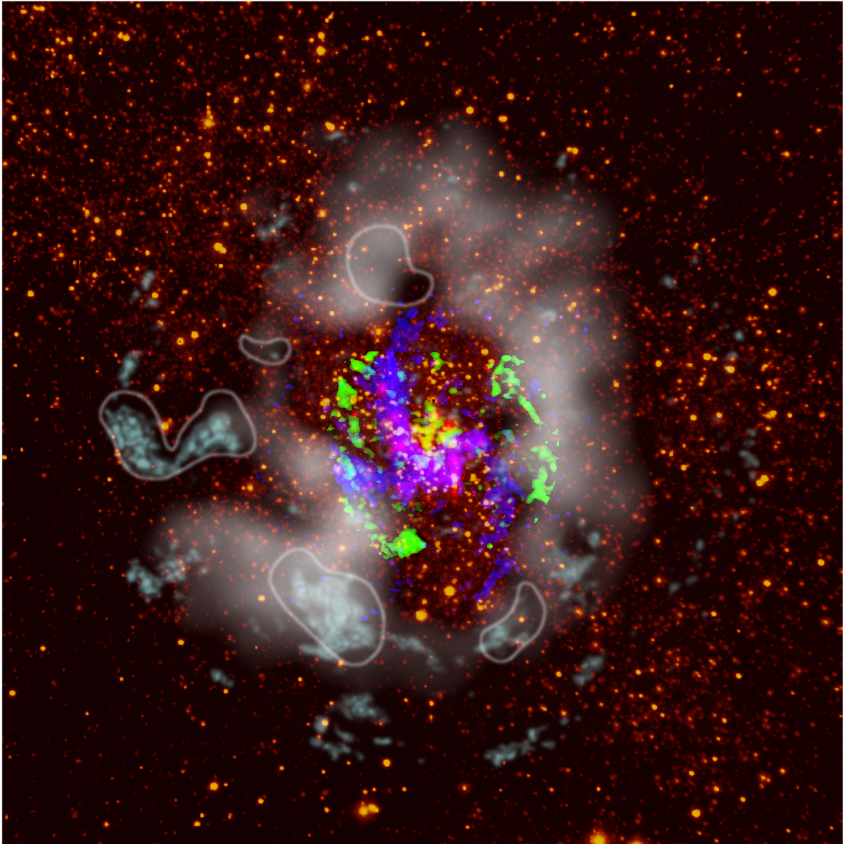}
  \caption{Combined near-infrared, sub-mm, and inverted X-ray image of the innermost 5.6 $\times$ 5.6 parsecs of the Galactic centre. Colors represent different structures and phases of matter: ionized gas (blue; based on 250 GHz continuum and H39$\alpha$ recombination line observations with ALMA, $0.75''$ beam; \citeauthor{2017A&A...603A..68M}, \citeyear{2017A&A...603A..68M}), molecular gas at the inner rim of the circumnuclear disk (green; CS$(5-4)$ line-emission observations with ALMA, $0.75''$ beam; \citeauthor{2017A&A...603A..68M}, \citeyear{2017A&A...603A..68M}), stars and warm dust (red; $3.8\,{\rm \mu m}$ L'-band observations with VLT NACO; \citeauthor{2012A&A...545A..70S}, \citeyear{2012A&A...545A..70S}). On the larger scales, the extended white ring marks the footprint of the circumnuclear disk that is seen as a depression in the X-ray diffuse emission, see \citet{2018MNRAS.474.3787M} for the detection and the detailed analysis. The stellar background consists of yellow point sources and it is the combination of ISAAC 1.19, 1.71, and 2.25\,${\rm \mu m}$ narrow-band images, see \citet{2013A&A...549A..57N} for details. The cyan patches along the circumnuclear disk stand for $N_2 H^{+}\,(1-0)$ molecular complexes \citep{2017A&A...603A..68M}, the largest of which coincide with the dark extinction regions in the near-infrared domain (encircled regions). These are dense molecular clouds that are candidates for star-forming regions in the central $\sim 10\,{\rm pc}$ from Sgr~A*. Figure adopted from \citet{2019JPhCS1258a2019E}, courtesy of \citet{2017A&A...603A..68M,2018MNRAS.474.3787M}.}
  \label{img_minispiral_multiwav}
\end{figure}

However, the medium in the central cavity is in general multi-phase, with the three Minispiral arms (also denoted as thermal Sgr~A~West) reaching larger gas densities of a few of $\sim 10^4\,{\rm cm^{-3}}$ \citep{2009ApJ...699..186Z,2010ApJ...723.1097Z} in comparison with the hot plasma in their surroundings. The gas in the arms is mostly ionized and has the electron temperature of $\sim 5\,000-13\,000\,{\rm K}$ \citep{2010ApJ...723.1097Z,2012A&A...538A.127K}. The Minispiral arms have a total gas mass of $\sim 100\,M_{\odot}$ and the Northern and the Western arms are considered to be ionized inner parts of the CND \citep{2005ApJ...622..346C}. The dynamics of the three arms can be fitted by three bundles of quasi-Keplerian ellipses with Sgr~A* at the focus \citep{2010ApJ...723.1097Z}. Molecular gas is detected in Sgr~A~West as well \citep{2017A&A...603A..68M}, mostly outside of the Minispiral arms that are predominantly ionized. In Fig.~\ref{img_minispiral_multiwav}, three different phases are depicted: ionized gas (blue; based on 250 GHz continuum observations with ALMA, $0.75''$ beam), molecular gas (green; CS$(5-4)$ line-emission observations with ALMA, $0.75''$ beam), stars and warm dust (red; $3.8\,{\rm \mu m}$ observations with VLT). 

Because of the presence of other bodies in the vicinity of Sgr~A*, the classical general-relativistic solutions derived for vacuum, i.e. Kerr-Newman metric in the most general form (which turns into the Kerr metric for the assumed zero electric charge and subsequently the Schwarzschild case for an uncharged non-rotating black hole), are basically not valid. However, since typical stars and gas clouds have negligible masses in comparison with Sgr~A*, they can be treated as test particles in most relevant cases. In many cases, it is sufficient to treat the motion of stars in the post-Newtonian framework \citep{2001A&A...374...95R}. When necessary, the motion of test particles can be calculated for the Kerr metric with relevant perturbation effects of the stellar cluster and the gaseous-dusty structures. In particular, general-relativistic treatment of self-gravitating disks and rings made up of stars and gas was analyzed \citep{2004CQGra..21R...1K,2010MNRAS.404..545S,2012MNRAS.425.2455S,2015MNRAS.451.1770W}. Such structures can be detected in the standard AGN model (standard thin Shakura-Sunyaev accretion disk) as well as in the Nuclear Star Cluster around Sgr~A* in the form of the Clock-Wise stellar disk \citep{2010ApJ...708..834B} between $0.05$ and $0.5$ parsecs. The analysis of the 3D dynamical state of the S cluster revealed that it consists of at least two nearly perpendicular stellar disks, i.e. the stellar distribution deviates significantly from isotropy \citep{2020ApJ...896..100A}. 

\begin{figure}[tbh!]
  \centering
  \includegraphics[width=0.49\textwidth]{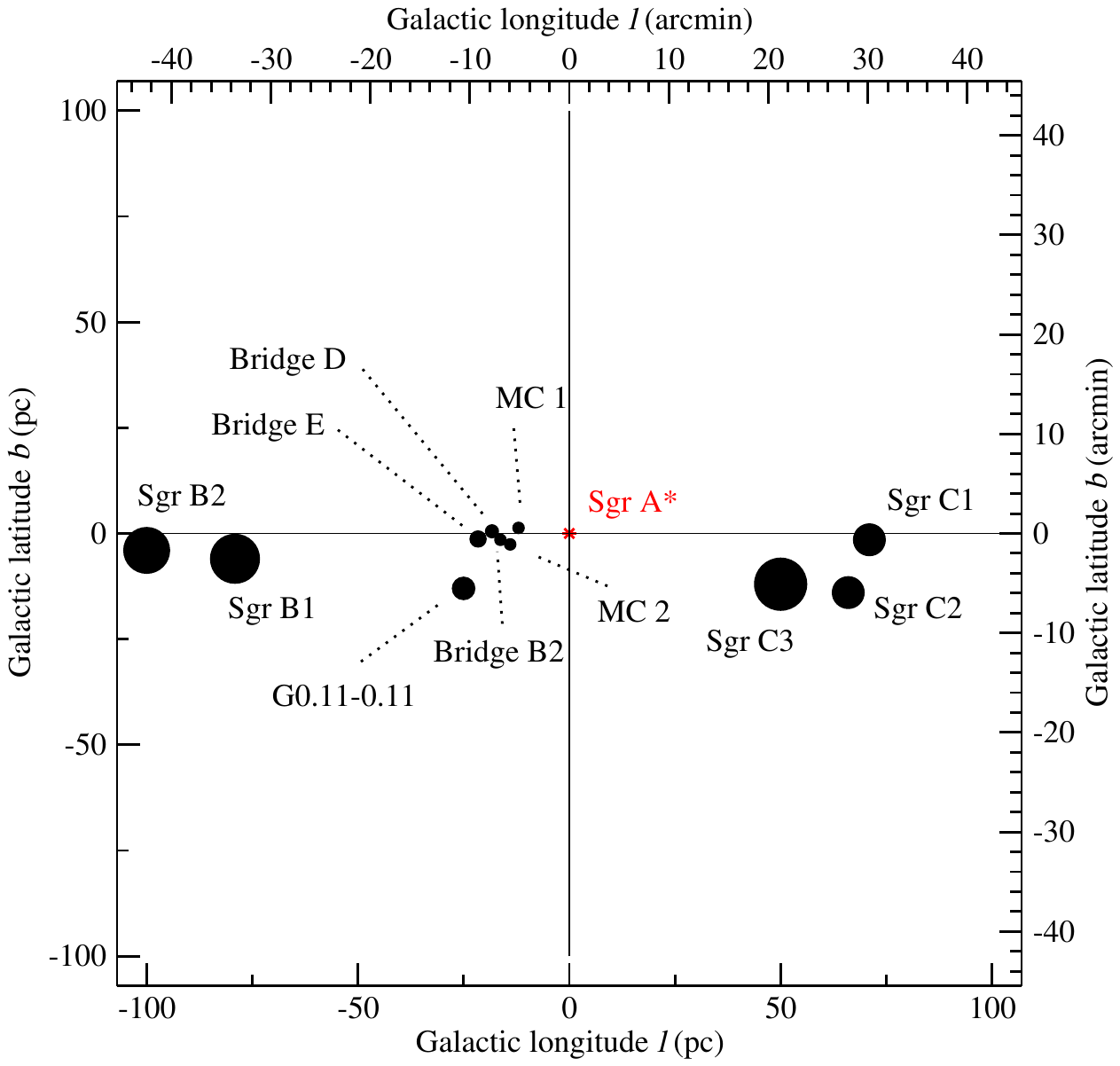}
  \includegraphics[width=0.49\textwidth]{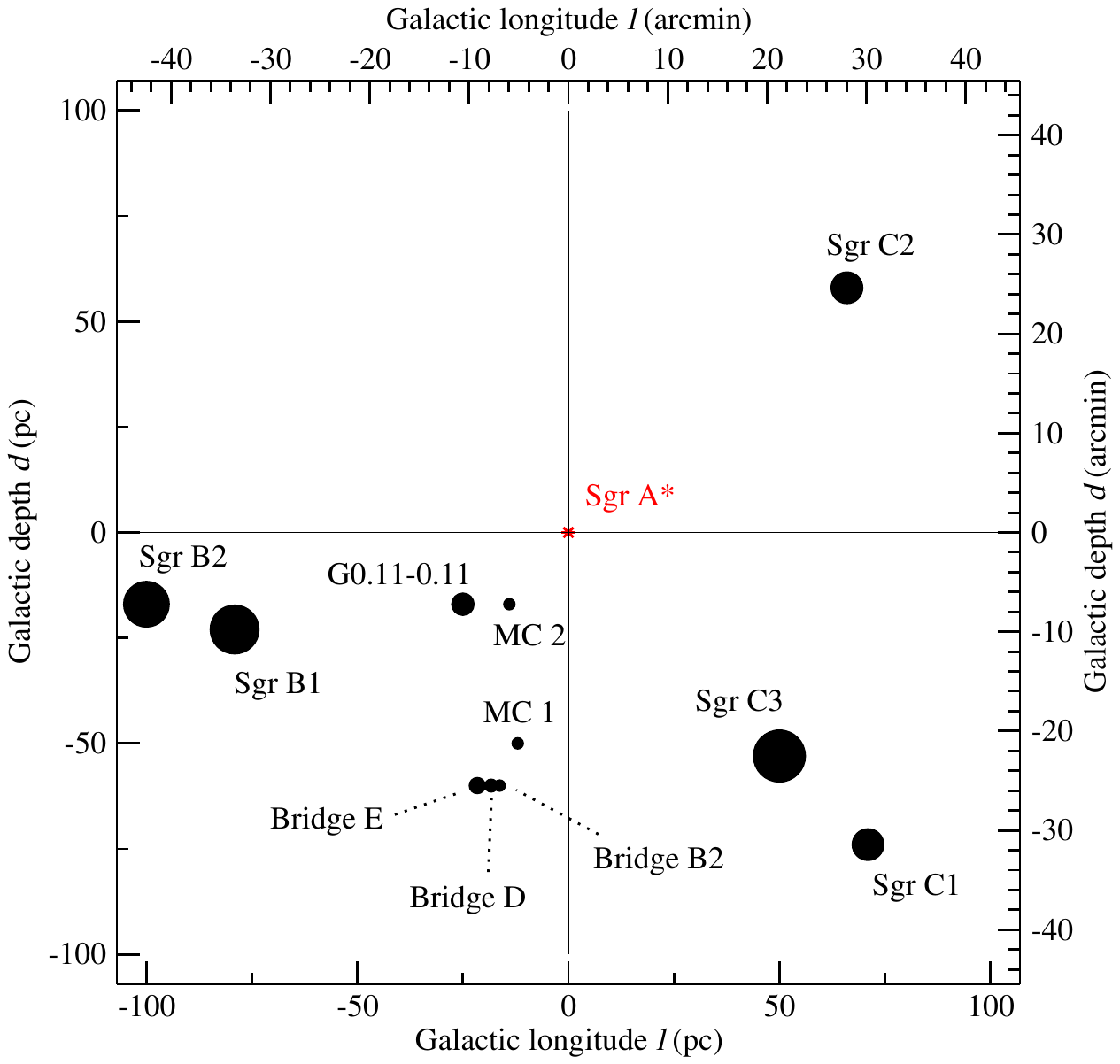}
  \caption{Positions of the molecular clouds around Sgr A* \citep{2011ApJ...735L..33M,2012A&A...545A..35C} can be 
  reconstructed with improved precision of future X-ray imaging polarimetry, 
  assuming that the primary radiation signal from the SMBH accretion has been scattered towards the observer
\citep{2015A&A...576A..19M}. Left panel is oriented along the edge-on view (as seen from Earth), whereas the right
panel shows the perpendicular view, i.e. from the perspective of the pole-on direction.}
  \label{img_minispiral_multiwav2}
\end{figure}

\citet{1989Natur.339..603K} suggested that the X-ray emission, and in particular the intensity of 6.7 keV iron line speak 
in favour of presence of interstellar medium shock-heated by an energetic explosion.
Although Sgr A* appears to be very quiet and underluminous in its present state, there is circumstantial evidence for much more vigorous activity in the relatively recent history over past few hundred years. The evidence has been accumulated for at least two high-luminosity phases during which the SMBH environment was illuminated by intense X-rays \citep{1998MNRAS.297.1279S,2009PASJ...61S.241I,2010ApJ...714..732P}. Going back in the history, the activity can be recognized in the early broadband (15 arcmin resolution) images from Granat satellite \citep{1993ApJ...407..606S}. Indeed, in order to probe the enormously crowded field surrounding Sgr A*, an excellent imaging capability is essential. But in addition to it, the imaging polarimetry can constrain the physical processes with much higher confidence. Especially the scenario of centrally illuminated distribution of clouds orbiting the SMBH automatically suggests a possibility of a polarized signal (the reflection component) and the polarization angle and degree can be used to trace the 3D positions of the clouds \citep{2014MNRAS.441.3170M}.

\begin{figure}[tbh!]
    \centering
    \includegraphics[width=0.7\textwidth]{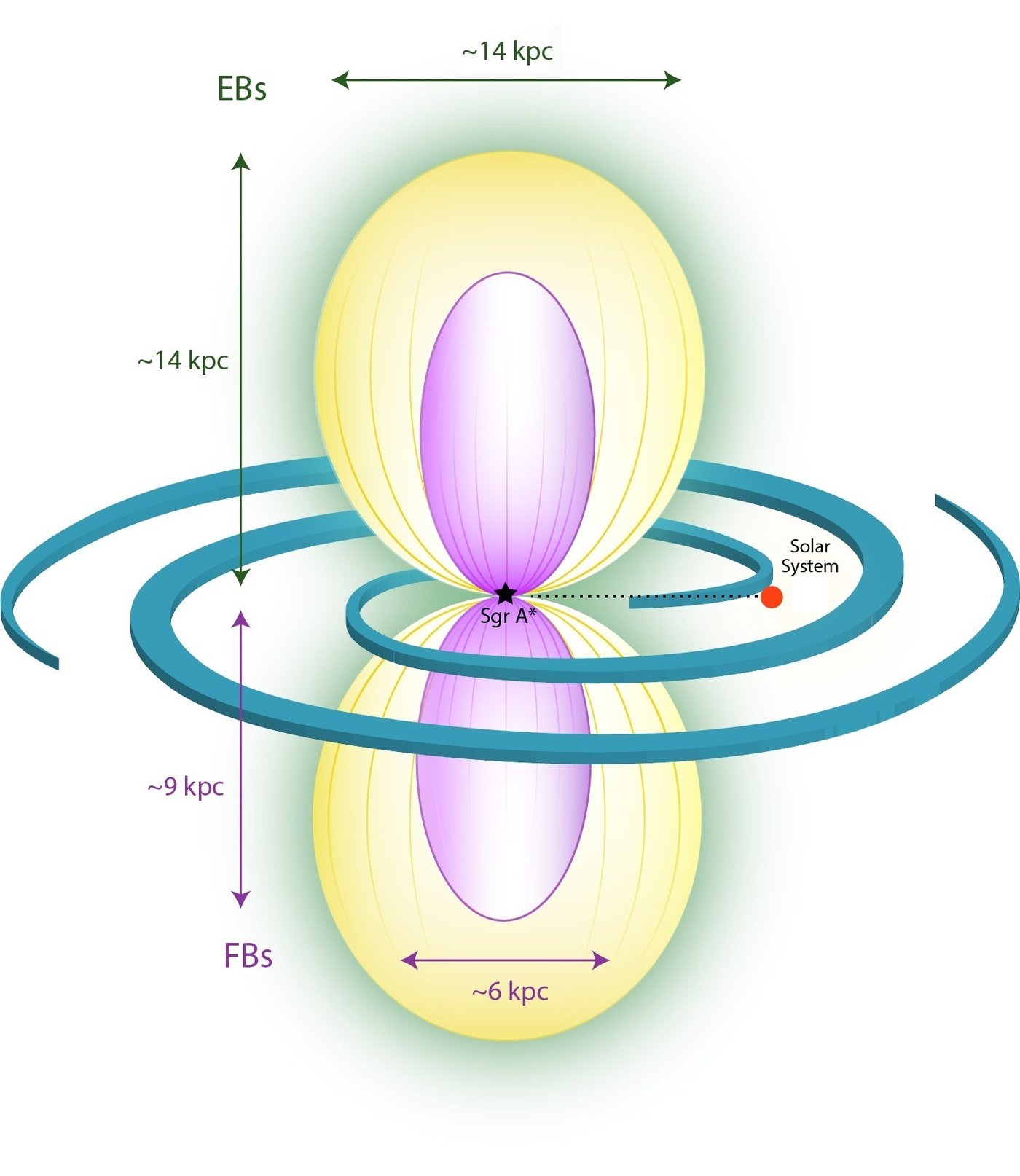}
    \caption{A schematic comparison of the size of the X-ray eROSITA bubbles (EBs) surrounding the previously discovered $\gamma$-ray Fermi bubbles (FBs) above and below the Galactic plane. The inflated bubbles seen in the X-ray domain have the length-scale comparable with the whole Galactic disk and they are clearly larger than the more compact $\gamma$-ray Fermi bubbles. For a detailed description of the observational analysis of eROSITA data; see \citet{2020Natur.588..227P}. Image credit: Max Planck Institute for Extraterrestrial physics.}
    \label{fig_erosita_fermi_bubbles}
\end{figure}

\begin{figure}
    \centering
    \includegraphics[width=0.6\textwidth]{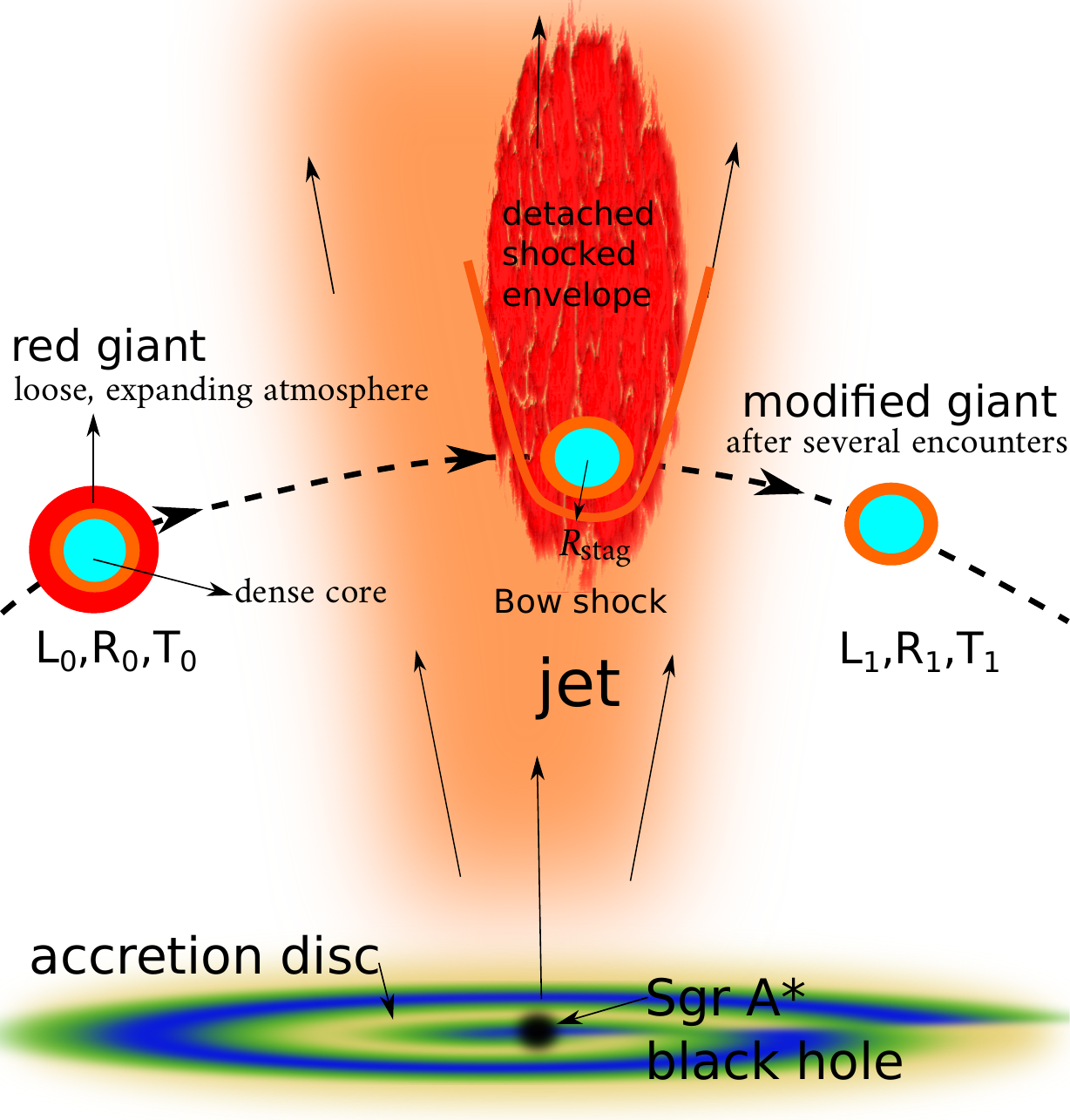}
    \caption{A scheme of the jet-ablation mechanism that likely operated in the Galactic center during the previous Seyfer-like state. As the red giant with a loose, expanding atmosphere enters the jet, its envelope is shocked and partially detached. After the interaction, the giant is truncated and has a higher effective temperature and a smaller infrared luminosity. For details, see \citet{2020ApJ...903..140Z}. Adopted from \citet{2020arXiv201112868Z}.}
    \label{fig_jet_RG}
\end{figure}

\begin{figure}
    \centering
    \includegraphics[width=0.7\textwidth]{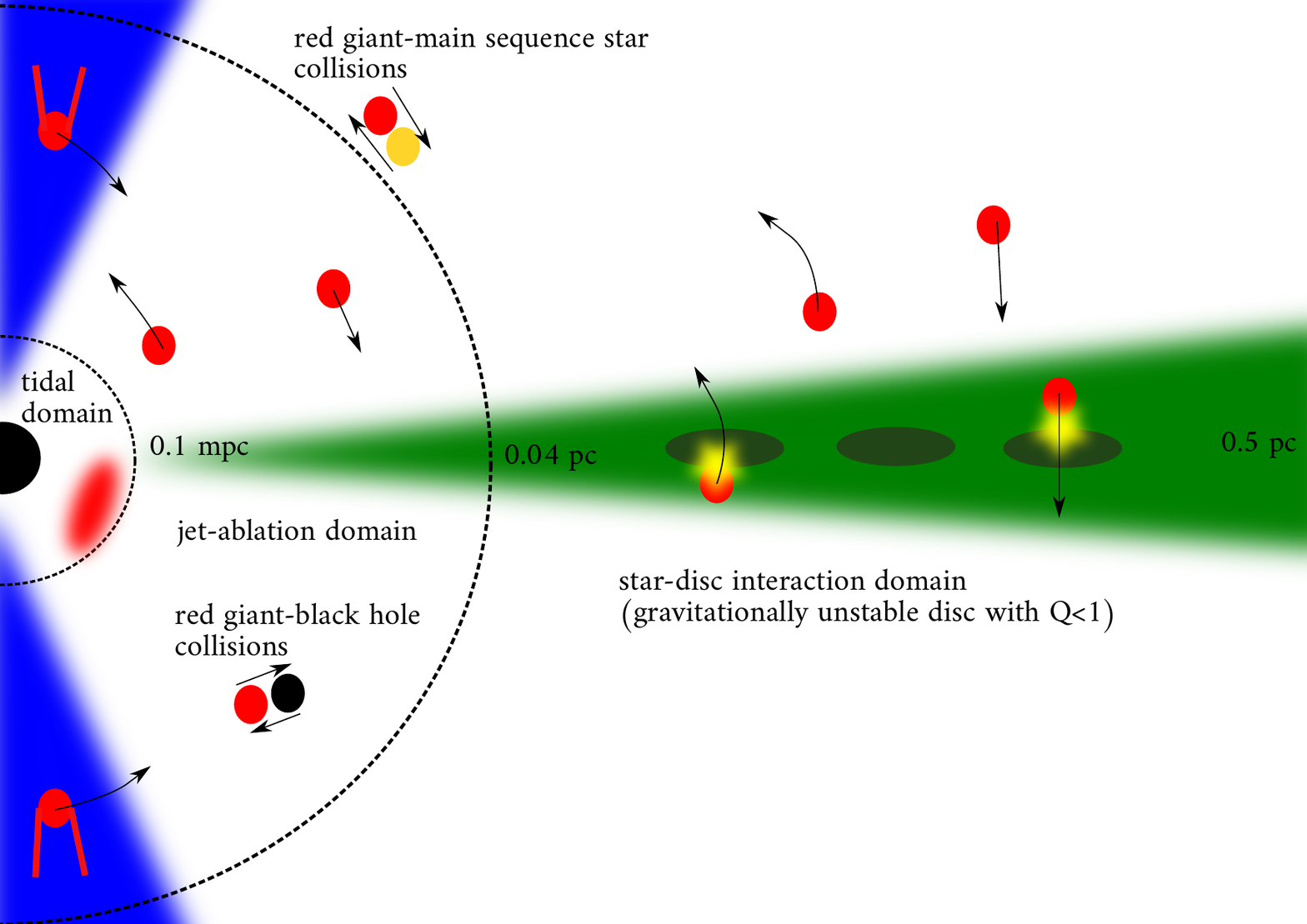}
    \caption{Illustration of different mechanisms that likely contributed to the observed dearth of bright red giants in the Galactic center. At smaller scales, the tidal disruption dominated, while further away the loose red-giant envelopes were ablated and removed via the collisions with the denser parts of the accretion disk. At intermediate scales, the jet-ablation likely contributed to the depletion of the brightest red giants.  Adopted from \citet{2020arXiv201112868Z}.}
    \label{fig_depletion_mechanisms}
\end{figure}

Every few million years the Galactic center is expected to significantly increase its activity by several orders of magnitude in terms of the bolometric accretion luminosity. During this Seyfert-like stage, Sgr~A* was surrounded by a dense accretion disk that could also have powered a bipolar jet with the kinetic luminosity of $10^{41}-10^{44}\,{\rm erg\,s^{-1}}$ for $\sim 0.1-1.0$ Myr, see \citet{2020ApJ...903..140Z} and references therein. One of the traces of such an enhanced activity could be large-scale $\gamma$-ray \textit{Fermi} bubbles \citep{2010ApJ...724.1044S} and even more extended \textit{eRosita} bubbles \citep{2020Natur.588..227P} that could have been inflated by the nuclear outflow or a collimated jet. The alternative theory of their inflation is the recent nuclear starburst accompanied by supernova explosions, but this seems to be currently ruled out based on energetic grounds \citep{2019ApJ...886...45B}. The X-ray eRosita bubbles with the length-scale of $\sim 14\,{\rm kpc}$ are larger  than the $\gamma$-ray Fermi bubbles ($\sim 9\,{\rm kpc}$), see Fig.~\ref{fig_erosita_fermi_bubbles} for the comparison of both large-scale structures. Another signature of the enhanced accretion and jet activity of Sgr~A* in the past million years could be missing bright red giants. Evolved late-type stars can form CO molecules in their atmospheres and inside $\sim 0.5\,{\rm pc}$, these is a drop in the CO bandhead absorption, which indicates that there is a decrease in the number of red giants close to Sgr~A* \citep{1990ApJ...359..112S}. More precisely, bright red giants exhibit a flat stellar surface-density distribution while fainter late-type stars have a cuspy surface-density profile \citep{2020A&A...641A.102S}, which indicates that a preferential depletion of larger, brighter red giants took place. The jet activity fits into this scenario of the preferential destruction of bright red giants since it can effectively ablate the upper layers of diluted and extended red-giant envelopes \citep{1969AcA....19....1P}, which would lead to its decreased infrared luminosity \citep{2020ApJ...903..140Z}, see Fig.~\ref{fig_jet_RG} for the illustration. The jet-ablation can complement other mechanisms of the red-giant depletion in the Galactic center: on smaller scales, the tidal disruption of red giants operates, while further away the red giants suffer ablation and mass loss while passing through the denser parts of the accretion disk. In addition, star-star and star-compact remnant collisions can contribute to the red-giant destruction, see Fig.~\ref{fig_depletion_mechanisms} for the illustration of different mechanisms.

\subsubsection*{Sgr~A* and hot accretion flows}

The compact radio source Sgr~A* is a very faint source in comparison with active galactic nuclei, with the bolometric (accretion) luminosity several orders of magnitude below its Eddington limit. With the measured bolometric luminosity of $L_{\rm Sgr~A*}\lesssim 10^{37}\,{\rm erg\,s^{-1}}$ \citep{1998ApJ...492..554N,2010RvMP...82.3121G,2017FoPh...47..553E} and the mass of $M_{\bullet}\simeq 4 \times 10^6\,M_{\odot}$, the Eddington ratio for Sgr~A* is,

\begin{equation}
  \lambda_{\rm Edd}=\frac{L_{\rm Sgr A*}}{L_{\rm Edd}}\lesssim \frac{10^{37}\,{\rm erg\,s^{-1}}}{5.2\times 10^{44}\,{\rm erg\,s^{-1}}}\approx 1.9 \times 10^{-8}\,,
  \label{eq_Eddington_ratio_SgrA*}
\end{equation}
i.e., Sgr~A* is eight to nine orders of magnitude below the Eddington limit for $4\times 10^6\,M_{\odot}$ black hole. From the Faraday rotation measurements of Sgr~A*, \citet{2007ApJ...654L..57M} inferred the upper and the lower limit of the accretion rate, $\dot{M}_{\rm acc}=2\times 10^{-7}\,{\rm M_{\odot}\,yr^{-1}}$ and $\dot{M}_{\rm acc}=2\times 10^{-9}\,{\rm M_{\odot}\,yr^{-1}}$, respectively. The source of the material are potentially the stellar winds of $\sim 100$ massive Wolf-Rayet (WR) of spectral types O and B that orbit the black hole in the innermost one parsec \citep{2009ApJ...690.1463L,2010ApJ...708..834B}. The gas from stellar winds is captured by the gravitational pull of the black hole within the Bondi radius $r_{\rm B}$ \citep{1952MNRAS.112..195B}, at which the gravitational potential of Sgr~A* overcomes the thermal pressure of the ionized plasma,

\begin{equation}
  r_{\rm B}\simeq \frac{2GM_{\bullet}}{c_{\rm s}^2}\approx 0.15\left(\frac{M_{\bullet}}{4\times 10^6\,M_{\odot}}\right)\left(\frac{\mu_{\rm HII}}{0.5}\right)\left(\frac{\gamma}{1.4}\right)^{-1}\left(\frac{T_{\rm p}}{10^7\,{\rm K}}\right)^{-1}\,{\rm pc}\,,
  \label{eq_Bondi_radius}
\end{equation}
where the mean atomic weight is set to $\mu_{\rm HII}=0.5$, which corresponds to the fully ionized plasma or HII region.

\begin{figure}[tbh!]
    \centering
    \includegraphics[width=0.99\textwidth]{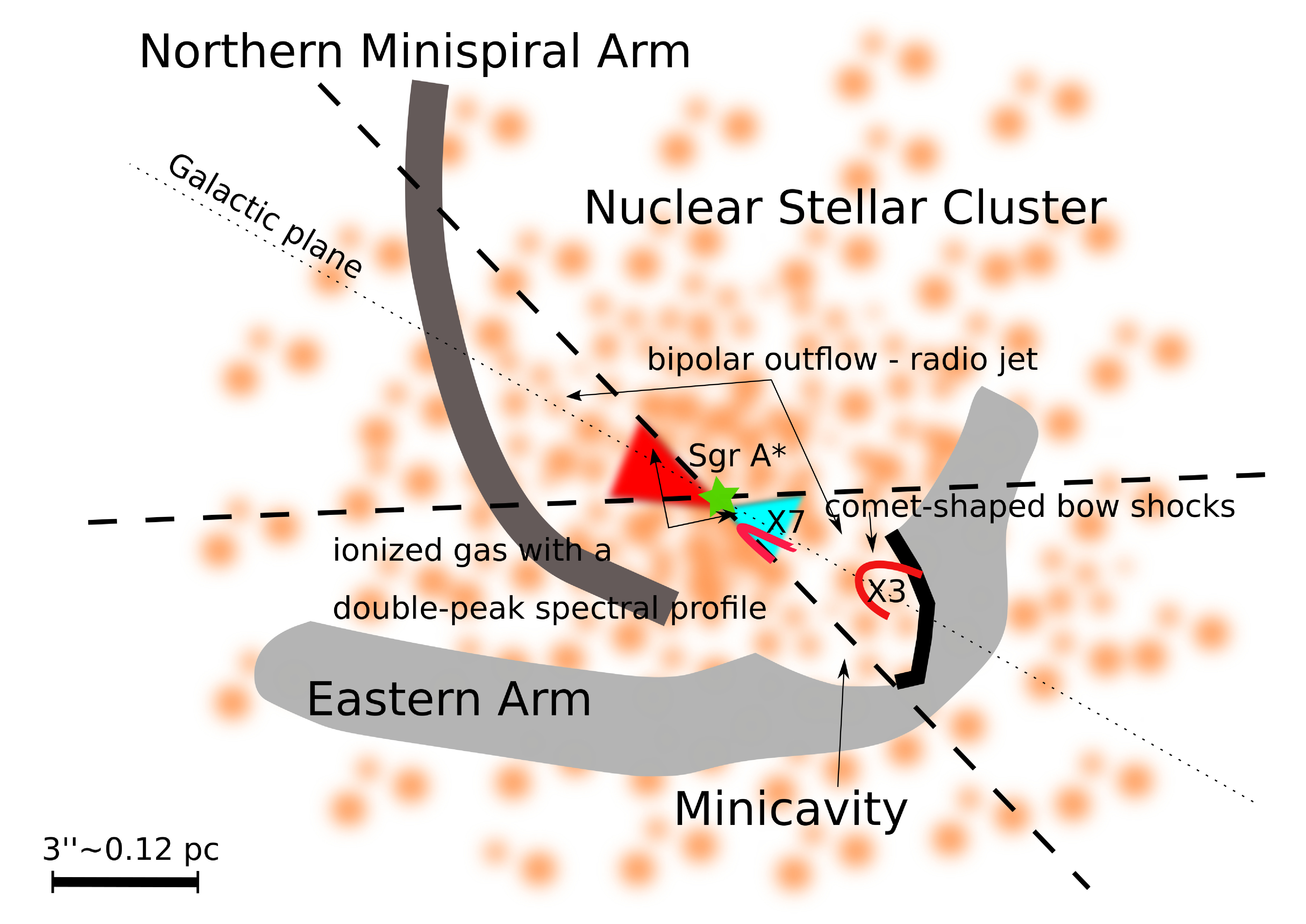}
    \caption{An illustration of the orientation of a bipolar outflow -- a radio jet with respect to the Galactic plane and other structures in the Galactic center according to \citet{2020MNRAS.499.3909Y}. In particular, we depict the parallel orientation of two prominent bow-shock sources X7 and X3, which are located $\sim 0.8''$ and $\sim 3.4''$ from Sgr~A* in projection, respectively; see also \citet{2010A&A...521A..13M,2019A&A...624A..97P,2021ApJ...909...62P}.}
    \label{fig_jet_illustration}
\end{figure}

Inside the Bondi radius expressed by Eq.~\eqref{eq_Bondi_radius}, the accretion flow towards Sgr~A* is currently well-explained in the framework of radiatively-inefficient accretion flows (RIAFs). One of the models of RIAFs is an optically thin advection dominated accretion flow (ADAF), which is characterized by \textit{advective cooling} in comparison with standard accretion disks (see Chapter~\ref{Radiation_processes}), which cool efficiently down radiatively. By advective cooling it is understood that the energy released via the flow viscosity is stored as entropy and essentially is transported inward by accretion. If we look at the flow element at the fixed radius from a rest frame, the material with larger entropy moves inward and is replaced by the lower-entropy material from an adjacent outer part of the flow. This way the ADAF flows can be become much hotter than standard thin disks, which leads to high-energy emission. The peak of the spectral energy distribution of Sgr~A* hot flow is in the sub-mm/mm domain, which is in contrast with the spectrum of the thin, optically thick disks typical of AGN that peak in the UV domain, which is known as the Big Blue Bump, see also Fig.~\ref{agn_sed} for comparison. The submm/mm bump of Sgr~A* spectrum divides the optically thick part of the spectrum towards lower frequencies and the optically thin part towards the higher frequencies. The optically thin part of the spectrum is characterized by order-of-magnitude flares in the near-infrared (NIR) and the X-ray domains. There are about $\sim 3-5$ flares in the NIR domain, while about one X-ray flare a day. In addition, there is a correlation between the two domains -- every X-ray flare has a NIR counterpart, while the opposite is not true, see \citet{2021ApJ...917...73W} for a detailed analysis and multi-wavelength modelling of the flares. 

Another property of hot flows is the radial dependency of the accretion rate, $\dot{M}_{\rm in}\approx \dot{M}_{\rm in}(r)$ \citep{2014ARA&A..52..529Y}. In other words, at the Bondi radius, the accretion rate is as large as $\dot{M}_{\rm B}\approx 10^{-3}\,{\rm M_{\odot}\,yr^{-1}}$, while Sgr~A* accretes less than $1\%$ of the Bondi rate as inferred from Faraday-rotation measurements \citep{2006ApJ...640..308M,2007ApJ...654L..57M}. Hence, most of the accreted matter at the Bondi radius is lost via the outflow, which can be collimated and form a bipolar outflow or a radio jet close to the Galactic plane \citep{2020MNRAS.499.3909Y}. This outflow can interact with the minispiral streamers, and the minicavity region is speculated to have been formed by the interaction with the nuclear outflow. In addition, there are several comet-shaped stellar bow-shock sources -- X7 \citep{2010A&A...521A..13M,2021ApJ...909...62P}, X3 \citep{2010A&A...521A..13M} and X8 \citep{2019A&A...624A..97P} -- whose axis of symmetry is oriented along the candidate collimated outflow, see also Fig.~\ref{fig_jet_illustration} for the illustration. Alternatively, this outflow could originate from the stellar cluster and not from the accretion flow. Either way, stellar bow shocks are powerful probes of the ambient circumnuclear medium in terms of the ambient-medium flow \citep{2014A&A...567A..21S,2018acps.confE..49Z, 2020A&A...644A.105H}.

\clearpage

\section{Accretion onto supermassive black holes}
\epigraph{\textit{``The key inspiration which determined my future life came with the series of lectures by Bohdan Paczy\'{n}ski. His lectures showed astrophysics as a dynamically developing new branch, with plenty of things to do. This was something completely new, like opening a window and letting in fresh air."}}{--- \textup{Bo\.{z}ena Czerny}}
\label{Radiation_processes}
  \subsection{Strong gravity and accretion tori}



The gravitational field is described by the Kerr metric, which can be written in the well-known Boyer--Lindquist coordinate system $(t,r,\theta,\phi)$ \citep{1973grav.book.....M,1983mtbh.book.....C}. This spacetime is asymptotically flat and it obeys the axial symmetry about the rotation axis and stationarity with respect to time; the singularity is hidden below the event horizon. The mass $M$ of the black hole is concentrated in the origin of the coordinate system. The spin parameter $a$ of the Kerr metric describes its rotation; the condition about the presence of the outer event horizon at a certain radius, $r=R_+$ (where the horizon encompasses the singularity) leads to the maximum value of the dimensionless spin rate: $|a\leq1|$. The solution can be then written in the form of the metric element \citep{1973grav.book.....M,1983mtbh.book.....C,1984ucp..book.....W}
\begin{equation}
{\rm d}s^{2} = 
  -\frac{\Delta\Sigma}{A}\,{\rm d}t^{2}
   +\frac{\Sigma}{\Delta}\,{\rm d}r^{2}
   +\Sigma\,{\rm d}\theta^{2}
+\frac{A\sin^2\theta}{\Sigma}\;
   \left({\rm d}\phi-\omega\,{\rm d}t\right)^{2},
\label{metric}
\end{equation}
where $\Delta(r)=r^{2}-2r+a^{2}$,
$\Sigma(r,\theta)=r^{2}+a^{2}\cos^2\theta$, $A(r,\theta)=(r^{2}+a^{2})^{2}-{\Delta}a^{2}\sin^2\theta$, $\omega(r,\theta)=2ar/A(r,\theta)$ (geometrical units are assumed with the speed of light $c$ and gravitational constant $G$ set to unity). 

\begin{figure}[tbh!]
\begin{center}\vspace*{-16em}
\includegraphics[width=\textwidth]{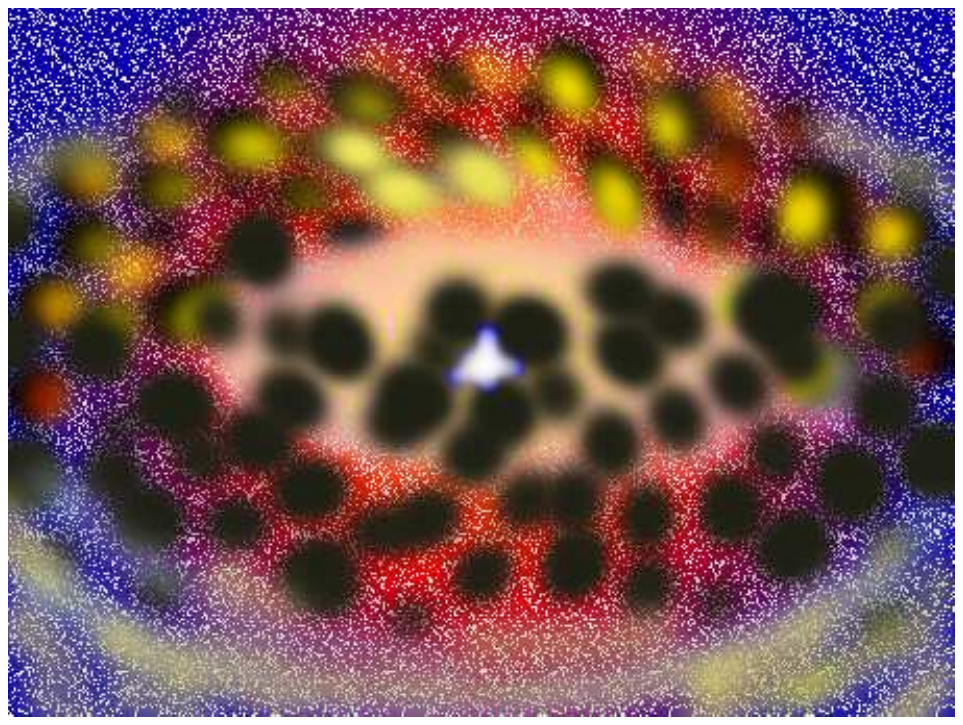}\vspace*{-8em}
\caption{An illustration of an active galactic nucleus. The main components of the system are shown: the supermassive black hole is hidden within the bright accretion region in the centre, surrounded by orbiting clouds and stars. The origin of clouds is probably due to disruption of the accretion torus by instabilities; they are irradiated from the centre and the primary X-rays are reprocessed on the clouds surfaces. The
intrinsic spectrum is determined mainly by the Comptonization, depending
sensitively on the ionization level. An almost clean line of sight to the 
central source temporarily emerges. 
This illustration has been adapted from \citeauthor{2001PASJ...53..189K}  \citep{2000MNRAS.318..547K}.}
\label{f0}
\end{center}
\end{figure}

\begin{figure}[tbh!]
\begin{center}
\includegraphics[width=0.7\textwidth]{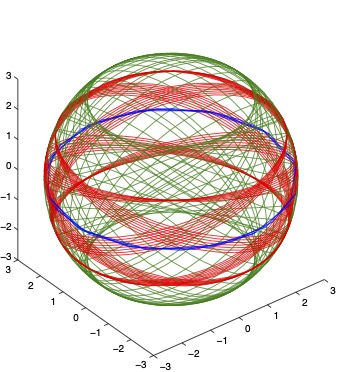}
\caption{Example of three spherical orbits of free particles moving along different inclination angles near a rapidly spinning Kerr black hole in the center of the coordinate system. Axes are labeled with dimensionless length units of $GM/c^2$; the blue circular trajectory resides in the equatorial plane (the plot was drawn by Ond\v{r}ej Kop\'a\v{c}ek).}
\label{spherical}
\end{center}
\end{figure}

Despite the fact that the Kerr metric is only axially symmetric (it does not obey spherical symmetry unless $a=0$) it is interesting to note that free spherical $R={\rm const}$ orbits are possible with the radius above the ISCO at an  arbitrary inclination with respect to the equatorial plane (even residing within the polar plane) as shown in Fig.\ \ref{spherical} \citep{1972PhRvD...5..814W,2021GReGr..53...10T}.

\begin{figure}[tbh!]
  \centering
  \includegraphics[bb = 150 270 470 770,width=0.5\textwidth]{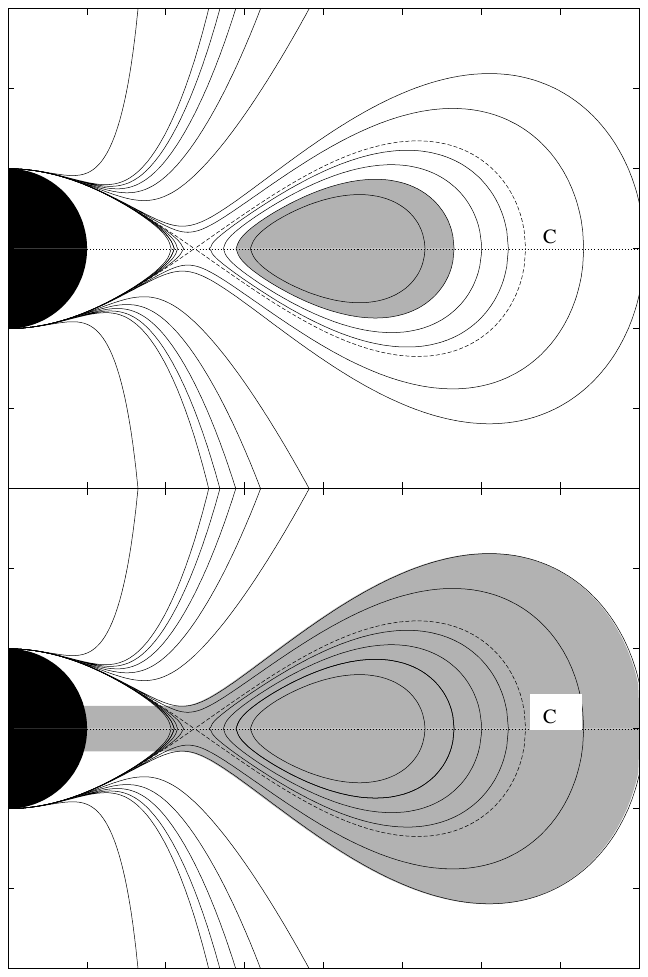}
  \caption{Schematic sketch of meridional sections across the equipotential surfaces $W={\rm const}$. Possible shapes are shown of non-magnetized polytropic tori with constant angular momentum density ($k=4/3$, $l={\rm const})$ near a black hole (denoted by black circle). The fluid can overflow over the critical, self-crossing surface {\sf C} (dashed line). Top panel: all material is confined within a closed, stable configuration; bottom panel: accretion of material overflowing across the critical surface.}
  \label{torus1}
\end{figure}

The plasma forms an accretion disk or a torus residing in the equatorial plane, so that the axial symmetry is maintained on the overall scale (Fig.\ \ref{f0}). In a non-magnetized (purely hydrodynamical) limiting case one can find the classical solution for the density distribution $\rho\equiv\rho(r,z)$ and pressure $P\equiv P(r,z)$, and geometrical shape $H\equiv H(r)$ of a non-gravitating polytropic torus, $P=K\rho^k$ \citep{1978A&A....63..221A,1980ApJ...242..772A}. Introducing enthalpy of the medium, $W(P)\equiv\int dP/\rho.$ By setting $P_{\rm in}=P_{\rm out}=0$ at the inner and the outer edges of the density distribution, \citet{1980ApJ...242..772A} find
\begin{equation}
 W_{\rm out}-W_{\rm in}=\int_{R_{\rm in}}^{R_{\rm out}}
  \,\frac{l(R)^2-l_{\rm kep}(R)^2}{R^3}\,dR=0,
\end{equation}
where $l_{\rm kep}(R)$ is the radial profile of the Keplerian angular momentum density in the equatorial plane. Several properties of this solution are worth mentioning \citep{1991MNRAS.250....7C,1924MNRAS..84..665V}: (i)~The level surfaces of functions $P,$ $\rho,$ and $W$ coincide; (ii)~If the torus boundary $P=0$ forms a closed surface, the torus center is defined by where $dP/dR=0,$  the pressure is maximum; (iii)~The shape of the torus can be found by integrating the vertical component of the Euler equation (Fig.\ \ref{torus1}).

Above a certain critical value, $W>W_c,$ the torus forms a stable configuration (see the shaded region), while for $W<W_c$ matter overflows onto the central object even if we neglect viscosity. This behaviour resembles the Roche lobe overflow in binary systems, however, here it is a consequence of the non-monotonic radial dependence of the Keplerian angular momentum in the relativistic regime near the black hole. Whereas the material inflowing from the atmosphere of a primary component of the binary system concentrates near the equatorial plane and naturally forms the torus, in the case of a single, isolated central black hole matching the inner (toroidal) structure to the outer reservoir of matter depends on many circumstances at the outer boundary region. This is also the case of supermassive black holes residing in nuclei of galaxies, which are fed by the interstellar medium from a surrounding (spheroidal) nuclear star cluster and the galaxy bulge; flattening of the structure is a parameter that can vary from disk-type equatorial inflow up to perfectly spherical (Bondi-type) solution \citep{2008ApJ...686..172S} at large radius. Let us note that the assumptions about constancy of $l(R)$ (angular momentum density along von Zeipel cylinders) and vanishing magnetization in Fig. \ref{torus1} are astrophysically unrealistic but they are useful to simplify calculations and allow an analytical insight.

Even in the case of a strongly magnetized environment the contribution of an astrophysically realistic magnetic energy to the space-time curvature is negligible. We can thus neglect its effect on the metric terms and assume a weak-field limit on the background of the Kerr black hole. In order to initiate the numerical code, we can employ an initially uniform magnetic field \citep{1974PhRvD..10.1680W}, which is described completely by two non-vanishing components of the four-potential,
\begin{eqnarray}
A_{t} &=& Ba\Big[r\Sigma^{-1}\left(1+\cos^2\theta\right) -1\Big], \label{mf1}\\
A_{\phi} &=& B\Big[{\textstyle\frac{1}{2}}\big(r^2+a^2\big)
 -a^2r\Sigma^{-1}\big(1+\cos^2\theta\big)\Big] \sin^2\theta \label{mf2},
\end{eqnarray}
in dimensionless Boyer--Lindquist coordinates and $B$ is the magnetic intensity of the uniform field far from the event horizon. The magnetic field and the associated electric component are generated by currents flowing in the accreted medium far from the black hole, as the latter does not support its own magnetic field (for further discussion, see below in the text). The set of two non-vanishing four-potential vector components defines the structure of the electromagnetic tensor, $F_{\mu\nu}\equiv A_{[\mu,\nu]}$; by projecting onto a local observer frame, one then obtains the electric and magnetic vectors. However idealized the initial configuration may be, the numerical solution rapidly evolves in a complex entangled structure, with field lines turbulent within the accreting medium and more organized in the empty funnels that develop outside the fluid structure.

  \subsection{Main components of the radiation spectrum from black hole accretion}
  \label{observed_radiation}

The AGN SED differs from normal galaxies mainly by the large luminosity across the electromagnetic spectrum from radio to gamma-rays, both extreme long and short wavelengths being present particularly when the AGN produces significant emission from the outflowing, relativistically fast jet. The large AGN luminosity (typical AGN luminosity is by more than eight orders of magnitude larger than the luminosity of Sgr~A* in our Galactic centre) can be well explained by accretion onto a supermassive black hole. Depending on the accretion mode (see the chapter \ref{AGN_across_mass} for more details), the total emission is either dominated by thermal emission from an accretion disk or by non-thermal emission from the accretion disk corona and/or relativistic jet.

\subsubsection*{Accretion disk thermal emission}

The thermal blackbody radiation is described by the Planck radiation law. The spontaneous radiative distribution of the entropy emerges radiation with the specific intensity $B$\, [W\,m$^{-2}$\,Hz$^{-1}$\,sr$^{-1}$], given by the Planck function:
\begin{equation}
\label{planck}
B(\nu,T) = \frac{2h\nu^3}{c^2}\frac{1}{{\rm e}^{\frac{h\nu}{kT}}-1},
\end{equation}
where $\nu$ is the emitted frequency and $T$ is the temperature of the blackbody surface. The blackbody emission is radiated by an optically thick matter, in which the thermal photons emerge multiple scattering with particles and the temperature possesses a value corresponding to the kinetic energy of the particles.

The peak frequency $\nu_{\rm peak}$ is the frequency at which $B(\nu,T)$ given by Eq.\ (\ref{planck}) reaches its maximum. It is proportional to the temperature:
\begin{equation}
h\nu_{\rm peak} = 2.82 kT,
\end{equation}
which was known before the Planck's law as the Wien displacement law. For frequencies well below the peak, its course is well described by:
\begin{equation}
B(\nu,T) \approx \frac{2\nu^2kT}{c^2} \quad \mbox{(Rayleigh--Jeans's law)},
\end{equation}
and for frequencies well above the peak by:
\begin{equation}
B(\nu,T) \approx \frac{2h\nu^3}{c^2}{\rm e}^{-\frac{h\nu}{kT}} \quad \mbox{\rm (Wien's law)}.
\end{equation}
It is straightforward to derive both, the Rayleigh--Jeans's approximations from Eq.~(\ref{planck}) by the Taylor expansion ${\rm e}^{\frac{h\nu}{kT}} = 1 + \frac{h\nu}{kT}+ \cdots$ (and the Wien's approximation by neglecting unity against the exponential in the denominator of Planck's law).

The energy flux of blackbody radiation is given by
\begin{equation}
 F = \sigma T^4 \quad \mbox{Stefan--Boltzmann's law},
\end{equation}
where $\sigma$ is the Stefan--Boltzmann constant $\sigma = \frac{2\pi^5k^4}{15c^2h^3}$.

An accretion disk does not emit blackbody radiation with a single temperature. The temperature at each position is determined by the particle's kinetic energy. This energy decreases with the radius and thus also the temperature drops. The radial dependence of the temperature for the accretion disk surface around a black hole can be determined by \citet{1973A&A....24..337S} standard-disk scenario:
\begin{equation}
\label{tempAD}
T(r)=\left(\frac{3GM\dot{M}}{8\pi\sigma r^{3}}\left[1-\left(\frac{r_{\rm in}}{r}\right)^{\frac{1}{2}}\right]\right)^{\frac{1}{4}},
\end{equation}
where $r_{\rm in}$ is the inner edge of the disk.

The total thermal emission from an accretion disk is the composition of contribution from multiple radii with different surface temperatures, known as the multicolor blackbody emission. This multicolor blackbody was successfully applied to thermal X-ray spectra of X-ray binaries \citep{Mitsuda1984}, confirming the general validity of the temperature radial dependence given by Eq.~\ref{tempAD}.
More sensitive X-ray data have later revealed deviation from the perfect blackbody emission, mostly caused by the Comptonization of thermal photons in the upper layers of the accretion disk \citep{Czerny1987, Ross1992, Shimura1993}. Taking into account the vertical structure of the disk leads to the spectral hardening and the local specific flux is diluted blackbody emission:
\begin{equation}
 F_{\nu} = \frac{\pi}{f^4_{\rm col}} B_{\nu} (f_{\rm col}, T_{\rm eff}),
\end{equation}
where $B_{\nu}$ is the Planck function (Eq.~\ref{planck}), $T_{\rm eff}$ is the effective temperature, and $f_{\rm col}$ is the hardening factor. The value of the hardening factor is a function of the accretion rate and may vary depending on the accretion state \citep[see, e.g.][]{Davis2005}.

Equation~(\ref{tempAD}) can be re-written to have a useful formula for estimating the temperature of accretion disks around black holes with various mass or accretion rate in terms of Eddington units:
\begin{equation}
\label{tempAD_rgedd}
T\left(\frac{r}{R_g}\right) \approx 6.3 \times 10^5\,{\textnormal K} \left({\frac{M}{10^8M_{\odot}}}\right)^{-\frac{1}{4}} \left(\frac{\dot{M}}{\dot{M}_{\rm Edd}}\right)^{\frac{1}{4}} \left(\frac{r}{R_g}\right)^{-\frac{3}{4}}f(r),
\end{equation}
where $R_g$ is the gravitational radius (defined by Eq.~\ref{rg}), and $f(r) = \left[1-\left(\frac{r_{\rm in}}{r}\right)^{\frac{1}{2}}\right]^\frac{1}{4}$.
For the same accretion rate in Eddington units and the inner edge of the disk, the temperature scales with mass as:
\begin{equation}
\label{tempAD_massdepend}
T \propto M^{-\frac{1}{4}}.
\end{equation}
Note that Eq.~\ref{tempAD_massdepend} is valid when the accretion
rate is expressed in units of Eddington values, otherwise the temperature would be 
$T \approx M^{-\frac{1}{2}}\dot{M}^{\frac{1}{4}}$.
However, for comparison of accreting black holes of different mass
but with a similar Eddington ratio, Eq.~\ref{tempAD_massdepend},
is the most useful and we can immediately estimate where 
the peak of AGN thermal emission should be based on the knowledge of measured accretion disk temperatures in X-ray spectra of XRB.
The thermal emission of accretion disks in AGN thus peaks in the UV range ($T \approx 10^5$\,K), while the temperature peak occurs at the soft X-ray band for stellar-mass black holes in X-ray binaries ($T \approx 10^7$\,K).

The observed spectrum of AGN has indeed typically a peak in the UV band, the big blue bump \citep{Richards2006}. Some UV spectra of quasars can be well fitted by the thermal disk emission \citep{2011MNRAS.415.2942C},
but it is not the case in general \citep{Elvis1994}. The UV and especially the range between far-UV and X-rays is heavily affected by the interstellar absorption, preventing astronomers from a detailed study of spectral properties of the thermal accretion disk emission in AGN.

\subsubsection*{Thermal bremsstrahlung radiation}

Thermal bremsstrahlung is radiated in a hot plasma due to the Coulomb collisions of the ions. The electrons are accelerated by the collisions to high velocities and emit radiation that can escape the plasma if it is optically thin. Therefore, the AGN corona produces not only up-scattered emission from the accretion disk but also its own thermal emission.
This is, however, very different from the thermal emission of the blackbody emission of the optically-thick accretion disk and is, therefore, sometimes not accurately referred to as a non-thermal component of AGN.

The temperature of the plasma is defined by the ion velocities, which have in thermal equilibrium the Maxwell--Boltzmann distribution. The temperature is proportional to the average kinetic energy of ions:
\begin{equation}
\frac{3}{2} kT = \left<\frac{1}{2}mv^2\right>,
\end{equation}
where $m$ and $v$ are the ion mass and the thermal speed, respectively. In the thermal equilibrium, the average kinetic energy is equal for ions and electrons. Therefore, the ions can be considered stationary with respect to the fast moving electrons. Different situations can happen in the collisions. The free-free transition occurs when the electron freely passes near the ion and is not kept. Such radiation yields the continuum spectrum without lines, and is called bremsstrahlung. 

The radiated power per unit photon energy and per unit volume is given by:
\begin{equation}
\epsilon_{\rm free-free} = 2\sqrt{\frac{2}{3\pi}}\,n_e\,n_i\,\alpha\,\sigma_T\,c\,Z^2\,\Theta^{-1/2}\,g(\Theta,E)\,e^{-E/kT}, 
\end{equation}
where $n_i$ is the ion density, $n_e$ is the electron density, $\alpha$ is the fine structure constant, $Z$ is the effective charge of the ion, $\Theta \equiv k_BT/m_ec^2$, and $g$ is the Gaunt factor, which is a dimensionless quantity of order of 1 and slightly changes with the spectral energy (frequency).

When the electron is captured by the ion and fills one of the ion vacancies, the transition is free-bound and the continuum emission is associated also with spectral-line emission corresponding to the transition. The process of free-bound transition is called radiative recombination. How often the ionization and recombination of atoms is happening depends mainly on the plasma density but also on the nature of equilibrium. More recombinations occur at the photoionization equilibrium than in the collisional ionization equilibrium since the temperature at the photoionization equilibrium is lower for comparable ionization.

\begin{figure}[tbh!]
\begin{center}
\includegraphics[width=0.48\textwidth]{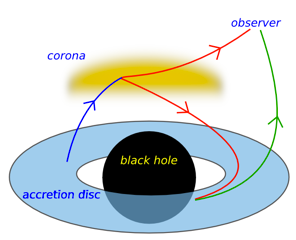}
\includegraphics[width=0.48\textwidth]{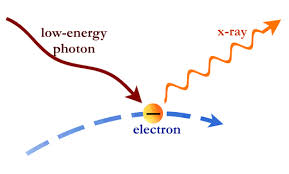}
\caption{Left: a schematic picture of the main radiation components of an accreting black hole: thermal radiation from the accretion disk, up-scattered in a hot corona and reflection emission. Light rays are distorted by strong gravity of the central black hole.
Right: a schematic view of the inverse Compton effect in the local frame frame where electrons have energy high enough compared to the ambient light, so the mechanism up-scatters the ambient photons.
}
\label{fig_inverse_compton}
\end{center}
\end{figure}

\subsubsection*{Non-thermal AGN emission}

The thermal emission of AGN accretion disks contributes mainly to the UV domain, but AGN are very bright X-ray sources as well and the thermal bremsstrahlung of the corona is not sufficient to produce the observed X-ray intensity. The X-ray emission can be, therefore, best explained by scattering of the thermal emission in the corona, as illustrated in Figure~\ref{fig_inverse_compton}. The very high temperature causes that the electrons in the corona possess very high velocities. Their kinetic energy is partly transported to the low-energy photons scattering on the electrons, and the photons thus gain energy and increase their frequencies. The resulting radiation dominates in X-rays, more energetic than UV. This effect is known as the {\textbf{inverse Compton scattering}}. Further, we will show that this effect can be understood in terms of the special-relativistic transformation in relations for Compton scattering.

The initial and final energy of a photon can be written in the units of electron rest energy:
\begin{equation}
e_i = \dfrac{h \nu_i}{m_e c^2}  \;, \qquad 
e_f = \dfrac{h \nu_f}{m_e c^2} \;,
\end{equation}
where $m_e$ is the mass of electron, $h$ is the Planck constant, and $\nu_i$, $\nu_f$ are the initial (before collision) and final (after collision) frequencies of photon, respectively. Based on the law of energy and momentum conservation, the final energy of the photon after the Compton scattering is:
\begin{equation}
\label{eq:final_photon_energy}
e_f = \dfrac{e_i}{1 + e_i \Lambda}	\;,
\end{equation}
where $\Lambda = 1 - \cos{\vartheta}$ and $\vartheta$ being an angle between the direction of photon before and after the collision.
The final frequency of the photon can be obtained by multiplying the equation \eqref{eq:final_photon_energy} by factor $m_e c^2$:
\begin{equation}
\label{eq:final_photon_frequency}
\nu_f = \dfrac{m_e c^2 \nu_i}{m_e c^2 + h \nu_i \Lambda}	\;.
\end{equation}

Equations~(\ref{eq:final_photon_energy}) and (\ref{eq:final_photon_frequency})
are valid in the electron rest frame. 
But now, let us assume that the electron is in the relativistic regime
with the Lorentz factor $\gamma = \frac{1}{\sqrt{1 - v^2/c^2}} = \frac{1}{\sqrt{1 - \beta^2}}$,
and rewrite the equation for Compton scattering in the laboratory frame.
We further denote laboratory frame by an index "L" and the rest frame by an index "R".
The final energy of the photon in the laboratory frame can be written as:
\begin{equation}
e^L_f = e^R_f \gamma \left( 1 + \beta \mu^R_f \right),
\end{equation}
where $\mu = \cos{\chi}$ and $\chi$ is the angle between the direction of electron and photon before the collision.
Using Eq.~\ref{eq:final_photon_energy}, we get:
\begin{equation}
e^L_f = \dfrac{e^R_i}{1 + e^R_i \Lambda^R} \gamma \left( 1 + \beta \mu^R_f \right) 
\end{equation}
Using transformations $\dfrac{1}{\gamma} \dfrac{1}{1 - \beta \mu^L_f} = \gamma \left( 1 + \beta \mu^R_f \right) $ and $e^R_i = e^L_i \gamma \left( 1 - \beta \mu^L_i \right)$, we get a formula for the final energy of the photon in the laboratory frame as:
\begin{equation}
\label{eq:final_photon_energy_inverseC}
e^L_f = e^L_i \, \dfrac{1 - \beta \mu^L_i}{1 - \beta \mu^L_f} \, \dfrac{1}{1 + e^L_i \gamma \left( 1 - \beta \mu^L_i \right) \Lambda^R}		\;. 
\end{equation}
Assuming photons with a low initial energy, so that the condition $\gamma e^L_i \ll 1$ is fulfilled, the equation can be simplified to:
\begin{equation}
e^L_f \approx e^L_i \, \dfrac{1 - \beta \mu^L_i}{1 - \beta \mu^L_f} 	\;, 
\end{equation}
which can be rewritten to:
\begin{equation}
e^L_f \approx e^L_i \gamma^2 \left(  1 - \beta \mu^L_f \right) \left(  1 + \beta \mu^R_f \right).
\end{equation}
From the last equation, it follows that the final energy of the photon in the laboratory frame must be larger than the initial energy and lower than the product of the initial energy and a factor of 4$\gamma^2$ that is larger than 1 and is the higher the higher velocity of electron is:
\begin{equation}
e^L_i ~ \lesssim ~ e^L_f ~ \lesssim ~ 4 \gamma^2 e^L_i		\;.
\end{equation}

Other source of the AGN non-thermal emission is the {\textbf{synchrotron emission}} from the relativistic jets. The synchrotron emission comes from the accelerated electrons in the magnetic field. Its power (in the observer's frame) increases with the second power of electron's energy and magnetic-field intensity. Therefore, it is strong especially in the black-hole jets  where the electrons are highly relativistic (see Chapter~\ref{Jets}). This causes a strong beaming effect and most of the energy is radiated into a narrow beam. Therefore, the synchrotron emission is the most dominant radiation component in blazars, i.e. in AGN with strong jets pointing towards us, while it is negligible for radio-quiet AGN. 
We refer to standard books, \citet[e.g.,][]{1979rpa..book.....R}, to read more about different mechanisms of thermal and non-thermal radiation.

Another important component of the AGN emission is reprocessing of the primary radiation. This happens very close to the black hole in the accretion disk, as it is illustrated in Fig.~\ref{fig_inverse_compton} (left panel), as well as at more distant structures, such as AGN tori. The reprocessing at tori usually produce a large fraction of infrared emission because the AGN radiation heats the dust there. The reflection at the innermost disk is particularly interesting in the X-ray domain, because the reprocessed emission there can even overcome the primary radiation that is declining at high energies. The most prominent features of the innermost X-ray reflection are the broad iron line and the Compton hump. These spectral features can be used to derive the basic parameter of the central black hole, its angular momentum (spin), as it is shown in the next chapter.


\begin{figure}[tbh!]
\begin{center}
\includegraphics[width=0.8\textwidth]{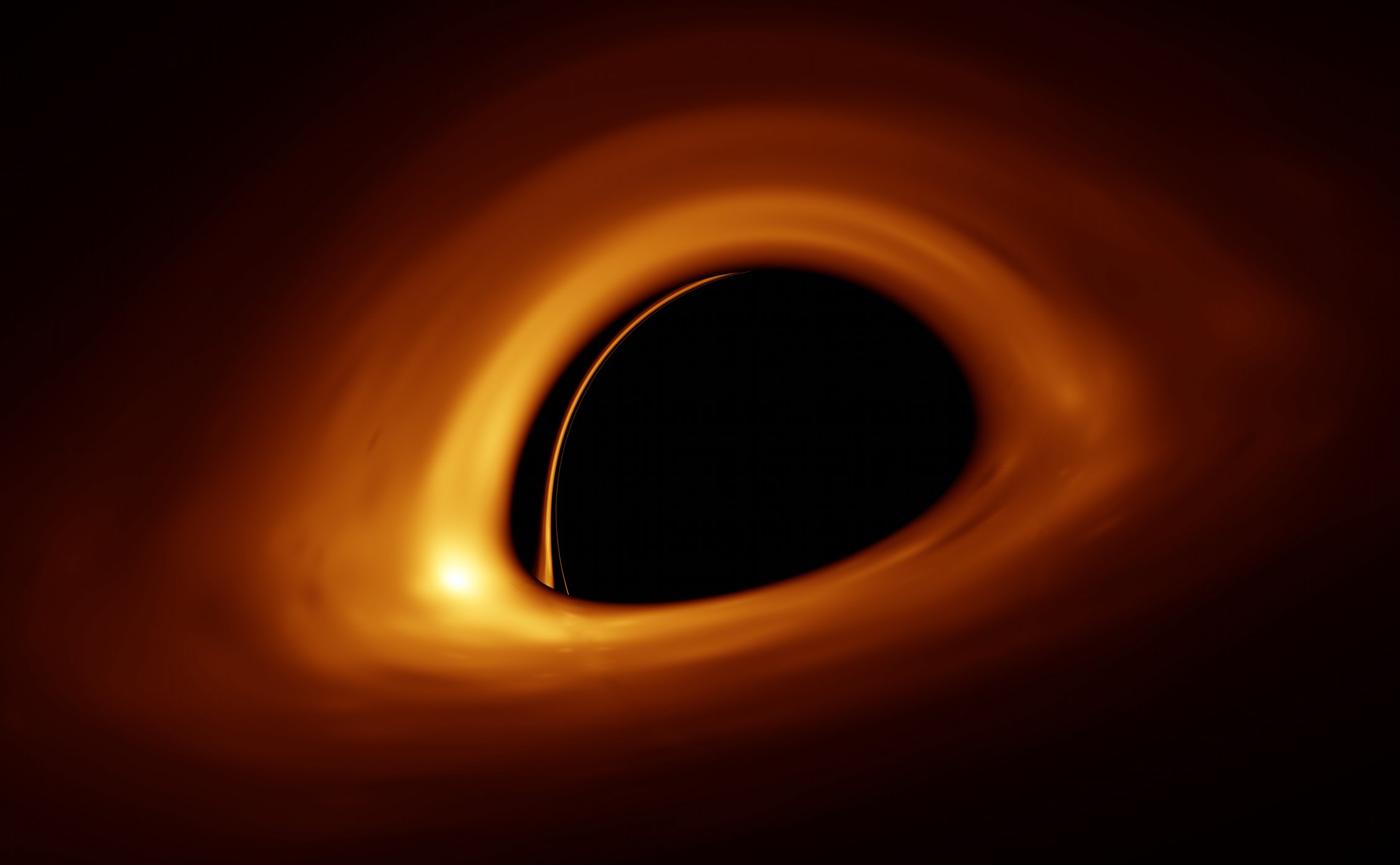}
\caption{A computer-generated image of a black hole accretion disk, as seen by a distant observer at $\theta_{\rm inc}=80$ degree inclination view angle \citep{2007ragt.meet...21B}. A black hole shadow develops in the central part of the image, surrounded by a set of narrow rings of photons encircling the black hole along indirect, highly bent rays.}
\label{fig11}
\end{center}
\end{figure}

\subsubsection*{Strong-gravity effects in radiation signal from accretion disk}
Accretion disks may be seen as axially symmetric and stationary at the zero
approximation. However, a variety of non-axisymmetric perturbations are known to develop, grow and influence their structure. This leads to the fluctuating
observed signal with characteristic imprints of the black hole strong gravity in continuum and spectral-lines \citep{2002apa..book.....F,2008bhad.book.....K}. 
Relativistic effects can be described within the
geometrical optics, where photons follow null geodesics
towards a distant observer. The radius of the emission site and
the viewing angle (inclination)
of the observer with respect to the disk plane are the 
main parameters for the signal arising near the inner edge, supplemented by the black-hole spin \citep{2000PASP..112.1145F,2018arXiv180110203K}. These imprints
can be characterized in several steps.

Each pixel within the detector plane corresponds to a photon ray arriving from the source,
namely, the accretion disk surface $z\equiv z(R)$. Photon trajectories are 
not straight lines in space, leading to a distorted image (Fig.~\ref{fig11}). 
Radiation intensity transforms from the local disk frame,
$I^{\rm R}(R,z(R);\,$$\nu^{\rm R},\mu^{\rm R}),$ to the observer's (laboratory) frame,
$I^{\rm L}(R,z(R);\,$$\nu^{\rm L},\mu^{\rm L})$, and propagates to the detector plane
(here, $\mu$ is the directional cosine of the emerging photon measured from the normal to the disk surface.).

The gravitational field of the central object and
the disk modify the shape of spectral features coming from the
disk \citep{1975ApJ...202..788C,1976ApJ...208..534C,1991ApJ...376...90L}; this is apparent especially at high
view angles and with fast-rotating black hole, where photons experience
large line-of-sight velocity and the radiation is produced mostly at very small
radii near the black hole. Finally, 
\citet{1989MNRAS.238..897L} studied continuum spectra produced
by self-gravitating disks, whereas
\citet{1995ApJ...440..108K}, \citet{Schee}, and \citet{1998MNRAS.301..721U} calculated the predicted
profiles of spectral lines within the
framework of exact general-relativistic solutions.
These approaches give an improved (consistent) picture
where the matter of the accretion flow generates radiation and, at the same time, the
accretion disk mass modifies its internal structure and acts back on the emerging photon rays as they travel toward the
observer.

\begin{figure}[tbh!]
\begin{center}
\includegraphics[width=0.6\textwidth]{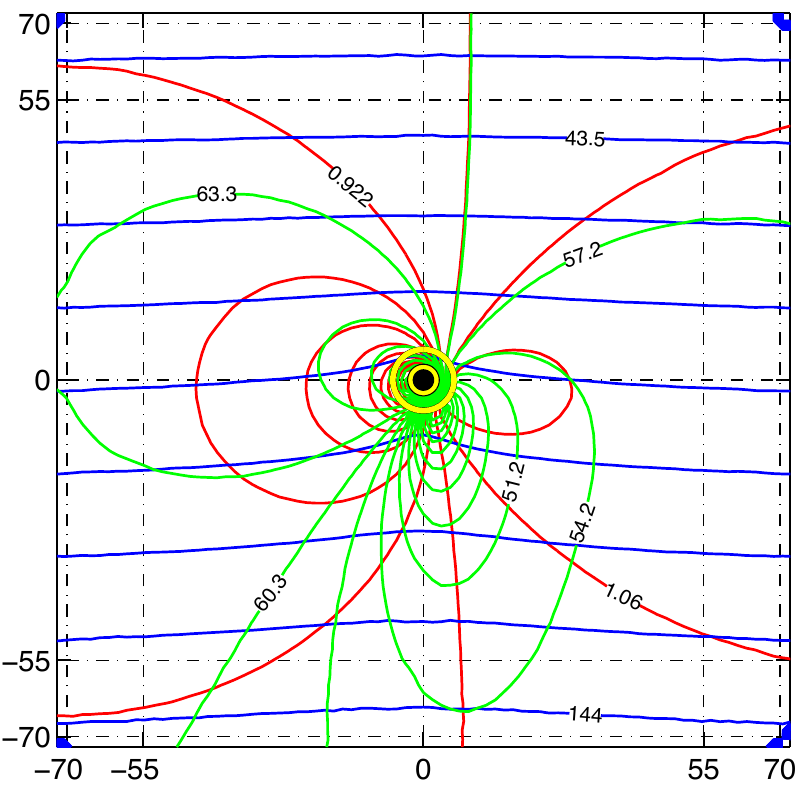}
\hfill
\includegraphics[width=0.33\textwidth]{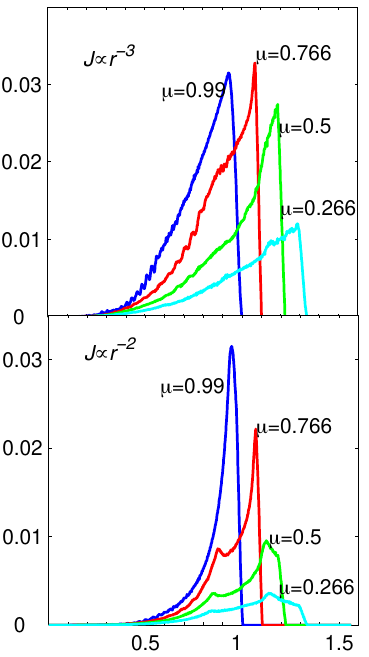}
\caption{Left panel: contour lines of three governing functions in the equatorial disk plane $(r,\phi)$ that determine effects of strong gravity of the central black hole
onto observed radiation from a Keplerian disk: (i) the overall (Doppler and gravitational) energy shift $g(R,\phi)$ (red); (ii) local emission angle (in degrees) with respect to
disk, affected by the relativistic aberration (green); (iii) time-delay (in geometrized units) due to light travel time from the disk plane to the distant observer located at large radius. Parameters of the plot: view angle $\theta_{\rm inc}=60$~deg, spin $a/M=0$ (Schwarzschild black hole -- the black circle). The grid extends up to $70 GM_\bullet/c^2$; ISCO radius $R_{\rm ms}=6GM_\bullet/c^2$ (yellow circle). Right panel: the resulting (computed) profiles of the relativistic spectral line. Effect of the overall broadening and skewness are apparent
(more profoundly for edge-on view angles).
Different power-law radial profiles of the intrinsic emissivity $J(r)\propto r^{-n}$ 
have been laid on the accretion disk with
several view angles $\mu\equiv\cos\theta_{\rm inc}$. 
Energy on the horizontal axis is normalized to the line intrinsic energy; 
the radiation flux on the vertical axis is in
arbitrary units. For a direct comparison 
with previous papers we selected the case of
a rapidly rotating Kerr black hole $a/M=0.998$ and an accretion disk
extending from the innermost stable circular orbit \citep[ISCO;
other parameters as in][]{1991ApJ...376...90L}.}
\label{fig12}
\end{center}
\end{figure}

In order to reveal the spectral profile, 
we calculate the total observed flux of radiation by collecting photons,
\begin{equation}
 F^{\rm L}(\nu^{\rm L})_{\mid\theta=\theta_{\rm obs}}=
 \int_{\lower 3pt\hbox{\footnotesize (Over observer's plane)}}
 \hspace*{-17ex}I^{\rm L}(\nu^{\rm L})_{\mid R\rightarrow\infty,\theta=
 \theta_{\rm obs}}\,dS.
\end{equation}
A typical double-horn skewed profile arises in the observed spectral line 
\citep{1989MNRAS.238..729F,1996MNRAS.282L..53M,2000MNRAS.312..817M}. 

Doppler factor, $g\equiv g(r;a/M,\theta_{\rm inc})\,\equiv\,1+z=E_{\rm obs}/E_{\rm em}$, 
characterizes the redshift function $z$ between the emitted (subscript ``''{\sf em}'') and observed (``{\sf obs}'') 
of photons originating at different sides of the source
(increase of the energy at the approaching side, and vice versa at the receding side of the disk). This relates to the change of radiation intensity by the corresponding 
factor $g^3$. The effect
tends to enlarge the width of the spectral features
and to enhance the observed flux at its high-energy tail. In addition to the special-relativistic effect, the overall
gravitational redshift of General Relativity contributes to the $g$-factor and affects 
mainly photons coming from the inner parts of the source in the deep potential well.
Finally, the gravitational lensing enhances the emerging flux from the far side, i.e., 
the section of the accretion disk at the upper
conjunction with the black hole; this leads to the enhancement 
of the observed flux around the centroid energy (see Fig.~\ref{fig12}).\footnote{See also https://astro.cas.cz/karas/papers/ky/default.htm for further details.} 

Let us remind the reader that the central (Kerr) black hole is assumed to be 
the only source of gravitational field in and there are no secondary bodies influencing 
gravitationally the inner disk. This may not be true in the case of binary
SMBHs that can warp the disk; furthermore, an alternative description of gravity,
if viable, could modify the standard picture described here.

Light rays of the geometrical optics approximation are determined by
Maxwell's equations for the electromagnetic field tensor. In the vacuum we can write
\begin{equation}
{F^{\mu\nu}}_{\!;\nu}=0, \quad {^\star F^{\mu\nu}}_{\!;\nu}=0,
\end{equation}
where the asterisk denotes a dual tensor. The associated electric field components
are projections onto observer's four-velocity,
$E^\alpha=F^{\alpha\beta}u_\beta$, and an electromagnetic wave is 
defined as a solution of the form
\begin{equation}
F_{\alpha\beta}=\Re e\left[u_{\alpha\beta}\;e^{\Im \psi(x)}\right].
\end{equation}

\begin{figure}[tbh!]
\begin{center}
\includegraphics[width=0.49\textwidth]{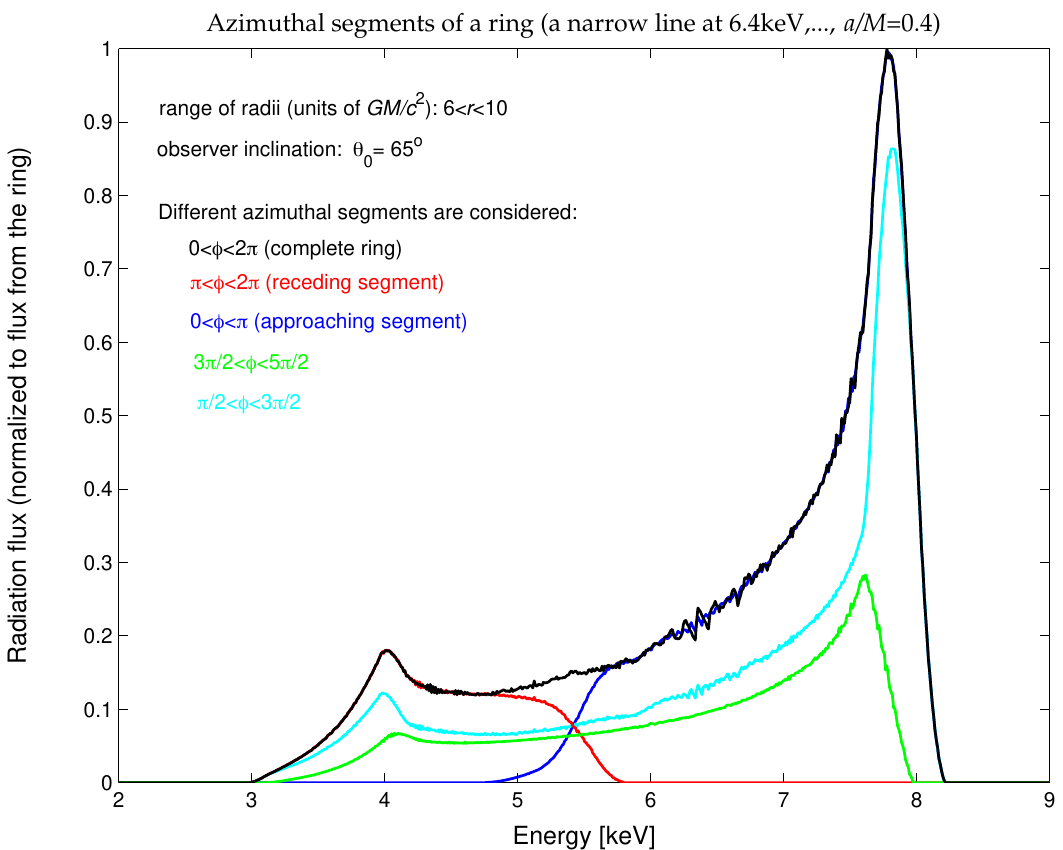}
\hfill
\includegraphics[width=0.47\textwidth]{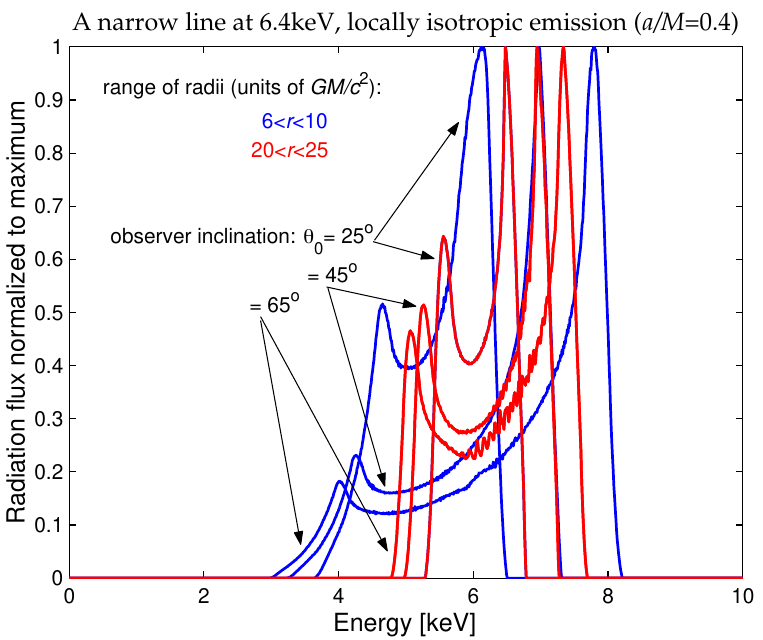}
\caption{Explaining the formation of relativistic spectral line profiles originating from a 
range of radii within a geometrically thin, planar, Keplerian accretion disk that resides in 
the equatorial plane of a 
moderately spinning Kerr black hole ($a/M_\bullet=0.4$). Left panel: intrinsically, the line has a narrow
($\delta$-type) profile centered around the rest-frame energy $6.4$\,keV. The composite profile (black) 
consists of contributions from four azimuthal segments in the disk, as indicated by different 
components and described in the figure legend. The approaching side of the disk creates the amplified
high-energy wing (blue), whereas the receding part is weaker and shifted toward the less energetic part
of the spectrum (red). Right panel: Examples of the line profile broadening that arises 
from different radii and observer inclinations, as indicated within the plot.}
\label{fig10a}
\end{center}
\end{figure}

\begin{figure}[tbh!]
\begin{center}
\includegraphics[width=0.9\textwidth]{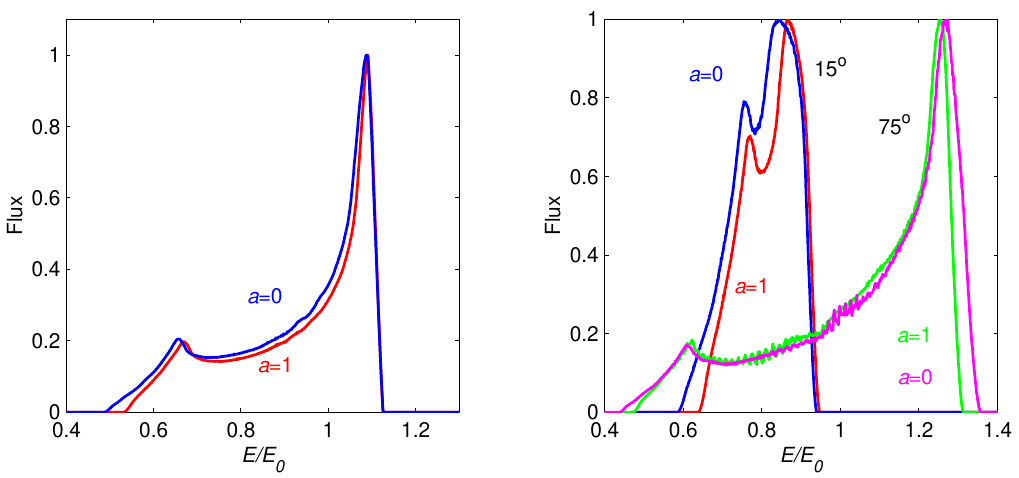}
\caption{The dependence on the black hole
spin and the inclination angle 
for the predicted profiles from a ring extending between ($6$--$10)\,GM/c^2$
\citep[other parameters as in][]{1991MNRAS.250..629K}.}
\label{fig-kojima}
\end{center}
\end{figure}

\begin{figure}[tbh!]
\begin{center}\vspace*{-14em}
\includegraphics[width=\textwidth]{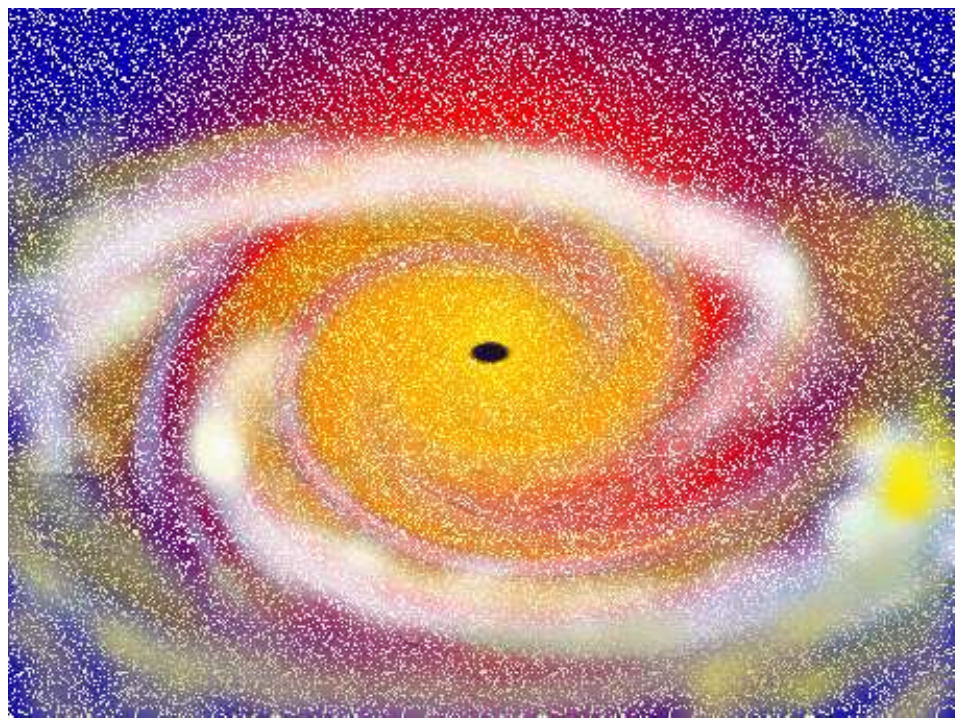}\vspace*{-10em}
\caption{An artistic sketch of spiral patterns on an
 accretion disk as another physically motivated origin of quasi-periodic modulation of the incoming radiation signal. Spiral-wave structures produce a rich variety of spectra and light curves, broader than what we can explain with the help of strictly localized
spots. This scheme is complementary to the idea of instant off-axis flares inducing the observed line variations and it enables us to restrict the geometry of the source by means of the reverberation technique. Figure adopted from \citeauthor{2001PASJ...53..189K}  \citep{2001PASJ...53..189K}.}
\label{fig10b}
\end{center}
\end{figure}

\begin{figure}[tbh!]
\begin{center}\vspace*{-4em}
\includegraphics[width=0.49\textwidth]{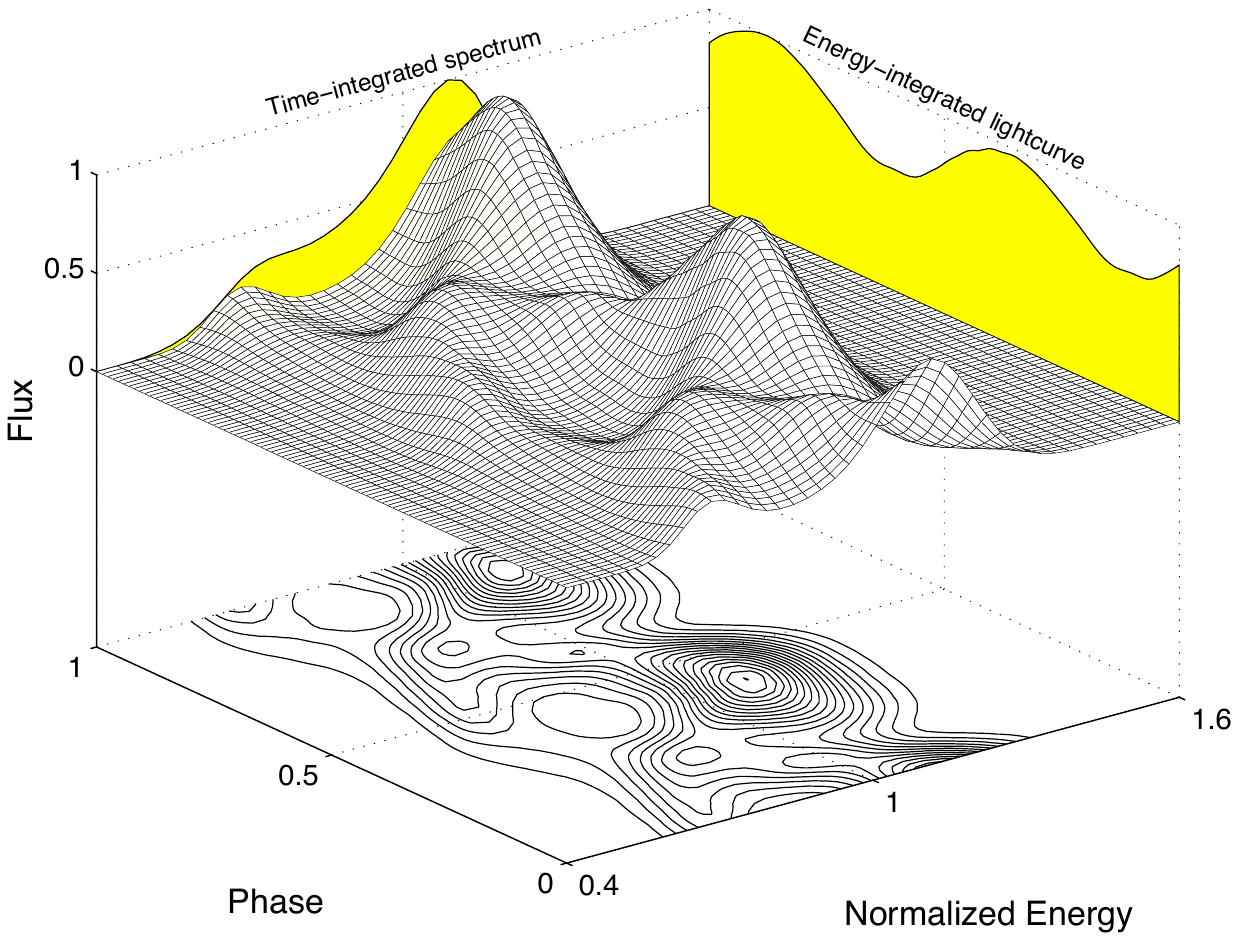}\hfill\includegraphics[width=0.49\textwidth]{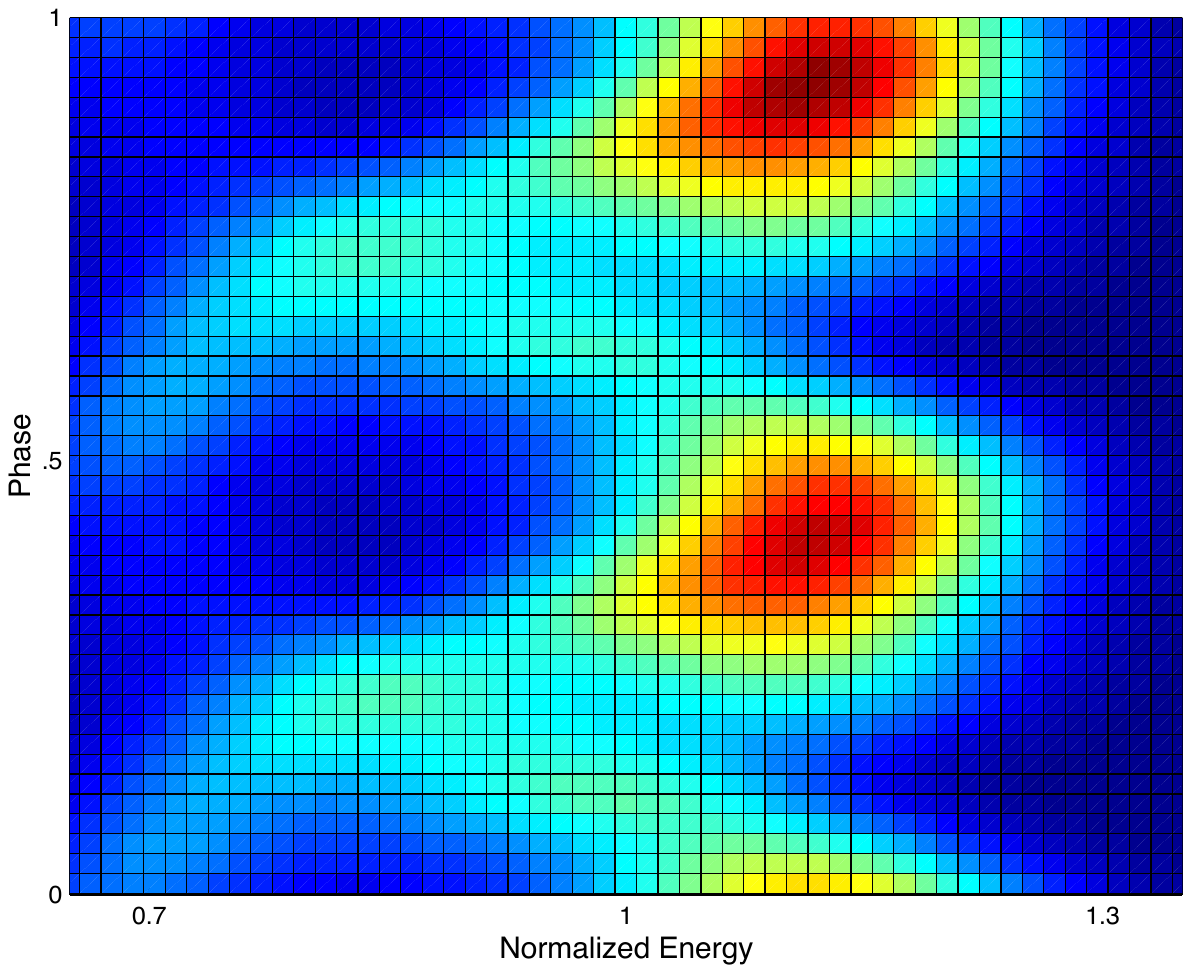}\vspace*{-4em}
\caption{A time-dependent model spectrum produced by a
non-axisymmetric pattern (spiral wave) orbiting within
a black hole accretion disk. Left panel shows the predicted radiation as detected by a distant observer
(moderate inclination angle $\theta_{\rm inc}=50$~deg). The energy-integrated light curve and time-integrated spectrum
are shown as projections of the full dynamical spectrum. 
In a typical situation of an AGN accretion disk, and especially for
small radii and intermediate to large inclination angles, most of the line
emission arises from only a small fraction of the orbit. 
Right panel: observed energy of the spectral line
vs.\ orbital phase of the pattern. The color scale of the diagram encodes different levels of the observed flux (red corresponding to the maximum enhancement, blue the minimum flux; \citeauthor{2006AN....327..961K} \citeyear{2006AN....327..961K}).
This can be compared with an early evidence in \citet{2004MNRAS.355.1073I}.}
\label{fig10}
\end{center}
\end{figure}

Two scales are introduced at this point. The phase $\psi(x)$ is a
a rapidly varying function, while the amplitudes $u_{\alpha\beta}$
vary slowly. We can thus define a wave vector, $k_\alpha\equiv
\psi_{,\alpha}$, which obeys parallel transported along the null geodesics,
$k_{\alpha;\beta}\,k^\beta=0$, $k_\alpha k^\alpha=0$. The propagation
law in the empty curved spacetime thus reads 
\citep{1967rta1.book.....E,1971grc..conf..104E}
\begin{equation}
DF_{\alpha\beta}-2\theta F_{\alpha\beta}=0,
\end{equation}
where $\theta\equiv-\frac{1}{2}{k^\alpha}_{;\alpha}$ describes the
expansion of null congruences, $D\,\equiv\,u^\alpha\nabla_\alpha$.
Superposition of the above-described influences is modeled and explained with 
Fig.\ \ref{fig10a}--\ref{fig-kojima}, where we
show the dependence of the predicted profiles on the parameters in comparison
with simulations in the original works \citep{1991ApJ...376...90L,1991MNRAS.250..629K,1992MNRAS.259..569K};
we notice an excellent agreement and give
the values that can be useful especially for tests of new numerical codes.
Furthermore, in Figs.\ \ref{fig10b}--\ref{fig10} we explore dynamical spectrum expected
to be seen from an orbiting spot or a similar kind of a feature of enhanced emission on
the accretion disk surface; time resolution offers a significant
improvement to constrain the line-producing region but the observation requires
much longer duration and sensitivity to capture enough photons.

\begin{figure}[tbh!]
\begin{center}
\includegraphics[width=0.75\textwidth]{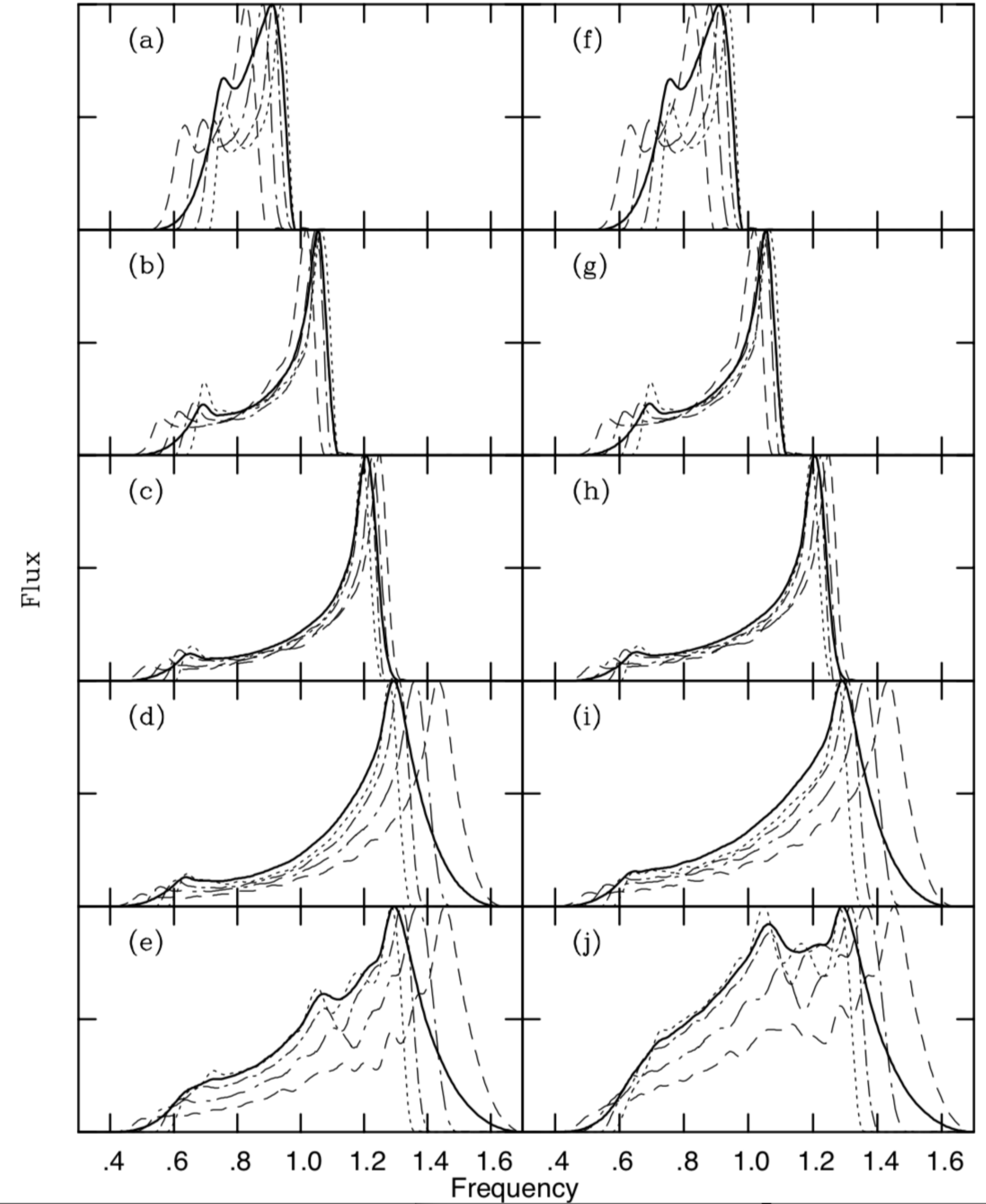}
\caption{Examples of synthetic emission-line profiles from an accretion disk,
taking into account a variety of relativistic effects (Doppler and a gravitational shift of 
frequency, lensing of the observed radiation flux due to gravity
of the central black hole) as well as the self-gravity of the accretion disk. 
The latter influence should be taken into account in realistic models
of AGN spectra (in case of the fluorescent iron line in the X-ray domain) because it must inevitably be present. The disk material 
orbits at a speed that is different from the Keplerian velocity due to self-gravity
of the disk matter. Different curves illustrate effects of the black-hole rotation $a/M_\bullet$, 
observer's inclination $\theta_{\rm inc}$, and different disk models (typically, $M_{\rm
 disk}/M_{\bullet}\approx10^{-2})$. The inclination angle varies over different panels
 from a small value (pole-on view -- top left, narrow line)
 up to $\theta_{\rm inc}\simeq80$\,deg (almost edge-on view -- bottom right,
 relativistically broadened and skewed profiles).
In each panel, contributions from different radial regions of the disk are indicated by 
different line styles, while the total flux from the entire disk is plotted by the solid line
(background subtracted and normalized to the maximum). Figure adapted
 from \citet{1995ApJ...440..108K}.}
\label{fig10-sg}
\end{center}
\end{figure}

Let us make an obvious but important remark, notably, that the above-mentioned idealized 
scheme must become more complex in realistic astrophysical circumstances. This leads
to additional degrees of freedom of the system and causes some kind of degeneracy
between parameters (they cannot be constrained in a unique way, instead, the best-fit
procedure can find during the minimization process different local minims). 
Hence, one needs more accurate observations and improved signal-to-noise
resolution in order to constrain the models. This complication is demonstrated in Fig. \ref{fig10-sg},
where we plot exemplary profiles of a spectral line influenced by additional diverse effects, including
the self-gravity, which have been ignored in previous models \citep{1998MNRAS.301..721U}.
Likewise, a geometrically thick (flaring) vertical profile of the accretion disk surface
and/or the presence of warps need to be taken into account especially 
at higher luminosity \citep{2003A&A...403...29C,2018ApJ...855..120T}, but these have been
neglected in the past mainly for the sake of simplicity of calculations.

In the presence of a refractive and dispersive media the approach of relativistic 
geometric optics in curved spacetimes was extended  \citep{1975ApJ...202..454A,1975A&A....44..389B} and then implemented numerically for practical 
computations \citep[e.g.][]{2018arXiv180404670P}. The effects of gravitational
lensing in the curved spacetimes combine with refraction
as light rays propagate through the medium.
\citet{2016ApJ...826...52K} have matched 
general relativistic magnetohydrodynamic (GRMHD) simulations with a Monte Carlo radiation transport code; this allows the authors to determine the X-ray flux irradiating the disk surface and the coronal electron temperature self-consistently, leading to 
self-consistent predictions for the relativistic Fe K$\alpha$ spectral profiles.
Having in mind the applications to present-day X-ray observations, the
energy shifts, gravitational lensing and time delays are the principal
effects which originate from general relativity and can be currently 
tested with data that are in our disposal. Polarimetric information is
more complete \citep{1975MNRAS.171..457R} and the relevant technology is 
available, however, it reaches beyond capabilities of present observational 
instruments available in the orbit \citep{2010xpnw.book.....B}. 

The relativistic polarization tensor is defined as
$J_{\alpha\beta\gamma\delta}\,\equiv\,\frac{1}{2}\langle 
F_{\alpha\beta}F_{\gamma\delta}\rangle$, and
its projections
$J_{\alpha\beta}=J_{\alpha\beta\gamma\delta}\,u^{\gamma}\,u^{\delta}
=\langle E_\alpha \bar{E}_\beta\rangle$ are introduced. Four parameters
$S_A$ are then given by $S_{\!_A}\equiv \frac{1}{2}(k_\alpha
u^\alpha)^2F_{\!_A}$, where $F_{\!_A}$ $(A=0,\,\ldots\,3)$ are 
constructed by projecting the polarization tensor \citep{1974ApJ...189...39A}.
These quantities satisfy the relations 
$J_{\alpha\beta}u^\beta=0$, $J_{\alpha\beta}k^\beta=0$, 
$\omega=u_\alpha k^\alpha$ and can be connected with the traditional 
definition of the Stokes parameters \citep{1959prop.book.....B,1960ratr.book.....C}.

The normalized Stokes parameters are defined $s_1\,\equiv\,S_1/S_0$,
$s_2\,\equiv\,S_2/S_0$, and $s_3\,\equiv\,S_3/S_0$. The degree of linear
and of circular polarization is  $\Pi_l=\sqrt{s_1^2+s_2^2}$,
$\Pi_c=|s_3|$, and the total degree of polarization
$\Pi=\sqrt{\Pi_l^2+\Pi_c^2}$. By propagation through an arbitrary, curved 
(but empty) spacetime,
the radiation flux obeys the relation, $F_{\!_{A,\,\rm
em}}\,dS_{_{\rm em}}=F_{\!_{A,\,\rm obs}}\,dS_{_{\rm obs}}$ from which
the redshift $z$ relates the photon energy at a point of emission with that
at a distant observer,
\begin{equation}
1+z=
\frac{(k_\alpha u_\alpha)_{\rm em}}{(k_\alpha u_\alpha)_{\rm obs}},\quad
S_{\!_A}=\frac{k_{\!_A}}{(1+z)^2dS}.
\label{eq:prop}
\end{equation}

For example, in the case of a thin rotating disk near Kerr black hole, 
the redshift factor $z$ and the
local emission angle $\vartheta$ are given by
\begin{equation}
1+z=\frac{r^{3/2}-3r^{1/2}+2a}{r^{3/2}+{a}-\xi}, \quad
\cos\vartheta=\frac{g\eta^{1/2}}{r};
\end{equation}
$\xi$ and $\eta$ are constants of motion. 

The gravitational field of a Kerr black hole (\ref{metric}) can be written in the form 
\citep{1973grav.book.....M}
\begin{equation}
ds^2=-\frac{\Delta}{\Sigma}\Big(dt-a\sin^2\theta\,d\phi\Big)^2
+\frac{\Sigma}{\Delta}\,dr^2+\Sigma\;d\theta^2
+\frac{\sin^2\theta}{\Sigma}\Big[a\,dt\!-\!\big(r^2+a^2\big)\,d\phi\Big]^2
\label{eq:metric}
\end{equation}
in Boyer--Lindquist spheroidal coordinates, where $\Delta(r)$
and $\Sigma(r,\theta)$ are known functions. The horizon occurs at 
$\Delta(r)=0$. Assuming sub-maximal rotation, $|a|\leq1$, the
outer radius is found at the dimension-less radius $r=1+(1-a^2)^{1/2}$ and it
hides the singularity from a distant observer. As a purely GR effect,
as soon as $a\neq0$, all particles and photons are dragged by the rotation of
the black hole (the frame-dragging effect).

Eq.~(\ref{eq:prop}) shows that the polarization properties of the disk
emission are modified by the photon propagation. As the reflecting 
medium has a disk-like geometry, a
substantial amount of linear polarization is expected because of Compton
scattering. In order to compute
the observable characteristics (i.e., different
polarization modes), one has to combine the reflected component with the
primary continuum. The polarization degree of the resulting signal
depends on the mutual proportion of the different components.
Future X-ray satellites will be essential to
study GR in the neighbourhood of accreting black holes with outstanding 
precision, including the polarimetric information.

It is worth noting that the precise shape and variability of the spectral features reflects
the intrinsic microphysics of the emitting medium and parameters of the black hole
(spin), but it also depends on the gravitational theory that determines 
the photon propagation. Therefore, this opens a promising way to verify (or reject) the predictions of
Einstein's theory vs.\ alternative viable theories \cite[for reviews, see][]{2012PhR...513....1C,2014LRR....17....4W}.
It has been proposed that the detailed features in radiation spectra 
from accreting black holes and the
morphological features shaping the black hole shadows bear
tiny signatures that could reveal departures from General Relativity 
\citep[][ and further references cited therein]{2013ApJ...773...57J,2015ApJ...802...63B,2017ApJ...842...76B}, 
although the effects are far too small to be seen 
at the available resolution of present-day data. In this context an exciting
new avenue has been recently opened with the prospects to apply the currently
available advanced technology of X-ray polarimetry during the forthcoming decade
\citep{2016SPIE.9905E..1QZ,2016SPIE.9905E..17W,2016SPIE.9905E..15S,2016APh....75....8K}.



  \subsection{X-ray reprocessing in accretion disk coronae}
%


The X-ray emission produced in AGN is widely accepted to originate via the inverse Compton scattering of thermal UV photons that are scattered in a hot plasma consisting of relativistically moving particles \citep{1976ApJ...204..187S, 1991ApJ...380L..51H, 1994ApJ...432L..95H}. Such spectrum can be described as a power law from the range $\approx 3\,kT_{\rm bb}$ to $kT_e$  \citep{1980A&A....86..121S}, where $kT_{\rm bb}$ is thermal black-body emission from the accretion disk and $kT_e$ is the electron temperature of the corona. For typical parameters of an AGN super-massive black hole, this range corresponds to energies from a few tens of eV up to a hundred of keV.

It was originally proposed that the AGN corona can originate 
due to vertically transported heat from the accretion disk by convection \citep{1977A&A....59..111B, 1977ApJ...218..247L}
or in magnetically confined loop-like structures similar to the observed structure of the solar corona \citep{1979ApJ...229..318G}.
The major difference between the solar and AGN corona consists in the amount of power that can be released into radiation. While the solar corona contributes to the total flux by less than one percent, the contribution of the X-ray corona in AGN can be any depending on the exact physical and geometrical conditions. 
The fraction of the corona emission mainly depends 
on the accretion state that is determined by the accretion rate.
Low-accreting sources are accompanied with much stronger corona
than highly accreting sources,
as seen mainly from the spectral-hardness changes
observed in accreting stellar-mass black holes in X-ray binaries \citep{2000A&A...354L..67M, 2006ARA&A..44...49R}.
But a similar trend is also observed for AGN.
The ratio between the optical and X-ray flux, the $\alpha_{\rm ox}$ parameter,
increases with the UV luminosity \citep{Steffen2006, Lusso2010, 2017A&A...607A..94M}.

\subsubsection*{Nature and geometry of the X-ray corona}

Despite large theoretical and observational attempts in the last decades,
the basic questions about the nature and geometry of the corona still remain open.
Different heating mechanisms have been proposed from convection of the heat through accretion disk surface to include magnetic reconnection and flares that instantaneously heat the corona. From different considered heating mechanisms, different geometries have been proposed. The corona might be extended, ``sandwiching'' the disk like its ionized plane-parallel atmosphere \citep{1977ApJ...218..247L, 1991ApJ...380L..51H, 1995MNRAS.277...70Z, 1999MNRAS.305..481R, 2000A&A...360.1170R, 2002ApJ...575..117L, 2005A&A...430..761M}, or ``patchy'', related with the single flares due to magnetic reconnection \citep{1979ApJ...229..318G, 1998MNRAS.299L..15D, 2004A&A...420....1C}.
Even for the sandwich-structured (also often called as two-phase corona), the importance of the magnetic dissipation of energy as the heating mechanism was noticed
\citep{1990ApJ...350..295S, 1997MNRAS.291L..23D, 2002ApJ...572L.173L}.

Another often considered geometry of the corona is the so-called ``lamp-post'' model \citep{Matt1991, 1996MNRAS.282L..53M, 2014arXiv1412.8627D, 2016ApJ...821L...1N}.
This model approximates the corona to be a point-like source above a black hole on its rotational axis. The advantage of this setup is mainly the simplicity of the model that can be used for calculating the spectral shape of the primary as well as the reflected radiation from black hole accretion disk. Thus, it can be easily used for fitting the observed X-ray spectra \citep{2004ApJS..153..205D,2016AN....337..362D}, on contrary to e.g. a patchy corona where many additional parameters would be present. The physical motivation of such a model is that it can well approximate a base of the jet \citep{2005ApJ...635.1203M} or a collision of shocks in an aborted jet \citep{2004A&A...413..535G}, or simply a very compact corona. 

The high compactness of the X-ray corona is suggested from X-ray observations. The first indication came from measuring steep radial-emissivity profiles of X-ray reflection spectra of several AGN, such as MCG\,-6-30-15 \citep{2002MNRAS.335L...1F, 2004MNRAS.348.1415V, 2007PASJ...59S.315M}, 1H0707-495 \citep{2009Natur.459..540F, 2010MNRAS.401.2419Z, Wilkins2011, 2012MNRAS.422.1914D}, IRAS\,13224-3809 \citep{2010MNRAS.406.2591P}, and Mrk 335 \citep{Wilkins2015}, as well as X-ray black hole binaries, such as XTE\,J1650-500 \citep{2004MNRAS.351..466M}, GX\,339-4 \citep{2007ARA&A..45..441M}, and Cyg\,X-1 \citep{2012MNRAS.424..217F}. The radial emissivity describes the function of the reflection emissivity on the radius from the centre. 
The measured indices reach values up to $\approx 7$ (i.e. the reflection emissivity $\approx r^{-7}$), which can be hardly explained without requiring a compact X-ray source and strong light-bending effects \citep{Svoboda2012, 2013MNRAS.430.1694D, 2014arXiv1412.8627D}.
Another independent indication of high compactness of the corona comes from the gravitational microlensing of quasars. Recent results from monitoring observations of lensed quasars \citep{Chartas2015} showed that the X-ray corona can indeed be as small as $30\,R_g$. In some cases, such as RXJ 1131-1231, the region producing the X-ray continuum is suggested to be even smaller, about $\sim 10\,R_g$ \citep{2010ApJ...709..278D}.

Nevertheless, any realistic corona cannot be obviously just a point-like source, but has to have some spatial extension. \citet{2016AN....337..441D} have investigated the minimum size of the corona to be still able to up-scatter enough photons to produce the observed level of Comptonized emission. The authors found that any plausible corona has to be at least a few gravitational radii in size and showed that some extreme cases of low height measurements, such as $h < 3\,R_g$ in 1H0707-495 \citep{2012MNRAS.422.1914D}, would require large super-Eddington intrinsic flux. Therefore, a strict lamp-post scenario of the corona with no spatial extension needs to be considered only as an approximation that allows us to make rough ideas of the corona compactness and how important could be the relativistic effects in spectra, such as light bending, gravitational redshift and aberration \citep{2004MNRAS.349.1435M, 2014arXiv1412.8627D}.

Moreover, if the corona represents the base of the jet, also the kinematic properties of the corona would affect the resulting X-ray spectrum including the reflection component \citep{2007ApJ...664...14F, 2013MNRAS.430.1694D}. A dynamic spatially extended corona was studied by \citet{Wilkins2015} who reported a dynamic evolution of the corona in a Seyfert I galaxy Mrk 335. The changes in flux, power-law photon index and reflection emissivity profiles were explained by changes of an X-ray primary source from an extended flat corona to a vertically collimated structure during high-flux states and finally to a compact region (explained as a failed aborted jet) during low-flux states when the flux dropped by an order of magnitude. Such low-flux states are not explained due to a sudden decrease of seed thermal photons (e.g. by a rapid drop of the accretion rate) but rather in the context of the light-bending scenario where a large fraction of radiation is captured by a black hole or bent from the direction towards the observer when corona gets too close to a black hole \citep{2004MNRAS.349.1435M,Dauser2014}.

On the other hand, when the low-flux state is due to the low accretion rate and a lack of the thermal seed photons, a large extended corona can be built on top of the accretion disk. Such geometry is often called as a ``two-phase'' or a ``two-temperature'' accretion flow \citep{1991ApJ...380L..51H, 2002ApJ...575..117L} where the cold phase is the accretion disk and the corona represents a hot accretion flow. If the accretion rate drops below a certain value, the accretion disk may evaporate and the whole accretion flow comprises only the coronal hot flow \citep{1999ApJ...527L..17L}. Such an accretion flow is advection-dominated (ADAF) and radiatively inefficient \citep{1994ApJ...428L..13N}. Observationally, it corresponds to a low-hard state in X-ray binaries when the X-ray spectrum is hard and the thermal emission is diminished.

In the two-phase accretion flow, when the disk becomes luminous, many photons are Comptonized in the corona, which implies cooling of the hot flow and thus significant weakening of the X-ray emission from the corona \citep{2012A&A...544A..87M}. Therefore, the Compton cooling prevents ADAF-like coronae to exist in high-luminosity states, which also naturally explains the observational fact that the large powerful coronae are not present in high-accreting sources \citep{2017A&A...607A..94M}.

The ADAF-like coronae are geometrically thick and optically thin. However, a two-temperature accretion flow may still exist when the heating mechanism is via magnetic dissipation through thin active regions on the thin disk \citep{1997MNRAS.291L..23D}. Most of the magnetic energy is acquired by protons that transfer the energy via Coulomb collisions to electrons and compensate the energy losses due to cooling via inverse Comptonization of photons. 
Such model would allow the existence of a plane-parallel corona above the accretion disk in standard AGN. The Comptonizing effects of the corona do not apply only to the thermal emission but also to the X-ray reflection \citep{2015MNRAS.448..703W,2017ApJ...836..119S}.



\subsubsection*{Measurements of the X-ray corona temperature}

The main physical characteristics of the corona is its temperature.
From the virial theorem, the temperature of particles near black holes can be
expressed as:
\begin{equation}
kT_{\rm vir} = \frac{GM_{\rm BH}m}{r} \approx \frac{mc^2}{r/R_g},
\end{equation}
and thus it does not depend on the black hole mass but only the mass of the particles.
The proton temperature is of order of a few MeV, while the electron temperature is
$T_{\rm vir} \approx 25(r/10 r_g)^{-1}$ keV. The radiation is produced mainly by scattering on electrons and thus the electron temperature determines the resulting spectrum. The X-ray power law is exponentially decreasing as:
\begin{equation}
A(E) = KE^{-\Gamma}e^{(-E/E_{\rm cut})},
\end{equation}
where $\Gamma$ is the power-law photon index and $E_{\rm cut}$ is the high-energy cutoff.
The high-energy cutoff is related to the electron temperature as:
\begin{equation}
E_{\rm cut} \approx 2-3\,kT_e,
\end{equation}
where the lower value is appropriate for the optical depth $\tau \leq 1$, 
while the higher value is for $\tau \gg 1$ \citep{2001ApJ...556..716P}.
The optical depth can be determined from the electron temperature and the measured photon index $\Gamma$.

The advancement of hard X-ray detectors onboard NuSTAR space mission \citep{2013ApJ...770..103H} lead to the first precise measurements 
of high-energy cutoffs in AGN \citep{Fabian2015}. The authors conclude that AGN coronae are hot and radiatively compact and the pair production and annihilation play important role in forming the spectral shape.


\begin{figure}[tbh!]
 \centering
 \begin{tabular}{cc}
 \includegraphics[width=0.8\textwidth]{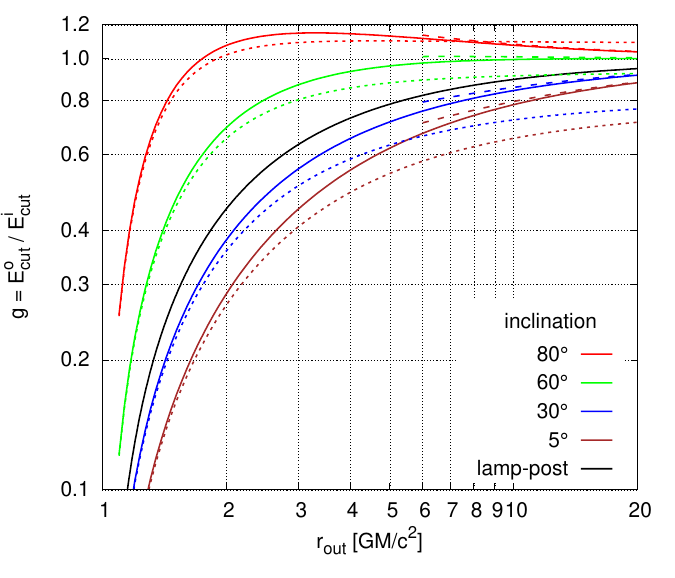} 
 \end{tabular}
\caption{The energy-shift factor plotted as function of an outer radius of a flat disk-like rotating corona, emitting a cutoff-powerlaw ($\Gamma = 2$) spectrum, for various inclination angles, assuming constant emissivity for extreme Kerr and Schwarzschild BH (solid and dashed lines, respectively) and a r$^{-3}$ emissivity profile for the case of the disk-corona around Kerr BH (dotted lines). The black solid line shows the energy-shift factor in the case of the lamp-post geometry (in this case, the r$_{\rm out}$ values on x-axis correspond to the height). Adapted from \citet{Tamborra2018}.}
\label{tamborra}
\end{figure}

\subsubsection*{Relativistic effects on the cutoff energy measurements}

In previous chapters, we discussed the relation between the spectral high-energy cutoff and the corona temperature. The observationally constrained cut-off energy can thus determine the physical nature of the corona. However, such measurements can be strongly affected by relativistic effects, namely by the gravitational redshift when the corona is compact, or by the Doppler shift when the corona is highly rotating, or by the combination of both. 

The relativistic redshift of the high-energy cutoffs was first considered by \citet{Fabian2015}, who employed the geometry of the corona from the lamp-post height constrained by the reflection models of the soft excess, the iron line, and the Compton hump. 
The energy shift of the emission from a point on the black-hole rotational axis at a given height $h$ from the centre can be calculated as:
\begin{equation}
\label{g_LP}
g \equiv E_{\rm observed}/E_{\rm intrinsic} = \sqrt{\dfrac{h^2-2h+a^2}{h^2+a^2}}.
\end{equation}
The observed power-law emission with a high energy cutoff has the form:
\begin{equation}
\label{f_obs_cutoff_lp}
F_{\rm obs} = NE^{-\Gamma}\exp\left(g\frac{-E}{E^{i}_{cut}}\right).
\end{equation}
The lamp-post corona represent, however, a static and dimensionless model. \citet{Tamborra2018} considered also the other extreme case -- a flat disk co-rotating corona, and noted that a realistic corona should be something in between. In such geometry, the energy shift $g$ is not a single value but is an integrated value over the corona extent. \citet{Tamborra2018} showed that the resulting profile of the corona emission can still be approximated by Eq.~\ref{f_obs_cutoff_lp} with a $g$-factor being not a precise value but the fit of the intrinsic integrated emission by a model defined by Eq.~\ref{f_obs_cutoff_lp}. Such g-factor is shown in Figure~\ref{tamborra} for different considered cases.

The energy shift of the cutoff energy in the disk co-rotating corona model is a function of an outer radius of such corona. If the corona is compact and the outer radius is within a few gravitational radii, then similarly to the lamp-post model, the  cutoff energy is more redshifted the more compact the corona is. However, the energy shift in the disk co-rotating corona model  
strongly depends also on inclination, as it is shown in Figure~\ref{tamborra}. When compared to the lamp-post model (black solid line in the plot), the larger redshift is obtained for inclinations lower than 30 degrees, and vice versa for higher inclinations. The higher redshift is mainly caused by integration of the corona emission down to the innermost stable circular orbit, but also the transverse Doppler shift contributes to the total redshift in the low-inclined systems. On the other hand, the Doppler boosting of emission coming from the matter moving rapidly towards the observer dominates for highly-inclined systems. As a consequence, the measured cutoff energy may become even blueshifted for inclinations as large as 80 deg.   


\clearpage

\section{Processes relevant for outflows and jets}  
\epigraph{\textit{``Science progresses best when observations force us to alter our preconceptions."}}{--- \textup{Vera Rubin}}
\label{Jets}
  \subsection[Strong-gravity]{Strong-gravity and electromagnetic fields}
\newcommand{\beq}{\begin{equation}}
\newcommand{\eeq}{\end{equation}}
\newcommand{\rd}{{\,\rm d}}
\newcommand{\ri}{{\rm{i}}}
\newcommand{\nab}{\mbox{\protect\boldmath$\nabla$}}
\newcommand{\BoldM}[1]{\mbox{\boldmath{${#1}$}}}



The electromagnetic field is governed by Maxwell's equations, 
the set of first-order differential equations for the electric and magnetic
intensity vectors. When expressed in the equivalent (perhaps more
elegant) four-dimensional tensorial formalism, the mutually coupled equations can be 
unified; the electric and the magnetic field components are then comprised
in a single quantity. The latter approach is particularly useful in the framework 
of the Special Theory of Relativity.
Whichever of the two equivalent formulations is preferred, one has to tackle
differential equations and, therefore, the appropriate initial and
boundary conditions must be specified in order to fully determine the
structure of the field. As the original literature, review articles and textbooks are numerous on the subject of astrophysical applications 
of magnetized plasma embedded in combined, mutually interacting gravitational and electromagnetic 
fields \citep[e.g.][]{1980panp.book.....M,2012bhae.book.....M} we will add to this the context of idealized
analytical solutions that allow us to gain some additional insight into 
physical mechanisms.

The cosmic environment is often ionized and the local electric field component is
thus efficiently neutralized in the frame comoving with the highly conducting medium
(force-free approximation). On the
other hand, magnetic fields are abundant throughout the interstellar and
intergalactic space and their intensity can grow to enormous magnitude
especially near rotating compact stars and pulsars \citep{2014MNRAS.445.2500G,2016smfu.book.....B}.

Maxwell's equations are not directly coupled with Newton's gravitational
law; one could thus conclude that there is no immediate relation
between the electromagnetic field of the body and its gravitational influence. Not so in
General Theory of Relativity. Energy density of the electromagnetic field, as of
any other field, stands explicitly in Einstein's equations for the
spacetime structure, contributing as a source of the gravitational
field. Einstein's and Maxwell's equations must be thus considered
simultaneously and this makes the whole system of equations
difficult to solve in full generality.
Fortunately, the energy density contained in realistic
electromagnetic fields turns out to be far too low to influence the
spacetime noticeably. Test-field solutions are adequate and accurate enough under such
circumstances. Here, by test-field solutions one means the solution of Maxwell's equations in a
fixed (albeit generally curved) spacetime, whose structure can be pre-determined by solving
Einstein's equations with no electromagnetic terms present.
Maxwell's equations are solved afterwards and, by assumption, they have
zero impact on the gravitational field. This is also an approximation which
we will adopt to start our discussion, but we will abandon it later to see the difference
and understand the interaction between Einstein--Maxwell fields. 

We first assume the spacetime of a
rotating black hole, namely, Kerr metric; see \citet{1973grav.book.....M}.
Different arguments can be put forward to demonstrate that 
magnetohydrodynamic effects often dominate the behaviour of cosmic plasma and 
must be taken into account even very close to the black hole event horizon. 
On the other hand, in the immediate vicinity of the black hole, in particular within 
the ergosphere of a fast rotating black hole, the gravitational effects must eventually
prevail. Also near compact stars the gravitational terms cannot be ignored, and so the
general-relativity MHD has to be employed. Here we will confine ourselves 
to the basic arguments in circumstances where gravitational effects cannot be ignored.  
The main intention is to describe the assumptions and to suggest 
selected references that would help entering the complex subject.
We will start with axially symmetric and stationary flows, so that
the system is further simplified by the presence of the additional symmetries. 

The aim of this chapter is to outline the basic equations governing the coupled
Einstein--Maxwell field and to mention different limits: the electro-vacuum case vs.
MHD case with plasma connected to the evolving magnetic field; and 
test (weak) electromagnetic fields vs.\ exact solutions of the full set, where the energy density of the 
electromagnetic component contributes to the space-time curvature. We will also mention 
the generating techniques that have been developed in order to find new solutions. These approaches
have been found particularly useful in adding the magnetic terms (``magnetizing'') to originally
unmagnetized black-holes and to reveal the impact of mutual interaction of Einstein--Maxwell fields.
Indeed, this interplay is interesting and non-trivial: asymptotic flatness is lost, singularities develop 
even outside the event horizon, solutions change their topological classification, etc.
We will give references to selected useful works but the limited space does not allow us to 
present a complete overview; we do not aim to give a comprehensive list of the 
literature. An extensive pedagogical exposition can be found in textbooks
\citep{2007coaw.book.....C,2001bhgh.book.....P,2013rehy.book.....R}.

Let us note that the influence of large-scale magnetic
fields upon the space-time metric can be characterized by
dimensionless parameter $\beta=B_0M$, where $B_0$ is the field strength
and $M$ is the mass of the object in geometrized units. Considering the
above-mentioned limit on the field strength near a one solar-mass
compact body, we obtain $\beta\approx10^{-5}B_{15}(M/M_{\odot})\ll1$, where
$B_{15}=B_0/(10^{15}{\rm{G}})$. Such magnetic terms contribute very
substantially to the spacetime metric on spatial scales of
$r\approx\beta^{-1}$ which in physical units, corresponds
to the length-scale of $\approx10^5$\,km. In the case of intergalactic fields
the magnetic intensity is small, typically below $1\,\mu$G,
but near the supermassive black hole the fields frozen into accreting plasma
become amplified by many orders. Specifying $M=10^9M_\odot$ and 
$B_0=10^4\,$G we have $\beta=10^{-7}$ and the critical length-scale is 
$r\simeq10^6R_{\rm g}$. Let us note that the transformation to physical units 
is achieved by relations $M= GM_{\rm phys}/c^2$ for the mass and 
$B_0= G^{1/2}B_{\rm phys}/c^2$  for the magnetic field parameter.

Historically, the field of astrophysical MHD has been covered in textbooks and review
articles; see, e.g., \citet{1976bah..book.....C}; \citet{1980panp.book.....M}; \citet{1983mfa..book.....Z};
\citet{1994ASIC..422.....L}. The applications range from solar physics to accretion disks 
in galactic nuclei; in our context, the original motivation
comes from the studies of particle acceleration near neutron stars where
strong electromagnetic fields and inertial effects are present \citep{1968Natur.219..145P,1969ApJ...157..869G,1982RvMP...54....1M}, and where the strength of the magnetic intensity is 
maintained and further enhanced by accretion. Magnetic fields
can be amplified by the dynamo action in the disk (\citet{1992ApJ...400..610B}; \citet{1993IAUS..157.....K}; \citet{2003ARA&A..41..555B}). 

The set of coupled Maxwell's and Euler's equations are the essential prerequisites 
to start with in the classical framework of non-vacuum equations. We can
follow a highly simplified exposition with
\begin{eqnarray}
 \BoldM{\nabla\times B} &=& \frac{4\pi}{c}\,
 \BoldM{j}+\frac{1}{c}\,
 \frac{\partial\BoldM{D}}{\partial t},\qquad
 \BoldM{\nabla\cdot B} \;=\; 0,\\
 \BoldM{\nabla\times E} &=& -\frac{1}{c}
 \frac{\partial \BoldM{B}}{\partial t},\qquad
 \BoldM{\nabla\cdot E} \;=\; 4\pi\rho_{\rm e}
 \label{mxw}
\end{eqnarray}
with 
$\rho_{\rm e}$ and $\BoldM{j}$ being density of {\em all\/} electric charges and 
currents (e.g., \citet{2005ragt.meet...71K}). Euler's equation adopts the appropriate form:
\begin{equation}
 \rho\,\frac{\partial\mbox{\protect\boldmath$v$}}{\partial t}
 +\rho\,\mbox{\protect\boldmath$v.\nabla v\;$}=
 -\mbox{\protect\boldmath$\nabla$}P+\;\mbox{\protect\boldmath$f$}
\label{eq:eul}
\end{equation}
with
\begin{equation}
 \mbox{\protect\boldmath $f$}=
 \mbox{\protect\boldmath $f$}_{\rm L}+
 \mbox{\protect\boldmath $f$}_{\rm g}=
 \rho_{\rm e}\mbox{\protect\boldmath $E$}+c^{-1}
 \mbox{\protect\boldmath $j\times B$}-\rho\mbox{\protect\boldmath $g$}
\end{equation}
being the Lorentz and gravitational terms, respectively. Mainly for the sake of
simplification the assumptions about axial symmetry and stationarity
are imposed. Although the general relativity effects have not been taken into account, 
a straightforward reformulation is possible
in which the form of the field equations remains almost identical even if
strong gravitational fields are present \citep[3+1 formalism;][]{1982MNRAS.198..345M,1986bhmp.book.....T}.

The primary distinction between the models with and without a black hole
consists in different boundary conditions imposed upon the
electromagnetic field, which threads the black hole horizon. The
relevant Maxwell equations describing the field outside the black hole
horizon can be solved by introducing appropriate imaginary currents
flowing on the surface of the horizon. These are defined in
such a way that the boundary conditions are satisfied. Currents
flowing along the field lines can thus close a circuit and the energy
extraction is then described in an analogous way as in the
discussion of magnetized disks or as in the theory of pulsar emission
\citep{1976MNRAS.176..465B,1986A&A...156..137C}.

Let us first employ the three-vector formalism in the classical form.
The above-given equations can be simplified by assuming the
force-free approximation,
\begin{equation}
 \rho_{\rm e}\mbox{\protect\boldmath $E$}
 +c^{-1}\mbox{\protect\boldmath$j\times B$}=0.
 \label{ff}
\end{equation}
The physical interpretation and consequences of the above relation
require a thorough discussion. Equation (\ref{ff}) tells us that
inertia of the material is neglected. In other words, the influence of
the Lorentz force acting on plasma in the comoving frame gets
neutralized immediately by induced electric currents; perfect
conductivity is thus assumed. A dimension-less condition for the validity
of the force-free approximation is $\rho\Gamma v^2/B^2\ll1$. This 
corresponds to a similar assumption of ideal MHD,
\begin{equation}
\mbox{\protect\boldmath$E^\prime\;$}\equiv\,
 \mbox{\protect\boldmath$E\;$}+\,c^{-1}\,
 \mbox{\protect\boldmath$v\times B\;$}=0;
 \label{mhd}
\end{equation}
where {\boldmath$E^\prime\;$} is the electric field in the local system
attached to plasma, and the two forms become equivalent if the current
density  is proportional to the velocity of the medium,
$\BoldM{j}=\rho_{\rm e}\BoldM{v}$. A~more general formula for the
current density that still satisfies the force-free assumption
(\ref{ff}) has a form $\BoldM{j}=\rho_{\rm{}e}\BoldM{v}+\mu\BoldM{B};$
$\mu$ is a scalar function to be determined. Both the force-free and
the perfect MHD fields are degenerate, i.e. {\boldmath$E\cdot
B\;$}$=0$. The approximation of ideal MHD can be understood as an
assumption about perfect electric conductivity of the material.
Substituting
\begin{equation}
 \BoldM{j}=\sigma\BoldM{E}^\prime
 \label{ohm}
\end{equation} for the vector of electric field from Ohm's law ($\sigma$
designates specific conductivity of the medium) and assuming perfect
conductivity $(\sigma\rightarrow\infty)$, we find
\begin{equation}
 \mbox{\protect\boldmath$\nabla\times\,$}(\mbox{\protect\boldmath$
 v\times B$}) = \frac{\partial\mbox{\protect\boldmath$B$}}{\partial t}.
 \label{freeze}
\end{equation}
Equation (\ref{freeze}) expresses the freezing of the magnetic field in
plasma material. The reason for this denomination is evident upon
realizing that the magnetic flux across an imaginary loop ${\ell}$
flowing together with the medium can be written as a sum of two terms,
the first one being determined by motion of the loop,
\begin{equation}
\int_{\cal S}\mbox{\protect\boldmath$\nabla\times\,$}(
  \mbox{\protect\boldmath$v\times B$})\mbox{\protect\boldmath$\cdot$}d
  \mbox{\protect\boldmath$\cal S$}=
  \oint_{\ell}(\mbox{\protect\boldmath $v\times B$})\,
  \mbox{\protect\boldmath$\cdot$}d\mbox{\protect\boldmath$\ell$}=
  -\oint_{\ell}\mbox{\protect\boldmath $B\cdot$}
  (\mbox{\protect\boldmath $v\times$}d\mbox{\protect\boldmath $\ell$}).
\end{equation}
The term $\partial${\protect\boldmath$B$}$/\partial t$ on the
right-hand side of (\ref{freeze}) corresponds to the change of the
magnetic flux due to the explicit time-dependence of
{\protect\boldmath$B$},
\begin{equation}
 \int_{\cal S}\frac{\partial\mbox{\protect\boldmath$B$}}{\partial t}
 \,\mbox{\protect\boldmath$\cdot\,$}d\mbox{\protect\boldmath$S$}.
\end{equation}
Equation (\ref{freeze}) thus expresses the fact that the magnetic flux
across any arbitrary closed loop remains constant. As we have seen
before, one can also understand this equation as a condition for the
electric field to vanish in the rest frame of plasma. 

\label{asf}
Now we can examine the basic relations valid of axially symmetric
magnetohydrodynamic equilibrium configurations under forces of (weak) gravity.
The relevant equations are capable of describing,
for example, magnetospheres and collimated outflows from
aligned rotators and magnetized accretion disks, as derived in detail
by \citet{1986A&A...156..137C} and \citet{1986ApJS...62....1L}. To remind the reader about the assumption,
we adopt the axial
symmetry and stationarity: we set $\partial/\partial\phi=0$,
$\partial/\partial{t}=0$ in all formulae. Starting equations are the mass
conservation law (the continuity equation); the momentum conservation law
(Euler equation, supplemented by the relation for the external
force $\BoldM{f}=\frac{1}{c}\,\BoldM{j\times B}-\rho\BoldM{g}$, where we
assume an electrically neutral plasma, $\rho_{\rm e}=0)$; and Maxwell's
equations with the perfect MHD constraint. The gravitational
acceleration \BoldM{g} is linked to density via the Poisson equation,
$\nabla^2\Phi=4\pi G\rho$, and, finally, it has to be 
supplemented by the first law of thermodynamics and the equation of state
to close the set of equations.

It follows from Faraday's law together with the conditions of axial symmetry and
stationarity ($\BoldM{\nabla\times E}=0$) that the toroidal part
$\BoldM{E}_{\rm T}$ of electric field vanishes,
\begin{equation}
 E_\phi^2=\BoldM{E}_{\rm T}\BoldM{\cdot E}_{\rm T} = 0.
 \label{et}
\end{equation}
The condition of perfect MHD implies the relation
for the poloidal flow velocity
\begin{equation}
 \BoldM{v}_{\rm P}=\xi\BoldM{B}_{\rm P},
 \label{xi}
\end{equation}
where $\xi(R,z)$ is a (yet undetermined) scalar function. It is
advantageous at this point to introduce into the Maxwell equations the
vector potential $\BoldM{A}$ and the scalar magnetic flux function,
$\Psi(R,z)\equiv RA_\phi$. Components of $\BoldM{B}_{\rm P}$ in terms
of $\Psi$ read $B_R=-\Psi_{,z}/R$, $B_z=\Psi_{,R}/R,$ where the coma
denotes partial differentiation. It is now evident from equation
(\ref{xi}) that
\begin{equation}
 4\pi\rho\xi=F_1(\Psi),
\end{equation}
where $F_1(\Psi)$ is an arbitrary function to be specified by the
boundary conditions and symmetries of the required solution. (We have
applied the Maxwell equation
$\BoldM{\nabla\cdot B}=\BoldM{\nabla\cdot B}_{\rm P}=0,$ and
the continuity equation.) We will see that
there is a set of such functions of $\Psi$ that determine a
specific solution. Each function can be identified with some conserved
quantity (derivation of $F_1$ utilizes the mass conservation 
equation). The existence of flux functions, which remain constant on
magnetic surfaces $\Psi={\rm const}$, is crucial in investigating axisymmetric
hydromagnetic flows (see \citet{1961hhs..book.....C}).

It follows from $\BoldM{v}_{\rm P}=\xi\BoldM{B}_{\rm P}$ [Eq.\
(\ref{xi})] that
\begin{equation}
 \BoldM{v\times B}=\BoldM{v_{\rm T}\times B_{\rm P}}+\BoldM{v_{\rm P}\times
 B_{\rm T}}=\frac{v_\phi-\xi B_\phi}{R}\,\BoldM{\nabla}\Psi.
\end{equation}
Curl of the last equation vanishes in accordance with the perfect MHD
condition and Faraday's law so that another stream function, $F_2,$ can be
introduced in the following way:
\begin{equation}
 \frac{v_\phi-\xi B_\phi}{R}=F_2(\Psi),\qquad
 \BoldM{E}=-c^{-1}F_2(\Psi)\BoldM{\nabla}\Psi.
\end{equation}
Further relations are obtained by projections of the Euler equation
and can be derived via straightforward but lengthy manipulations.
The toroidal part reads
\begin{equation}
 \BoldM{B}_{\rm P}\BoldM{\cdot\nabla}\left(RB_\phi-F_1Rv_\phi\right).
\end{equation}
For analogous reasons as those that have been presented with equation
(\ref{xi}), the term in parentheses is also a function of $\Psi$ only,
say $F_3(\Psi)$. Other two independent relations can be obtained by
projecting the Euler equation into the poloidal plane. The projection
along $\BoldM{B}_{\rm P}$ yields the Bernoulli equation
\begin{equation}
 {\textstyle{\frac{1}{2}}}\,v^2+\int_{\Psi={\rm const}}\frac{{\rm d}P}{\rho}
 +\Phi-Rv_\phi F_3=F_4(\Psi).
 \label{eul4}
\end{equation}
Compared to the hydrodynamical form of the Bernoulli integral, in which
electromagnetic effects are not considered, the additional term
$Rv_\phi F_3$ corresponds to the electromagnetic (Poynting) energy
transport. The projection of the poloidal component of the Euler
equation to the direction parallel to $\BoldM{\nabla}\Psi$ serves as a 
Grad--Shafranov master equation that can close the set of independent
equations
(for the application to black hole magnetospheres, see e.g. \citet{1977MNRAS.179..433B};
\citet{2000ApJ...541..257B}; \citet{2003PhRvD..67l4026I}; \citet{2013ApJ...765..113C}). This
is a non-linear differential equation for $\Psi,$ the explicit form of
which naturally depends on the equation of state and on stream
functions $F_k$. For example, we set $F_1=0$ if no poloidal flow of
material is required {\it a priori\/} (the case of disks). Force-free
approximation to the Grad--Shafranov equation is equivalent to the
self-consistent form of the pulsar equation from the astrophysical
literature (\citet{1973ApJ...179..269C}, \citet{1973ApJ...182..951S}). Within 
the general relativity framework, its applications to rotating
compact stars and black holes have been examined by various authors
(see e.g. \citet{2005MNRAS.358..998K}). On the other hand, laboratory
plasma in tokamaks are often described within the approximation of a vanishing
material flow, $F_1=F_2=0,$ and negligible gravity, $\Phi=0$. 

Electromagnetic forces act on charged particles and may substantially
modify the structure of accretion disks. These have been originally explored  
in a series of papers by \citet{1982RvMP...54....1M}, \citet{1983ApJ...266..188M}. The inclusion of
electromagnetic effects makes the disk theory much more complex. A
simplified approach is possible in terms of self-similar solution of
axially symmetric MHD equations \citep{1989ApJ...342..208K}. 


Let us consider a magnetized accretion disk in the equatorial plane and assume that the
magnetic field is frozen into the disk. The toroidal part of the field
arises from the dragging of the magnetic field by the disk material. It
follows that $\BoldM{B}_{\rm T}=B_\phi\,\BoldM{e_\phi},$
$\BoldM{B}_{\rm P}=\BoldM{B}-\BoldM{B}_{\rm T}=
B_R\,\BoldM{e_R}+B_z\,\BoldM{e_z}$. 


Now the basic relations adopt the following form. The Maxwell equation
$\BoldM{\nabla\cdot B}=0$ together with the consequence of axial
symmetry, $\BoldM{\nabla\cdot B_{\rm T}}=0,$ yield
$\mbox{\protect\boldmath$\nabla\cdot B_{\rm P}$}=0$. This means that
both the poloidal and the toroidal components can be separately
associated with unending field lines. The value of {\protect\boldmath$
E_{\rm P}$} follows from the force-free condition,
\begin{equation}
 0=\BoldM{E}^\prime=\BoldM{E}_{\rm P}+c^{-1}(
 \BoldM{\Omega\times r})\BoldM{\times B},
 \label{eq:ep}
 \end{equation}
where $\BoldM{\Omega}=\Omega^{\rm F}\,\BoldM{e_\phi}$
means the angular velocity of
each field line and {\protect\boldmath$r$} is the radius vector.
Charged particles move along field lines.
Using $\BoldM{\nabla\times E_{\rm P}}=0$  and
$\BoldM{\nabla\cdot B_{\rm P}}=0$
we find
\begin{equation}
 \BoldM{B_{\rm P}\cdot\nabla}\Omega^{\rm F}=0.
\end{equation}
In other words, the angular velocity of each field line remains
constant along its curve; thus $\Omega^{\rm F}$ does not change along
poloidal field lines. This result is called Ferraro's law of
iso-rotation \citep{1961itmf.book.....F}.

The light surface is the locus of points where  the velocity of the
field lines approaches the speed of light. Charged  particles cannot
corotate with field lines beyond the light surface; instead, they are
forced to move away and this is the basis of particle acceleration
around pulsars and possibly formation of jets in extragalactic
sources. On the other hand, the accretion disk itself can serve as a source of particles. Assuming
the perfect MHD condition inside the disk, we obtain for the particle
density
\begin{equation}
 n = \frac{1}{4\pi q_{\rm e}}\mbox{\protect\boldmath$\nabla\cdot E$}\;=
 -\frac{1}{4\pi q_{\rm e}c}\;\mbox{\protect\boldmath$\nabla\cdot$}(
 \mbox{\protect\boldmath$v\times B$})
 = -\frac{1}{2c}\Omega B_\phi.
\end{equation}
Non-zero charge-density generates an electric field, which pulls charged
particles out of the disk.

Magnetic field lines threading the disk exert a torque on its material,
$\BoldM{G}=\BoldM{R\,\times}\,(\BoldM{j\times B})$,
and are thus a source of effective viscosity. Such a disk does not
radiate (remember that we are considering axisymmetric stationary
configurations). Consider now a circle of radius $R$ centered on the symmetry axis. 
Ampere's law yields $B_{\rm T}={2J}/{cR}$.
We have already mentioned that in the force-free region currents flow along
magnetic surfaces, but in the disk and in the far region the force-free
condition is violated and dissipation occurs. The density of the
electromagnetic energy flowing through the force-free region is given by
$\BoldM{P}=c\,\BoldM{E\times B}\approx c\,\BoldM{E_{\rm P}\times B_{\rm T}}$.
Substituting for {\protect\boldmath$E_{\rm P}$} from
Eq.\ (\ref{eq:ep}) we estimate the magnitude of this vector as
${\cal P}\approx\Omega RB_{\rm P}B_{\rm T}$. Finally,
$B_{\rm P}$ and $B_{\rm T}$ are to be determined in accordance
with the boundary conditions.

Magnetohydrodynamic equations are only applicable if the medium
can be treated as a continuum. This assumption does not hold in case of
highly diluted, in which case the
mean free path of particles is comparable with
other characteristic length-scales of the system;
small-scale turbulence may save the MHD approximation
even at low densities but large-scale magnetic fields break it 
in regions of strong magnetic dominance \citep{1962clel.book.....J}. There
are various possibilities for extending the validity of the description
in terms of continuum beyond the ideal limit towards to include effects
of finite resistivity. For example, unequal directional action of the magnetic 
field on charged particles can be accommodated by replacing the scalar
conductivity $\sigma$ in Ohm's law by a tensorial quantity. Eventually,
the extreme situations of highly rarefied magnetized
plasma can be adequately described only by adopting the kinetic theory.

In the guiding-center
approximation one assumes that local gyrations and slow-drift motions can
be neglected \citep{Lehnert1965}. This restriction requires the radius of the gyrations be much
less than a characteristic length-scale of the field inhomogeneities,
and the period of the gyrations to be much less than a characteristic time-scale of
the field evolution. If the conditions are satisfied one can average out
gyrations and drifts. Instead of the exact trajectory, one follows
an imagined path of the center of gyrations. Flow lines of plasma coincide with the
common direction of electric and magnetic fields in the plasma
co-moving frame. The equation of motion of a particle of rest mass $m_0$ and electric
charge $q$ then reads \citep{1991JMP....32..714K}
\begin{equation}
\frac{d}{dt}\left(\Gamma m_0\BoldM{v}\right)=
 \Gamma m_0\BoldM{g}+
 \frac{q}{c}\,(\BoldM{E}+\frac{1}{c}\BoldM{v\times B}),
 \label{vb1}
\end{equation}
where $\Gamma$ is the Lorentz factor and $\BoldM{g}$ is the gravitational acceleration.

\subsection{General Relativity framework}
Let us now generalize our discussion to the case of General Relativity (GR)
framework for axially symmetric MHD flows. In this section, very naturally, we will employ the standard 
GR notation with geometrized units ($c=G=1$) and the signature of metric
$-\mbox{+++}$. First we will still maintain the approximation
of electromagnetic test fields (albeit in the curved background), so that the
spacetime metric is not electromagnetically influenced. The set of equations of perfect
magnetohydrodynamics can be written in the following form \citep{1989LNM..1385.....A,1967rhm..book.....L}. 

Let us note that the environment of diluted cosmic plasma, accretion flows in particular, 
has been permeated by magnetic fields of diverse magnitude and topology. In the context of
black-hole accretion the field lines can be either turbulent and highly entangled (on length-scales
$\ell\ll R_{\rm g}$, or they can be organized on much larger scales, as various simulations indicate
in the evacuated regions along the symmetry axis. Also time-scales vary wildly, from the fields that
change explosively (compared to the light-crossing time across the event horizon, $\tau_\ell\simeq\ell/c$)
to steady systems that evolve adiabatically over much longer (viscous) period. In following, to reveal 
effects connected with the overall geometry of the system and associated effect of gravitation, we
will concentrated on large-scale, organized, slowly evolving fields.

Conservation of the particle number reads
$ \left(\rho_0u^\alpha\right)_{;\alpha}=0$, with $\rho_0=mn; \label{mc}$, where
$m$ is the particle rest mass, $n$ numerical density, $u^\alpha$
four-velocity. Here we do not consider a possibility of creation of
pairs, which would break this conservation law.
Normalization condition for four-velocity is written
$ u^\alpha u_\alpha=-1$. As originally shown by \citet{ZnajekPhd}, 
by employing the explicit form of the energy-momentum conservation
$ {T^{\alpha\beta}}_{;\beta}=0$, and the definition of
energy-momentum tensor in terms of material density $\rho$ and pressure $P$, 
one finds \citep{PhinneyPhd}
\begin{equation}
T^{\alpha\beta}=T^{\alpha\beta}_{\rm matter}+T^{\alpha\beta}_{\rm EMG},
\end{equation}
where the right-hand side contains the source
term divided in two components,
\begin{eqnarray}
T^{\alpha\beta}_{\rm matter}&=&(\rho+p)u^\alpha u^\beta+pg^{\alpha\beta},\\
T^{\alpha\beta}_{\rm EMG}&=&\frac{1}{4\pi}\left(F^{\alpha\mu}F^\beta_\mu-
 \frac{1}{4}F^{\mu\nu}F_{\mu\nu}g^{\alpha\beta}\right),
\end{eqnarray}
with $F_{\mu\nu}=A_{\nu,\mu}-A_{\mu,\nu}$. 

The axial symmetry and stationarity guarantee the existence of two
Killing vectors, $k^\alpha=\delta^\alpha_t$ and
$m^\alpha=\delta^\alpha_\phi$, which satisfy relations
\begin{equation}
 0=k_\alpha {T^{\alpha\beta}}_{;\beta}=\left(k_\alpha T^{\alpha\beta}
 \right)_{;\beta}, \label{ka}
\end{equation}
\begin{equation} 0=m_\alpha {T^{\alpha\beta}}_{;\beta}=\left(m_\alpha T^{\alpha\beta}
 \right)_{;\beta}.
 \label{ma}
\end{equation}
Let us note that
the electromagnetic field may or may not conform to the same symmetries
as the gravitational field. Naturally, the problem is greatly simplified
by assuming axial symmetry and stationarity for both fields. 

In the fluid rest frame, the electric field is assumed to vanish
completely: $F_{\alpha\beta}u^\beta=0$. It follows 
\citep{ZnajekPhd,1977MNRAS.179..433B,1992ApJ...386..455H,1992A&A...263..401K,2017ApJ...836..193P}
%
%
\begin{equation}
 \frac{A_{t,r}}{A_{\phi,r}}=\frac{A_{t,\theta}}{A_{\phi,\theta}}\equiv-
 \Omega^{\rm F}. \label{omf}
\end{equation}
The latter relation implies
${\Omega^{\rm F}_{,r}}/{\Omega^{\rm F}_{,\theta}}={A_{\phi,r}}/{A_{\phi,\theta}}$.
Indeed, Eq.\ (\ref{omf}) is an analogy of Eq.\ (\ref{eq:ep}) with $\Omega^{\rm F}$ playing the role of  
angular velocity of
magnetic field lines. A more detailed exposition and derivation of these relations
can be found in \citet{PhinneyPhd}, where it is shown that the latter equation 
implies that the two Jacobian matrices
\begin{equation}
\frac{\partial(A_t,A_\phi)}{\partial(r,\theta)}=0,\qquad\qquad
\frac{\partial(\Omega^{\rm F},A_\phi)}{\partial(r,\theta)}=0
\end{equation}
vanish. Therefore, the mentioned functions are not independent: 
$A_t\equiv A_t(A_\phi)$ and $\Omega^{\rm F}\equiv
\Omega^{\rm F}(A_\phi)$. The flow stream-lines and the magnetic
field-lines lie in the level surfaces of $A_\phi,$ i.e.
$\vec{u}\cdot\nabla A_\phi=\vec{B}\cdot\nabla A_\phi=0,$ where
$\vec{B}=(^*{F})\cdot\vec{u}$ (the arrow denotes two-component
space-like vectors defined in the $(r,\theta)$-plane). 

In order to find the possible geometry of flow lines, one can introduce
the stream function $k(r,\theta)$ satisfying \citep{PhinneyPhd,1991PhRvD..44.2295N}
\begin{equation}
 \frac{u^r}{A_{\phi,\theta}}=-\frac{u^\theta}{A_{\phi,r}}\equiv
 \frac{k(r,\theta)}{4\pi\sqrt{-g}\rho_0};
 \label{k}
\end{equation}
%
%
apparently, the functional dependence is constrained to $k(r,\theta)\equiv k(A_\phi)$. Two additional stream functions can be
obtained by inserting the explicit form of $T^{\alpha\beta}$ into
equations (\ref{ka})--(\ref{ma}). 
%
Finally, we are still left with the two equations,
${T^{r\beta}}_{;\beta}=0$ and ${T^{\theta\beta}}_{;\beta}=0,$ but also
these relations are not independent
and they can be constrained
by contracting ${T^{\alpha\beta}}_{;\beta}=0$ with
any poloidal four-vector which is linearly independent of poloidal
projection of $u_\alpha$. The result is a non-linear second-order
differential equation which is a generalization of the Grad--Shafranov
equation within general relativity \citep[as outlined by][]{2005MNRAS.358..998K}.

So far our discussion has been restricted to weak,
electromagnetic test-fields in a given, fixed background spacetime; we
have neglected the influence of the electromagnetic field on the
spacetime metric. This approach was employed by a number of authors to
address the problem of electromagnetic effects near a rotating (Kerr)
black hole \citep{1986bhmp.book.....T,1989PhR...183..137W,1990ApJ...363..206T,2004MNRAS.350..427K}. 
The Kerr solution is a suitable and astrophysically accurate approximation of the 
space-time that captures cosmic environment relevant for the formation of jets
\citep{1997A&A...319.1025F}.
On the other hand, self-consistent solutions of coupled
Einstein--Maxwell equations for black holes immersed in electromagnetic
fields have been studied only within stationary, axially symmetric,
vacuum models \citep[e.g.,][]{1989JMP....30.1310G,1987ZhETF..92.1921D,1986pfvb.book.....G,1976JMP....17..182E,1981JMP....22.1828H,1991JMP....32..714K}. 

It turns out that the test electromagnetic field
approximation is fully adequate for modelling astrophysical sources, 
however, the long-term evolution of magnetospheres of
rotating black holes, consequences of non-ideal MHD and the effects
of oscillatory motion of the central body are 
still open to further work \citep{1992MNRAS.254..192O,1989ApJ...337...78P,2001MNRAS.322..723R,2004MNRAS.352.1161R,1996ApJ...460..199K}.


Electromagnetic fields play an important role in astrophysics. Near
rotating compact bodies, such as neutron stars and black holes, the
field lines are deformed by an interplay of rapidly moving plasma and
strong gravitational fields. Here we will illustrate purely
gravitational effects by exploring simplified vacuum solutions in which
the influence of plasma is ignored but the presence of strong gravity is
taken into account. In fact, we will restrict ourselves to purely
electro-vacuum solutions for which we we will outline the elegant formalism of null
tetrads.  We do not derive new solutions or technique in these lectures,
instead, we summarise useful relations with the focus towards the role
of large-scale organised fields interacting with rotating black holes. 

Let us note that, with exact solutions 
of the coupled Einstein--Maxwell electro-vacuum fields, an aligned magnetic flux becomes 
expelled from a rotating black hole as an interplay between the shape of
magnetic lines of force (which become pushed out of the horizon) and
the concentration of the magnetic flux tube toward the rotation axis (which 
becomes more concentrated for strong magnetic fields because of their
own gravitational effect). This is, however, important only for {\em very strong\/}
magnetic fields. In the the latter case the classical definition of the black
hole \citep{1973blho.conf....1H} has to be modified by taking into account the fact 
the spacetime is not asymptotically flat; instead, the electromagnetic effects 
contribute to the gravitational field and they may even produce singularities
outside the event horizon (see \citet{1976JMP....17..182E}; \citet{1991JMP....32..714K}, and
further references cited therein).

Now we are in a position to discuss weak {\em test} electromagnetic fields near black holes, which will then lead us
to the methods to generate {\em exact} stationary (axially symmetric) solutions.
We start with Einstein's equations in the form of a set of coupled partial differential 
equations \citep[e.g.,][]{1983mtbh.book.....C},
\beq 
R_{\mu\nu}-\textstyle{\frac{1}{2}}Rg_{\mu\nu}=8\pi T_{\mu\nu}
\eeq
($c=G=1$), where the right-hand side (the source term $T_{\mu\nu}$) is assumed to be of purely electromagnetic origin,
\beq
T^{\alpha\beta}\equiv T^{\alpha\beta}_{\rm EMG}=\frac{1}{4\pi}\left(F^{\alpha\mu}F^\beta_\mu- \frac{1}{4}F^{\mu\nu}F_{\mu\nu}g^{\alpha\beta}\right).
\eeq
The sources are constrained by the conservation relation,
${T^{\mu\nu}}_{;\nu}=-F^{\mu\alpha}j_{\alpha}$, where ${F^{\mu\nu}}_{;\nu}=4\pi j^\mu$, 
${^\star F^{\mu\nu}}_{;\nu}=4\pi\mathcal{M}^\mu$, and the dual tensor is denoted
by $^\star F_{\mu\nu}\equiv\frac{1}{2}{\varepsilon_{\mu\nu}}^{\rho\sigma}F_{\rho\sigma}$.

We assume that the electromagnetic test-fields reside in a curved
background of a rotating black hole (Kerr metric; see \citet{1986bhmp.book.....T}; \citet{1986pfvb.book.....G}). 
In other words, we study magnetised spacetimes that possess a horizon within
classical General Relativity. Let us note that higher-dimensional black holes and black rings 
in external magnetic fields have attracted a lot of interest in recent years and they have been 
explored by, e.g., \citet{2005JHEP...05..048O} and \citet{2006PhRvD..73f4008Y}. Also an 
extension to the case of naked singularity has been discussed recently 
\citep{2013CQGra..30t5007A}.

The presence of Killing vectors corresponds to symmetries of the
spacetime \citep{1983mtbh.book.....C,1984ucp..book.....W}, namely, stationarity and
axial symmetry.  Killing vectors satisfy the relation
\beq
\xi_{\mu;\nu}+\xi_{\nu;\mu}=0,
\eeq
where the coordinate system can be selected in such a way that the symmetry generating vector
adopts a simple form with the only non-vanishing component: 
$\xi^\mu=\delta^\mu_\rho$. 
In a vacuum spacetime, Killing vectors generate a  test-field solution
of Maxwell equations. One can check that a sequence of relations is obeyed:
\beq
0=\xi_{\mu;\nu}+\xi_{\nu;\mu}=\xi_{\mu,\nu}-\Gamma^\lambda_{\mu\nu}\xi_\lambda+\xi_{\nu,\mu}-\Gamma^\lambda_{\mu\nu}\xi_{\lambda}=g_{\mu\nu,\rho}.
\label{kill}
\eeq
Eq.\ (\ref{kill}) states that the metric tensor, due to the symmetry, 
is independent of the selected $x^\rho$ coordinate. In fact, the 
electromagnetic field may or may not conform to the same symmetries
as the gravitational field. Naturally, the problem is greatly simplified
by assuming axial symmetry and stationarity for both fields.

We define the electromagnetic field by
associating it with the Killing vector field, $F_{\mu\nu}=2\xi_{\mu;\nu}$.
Then it turns out that $F_{\mu\nu}=2\xi_{\mu;\nu}=-2\xi_{\nu;\mu}=-F_{\nu\mu}$,
and hence the symmetry vector generates the electromagnetic test field tensor in the form \citep{1984ucp..book.....W}
\beq
F_{\mu\nu}=\xi_{\mu;\nu}-\xi_{\nu;\mu}\equiv\xi_{[\mu;\nu]}.
\eeq
By employing the Killing equation and the definition of Riemann tensor, i.e., the relations
$\xi_{\mu;\nu;\sigma}-\xi_{\mu;\sigma;\nu}=-R_{\lambda\mu\nu\sigma}\xi^\lambda$, and
$R_{\lambda[\mu\nu\sigma]\mathrm{cycl}}=0$,
we find
\beq
\xi_{\mu;\nu;\sigma}=R_{\lambda\sigma\mu\nu}\xi^\lambda,\qquad
{\xi^{\mu;\nu}}_{;\nu}={R^\mu}_\lambda\xi^\lambda.
\eeq
The right-hand side vanishes in vacuum, hence
${F^{\mu\nu}}_{;\nu}=0$.
It follows that the well-known field invariants are given by relations
\beq
\mbox{\boldmath$E. B$}=\textstyle{\frac{1}{4}}\,{^\star}F_{\mu\nu}F^{\mu\nu},
\qquad
B^2-E^2=\textstyle{\frac{1}{2}}F_{\mu\nu}F^{\mu\nu}.
\eeq

\begin{figure}[tbh!]
\begin{center}
\includegraphics[width=0.6\textwidth]{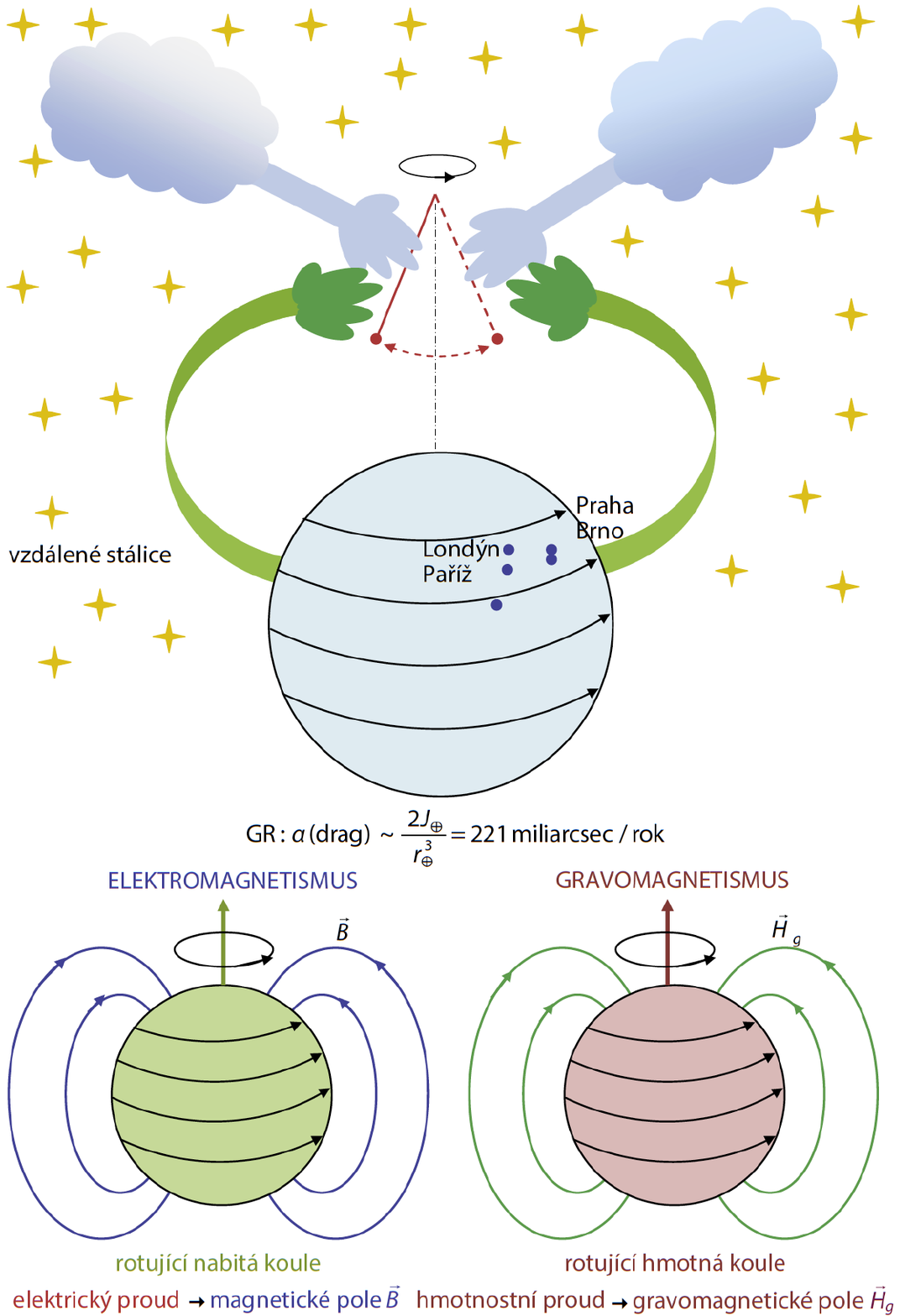}
\caption{Artist's conception of the Mach's principle and its link to the frame-dragging in General Relativity. A heuristic approach pictures the averaged motions of matter in the Universe as the basis of local inertial systems. Particles as well as fields ``feel'' gravitational effect by matter in rotation. This is the Lense--Thirring effect that influences the motion of Foucault pendulums, which is shown at the top of the plot. In the bottom part, according to the General Relativity, in analogy with the electromagnetic field of a rotating charged sphere, a rotating body creates the gravitomagnetic field in its surroundings \citep{1995mpfn.conf.....B,1982sils.book.....P}. The analogy demonstrates an inspiring role that the Mach's principle plays in gravitational physics, even if it is only partially incorporated in GR. This drawing has been adopted from the Czech review article (J.\ Bi\v{c}\'ak \citep{bicak2016}) dedicated to the centenary anniversary of Ernst Mach's death (*1838 $\dagger$1916); figure by Jana Musilov\'a was presented by Ji\v{r}\'{\i} Bi\v{c}\'ak during ``Mach, Relativity and Cosmology'' conference in Brno (2016).
The original drawing with a Czech legend indicates distant stars (yellow), a rotating charged sphere in electromagnetism (green), and a rotating massive sphere according to gravitomagnetism (red).}
\label{fig-mach}
\end{center}
\end{figure}

We start from the axial and temporal Killing vectors, existence of which is guaranteed in any axially symmetric and stationary spacetime,
\beq
\xi^\mu=\frac{\partial}{\partial t},\qquad \tilde{\xi}^\mu=\frac{\partial}{\partial \phi}.
\eeq
In the language of differential forms (e.g.\ \citet{1984ucp..book.....W}),
\beq
\underbrace{\textstyle{\frac{1}{2}}F_{\mu\nu}\rd x^\mu\wedge\rd x^\nu}_{\mbox{${\bf F}$}}=\underbrace{\xi_{\mu,\nu}\rd x^\mu\wedge\rd x^\nu}_{\mbox{${\bf d}${\protect\boldmath$\xi$}}}.
\eeq
The above-given equations allow us to introduce the magnetic and electric charges in the
form of integrals over the horizon. Magnetic~charge is defined by relation
\begin{equation}
4\pi \mathcal{M}=\int_\mathcal{S}{\bf F}=\int_\mathcal{S}\mbox{${\bf d}${\protect\boldmath$\xi$}},
\end{equation}
which is identical to zero as expected, while the electric charge is defined
\begin{equation}
4\pi Q=\int_\mathcal{S}\mbox{\protect\boldmath$^\star F$}=\int_\mathcal{S}\mbox{{\protect\boldmath$^\star$}${\bf d}${\protect\boldmath$\xi$}},
\end{equation}
which evaluates to $Q=-2M=4J$, where $M$ has a meaning of mass and $J$ is angular momentum of the source.
Here, integration is supposed to be carried out far from the 
source, i.e.\ in spatial infinity of Kerr metric in our case.

In an asymptotically flat spacetime, $\partial_\phi$ generates a uniform
magnetic field, whereas the field vanishes asymptotically for
$\partial_t$. These two solutions are known as the Wald's field \citep{1974PhRvD..10.1680W,1975PhRvD..12.3037K,1980PhRvD..22.2933B,2014ApJ...788..186N},
\beq
F=\textstyle{\frac{1}{2}}B_0\left(\rd\tilde{\xi}+\frac{2J}{M}\rd\xi\right).
\eeq
Magnetic flux surfaces are given by integration,
$4\pi \Phi_{\mathcal{M}}=\int_\mathcal{S}\mbox{\protect\boldmath$F$}=\mbox{const}$.
Magnetic field lines in the axisymmetric case are then
\beq
\frac{{\rd{r}}}{{\rd\theta}}=\frac{B_r}{B_\theta}.
\label{l1}
\eeq
This definition is in agreement with the natural expectation that the magnetic lines should reside in the surfaces of constant magnetic flux.

This brings us to the spin-coefficient formalism \citep{1962JMP.....3..566N,1973JMP....14..874G} 
as a special form of the tetrad formalism.
Although this should be a topic of an independent 
text beyond the scope of the present chapter, we refer the reader to Appendix \ref{appa} for an introduction
and the basic notation related to the highly useful 
method of spin coefficients.


An explicit form of the solutions can be constructed and illustrated
by plotting the lines of force where the Schwarzschild case is among the simplest examples, where the effects of
strong gravitational field persist \citep[][ for further references]{2014ragt.conf...75K}.

\begin{figure}[tbh!]
\begin{center}
\includegraphics[width=0.45\textwidth]{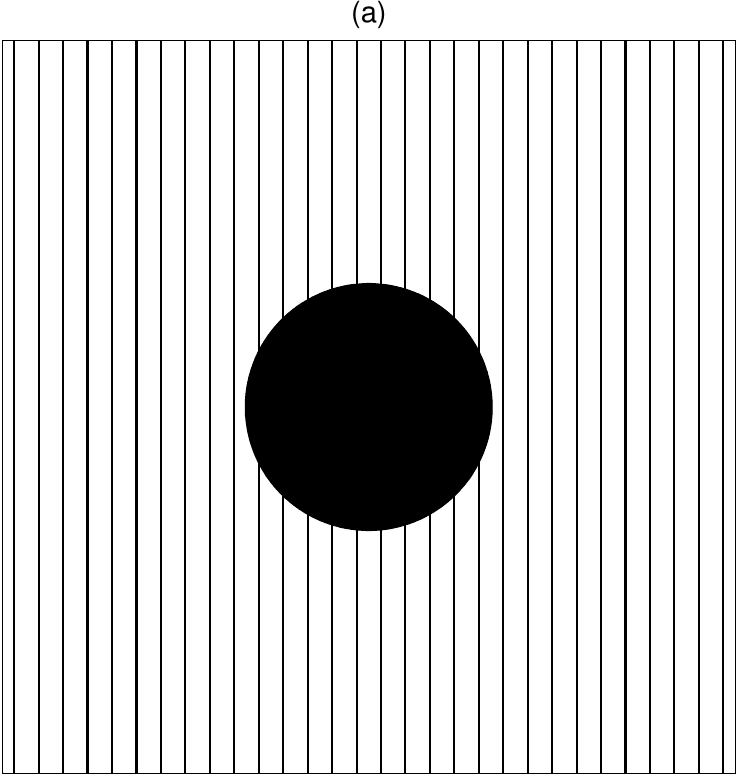}\hfill\includegraphics[width=0.45\textwidth]{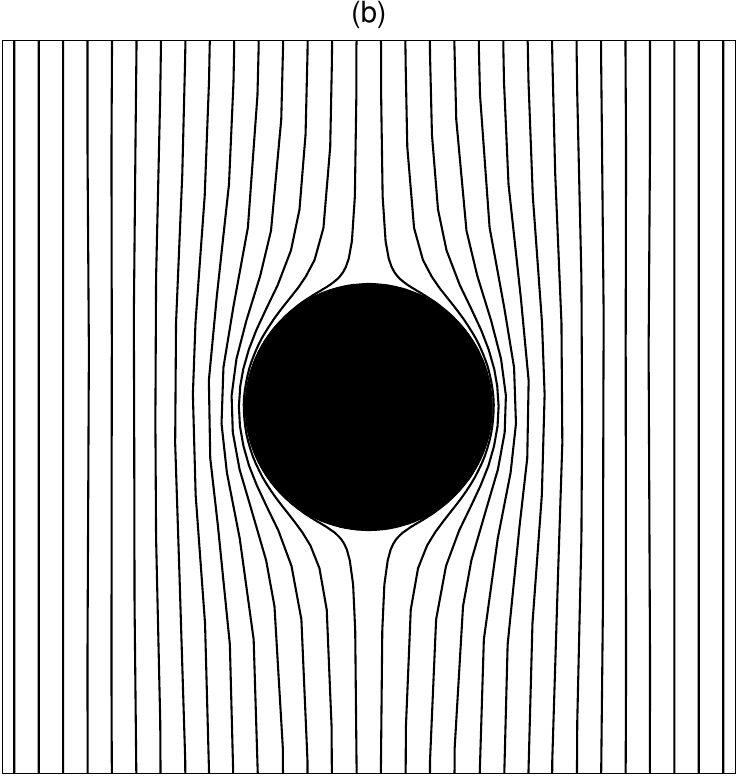}
\caption{The poloidal ($\phi=$\,const) section across a system of the black hole (black circle) embedded in an 
axially symmetric (weak) uniform magnetic field: (a)~$a=0$ (static black hole -- Schwarzschild solution for the
space-time geometry) vs.\ (b)~$a=M$ (maximally rotating Kerr black hole). In the left panel (a), 
magnetic lines of force are plotted for 
the non-rotating case in Schwarzschild coordinates, 
where the gravitational effects on the homogeneous field lines do not show up in the projection.
On the other hand, the right panel (b) does show the effect of the expulsion due to fast rotation that influences
the spacetime and the magnetic field.}
\label{fig1}
\end{center}
\end{figure}

\begin{figure}[tbh!]
\begin{center}
\includegraphics[width=0.45\textwidth]{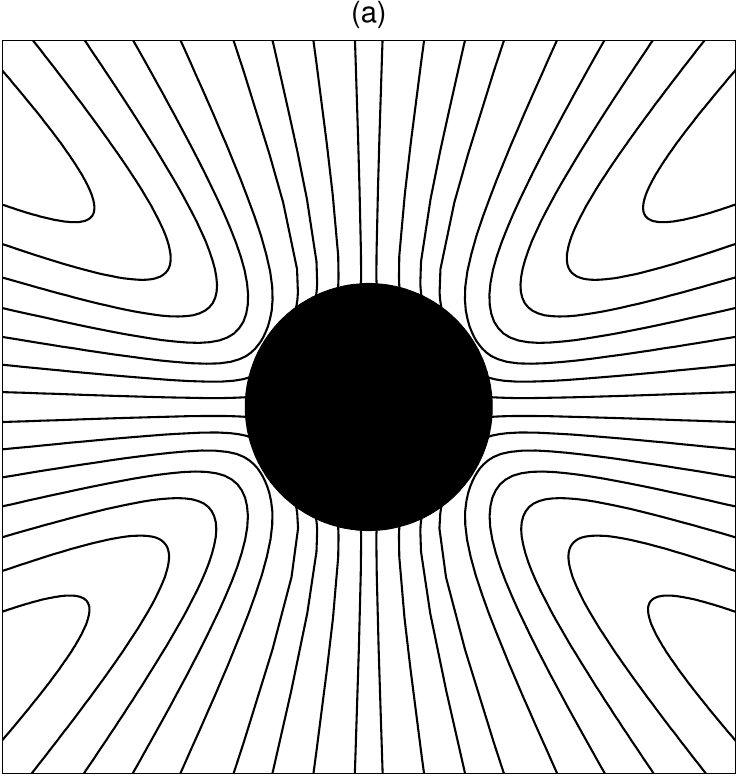}\hfill\includegraphics[width=0.45\textwidth]{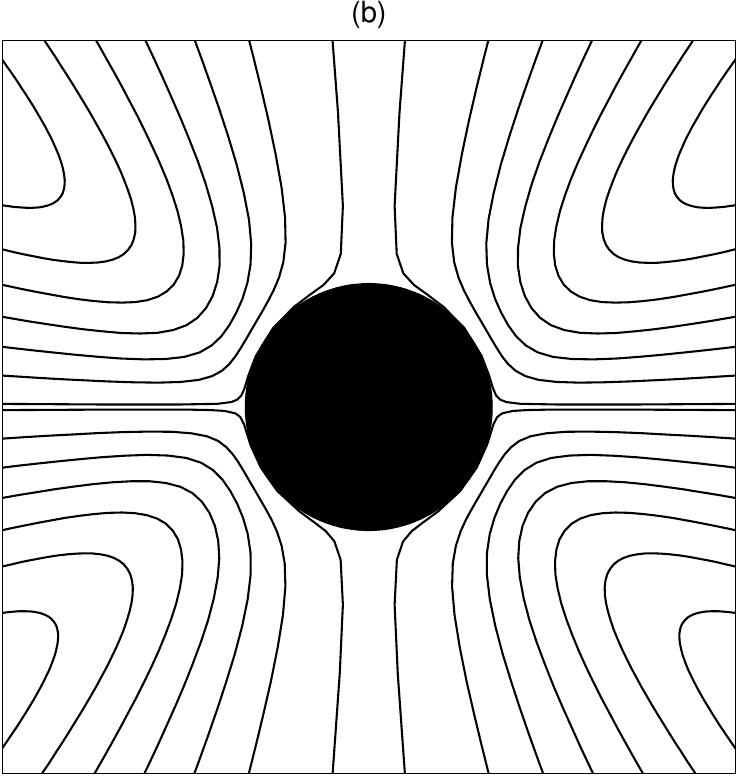}
\caption{Field lines of the gravito-magnetically induced electric field produced by the aligned, asymptotically uniform (test) magnetic field in vacuum. (a)~The case of a fast-rotating Kerr black hole ($a=0.95M$); (b)~the maximally rotating black hole ($a=M$). Let us notice the specific form of the induced electric field which becomes radial at the larger distance. 
Figure adapted from \citet{2000EJPh...21..303D}.}
\label{fig2}
\end{center}
\end{figure}


Lorentz force acts on electric/magnetic monopoles residing at rest with
respect to  a locally non-rotating frame (fig.\ \ref{fig-mach}),
\beq
\frac{\rd u^\mu}{\rd\tau}\propto \,^{\star}\!F^\mu_\nu\,u^\nu,\qquad \frac{\rd u^\mu}{\rd\tau}\propto F^\mu_\nu\,u^\nu.
\eeq
Magnetic lines are defined in the form \citep{1973oeco.book.....C}
\beq
\frac{\rd r}{\rd\theta}=-\frac{F_{\theta\phi}}{F_{r\phi}},\qquad \frac{\rd r}{\rd\phi}=\frac{F_{\theta\phi}}{F_{r\theta}},
\eeq
whereas the magnetic flux (in an axially symmetric case) reads
\beq
\Phi_{\rm m}=\pi B_0\left[r^2-2Mr+a^2+\frac{2Mr}{r^2+a^2\,\cos^2\!\theta}\left(r^2-a^2\right)\right]\sin^2\!\theta=\mbox{const}.
\eeq
The shape of field lines in the poloidal projection is associated with the coordinates, whereas the magnetic flux is an 
invariantly defined scalar quantity that characterizes the system. 
Notice: $\Phi_{\rm m}=0$ for $r=r_+$ and $a\rightarrow M$. The axisymmetric flux is expelled out of the horizon 
-- a somewhat surprising but well-known effect that resembles the Meissner effect in 
physics of (super-)conducting magnetized bodies \citep{2000NCimB.115..739B,2014PhRvD..89j4057P,2015PhRvD..92j4006B}. The analogy,
however, is not complete. While the true Meissner effects concerns the magnetic field expulsion out of superconducting medium, its black-hole variant describes a strong gravity influence on the field lines in vacuum spacetime with a specific high degree of symmetry. 

Let us note that the mechanism operates not only in
case of asymptotically uniform magnetic fields but it similarly influences higher multipoles.
Apparently, the effect does not operate in oblique (non-axisymmetric) or plasma filled (force-free) or non-stationary 
magnetospheres \citep{2016ApJ...816...77P,2016PhLB..760..112G,2018arXiv180505952E}. There has been a continued discussion whether the process
of expulsion might operate in the case of Kerr black hole in extreme rotation ($a/M=\pm1$); 
it turns out, however, that the mechanism is relevant just for electro-vacuum fields and non-ideal MHD environments.
In particular, the force-free condition (typically satisfied in highly conducting cosmic environments) leads
to magnetic field lines being rapidly accreted together with the infalling matter, so that
the magnetic flux tubes penetrate the horizon in an almost radial direction.

On the other hand, we will demonstrate further below that the electro-vacuum
case is not limited to weak magnetic fields; it
can be reproduced within the {\em exact electro-vacuum} axisymmetric solutions of the coupled Einstein--Maxwell equations
if the notion of the black hole is appropriately extended.
This is in agreement with the view of the GR Meissner effect
as a purely geometrical effect that influences the vacuum structure of the magnetic field lines around the event horizons, however, it neither enhances nor
counter-acts the competition from MHD processes when
a conducting (non-vacuum) environment is present.

The electric fluxes and field lines can be introduced in an analogous
manner by interchanging the electromagnetic field tensor
and its dual tensor, and simultaneously the magnetic charge by the electric charge, and vice versa
wherever they appear in the formulae. 
The induced electric field vanishes in the non-rotating case. Based
on analogy with a conducting sphere in rotation (within the classical 
electromagnetism), one could 
expect a quadrupole-type component, however, here the leading term of the
electric field arises due to gravo-magnetic interaction, which is a purely
general-relativity effect. This electric field falls off radially
as $\propto r^{-2}$ (Fig.\ \ref{fig2}).

Magnetic field lines reside in surfaces of constant magnetic flux, and
this way the lines of force are defined in an invariant way (see
Fig.~\ref{fig1}). An electric field is induced by the gravito-magnetic
influence of the  black hole. The resulting field lines are shown in
Fig.~\ref{fig2}. An asymptotic form of the electric field-lines reads
\begin{eqnarray}
\frac{\rd r}{\rd\lambda}&=&\frac{B_0aM}{r^2}\left(3\cos^2\theta-1\right)\;+\;\frac{3B_{\perp}aM}{r^2}\,\sin\theta\,\cos\theta\,\cos\phi+\mathcal{O}\left(r^{-3}\right),\\
\frac{\rd\theta}{\rd\lambda}&=&\mathcal{O}\left(B_{\perp}r^{-3}\right),\qquad\frac{\rd\phi}{\rd\lambda}\;=\;\mathcal{O}\left(B_{\perp}r^{-3}\right).
\end{eqnarray}

\begin{figure}[tbh!]
\begin{center}
\includegraphics[width=0.53\textwidth]{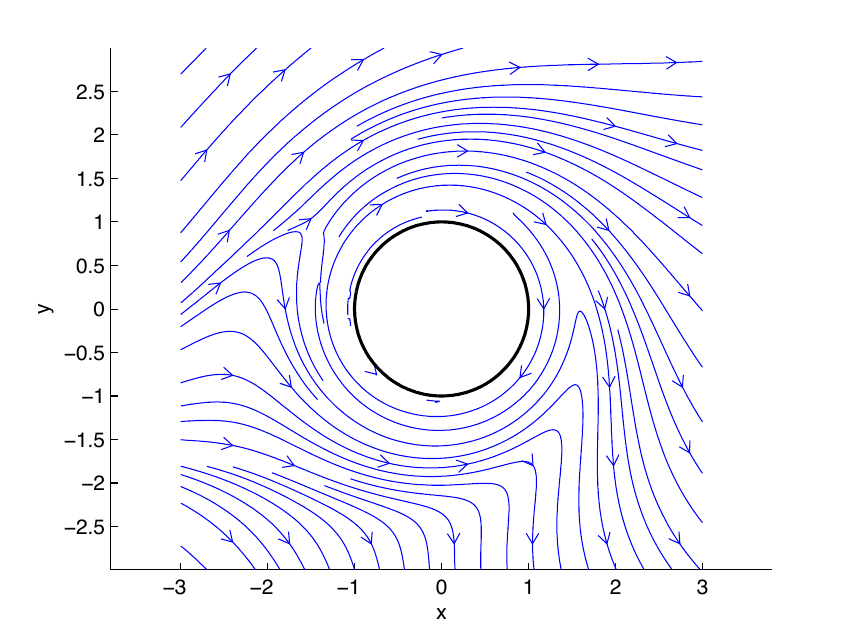}
\includegraphics[width=0.46\textwidth]{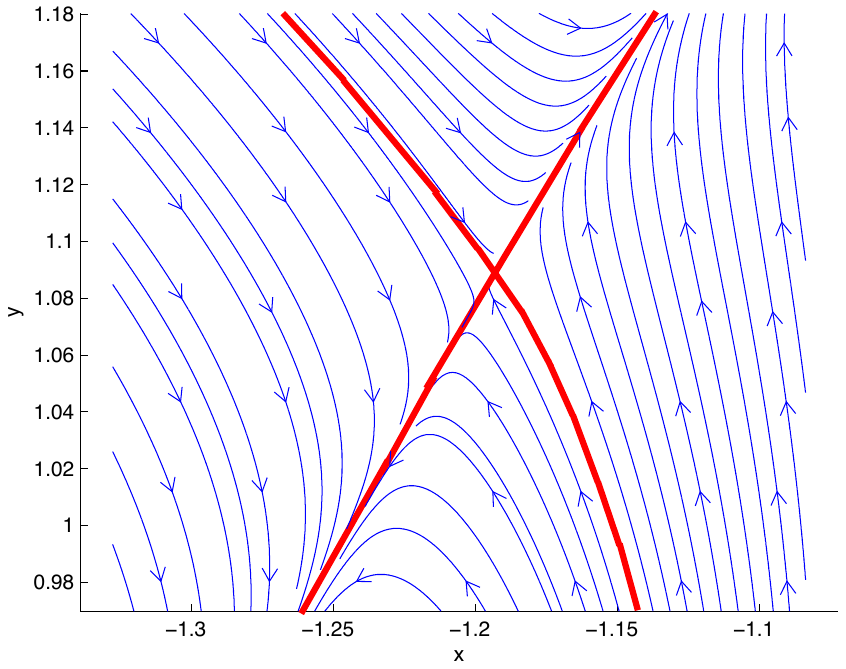}
\caption{Magnetic field lines of an asymptotically uniform
(weak electro-vacuum) magnetic field perpendicular to the black hole rotation axis \citep{2009CQGra..26b5004K,2014AcPol..54..398K}. Left: the equatorial plane is shown ($xy$ Cartesian coordinates in units of $GM/c^2$, the horizon radius is indicated by circle). The magnetic intensity vanishes in two neutral X-type points near in the vicinity of a  rapidly spinning Kerr black hole ($a/M=0.99$). Right: an enlarged detail of the structure around the magnetic null point.}
\label{fig-null}
\end{center}
\end{figure}

As mentioned above, the magnetic field aligned with the rotation axis produces an asymptotically
radial electric field, rather than a quadrupole field, as could be expected under
similar circumstances in the classical electrodynamics. This difference is
due to the combined gravito-magnetic effect of rotation and strong gravity of the black hole.

Magnetic fields misaligned with respect to the rotation axis exhibit very complex structure.
Frame-dragging acts on the field lines and distorts them in the sense of black hole rotation. Non-zero magnetic flux still enters into the horizon and, naturally, 
the same magnetic flux has to emerge out of the black hole. Magnetic lines of an inclined (oblique) field
are progressively wound up and the conditions suitable for magnetic reconnection are created \citep[magnetic
neutral points emerge in the ergosphere; cf.][]{2012CQGra..29c5010K}. Let us emphasize that these
effects occur due to purely geometrical action of strong (rotating) gravity, however, the presence of
conducting (force-free) plasma of an accretion disk can lead to a similar structure of anti-parallel, 
rotationally distorted field lines \citep{2008ApJ...682.1124K}. The latter authors demonstrate the possibility 
of magnetically driven energy extraction from the black hole through the reconnection {\em within\/} the ergosphere. 
 
We notice that magnetic null points emerge near the black hole,
suggesting that magnetic reconnection can be initiated by the purely
gravitomagnetic effect of the rotating black hole (Fig.\ \ref{fig-null}). Indeed, this new
mechanism has been proposed \citep{2009CQGra..26b5004K} in 
the context of particle acceleration processes in the ergosphere 
of rapidly rotating magnetized black holes. The induced electric component does
not vanish in the magnetically neutral point, and hence it
can accelerate the particles efficiently.

\begin{figure}[tbh!]
\begin{center}
\includegraphics[width=0.6\textwidth]{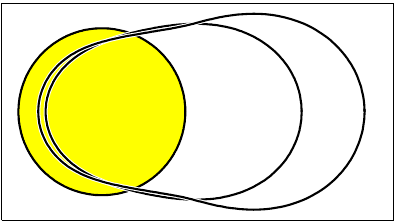}
\caption{A cross-sectional area for the capture of magnetic flux by a rotating black hole. The three oval-shape curves correspond to different values of the black-hole angular momentum: spin $a=0$ (the cross-section has a form of circle of radius $2M$; its projection coincides with the black-hole horizon, indicated by yellow color); $a=0.95\,M$; and an extremely rotating Kerr black hole, $a=M$. The enclosed area encompasses the field lines of the asymptotically perpendicular magnetic field which eventually enters into the black hole horizon. This area grows with the black hole spin and its shape is distorted by the gravitomagnetic interaction (figure adapted from \citeauthor{2000EJPh...21..303D}, \citeyear{2000EJPh...21..303D}).}
\label{fig4}
\end{center}
\end{figure}

As mentioned above, the magnetic field lines thread the horizon and become captured by the black hole if the orientation is
mis-aligned with the rotation axis. The capture of magnetic 
field lines of the perpendicular field is seen in Fig.~\ref{fig4}, where we plot the black 
hole effective cross-sectional area. The oval-shaped region is formed
by end-points of those field lines at large spatial distance from the black hole 
($r/M\rightarrow\infty$) which eventually cross the horizon and enter into the
black hole. Starting from the circular shape for $a/M=0$ static case, the 
area in mention is increasingly deformed as the spin grows.

\begin{figure}[tbh!]
\begin{center}
\includegraphics[width=0.49\textwidth]{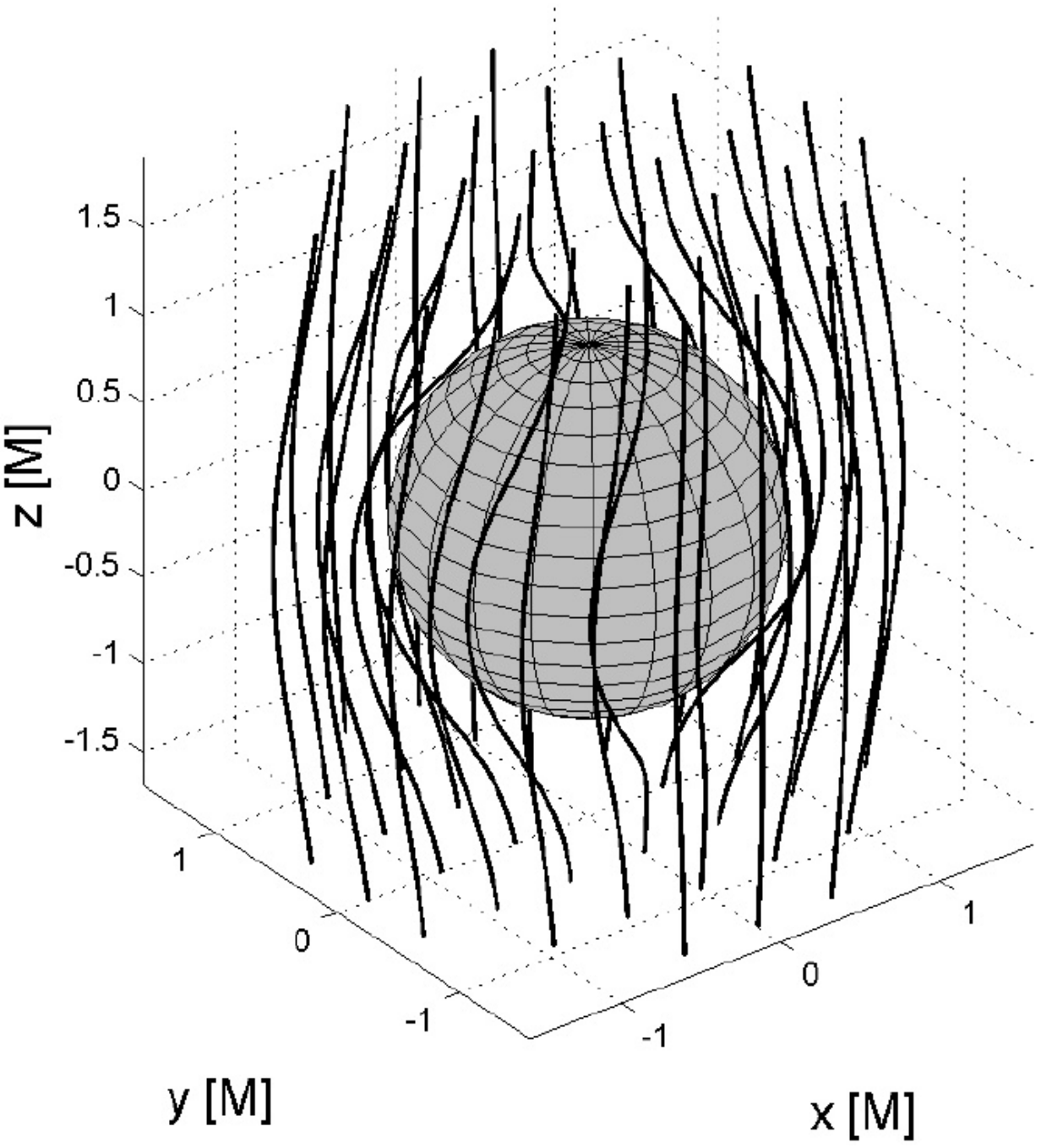}\hfill\includegraphics[width=0.49\textwidth]{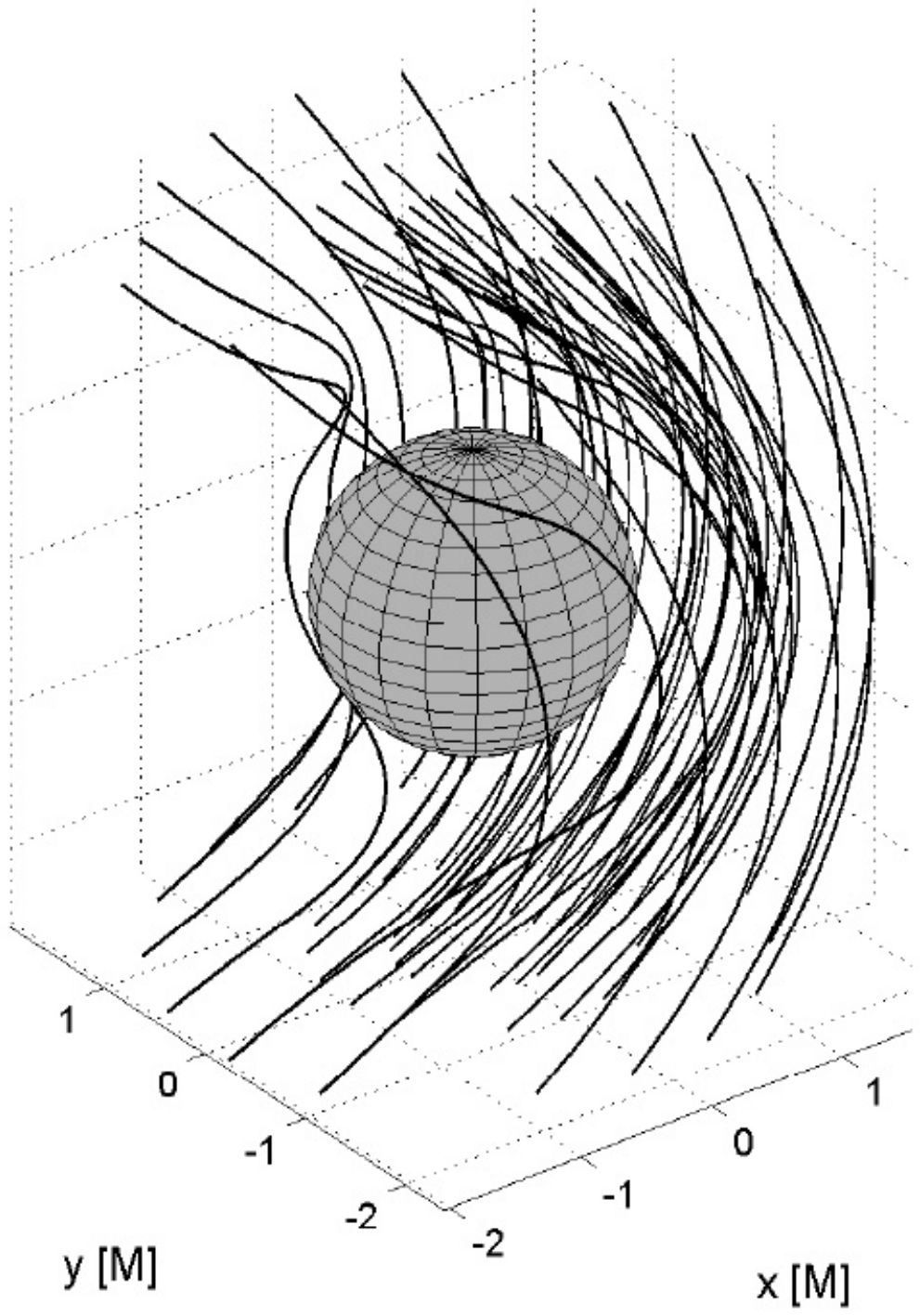}
\caption{Two examples of organized (electro-vacuum) magnetic fields near a rotating black hole in 3D projection. Left: the case of an 
asymptotically uniform magnetic field aligned with the rotation axis (parallel to the direction of $z$-axis). 
Right: the effect of 
linear (translatory) boost, where the black hole moves along the $x$-axis (figure from \citeauthor{2012JPhCS.372a2028K},
\citeyear{2012JPhCS.372a2028K}).}
\label{fig8}
\end{center}
\end{figure}

We can conclude that the frame/dragging effects on inclined magnetic 
magnetic fields lead to the emergence of 
magnetic neutral points and the associated reconnection. At the same time
the magnetic flux is not rotationally expelled out of the black hole, so that
the energy extraction via the Blandford-Znajek mechanism is possible even
from a maximally rotating Kerr black hole. Finally, Fig.~\ref{fig8} shows the 
deformation imposed on the structure of the magnetic lines by the 
translatory boost onto the magnetic field. Again, this exhibits the purely geometrical
effect which the black hole gravity has on the vacuum magnetic fields.

Surface charge can be formally defined by the radial component of electric field in non-singular coordinates \citep{1986bhmp.book.....T},
\begin{eqnarray}
\sigma_{\mathrm{H}}&=&\frac{B_0a}{4\pi\Sigma_+}\left[r_+\sin^2\theta-\frac{M}{\Sigma_+}\left(r_+^2-a^2\cos^2\theta\right)\left(1+\cos^2\theta\right)\right]\\
&&+\frac{B_{\perp}a}{4\pi\Sigma_+}\,\sin\theta\,\cos\theta\left[\frac{Mr_+}{\Sigma_+}+1\right]\left[a\sin\psi-r_+\cos\psi\right],
\end{eqnarray}
with
\beq
\psi=\phi+\frac{a}{r_+-r_-}\,\ln\frac{r-r_+}{r-r_-}\,\propto\ln(r-r_+).
\eeq
For $a\ll M$,
\beq
\sigma_{\mathrm{H}}=\frac{a}{16\pi M}\left[B_0\left(1-3\cos^2\theta\right)+3B_{\perp}\sin\theta\,\cos\theta\,\cos\psi\right].
\eeq
It should be noted that $\sigma_{\mathrm{H}}$ does not represent any
kind of a real charge distribution. Instead, it is introduced  by analogy with junction conditions for Maxwell's equations 
in classical electrodynamics. The
classical problem was treated in original works by Faraday, Lamb,
Thomson and Hertz, and more recently in \citet{1949RSPSA.199..413B} and  \citet{1950RvMP...22....1E}. 
It is quite enlightening to pursue this similarity
to greater depth (see e.g.\ \citet{2000PhyS...61..253K}, and references cited
therein) despite the fact that this is purely a formal analogy,
as pointed out by \citet{2001bhgh.book.....P}.

So far we discussed test-field solutions of Einstein equations which reside in a
prescribed (curved) spacetime. In the rest of this chapter we will briefly outline a mathematical procedure
to construct {\em exact\/} solutions of mutually coupled Einstein--Maxwell
equations \citep{1980esef.book.....K}. Because this task can be rather complicated, astrophysically realistic results
can be only obtained by numerical approaches. However, an important insight can be
gained by using the simplified analytic solutions. We defer a brief mathematical exposition to this latter approach to Appendix \ref{appb}.

\begin{figure}[tbh!]
\begin{center}
\includegraphics[width=0.9\textwidth]{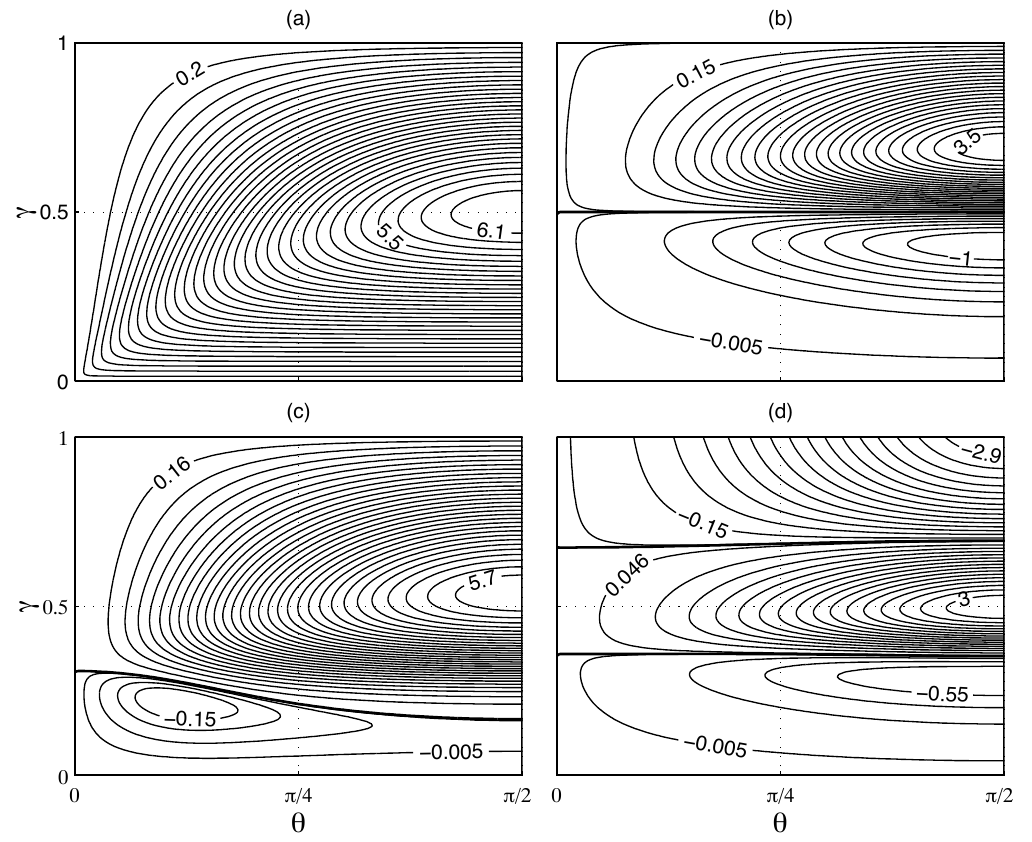}
\end{center}
\caption{Contours of magnetic flux across a cap on the horizon (latitude angle $\theta$ is measured from the rotation axis) of a magnetized black hole: (a)~$a=e=0$; (b)~$a=1$, $e=0$; (c)~$a=0.2$, $e=0$; (d)~$a=-e=1/\protect\sqrt{2}$ (electric charge and spin of the black hole). Here, $\gamma\equiv\left(1+\beta\right)^{-1}$, $\beta\equiv B_0M$. The horizontal separatrices (thick lines) correspond to extreme configurations, where the Meissner-type effect causes a complete expulsion of the magnetic field out of the black hole horizon. This figure (adapted from \citeauthor{2000PhyS...61..253K} \citeyear{2000PhyS...61..253K}) illustrates strong-gravity effects on magnetic fields that do not occur in the weak-magnetic (test) field approximation, namely, the expulsion of the magnetic flux as a function of the intensity of the imposed magnetic field (notice the horizontal lines of separatrices, where
the magnetic flux vanishes). Let us emphasize that, unlike the original discussion of the Meissner effect on weak magnetic fields, here we consider {\em exact, asymptotically non-flat, electro-vacuum\/} spacetimes of coupled Einstein--Maxwell fields.}
\label{fig5}
\end{figure}

It can be shown that the only static magnetized Kerr-Newman black hole is the one with the vanishing total charge 
$Q_{\rm{H}}$ and total angular momentum $J_{\rm{H}}$ \citep[as defined by Komar's integral relation;][]{1991JMP....32..714K}. Let us remind the reader that
these solutions are not asymptotically flat because the magnetic field does not disappear at large spatial distance from the black hole.
Instead, the spacetime metric approaches the magnetic universe \citep{1965PhRv..139..225M}. 

Fig.\ \ref{fig5} shows the magnetic flux across the black hole hemisphere in the exact magnetized black hole solution.
Four panels capture the topologically different configurations arising for different combinations of spin parameter $a$ and the electric
charge $e$. By setting $\theta=\pi/2$, one obtains the flux across the entire hemisphere.
The top-left panel (a) corresponds to the case of magnetized non-rotating (Schwarzschild--Melvin) metric
($a=e=0$) while (b) and (d) show two extreme configurations ($a^2+e^2=1$). In Fig.\ \ref{fig5}a the flux reaches
maximum of $F=2\pi$ for $\gamma=0.5$ \citep[see also][]{1986pfvb.book.....G,2000PhyS...61..253K}. 
In other words, the flux is not a monotonic function of the
magnetic-field parameter and it decreases when the field strength
exceeds a certain critical value, depending on the interplay between the magnetic intensity, distribution of the flux across
the horizon, and its surface. In terms of its $\beta$-dependence, the flux
first concentrates to symmetry axis $\theta=0$ when $\beta$ increases
from zero to unity, but than it spreads away from the axis of rotation. Let us note that this is an exemplary
behaviour that occurs in exact solutions with strong magnetic fields but it diseappears in weak test fields.

  \subsection{Particle acceleration and jet launching near SMBH}
Observations reveal that collimated outflows of matter are
a generic phenomenon connected with certain types of astronomical
objects. These jets exist on very different length scales and they are
associated with sources ranging from stars to galactic nuclei -- over 
nine orders of magnitude in the mass of the central source
\citep[see][for authoritative reviews]{1984RvMP...56..255B,2018arXiv181206025B}.
The nature of jets and the associated mechanism of jet formation, acceleration 
and disruption are very diverse, but they also exhibit common properties \citep{2015ASSL..414.....C}. Unifying
schemes have thus been proposed, initially motivated by observations. 
It has been speculated about
analogies between electromagnetic processes that accelerate particles
near pulsars \citep{1984ApJ...282..154B} versus processes in magnetospheres of
supermassive black holes
\citep{1977MNRAS.179..433B,1992ApJ...386..455H}.
It is tempting to ascribe (part of) the differences, at least
in case of AGN jets, to an unequal orientation of these objects with respect to the observer.
The interest in this subject has been amplified by the discovery of
relativistic motion in microquasars in our Galaxy \citep{1994Natur.371...46M,1995Natur.375..464H,1995Natur.374..141T}
which have their well-known counterpart in extragalactic superluminal
jets \citep{1990psrj.book.....Z}. Excellent original articles reviewing
the developments of our knowledge about jets are available, both for relevant stellar-scale objects
in the Galaxy \citep{1991bja..book..484P,1994SSRv...76..368G}
and for extragalactic jets \citep{1984RvMP...56..255B,1995PASP..107..803U}.

Different mechanisms have been proposed by which a black hole and its accretion disk can launch jets
\citep{1986bhmp.book.....T}.
Firstly, an outflowing wind from the gaseous disk may create a bubble in an infalling rotating cloud; 
hot gas then makes the orifices and the jet is shot out. Secondly, the surface of a geometrically thick 
rotating accretion disk may form funnels and collimate the wind. Thirdly, magnetic field lines are anchored 
in the disk material, thus spinning due to the overall rotation which pushes plasma to create jets. Finally, 
magnetic lines thread the hole, which forces them to spin and push plasma outwards.
The jets are presumably formed in the innermost regions of the source
(within a few or a few tens of gravitational radii, $R_g$, from the centre) and
then they emanate outwards along the rotation axis of the central object.
The initial phase of the jet formation is sometimes called 
pre-collimation in the literature, in distinction to processes of
subsequent collimation that operate in more distant regions (magnetic fields play
most probably a major role in maintaining collimated outflows on
their course; \citep{1986A&A...156..137C,1986ApJS...62....1L,1982MNRAS.199..883B}.
If we are interested in the contribution to the total radiation
that originates near the compact object, this still remains
beyond current observational limits (which are at about $10^3R_{\rm g}$
for extragalactic sources). However, this has recently been changed by the 
Event Horizon Telescope observations \citep{2019ApJ...875L...1E} that have utilized the mm-VLBI technique \citep{2017A&ARv..25....4B}. 


Several mechanisms of the jet initial pre-collimation have been proposed to operate
at the launching site near SMBH, where strong gravity effects and the relativistic frame dragging 
operate: \citet{1980ApJ...241L...7A} and \citet{1981MNRAS.197..529S}
consider collimation inside a funnel of a luminous thick
accretion disk. Material of the jet is in mutual interaction with
the disk radiation which determines its terminal speed
\citep{1987slrs.work..301P}. Since early 1980s the idea of the disk-like accretion flows
at high accretion rates has evolved to an intricate form but the
general scheme of jets
flowing along the disk axis under radiative drag remains viable.
The model has been advanced by detailed quantitative
calculations of the jet acceleration both within the framework of the
hydrodynamic \citep{1992ApJ...394..459L}
 and the test particle approximations \citep{1989ApJ...340..162M,1991A&A...252..835V}.
These calculations confirm the importance of the disk radiation
on the jet dynamics.
\citet{1992CQGra...9.1303D} constrained the rate of change of energy
and angular momentum of an interesting special family of particle
trajectories that spiral along
the axis. They do find collimation but the physical nature of
dissipative processes that cause the loss of particle energy and angular
momentum remains unclear.
Purely geodesic motion in the Kerr geometry does not contribute to
collimation except when the existence of a hypothetical naked singularity in
the centre is accepted \citep{1993MNRAS.263..545B,2018HabT........57D}. 

\begin{figure}[tbh!]
\begin{center}
\includegraphics[angle=0,width=0.5\textwidth]{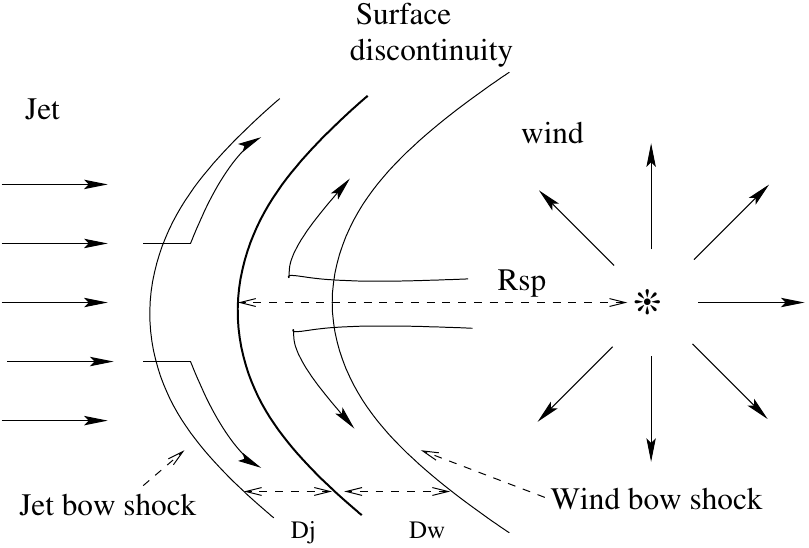}
\hfill
\includegraphics[angle=90,width=0.43\textwidth]{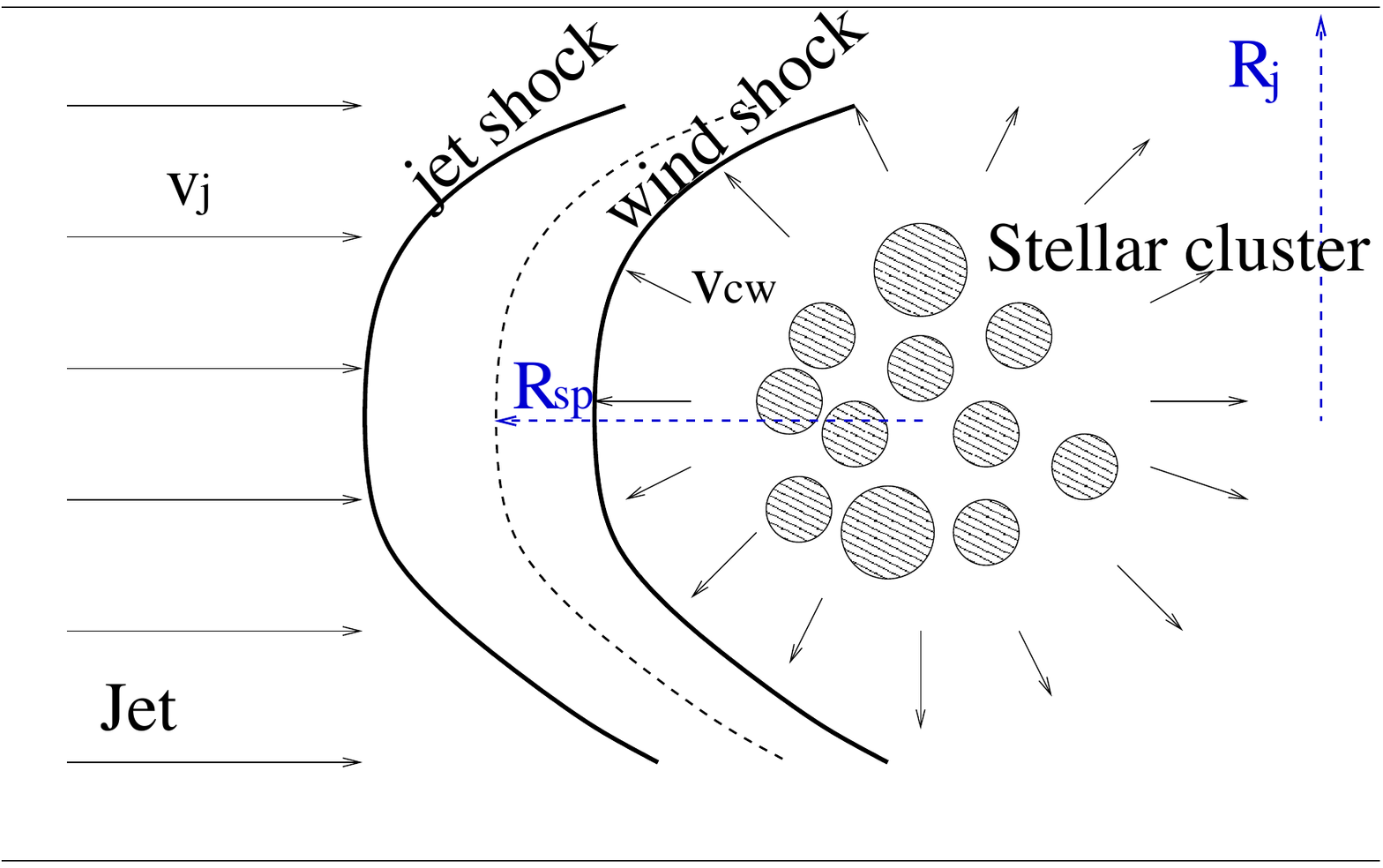}
\caption{The particle acceleration operates across the shock discontinuity when a 
relativistic jet interacts with a passing star. 
Left panel: a double bow-shock structure arises due to a stellar wind interacting with a supersonic jet. 
The jet and the wind shocked regions are separated by a contact discontinuity. The shocked matter flows downstream, i.e., away from the shock apex. $D_j$ and $D_w$ denote the size and thickness of the jet and the wind bow-shock downstream regions, respectively.
Right panel: the efficiency of the mechanism increases
dramatically as a dense star-cluster happens to cross the jet flow line 
(\citeauthor{2013MNRAS.436.3626A} \citeyear{2013MNRAS.436.3626A}, 
\citeyear{2017bhns.work....1A}).}
\label{fig9}
\end{center}
\end{figure}

Furthermore, stars can occasionally transit across the jet and produce shocks \citep{2015ApJ...807..168B}.
The interaction provides a scenario
to address non-thermal processes. A double 
bow-shock structure forms and particles become accelerated 
via DSA mechanism. Individual encounters have a limited 
effect, however, dense clusters of massive stars can truncate the jet
as the cluster crosses the jet line near the jet launching region.
\citet{2017bhns.work....1A} considered the effects of interaction of jets in Active Galactic Nuclei
when they encounter multiple obstacles, 
namely, stars in Nuclear Star Cluster surrounding the nucleus
and globular clusters passing across the inner jet (Fig.~\ref{fig9}). 

It may be interesting to note that the interaction between the accretion
disk, jet, and star of the Nuclear Star Cluster can greatly influence that 
mode of accretion, the jet emanation, and the radiation produced by the two components.
Collisions of the star with an accretion disk can (quasi-)periodically modify the
radiation flux from the source and give us information about the evolution
of the orbit. Its orientation with respect to
the observer will be affected by general relativistic effects (periastron
shift, Lense--Thirring precession) and corresponding modulation with specific
periodicities may provide us with the evidence of the black hole in the core
\citep{1994ApJ...422..208K}.
The model of a star orbiting a SMBH has been considered
by several authors in different situations.
A mechanism of the tidal
capture of a star by a black hole has been discussed
\citep{1988Natur.331..687H,1988Natur.333..523R,1992MNRAS.255..276N}.

Star-disk direct hydrodynamical collisions offer another possibility \citep{1991MNRAS.250..505S,1994MNRAS.267..557P}.
\citet{1983ApJ...273...99O} proposed that the collisions could enhance the viscous
drag on the accretion disk. \citet{1983Ap&SS..95...11Z} calculated a temperature
profile of a bright spot which is created in the place were the star crashes
through the disk. She estimates that the maximum local intensity is in
the UV band. The collisions are highly supersonic: Mach number of the order $10^2$--$10^3$. 
Assuming that the core of
AGN is embedded in a dense star cluster, \citet{1992AIPC..254..564Z} 
estimate that the amount of the disk material swept out of the disk
by stars' passages is enough to form the
gas clouds of the broad line region. \citet{1991MNRAS.250..505S} calculate
time-scales for the evolution of the orbital parameters in the Newtonian
approximation: star-disk 
collisions result in circularization of the orbit, the inclination
is reduced and the star becomes a part of the disk.
\citet{1993MNRAS.265..365V} have
obtained similar results for a star moving in the field of a Kerr
black hole. In contrast to the Newtonian case,
the probability that the star will be captured by the black hole is now higher
because subsequent collisions can
set the star in an unstable orbit which ends in the black hole.
Let us note that the effective cross-section for star-disk interaction 
can be significantly enlarged in case of a magnetic star. This may then 
lead an increased drag; for a textbook account of the
subject, see \citet{1987anz..book.....L} and \citet{2003ApJ...588..400R}.



\clearpage

\section{Unification of accreting black holes across the mass scale}
\epigraph{\textit{``It's a bit like playing chess with nature as your opponent. You make a good move, but nature counters with half a dozen new ones."}}{--- \textup{Eleanor Margaret Burbidge}}
\label{AGN_across_mass}
   
%
%
\subsection{Fundamental plane of black hole activity}

Processes related to black hole accretion,
which we described in previous sections, are not 
exclusively related to AGN physics only.
Black holes themselves are very simple objects
characterised by their mass and rotational momentum
(electric charge is probably unimportant for astrophysical
black holes in practice). That's why it is widely believed that similar processes
occur around black holes across their mass range
from a few masses of Sun in stellar-collapsed black holes 
in X-ray binaries (XRBs) up to billions of Solar masses
in the most massive AGN.

\begin{figure}[tbh!]
\begin{center}
\includegraphics[width=0.8\textwidth]{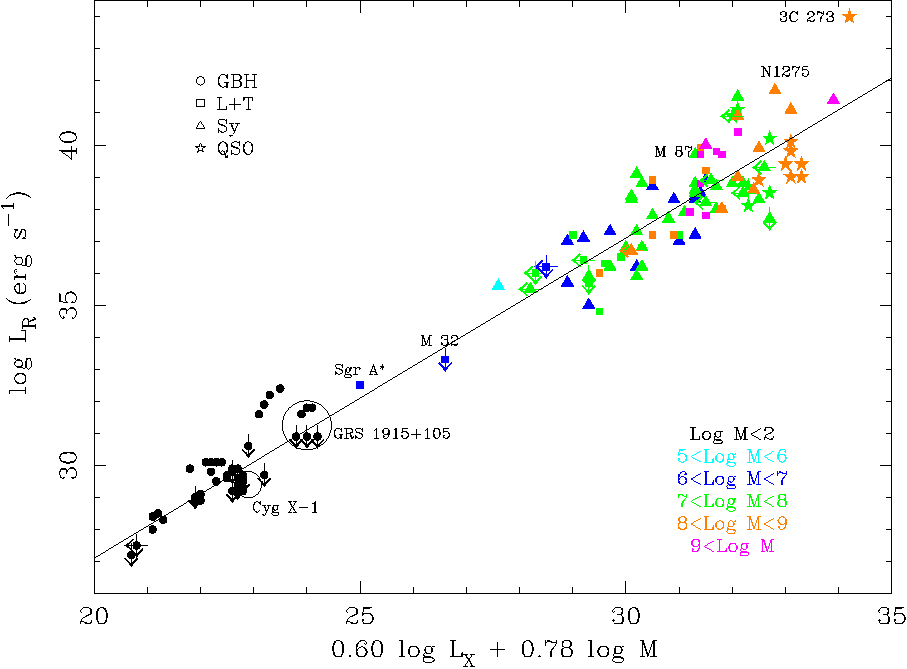}
\end{center}
\caption{Fundamental plane of black hole activity showing a relation between the radio and X-ray luminosity with mass for different kinds of black holes \citep[from][]{Merloni2003}.
}
\label{fundamental_plane}
\end{figure}

The unification between the supermassive
and stellar-mass black holes became first evident from an empirical extension of the XRB relation between the radio and X-ray luminosity
to AGN,
the so called fundamental plane of black-hole activity \citep{Merloni2003}.
Figure~\ref{fundamental_plane} shows the dependence of the radio luminosity on the X-ray luminosity and the black-hole mass for different kinds of objects, including XRBs (circles),
low-accreting supermassive black holes (Sgr A* and M\,32,
shown by squares) as well as highly-accreting Seyferts (triangles) 
and quasars (stars). The color denotes the central black-hole mass
(black represents mass corresponding to stellar masses, i.e. lower than 100\,M$_\odot$, while the other colors represent ranges for masses of the super-massive black holes in galactic centres). The found correlation from the multivariate regression analysis by \citet{Merloni2003} was:
\begin{equation}
\log L_{\rm R} = (0.6 \pm 0.11) \log L_{\rm X} + (0.78^{+0.11}_{-0.09}) \log M + 7.33^{+4.05}_{-4.07}. 
\end{equation}

The radio luminosity plays an important indicator for the jet presence,
while X-rays are connected with the accretion.
The fundamental plane of the black-hole activity thus observationally demonstrated the relation between the accretion physics and the jet launching mechanisms. Moreover, it showed that the relation is similar for XRBs and AGN, i.e. for objects with entirely different masses, suggesting that the jet models are scale-invariant, as proposed, e.g., by \citet{Heinz2003}.
Moreover, the found correlation is quantitatively consistent with the advection-dominated accretion flow (ADAF) rather than the standard accretion disk scenario. This indicates that the jet is a prominent feature mainly during this kind of accretion flow that is related with a sub-Eddington accretion rate, and the fundamental plane is appropriate for sources in this so-called hard state.

The fundamental plane has shown similarities between XRBs and AGN
and has opened the way to the unification of black-hole activity
across the mass range. Understanding what is common and what is inherently
different between these two classes will allow us to use the knowledge
of one black-hole kind to the other and better understand the black-hole physics in general. What is clearly different between the XRBs and AGN
is the characteristic time scale, at which the accretion flow evolves. While for XRBs several changes of their accretion modes have been detected (and different associated spectral states have been identified, see next Section for more details), AGN probably remain in the same accretion state during monitoring over years and decades. Several observational phenomena that are related with the XRB spectral states are, however, observed for AGN as well, including large variability and outflows in the form of jets and winds.
Also, in the other way round, some XRB spectra revealed the features
more typical for AGN. An example could be V404\,Cyg in its recent outburst,
where \citet{Motta2017} revealed the presence of winds and large obscuration
analogical to AGN spectra. The next sections are devoted to describe the spectral states of XRBs and to discuss possible analogy of spectral states in AGN.
\subsection{Spectral states: AGN vs.\ microquasars?}

The XRB spectral state can be quite well defined 
by the intensity and spectral hardness (ratio between hard and soft X-ray flux)
in the so-called hardness-intensity diagram (HID) \citep{Fender2004}. 
The soft X-ray band is dominated by the accretion-disk thermal emission.
Thus, sources in the ``high/soft'' state (HSS) produce 
most emission in the form of thermal radiation. 
Oppositely, sources lacking the accretion-disk thermal emission 
are dimmer and harder and the state is referred to as ``low-hard'' state (LHS). 
It is believed that sources
in the LHS have their accretion disk truncated 
at a further radius 
while in HSS the disk innermost edge coincides with the innermost
stable circular orbit at 1--6\,$GM/c{^2}$ depending on the black hole spin.
The X-ray corona is weak in HSS, while it is strong in LHS.

The spectral states of X-ray binaries evolve very rapidly \citep{Dunn2010},
they can change in order of hours to days.
An intriguing property is an observed hysteresis in HID. 
During a typical outburst,
the source gets brighter by several orders of magnitude,
but its hardness does not change significantly \citep[see, e.g.,][]{Fender2012}.
Then, either the outburst fails and the source goes back to quiescence,
or the source very rapidly moves to the HSS 
and thermal emission starts to prevail in the spectrum.
The transition back from soft to hard always happens
when the luminosity is much lower. 
Several explanation of this hysteresis behaviour were proposed
\citep{Chakrabarti1995, SmithD2002, Meyer-Hofmeister2005, Petrucci2008, Contopoulos2015},
but none has been widely accepted as the unique scenario.

With the XRB spectral states, several observational phenomena are related.
Large variability (rms of order of a few tens of percent) is observed in the hard state
while it diminishes in the soft state \citep{Homan2005}. 
Interesting timing features occur 
during the spectral transitions when the variability suddenly drops. 
The power spectra reveal different kinds of quasi-periodic oscillations \citep[see, e.g.,][]{Belloni2010}.
Also the jet production is closely associated with the spectral states \citep{Fender2004}. 
A persistent jet is present in the hard state while it disappears after the transition
to the HSS.
\citet{Ponti2012} showed that XRBs in the soft state produce equatorial winds  
instead of the jets.
Only sources in the so-called ``very-high state'' (or ``steep-powerlaw state'') can still 
produce blobs of relativistically accelerated matter ejected from the source in the polar (jet) direction. 
These ballistic jets are highly relativistic
(have very large Lorentz factor) and unstable (they never built a persistent jet). 
\citet{Narayan2012} showed that the strength of the ballistic jets is correlated
with the BH spin, while the correlation between the spin and the radio power of persistent jets
was found to be dubious \citep{Fender2010}.

The spin - jet power relation has significant implication also for AGN physics. From large extragalactic surveys, it was shown that $\sim 10\%$ of AGN are radio-loud \citep{Kellermann1989}, while the majority is radio-quiet. There have been long discussions what physical processes lead to this distinction, the spin paradigm being one of the most favourite physical interpretation \citep{Moderski1998, Sikora2007}. However, in analogy to XRBs,
it was suggested by \citet{Koerding2006a} and \citet{Svoboda2017} 
that the observed radio dichotomy may be explained by AGN spectral states similar to those in XRBs.

  Prominent jets extending to extra-galactic space
are observed in several nearby AGN, such as M87 \citep{Junor1999}.
Jet emission represents a significant contribution in radiation
observed from low-luminosity AGN,
for which the radio loudness was found to be anti-correlated with the Eddington ratio 
\citep[see, e.g.,][]{Ho2002, Panessa2007, Sikora2007, Ishibashi2014},
qualitatively confirming the possible analogy with the jet presence in the XRB ``low-hard'' state.
Indeed, it was proposed in the past that different kinds of AGN may correspond 
to different spectral states similar to XRBs 
\citep[see, e.g.,][]{Koerding2006a, 2006Natur.444..730M, Sobolewska2011}.

Direct comparison of AGN with XRBs is difficult mostly because of the largely different
mass ($M \approx 10^{5-10}\,M_{\odot}$) that determines both the size and the time scales.
The size scale affects the temperature of the accretion disk 
because the disk inner edge is located further from the BH singularity in AGN than in XRBs.
The AGN thermal emission thus dominates in the UV band 
\citep{Malkan1982, Laor1989}, which is, however, difficult
to observe due to large interstellar and intergalactic absorption.
In addition, AGN are not isolated. They are surrounded by the whole galaxy 
whose hot stars significantly contribute to the total UV emission coming from the
active galaxy.
Because of the observational difficulties, the UV spectrum
is one of the least understood part of AGN spectral energy distribution (SED). 
Although the apparent excess (Big Blue Bump) is usually associated
with the thermal emission from the accretion disk. 
It was, however, noticed already by \citet{Elvis1994} that the quasar spectra
cannot be generally described by the thermal blackbody emission of a thin accretion disk.
This lead to the hypothesis 
that AGN UV and soft X-rays need to be explained by an additional Comptonization
component that is unique for AGN and not present for XRBs \citep{Done2012, Done2014a}.

The other complication related to the mass is the different time scale of AGN and XRBs.
A complete cycle of an XRB outburst from quiescence
through hard state to soft state is of the order 
of a few hundred days \citep{2006ARA&A..44...49R, Dunn2010}.
AGN with significantly larger black holes 
have the time scales proportionally larger.
It is therefore impossible to wait for a spectral change in an AGN.
Even the shortest independently estimated timescales for an AGN phase is $10^5$ years
\citep{Schawinski2015}.
The only chance is thus to detect a source during the very fast transition,
such as the ``hard-to-soft'' transition, that takes only hours in XRB
and that would correspond to years in AGN.
Some changing-look AGN were indeed proposed to be candidates
of spectral changes \citep{LaMassa2015, MacLeod2016, McElroy2016}, 
but in some cases other interpretations have also been proposed \citep[see e.g.][]{Merloni2015}.
Most AGN are not variable in X-rays by a factor larger than a few over decades
\citep{Strotjohann2016}, suggesting that they remain in the same spectral state.
Only a large homogeneous AGN sample can populate the HID
with a sufficient number of sources.

\begin{figure}[tb!]
 \includegraphics[width=0.5\textwidth]{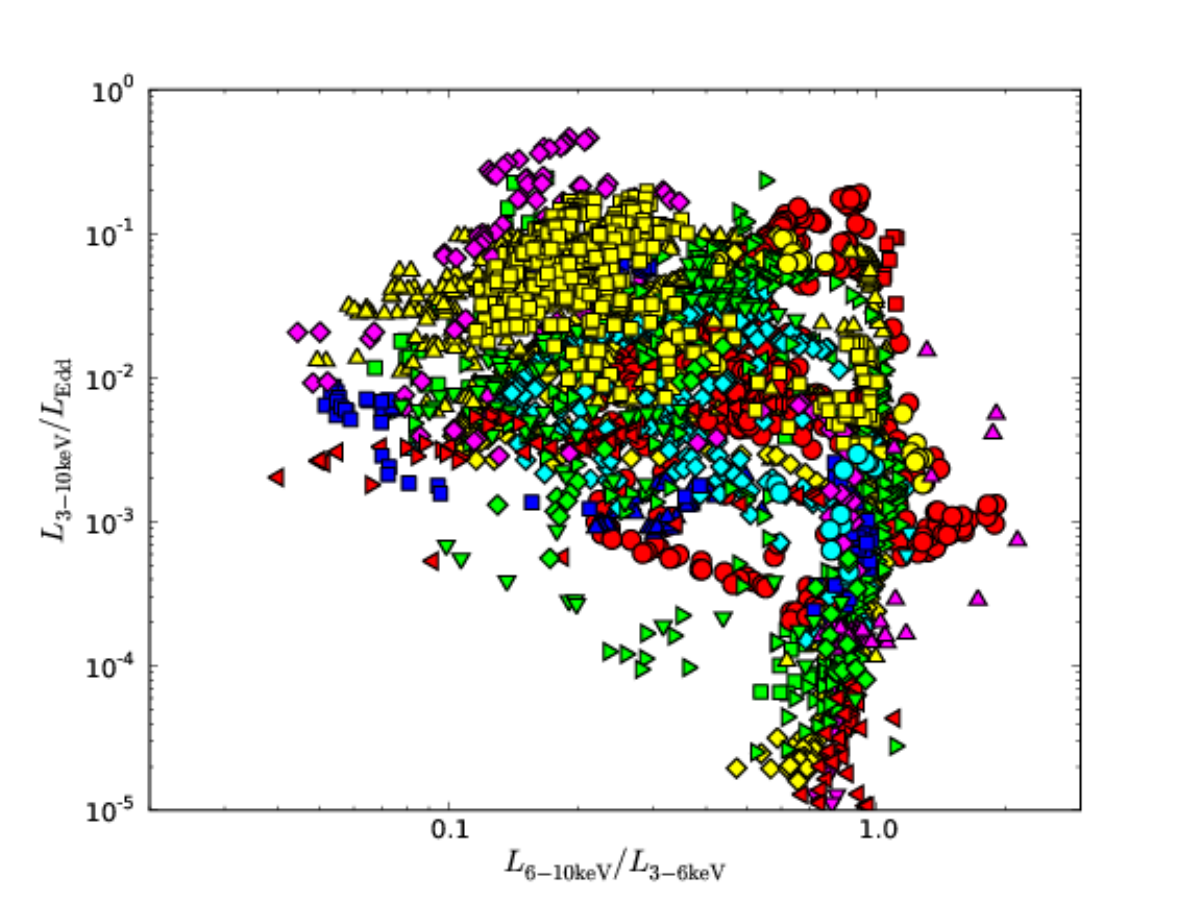}
 \includegraphics[width=0.47\textwidth]{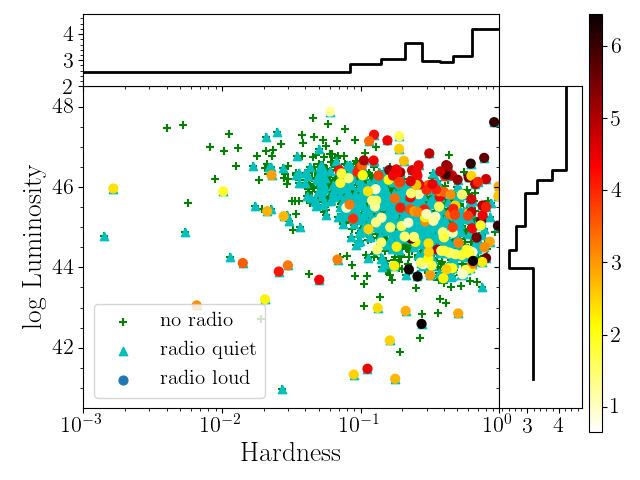}
\caption{\small{Comparison of hardness-intensity diagrams for an XRB sample in the left panel 
\citep{Dunn2010} vs. AGN sample on the right \citep{Svoboda2017}. Different colors in the XRB plot represent
different sources. The color in AGN plot denotes the relative radio loudness. The radio loudness
per hardness bin is shown in the upper histogram (for radio-loud sources only), 
revealing the increase of radio loudness towards hard states. 
The side histogram shows the relative radio loudness per luminosity bin.}
}
\label{agn_states}
\end{figure}

A seminal work on an AGN HID was done by \citet{Koerding2006a}.
They generated a disk-fraction luminosity diagram for a large sample
of quasars from the {{Sloan Digital Sky Survey}} \citep{2006ApJS..162...38A}
and from archival X-ray measurements from the {{ROSAT All-Sky Survey}} \citep{1999A&A...349..389V}.
They showed that radio-loud AGN lack the thermal emission and their luminosity is dominated 
by X-rays from non-thermal processes.
However, the SDSS sample does populate only high-luminosity part of the diagram
with the sensitivity limit ($L \approx 10^{44}$\,erg\,s$^{-1}$),
not including the low-luminosity AGN. Their results could also be affected 
by the limited bandpass of the ROSAT/PSPC detector,
and by the non-simultaneity of the data if the observed flux is significantly variable.
However, a recent work by \citet{Svoboda2017} revealed the same trend for AGN sample
based on simultaneous UV and X-ray observations obtained by {\xmm}.
The sensitivity of {\xmm} measurements allowed them to study low-luminosity sources as well.
They found that their radio loudness is also related with the UV/X-ray spectral hardness,
but the precise measurement of the spectral hardness is compromised
by generally unknown host-galaxy contamination that affects more UV than X-rays.
Therefore, the low-luminosity sources appear softer than would be their intrinsic nuclear emission.
This is apparent from the comparison between AGN and XRB HID (see fig.~\ref{agn_states}),
and was shown by \citet{Svoboda2017} that the effect of host galaxy
is indeed the most likely reason for this discrepancy.

The host-galaxy decomposition is difficult
in the absence of very high spatial-resolution observations
that are available only for the nearest galaxies.
One possible way is a detailed modelling of the AGN SED 
\citep[see e.g.][and references therein]{Ho1999, Ho2008}.
This has been done so far for only nearby AGN,
and there is not yet a fully accepted consensus on the UV emission.
While \citet{Ho2008} concluded that these sources completely lack the UV emission from the nucleus,
\citet{Maoz2007} found a significant contribution that can be associated with the nuclear activity.
Other possible method could be employed for data with available long-term monitoring.
While the AGN flux is inherently variable, the host-galaxy emission can be supposed
as constant contribution to the total flux \citep{Noda2016}.
The proper decomposition of the host galaxy from AGN emission
is crucial for a more detailed study of low-luminosity
(and thus very often low-accreting) sources.

Based on the qualitative agreement between the XRB and AGN spectral states, it was suggested by \citet{Koerding2006a} and \citet{Svoboda2017} 
that AGN spectral states may explain the observed radio dichotomy,
i.e. a distinction between radio-quiet and radio-loud AGN \citep{Kellermann1989}.
Previously, it was proposed that the black hole spin is the important parameter
\citep{Moderski1998, Sikora2007}.
However, we know from XRBs that the jet production does not occur in the HSS
(when the matter gets to the innermost stable circular orbit), 
but the persistent jets only appear in the LHS
(when the accretion disk is believed to be truncated). 
In this sense, the spectral state (accretion mode) might play a prime role
whether the jet is launched or not, and then the effect of the spin will be imprinted in the power of the jet.

If the jet is related to the spectral state,
it should be then possible to track the radio-morphology evolution in the HID.
The previous morphological classifications, such as one by \citet{Fanaroff1974},
distinguishing core- and lobe- dominated radio emission,
may be related with the interstellar medium in the host-galaxy 
rather than to the intrinsic jet evolution \citep{Gendre2013, Miraghaei2017}.
This implies that high-resolution and high-sensitive data from jet cores
will be essential in understanding the fundamental physical parameters of black hole activity \citep[see, e.g.,][]{Panessa2013}. 
The planned extension of the VLA Sky Survey (VLASS), ASKAP/EMU in the southern hemisphere or APERITIF/WODAN in the north 
\citep[for a review see e.g.][]{Norris2013},
and finally the SKA radio interferometer \citep{Dewdney2009},
will allow more detailed studies of AGN jet morphologies.

\subsection{Very high energy particles and photons}
The Earth upper atmosphere is continuously bombarded by very-high energy particles and gamma-ray photons, which have been originally discovered in an indirect manner, via balloon experiments performed at the beginning of 20th century \citep{Hess:262750,1939RvMP...11..288A}. According to the widely accepted scenario for explaining the cosmic-ray production, energetic nucleonic particles are produced via diffusive-shock acceleration mainly in supernova remnants. The particles remain trapped in the tangled Galactic magnetic field for $~10^7$ years (at GeV energies) until they eventually escape or interact with other nuclei in the interstellar medium.

\begin{figure}[tb!]
\begin{center}
\includegraphics[width=0.8\textwidth]{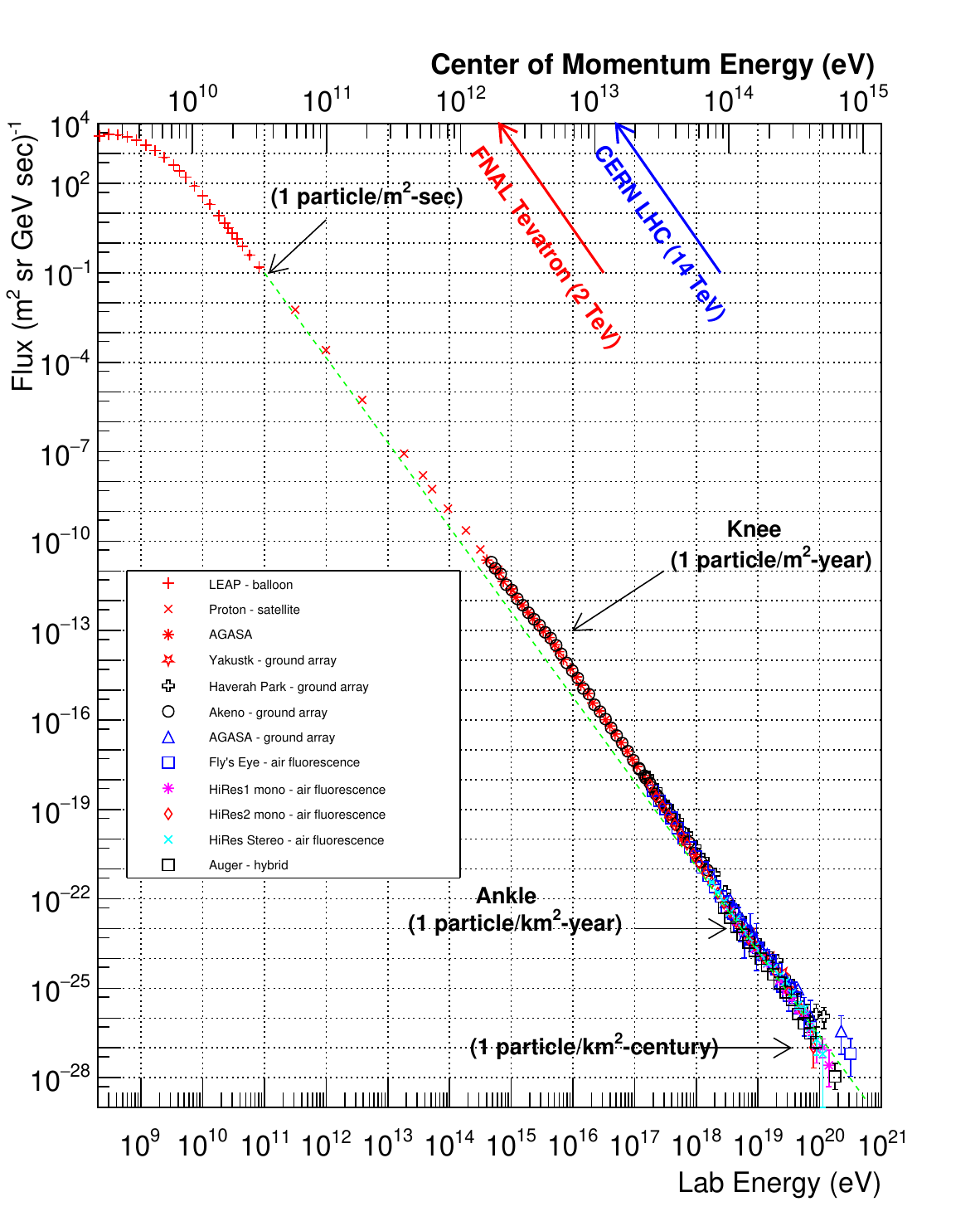}
\end{center}
\caption{\small{The spectrum of cosmic ray particles of different chemical composition is plotted against several limits of particle accelerator experiments. Relatively small departures from the uniform power-law dependence (the ``knee'' and the ``ankle'') provide important clues about the acceleration mechanisms and the composition of cosmic rays (figure from \citet{2008PhDT........19H}).}}
\label{vhe-spectrum}
\end{figure}

Highly energetic electrically charged particles of cosmic origin are usually denoted as cosmic rays, and despite a great progress their exact origin still remains very puzzling \citep{2012fura.book.....F}. Charged particles of cosmic rays and cosmic gamma rays are mutually connected by physical mechanisms of their origin and propagation. To date there are multiple candidates for the sources within the Galaxy and further out, however, it is believed that violent processes in the vicinity of compact stars and supermassive black holes play the main role in their production. At the very highest energies ($>10^{19}$~eV) these particles are mostly extra-galactic in origin, and possibly correlated with the structure of active galaxies \citep{2017Sci...357.1266P,2019FrASS...6...24K}.

Active Galactic Nuclei are thus promising potential sources of 
Ultra High Energy Cosmic Rays (UHECRs), which are
particles having an energy of $\sim$100~EeV.
These mysterious messengers arrive on Earth from outside the Galaxy, 
but their origin remains unknown \citep{2017Sci...357.1266P}. 
Radiogalaxies are a subclass of AGN where jets are detected 
at radio frequencies, which in turn are classified as type I and
type II Faranoff--Riley (FR) galaxies. In this context high resolution radio 
data provide a very powerful piece 
of information to 
study particle acceleration and to determine the nature of amplified
magnetic field. 
In both cases, jets moving in the intergalactic medium are believed to
produce shock waves where particles accelerate via the
diffusive shock acceleration mechanism \citep[DSA;][]{1978MNRAS.182..147B}. 
In fact, DSA appears to be 
the best established process to accelerate
particles in astrophysical sources where shock waves are present:
particles diffuse back and forth across
the shock front  and gain energy in each crossing.
Therefore, extended time is required to accelerate the most energetic cosmic 
rays unless the magnetic field around the shock is amplified.
The amplified turbulent field 
scatters particles, so that they cross the shock more frequently and
achieve high energy in the available time. 
The current state of the art, however, is a mainly phenomenological 
scenario where the acceleration process finishes 
at the moment when particles start radiating their energy (or when they can escape from
the source). The underlying assumption
behind this scenario is that the magnetic field persists over long distances 
all the way downstream of the shock. However, it has been shown
\citep{2015ApJ...806..243A} that the magnetic field must be highly discrete
to explain the thin radio emission 
in the jet termination region of some quasars. 
This result has important consequences for determining 
the maximum energy of particles accelerated up to about 1~TeV and the problem of UHECR
 acceleration remains open. 

The observed energy spectrum of high-energy cosmic rays is a power-law that exhibits a monotonic fall-off \citep{2008PhDT........19H,2009astro2010S.292S}, as seen in fig.\ \ref{vhe-spectrum}. The energy spectrum is an almost precise, uniform power-law decay.\footnote{See https://www.physics.utah.edu/\~\,whanlon/spectrum.html. Here, the energy range for the ground-based experiments has been corrected thanks to the updated information from W. Hanlon, with kind permission. The cosmic ray primary particle energy is measured in the laboratory frame, whereas the particle energy in accelerators is expressed in the center of momentum frame.} In order to gain sufficient Signal-to-Noise ratio at the highest energies one needs enormous collecting area of the instrument.  Multiplying the flux by a power-law proportionality with the energy power about 2.7 flattens the spectrum almost precisely below $10^{16}$~eV. There are additional features at higher energies that we can call the ``knee'' and the ``ankle''. In fact, the cosmic ray spectrum resembles a leg; we can imagine the knee (in the second break of the power law), and the ankle (in the third break in the power law) in the plot shape. 

Ultra-high energy photons may arise by a number of processes and they likely contribute to the total flux of cosmic rays. In 1966, Greissen, Zatsepin, and Kuzmin \citep{1966PhRvL..16..748G,1966JETPL...4...78Z} predicted that the interaction of charged particles with the low-energy relic photons of cosmic microwave background radiation will lead to the suppression of the cosmic ray flux at energy above $\simeq5\times10^{19}$~eV for sources beyond the distance of a few Mpc (GZK limit). However, see
\citet{1995ApJ...441..144B}.

Very-high energy gamma-rays arise in non-thermal processes by interactions of accelerated, electrically charged particles with the surrounding environment in rather extreme physical conditions, such as supernova remnants, and the immediate vicinity of black holes in AGN. The radiation processes leading to gamma rays are associated with accelerated charged particles, although gamma-rays could be potentially created also by more exotic mechanisms, such as the decay of dark matter particles.

Below GeV energies the cosmic rays are predominantly of solar origin, however, the cosmic ray spectrum extends much up to $\sim10^{20}$~eV. Because these are electrically charged particles, cosmic rays become deflected by the Galactic and extragalactic magnetic fields, which randomize their direction at the point of entry in the Earth atmosphere. On the other hand, unlike the charged cosmic rays, gamma-ray photons are not affected by magnetic fields, and so they can be traced back to their original sources at least in principle. It has been proposed that TeV photons could originate from the decay of pions that arise in interactions of protons $\gtrsim10\,$TeV energies \citep{2007ApJ...665L.131G}.

\begin{figure}[tb!]
\begin{center}
\includegraphics[width=0.7\textwidth]{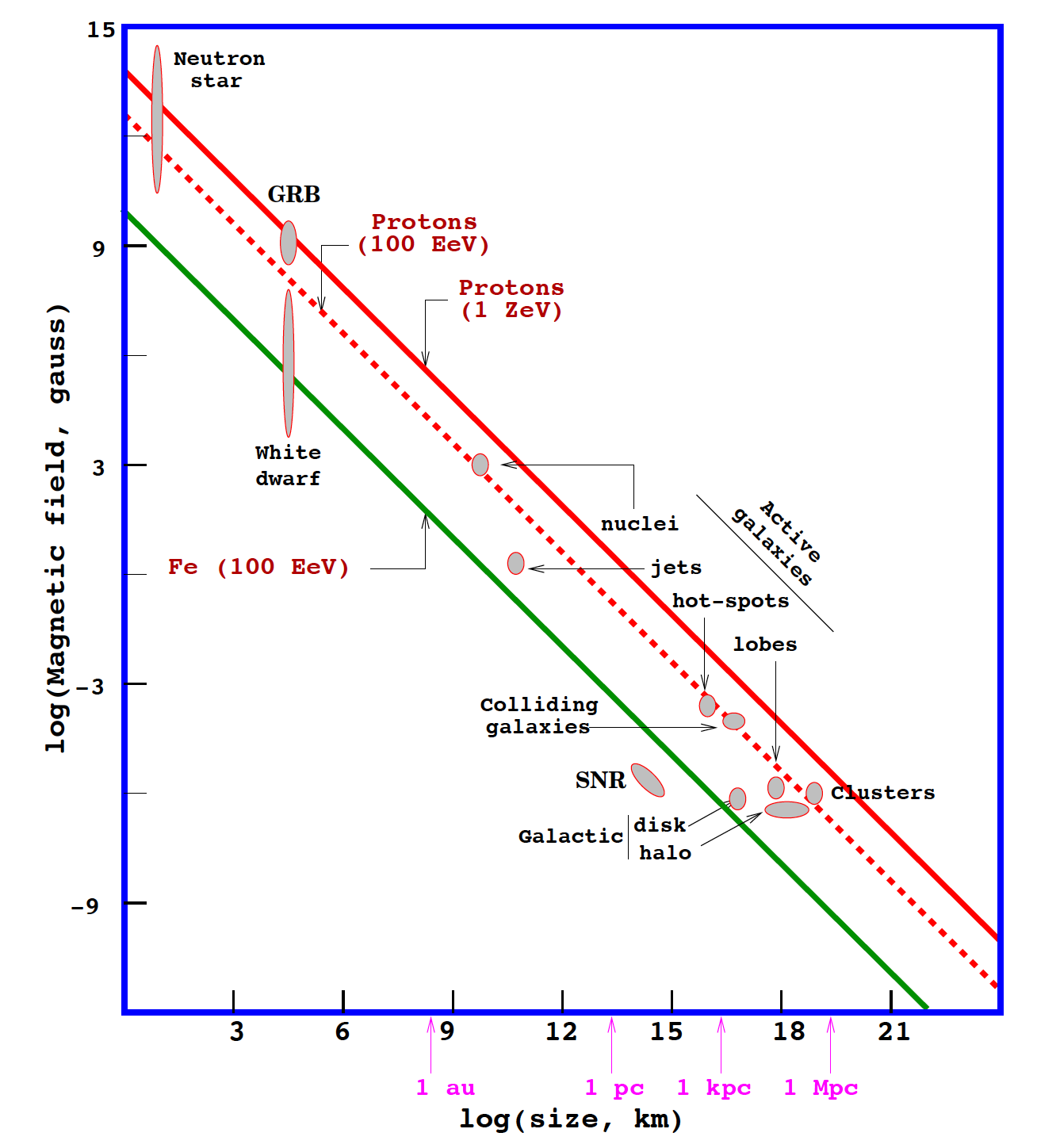}
\end{center}
\caption{\small{The Hillas diagram constrains the possible sites of cosmic rays depending on the characteristic size of the source and the magnetic field intensity confined within its volume. Diagonal lines indicate the maximum energy above which the source is unable to accelerate cosmic rays of a given chemical composition. In particular, the attainable energy for protons of $10^{20}$~eV (so far the highest value reported observationally) is shown by two red lines corresponding to different values of energy $E$ (or the shock velocity $\beta$; (see the text for further details). An analogous line for the case of iron atoms would be placed parallel towards bottom left (green). Various cosmic sources can produce cosmic rays above the limits of what current laboratory accelerators achieve (the position of the Large Hadronic Collider (LHC) would be placed around the size 1~km and the magnetic intensity $\sim10^5$~G. Figure drawn by Murat Boratav.}}
\label{hillas}
\end{figure}

Neutrinos are a very interesting aspect of cosmic ray production even if they are not considered as a separate component of this phenomenon (we do not have a neutrino spectrum). They propagate along a straight path and, thanks to their small cross-section, neutrinos can traverse enormous matter depth without experiencing any interaction or absorption. They can thus serve as excellent messengers from extragalactic sources, where they are produced by hadronic processes due to the acceleration of matter to high energies \citep{2019FrASS...6...24K}. Neutrinos are also produced by cosmic rays interacting with the cosmic microwave background. On the Earth, neutrinos can emerge in cosmic ray showers and they can be detected via methods of particle physics, such as Pierre Auger Observatory \citep{1998APh.....8..321C}, and ANTARES and the Ice-Cube experiments \citep{2016ApJ...823...65A,2017ApJ...850L..35A}. The neutrino trajectory between the source and the observer is straight and unperturbed for the neutrinos produced within the Galactic and the extragalactic sources. This has enabled to associate the neutrino with specific AGN and study the radio/gamma-ray properties of blazars, both BL Lac sources \citep[TXS 0506$+$056; ][]{2019A&A...630A.103B} as well as FSRQs \citep[PKS 1502$+$106; ][]{2021MNRAS.503.3145B}. However, the cosmic neutrino background may be affected by the gravitational lensing within the galaxy cluster and galaxy halos since a fraction of cosmic neutrinos has likely become already non-relativistic with larger deflection angles. This can then induce anisotropies in the cosmic neutrino background with the features including Einstein rings that, however, behave chromatically for massive neutrinos depending on the cosmic neutrino momentum spectrum \citep{2020JCAP...04..054Y}. In terms of the Galactic neutrinos that originate within core-collapse supernovae, there is a curious possibility of the supernova neutrino lensing by the Sgr~A* SMBH in case the supernova is nearly aligned along the Earth--Galactic center connection \citep{2007APh....28..348M}. However, these events are extremely rare: two orders of magnitude enhancement of neutrino flux is likely to occur once every few billion years, and one order of magnitude enhancement can occur once every 250 million years. A more probable event is the detection of the difference in the arrival time by as much as 20 seconds in case the supernova is not perfectly aligned with the Galactic center, but this is expected to occur still only once 29 million years \citep{2007APh....28..348M}.     

Besides supernova remnants, which appear to be highly probable sources in the Milky Way, very-high energy gamma rays can be efficiently generated in pulsar wind nebulae, in pulsar magnetospheres, and jets in microquasars. Among extragalactic objects, nuclei of active galaxies and starburst galaxies are likely sources. Gamma-ray bursts (GRBs) are expected to produce very-high energy gamma-rays in shocks via the synchrotron and inverse Compton processes in jets.

In 1984, Hillas \citep{1984ARA&A..22..425H} formulated an ingenuous and general criterion for the acceleration of charged cosmic ray particles of atomic number $A$ up to high energy $E_{\rm max}$ (see fig.\ \ref{hillas}). Namely, the characteristic condition that connects the intensity of the magnetic field $B$ and the acceleration region size $L$ is determined by the Larmor radius, which can be visualized in the Hillas relation \citep{2002APh....18..229O,2010EPJC...65..649C,2015SPPhy.161..283D,2018ipap.book.....D,2018mmea.book.....A},
\begin{equation}
E_{\rm max}\simeq10^{18} Z\beta \frac{L}{\rm kpc}\frac{B}{\mu\rm G}\;{\rm eV},
\end{equation}
where $\beta c$ is the characteristic velocity. It turns out that
only the most powerful AGN are able to accelerate protons to ultra-high energies, nevertheless, the acceleration of heavier nuclei is possible in more abundant Seyfert galaxies \citep{2010PhyU...53..691P}.
The Hillas limit is thus an upper limit on the energy that particles can achieve, quite independent of the particular acceleration mechanism although the shock acceleration appears to be among the most promising scenarios.
  
\clearpage
\section{Appendix}
\epigraph{\textit{``The beauty of mathematics only shows itself to more patient followers."}}{--- \textup{Maryam Mirzakhani}}
\subsection{Mathematical formalism of spin coefficients}
\label{appa}
Tensorial variables are projected onto a complete vector basis at each point of the curved spacetime \citep{1962JMP.....3..566N,1973JMP....14..874G}.
It has proven to be extremely useful choice with a far-reaching impact, also in the context of gravitational waves
\citep{1973ApJ...185..635T}, however,
here we confine ourselves to electromagnetism near black holes in vacuum
\citep{1975PhRvD..12.3037K,1980PhRvD..22.2933B}.
Besides other features the spin-coefficient formalism is attractive especially because it leads to a set of first-order equations.

The vectorial basis can be conveniently chosen as a 
complex null tetrad, 
$l^\mu$, $n^\mu$, $m^\mu$, $\bar{m}^\mu$, satisfying the conditions
$l_\nu n^\nu=1$, $m_\nu\bar{m}^\nu=-1$
(all other combinations vanish). A natural correspondence with an orthonormal tetrad reads
\begin{equation}
e_{(0)}=\frac{l+n}{\sqrt{2}},\quad
e_{(1)}=\frac{l-n}{\sqrt{2}},\quad
e_{(2)}=\frac{m+\bar{m}}{\sqrt{2}},\quad
e_{(3)}=\frac{m-\bar{m}}{\Im\,\sqrt{2}}.
\end{equation}
Although this seems to be more a topic of theoretical physics than astrophysics, we will
briefly expose the relevant mathematics; it has proved to be very useful in deriving and
elucidating the gravitational effects on electromagnetic fields in their mutual interaction. Let us emphasize that 
our overview is highly incomplete and focused only to
a few selected aspects.

Let us note that the adopted definition of the null tetrad is not unambiguous; the following three transformations maintain
the tetrad properties:
\begin{eqnarray}
l&\rightarrow& l,\quad m\rightarrow m+al,\quad n\rightarrow n+a\bar{m}+\bar{a}m+a\bar{a}l;\\
n&\rightarrow& n,\quad m\rightarrow m+bm,\quad l\rightarrow l+b\bar{m}+\bar{b}m+b\bar{b}n;\\
l&\rightarrow& \zeta l,\quad n\rightarrow \zeta^{-1}l,\quad m\rightarrow e^{\Im\psi}m;
\end{eqnarray}
with $\zeta$, $\psi\in\Re$.

Instead of six real components of $F_{\mu\nu}$, the framework of the null tetrad formalism describes the electromagnetic field by three independent complex quantities,
\begin{eqnarray}
\Phi_0&=&F_{\mu\nu}l^\mu m^\nu,\\
\Phi_1&=&\textstyle{\frac{1}{2}}F_{\mu\nu}\left(l^\mu n^\nu+\bar{m}^\mu m^\nu\right),\\
\Phi_2&=&F_{\mu\nu}\bar{m}^\mu n^\nu.
\end{eqnarray}
It can be checked that the backward transformation has a form
\begin{equation}
F_{\mu\nu}=\Phi_1\left(n_{[\mu}l_{\nu]}+m_{[\mu}\bar{m}_{\nu]}\right)+\Phi_2l_{[\mu}m_{\nu]}+\Phi_0\bar{m}_{[\mu}n_{\nu]}+c.c.
\end{equation}
The Newman--Penrose formalism defines the following differential operators:
\begin{equation}
D\equiv l^\mu\partial_\mu,\quad \delta\equiv m^\mu\partial_\mu,\quad \bar{\delta}\equiv\bar{m}^\mu\partial_\mu,\quad \Delta\equiv n^\mu\partial_\mu.
\end{equation}
Furthermore, \citet{1962JMP.....3..566N} introduce a set of spin coefficients (also called Ricci rotations symbols), which have been customarily denoted as follows,
\begin{eqnarray}
\alpha&=&-\textstyle{\frac{1}{2}}\left(n_{\mu;\nu}l^\mu\bar{m}^\nu-\bar{m}_{\mu;\nu}m^\mu\bar{m}^\nu\right),\\
\beta&=&\textstyle{\frac{1}{2}}\left(l_{\mu;\nu}n^\mu m^\nu-m_{\mu;\nu}\bar{m}^\mu m^\nu\right),\\
\gamma&=&-\textstyle{\frac{1}{2}}\left(n_{\mu;\nu}l^\mu n^\nu-\bar{m}_{\mu;\nu}m^\mu m^\nu\right),\\
\epsilon&=&\textstyle{\frac{1}{2}}\left(l_{\mu;\nu}n^\mu l^\nu-m_{\mu;\nu}\bar{m}^\mu l^\nu\right),\\
\kappa&=&l_{\mu;\nu}m^\mu l^\nu, \qquad ~~\lambda=-n_{\mu;\nu}\bar{m}^\mu \bar{m}^\nu,\\
\rho&=&l_{\mu;\nu}m^\mu \bar{m}^\nu, \qquad \mu=-n_{\mu;\nu}\bar{m}^\mu m^\nu,\\
\sigma&=&l_{\mu;\nu}m^\mu m^\nu, \qquad \nu=-n_{\mu;\nu}\bar{m}^\mu n^\nu,\\
\tau&=&l_{\mu;\nu}m^\mu n^\nu, \qquad ~\pi=-n_{\mu;\nu}\bar{m}^\mu l^\nu.
\end{eqnarray}
Despite that this is large number of variables the notation has turned out to be extremely
practical. The best way is to demonstrate the usefulness of the formalism with examples.

We can write the null tetrad for Schwarzschild metric \citep[see][]{1975PhRvD..12.3037K,1980PhRvD..22.2933B}.
The metric is then written in the form
\begin{equation}
\rd s^2=\left(1-\frac{2M}{r}\right)\rd t^2-\left(1-\frac{2M}{r}\right)^{-1}\rd r^2-r^2\rd\theta^2-r^2\sin^2\theta\rd\phi^2.
\end{equation}
The appropriate null tetrad is then given by
\begin{eqnarray}
l^\mu&=&\left([1-2M/r]^{-1},1,0,0\right),\\
n^\mu&=&\left(\textstyle{\frac{1}{2}},\textstyle{\frac{1}{2}}[1-2M/r],0,0\right),\\
m^\mu&=&\frac{1}{\sqrt{2}\,r}\left(0,0,1,\Im\sin^{-1}\theta\right).
\end{eqnarray}
An arbitrary type-D spacetime (e.g.\ the Schwarszchild metric) allows to
set $\kappa=\sigma=\nu=\lambda=0$. In particular, for the Schwarzschild
metric the explicit form of non-vanishing spin coefficients is:
\begin{equation}
\rho=-\frac{1}{r},\quad \mu=-\frac{1}{2r}\frac{1}{1-2M/r},\quad
\alpha=-\beta=-\sqrt{2}\,r\cot\frac{\theta}{2},\quad \gamma=\frac{M}{2r^2}.
\end{equation}

Maxwell's equations now adopt an elegant form
\begin{eqnarray}
(D-2\rho+2\epsilon)\Phi_1-(\bar{\delta}+\pi-2\alpha)\Phi_0&=&2\pi J_l,\\
(\delta-2\tau)\Phi_1-(\Delta+\mu-2\gamma)\Phi_0&=&2\pi J_m,\\
(D-\rho+2\epsilon)\Phi_2-(\bar{\delta}+2\pi)\Phi_1&=&2\pi J_{\bar{m}},\\
(\delta-\tau+2\beta)\Phi_2-(\Delta+2\mu)\Phi_1&=&2\pi J_n
\end{eqnarray}
with $J_l=l_\mu(j^\mu+\Im\mathcal{M}^\mu)$, $J_m=m_\mu(j^\mu+\Im\mathcal{M}^\mu)$,
$J_{\bar{m}}=\bar{m}_\mu(j^\mu+\Im\mathcal{M}^\mu)$,
and $J_n=n_\mu(j^\mu+\Im\mathcal{M}^\mu)$.
These are four equations for three complex variables.

Furthermore, \citet{1973ApJ...185..635T} derived the following general form of Maxwell's equations:
\begin{eqnarray}
\Big[(D&\!\!-\!\!&\epsilon+\bar{\epsilon}-2\rho-\bar{\rho})(\Delta+\mu-2\gamma)\nonumber\\ 
&&-(\delta-\beta-\bar{\alpha}-2\tau+\bar{\pi})(\bar{\delta}+\pi-2\alpha\Big]\Phi_0=2\pi J_0,\\
\Big[(D&\!\!+\!\!&\epsilon+\bar{\epsilon}-\rho-\bar{\rho})(\Delta+2\mu)\nonumber\\
&&-(\delta+\beta-\bar{\alpha}-\tau+\bar{\pi})(\bar{\delta}+2\pi\Big]\Phi_1=2\pi J_1,\\
\Big[(\Delta&\!\!+\!\!&\gamma-\bar{\gamma}+2\mu+\bar{\mu})(D-\rho+2\epsilon)\nonumber\\
&&-(\bar{\delta}+\alpha+\bar{\beta}-\bar{\tau}+2\pi)(\delta-\tau+2\beta\Big]\Phi_2=2\pi J_2
\end{eqnarray}
with
\begin{eqnarray}
J_0&=&(\delta-\beta-\bar{\alpha}-2\tau+\bar{\pi})J_l-(D-\epsilon+\bar{\epsilon}-2\rho-\bar{\rho})J_m,\\
J_1&=&(\delta+\beta-\bar{\alpha}-\tau+\bar{\pi})J_{\bar{m}}-(D+\epsilon+\bar{\epsilon}-\rho-\bar{\rho})J_n,\\
J_2&=&(\Delta+\gamma-\bar{\gamma}+2\mu+\bar{\mu})J_{\bar{m}}-(\bar{\delta}+\alpha+\bar{\beta}+2\pi-\bar{\tau})J_n.
\end{eqnarray}
Clearly this is an extremely useful form: the above-given
differential equations are {\em entirely decoupled!}

Within the Schwarzschild metric the equations are simplified to
\begin{eqnarray}
\left[\frac{\partial}{\partial r}+\frac{2}{r}\right]\Phi_1+\frac{1}{\sqrt{2}r} {^\star}\bar{\partial}\Phi_0&=&2\pi J_l,\\
-\frac{1}{\sqrt{2}r} {^\star}{\partial}\Phi_1+\frac{1}{2}\left[\left(1-\frac{2M}{r}\right)\frac{\partial}{\partial r}+\frac{1}{r}\right]\Phi_0&=&2\pi J_m,\\
\left[\frac{\partial}{\partial r}+\frac{1}{r}\right]\Phi_2+\frac{1}{\sqrt{2}r} {^\star}\bar{\partial}\Phi_1&=&2\pi J_{\bar{m}},\\
-\frac{1}{\sqrt{2}r} {^\star}{\partial}\Phi_2+\frac{1}{2}\left(1-\frac{2M}{r}\right)\left[\frac{\partial}{\partial r}+\frac{2}{r}\right]\Phi_1&=&2\pi J_n,
\end{eqnarray}
where the {\sf edth} operator acts on a spin weight $s$ quantity $\eta$,
\begin{equation}
^\star\partial\eta=-\left\{\sin^s\theta\left[\frac{\partial}{\partial\theta}+\frac{\Im}{\sin\theta}\frac{\partial}{\partial\phi}\right]\sin^{-s}\theta\right\}\eta.
\end{equation}
Spin weight is defined by the transformation property $\eta\rightarrow e^{\Im s\psi}\eta$ under the transformation $m\rightarrow e^{\Im\psi}m$.
$\Phi_0$, $\Phi_1$, $\Phi_2$ have spin weights $s=1$, $0$, $-1$, respectively.


\subsection{The approach of Ernst potential}
\label{appb}
Let us first assume a static spacetime metric in the form
\citep{1968PhRv..167.1175E,1968JMP.....9.1744H}
\begin{equation}
\rd s^2=f^{-1}\left[e^{2\gamma}\left(\rd z^2+\rd\rho^2\right)+\rho^2\rd\phi^2\right]-f\left(\rd t-\omega\rd\phi\right)^2
\end{equation}
(with $f$, $\omega$, and $\gamma$ being functions of $z$ and $\rho$ only)
and the coupled Einstein--Maxwell equations under the following constraints:
(i)~an electrovacuum case containing a black hole, (ii)~axial symmetry and stationarity
 \citep[see][, and further references cited therein]{1996PhRvD..53.1853A,1976JMP....17..182E,1990CQGra...7..391K,2015PhRvD..92j4006B}. 
Let us note that the system is {\it not\/} assumed to be asymptotically flat; instead, the energy in the magnetic field (extending to spatial infinity) ensures that the space-time at infinity does not go over to asymptotical flatness.

As explained in textbooks \citep{1973grav.book.....M} and in the above-mentioned works, 
one can proceed conveniently in one of the 
following ways to find the three unknown metric functions:
\begin{itemize}
\item The standard approach that proceeds from the second-order derivatives of the metric tensor: $g_{\mu\nu}\rightarrow\Gamma^\mu_{\nu\lambda}\rightarrow R^\alpha_{\beta\gamma\delta}\rightarrow G_{\mu\nu}$;
\item The formalism of exterior calculus: $e^\mu_{(\lambda)}\rightarrow\omega_{\mu\nu}\Omega_{\mu\nu}\rightarrow R^{\hat{\alpha}}_{\hat{\beta}\hat{\gamma}\hat{\delta}}\rightarrow G_{\hat{\mu}\hat{\nu}}$;
\item The approach of the variation principle: $\mathcal{L}=-\frac{1}{2}\rho f^{-2}\nab f\!\cdot\!\nab f+\frac{1}{2}\rho^{-1}f^2\nab\omega\!\cdot\!\nab\omega$.
\end{itemize}
Here we denoted the nabla operator, $\nab\!\cdot\!\left(\rho^{-1}\vec{e_\phi\times}\nab\varphi\right)=0$, $\forall\varphi\equiv\varphi(\rho,z)$.

Now, the vacuum field equations (without electromagnetic field) can be written in an elegant form
\citep{1968PhRv..167.1175E,1968JMP.....9.1744H}
\begin{equation}
f\nab^2f=\nab f\cdot\nab f-\rho^{-2}f^4\nab\omega\cdot\nab\omega,
\nab\cdot\left(\rho^{-2}f^2\nab\omega\right)=0.
\end{equation}
We will briefly summarize the formalism (originally developed in the late 1960s) and employ it for exact 
magnetized black-hole solutions. Our aim here is to explore the transition from weak magnetic fields
to the strong-field case that would be more appropriate to describe highly magnetized objects.

Let us define functions $\varphi(\rho,z)$, $\omega(\rho,z)$ by
the prescription
\begin{equation}
\rho^{-1}f^2\nab\omega=\vec{e_\phi\times}\nab\varphi,\qquad
f^{-2}\nab\varphi=-\rho^{-1}\vec{e_\phi\times}\nab\omega.
\end{equation}
By applying $\nab\cdot$ operator on the both sides of the last equation,
the relation for $\varphi$ comes out,
$\nab\cdot\left(f^{-2}\nab\varphi\right)=0$.
Let us further define $\mathcal{E}\equiv f+\Im\varphi$. Then, both field equations can be written in the form
\begin{equation}
(\Re\mathcal{E})\nab^2\mathcal{E}=\nab\mathcal{E}\cdot\nab\mathcal{E}.
\end{equation}
Now we can proceed to adding the electromagnetic field, for which
\begin{equation}
\mathcal{L}^\prime=\mathcal{L}+2\rho f^{-1}A_0\left(\nab A\right)^2-2\rho^{-1}f\left(\nab A_3-\omega\nab A_0\right)^2.
\end{equation}
Functions $f$, $\omega$, $A_0$, and $A_3$ are constrained by the variational
principle. Define $\Phi\equiv\Phi(A_0,A_3)$, $\mathcal{E}\equiv f-|\Phi|^2+\Im\varphi$:
\begin{equation}
\begin{array}{ll}
&\left(\Re\mathcal{E}+|\Phi|^2\right)\nab^2\mathcal{E}=\left(\nab\mathcal{E}+2\bar{\Phi}\nab\Phi\right)\cdot\nab\mathcal{E},\\
&\left(\Re\mathcal{E}+|\Phi|^2\right)\nab^2\Phi=\left(\nab\mathcal{E}+2\bar{\Phi}\nab\Phi\right)\cdot\nab\Phi.
\end{array}
\end{equation}
Let us assume $\mathcal{E}\equiv\mathcal{E}(\Phi)$ to be an analytic function \citep{1968PhRv..167.1175E,1968JMP.....9.1744H}
that satisfies the constraint
\begin{equation}
\left(\Re\mathcal{E}+\Phi^2\right)\frac{\rd^2\mathcal{E}}{\rd\Phi^2}\nab\Phi\cdot\nab\Phi=0.
\end{equation}
Assume further a linear relation, $\mathcal{E}=1-2\Phi/q$, $q\in\mathrm{C}$,
and introduce a new variable $\xi$, defined by the relation
\begin{equation}
\mathcal{E}\equiv\frac{\xi-1}{\xi+1},\qquad\Phi=\frac{q}{\xi+1},
\end{equation}
\begin{equation}
[\xi\bar{\xi}-(1-q\bar{q})]\nab^2\xi=2\bar{\xi}\nab\xi\cdot\nab\xi.
\end{equation}


An interesting aspect of the adopted formalism is that it allows one to generate new exact solutions based on the previously known
solutions (the physical interpretation  of each newly generated spacetime needs to be explored, especially the regularity
and the presence of singularities have to be checked).
We introduce new variables by relations
$\xi_0\rightarrow\xi=(1-q\bar{q})\xi_0$ and
$[\xi_0\bar{\xi}_0-1]\nab^2\xi_0=2\bar{\xi}_0\nab\xi_0\cdot\nab\xi_0$, i.e.
\begin{equation}
(\Re e\;\mathcal{E}_0)\nab^2\mathcal{E}_0=\nab\mathcal{E}_0\cdot\nab\mathcal{E}_0,\qquad\mathcal{E}_0\equiv\frac{\xi_0-1}{\xi_0+1}.
\end{equation}
where $\mathcal{E}_0$ has a meaning of an ``old'' vacuum solution.

Let $(\Phi,\mathcal{E},\gamma_{\alpha\beta})$ be a solution of Einstein--Maxwell electrovacuum equations with an anisotropic Killing vector field. Then there is another solution $(\Phi^\prime,\mathcal{E}^\prime,\gamma^\prime_{\alpha\beta})$, related to the old solution by transformation that satisfies one of the following forms \citep{1980esea.book.....K},
\begin{eqnarray*}
\mathcal{E}^\prime&=&\alpha\bar{\alpha}\mathcal{E},\quad\Phi^\prime=\alpha\Phi,\quad\mbox{\ldots dual rotation,~}^{\star}F_{\mu\nu}\rightarrow\sqrt{{\alpha}/{\bar{\alpha}}}\,^{\star}F_{\mu\nu},\\
\mathcal{E}^\prime&=&\mathcal{E}+\Im b,\quad\Phi^\prime=\Phi,\quad\mbox{\ldots calibration, no change in~}F_{\mu\nu},\\
\mathcal{E}^\prime&=&\mathcal{E}-2\bar{\beta}\Phi-\beta\bar{\beta},\quad\Phi^\prime=\Phi+\beta,\quad\mbox{\ldots calibration \ldots},\\
\mathcal{E}^\prime&=&\mathcal{E}(1+\Im c\mathcal{E})^{-1},\quad\Phi^\prime=(1+\Im c\mathcal{E})^{-1},\\
\mathcal{E}^\prime&=&\mathcal{E}{\underbrace{(1-2\bar{\gamma}\Phi-\gamma\bar{\gamma}\mathcal{E})}_{\Lambda=1-B_0\Phi-{\frac{1}{4}}B_0^2\mathcal{E}}}^{-1},\quad\Phi^\prime=(\Phi+\gamma\mathcal{E})(1-2\bar{\gamma}\Phi-\gamma\bar{\gamma}\mathcal{E})^{-1}.
\end{eqnarray*}
\begin{equation}
\mathcal{E}\rightarrow\mathcal{E}^\prime=\Lambda^{-1}\mathcal{E},\qquad f\rightarrow f^\prime=|\Lambda|^{-2}f,\qquad\omega\rightarrow\omega^\prime,
\end{equation}
\begin{equation}
\Phi\rightarrow\Phi^\prime=\Lambda^{-1}(\Phi-\textstyle{\frac{1}{2}}B_0\mathcal{E}),\ \nab\omega^\prime=|\Lambda|^2\nab\omega+\rho f^{-1}(\bar{\Lambda}\nab\Lambda-\Lambda\nab\bar{\Lambda}).
\end{equation}


While the calibration transformations are not of interest for us here, the relations involving the magnetic field $B_0$
are relevant and they define a non-trivial magnetization procedure. To illustrate the mechanism of the above-mentioned solution generating technique we can give three elementary examples where the well-know spacetime have been reproduced. In fact, it is possible to start from the most trivial set-up, i.e.\
the Minkowski spacetime. From this seed the outcome of the generating method leads us to the magnetic Melvin universe.
\begin{equation}
\rd s^2=\left[\rd z^2+\rd\rho^2-\rd t^2\right]+\rho^2\rd\phi^2.
\end{equation}
with the metric functions
$f=-\rho^2$, $\omega=0$, $\Phi=0$, $\mathcal{E}=-\rho^2$, $\varphi(\omega)=0$,
new metric functions
$f^\prime=-\Lambda^{-2}\rho^2$, $\omega^\prime=0$, $\Phi^\prime=\textstyle{\frac{1}{2}}\Lambda^{-1}B_0\rho^2$,
the associated components of the magnetic field $B_z=\Lambda^{-2}B_0$, $B_\rho=B_\phi=0$, and the 
generated line element
\begin{equation}
\rd s^2=\Lambda^2\left[\rd z^2+\rd\rho^2-\rd t^2\right]+\Lambda^{-2}\rho^2\rd\phi^2.
\end{equation}
It is the $\Lambda$ function that leads to asymptotically non-flat (cosmological) behaviour of
the new solution, where gravity of the magnetic field is in balance with the Maxwell pressure. 
Cylindrical symmetry is maintained along $z$-axis.


Next, we start from the Schwarzschild black hole which as a result of the application 
of the generating technique produces the Schwarzschild-Melvin black hole, i.e.
the spacetime that resembles the non-rotating black hole near its event horizon, however, the solution lacks 
the property of asymptotical flatness and reach the
above-mentioned Melvin's universe at large distance.
\begin{equation}
\rd s^2=\left[\left(1-\frac{2M}{r}\right)^{-1}\rd r^2 - \left(1-\frac{2M}{r}\right)\rd t^2+r^2\rd\theta^2\right]+r^2\sin^2\theta\rd\phi^2,
\end{equation}
with
$f=-r^2\sin^2\theta,\quad\omega=0$, $\rho=\sqrt{r^2-2Mr}\,\sin\theta$,
$B_r=\Lambda^{-2}B_0\cos\theta$, $B_\theta=-\Lambda^{-2}B_0(1-2M/r)\sin\theta$,
and
\begin{equation}
\rd s^2=\Lambda^2\Big[ \quad...\quad\Big]+\Lambda^{-2}r^2\sin^2\theta\rd\phi^2
\end{equation}
(the term within the brackets remains unchanged from the original metric form). 
The following limits hold for the magnetized Schwarzschild-Melvin black hole solution:
(i) $B_0=0\rightarrow$ Schwarzschild solution,
(ii) $r\gg M\rightarrow$ Melvin solution,
(iii) $|B_0M|\ll1\rightarrow$ Wald's test field in the region $2M\ll r\ll B^{-1}_0$.


As an even more general example we mention the result of the magnetizing technique when applied to the rotating, electrically charged black hole. The outcome in this case 
is the spacetime of magnetized Kerr--Newman black hole.
\begin{eqnarray*}
g&=&|\Lambda|^2\Sigma\left(\Delta^{-1}\rd{r}^2+\rd{\theta}^2-\Delta{A^{-1}}\rd{t}^2\right)\\
&&+|\Lambda|^{-2}\Sigma^{-1}A\sin^2\theta\left(\rd{\phi}-\omega\rd{t}\right)^2,
\end{eqnarray*}
$\Sigma=r^2+a^2\cos^2\theta$, $\Delta=r^2-2Mr+a^2+e^2$, $A=(r^2+a^2)^2-{\Delta}a^2\sin^2\theta$ are functions from the Kerr-Newman metric.

The characteristic function $\Lambda=1+\beta\Phi-\frac{1}{4}\beta^2\mathcal{E}$ of the magnetized solution 
is given in terms of the Ernst complex potentials $\Phi(r,\theta)$ and $\mathcal{E}(r,\theta)$:
\begin{eqnarray*}
\Sigma\Phi
 &=& ear\sin^2\theta-{\Im}e\left(r^2+a^2\right)\cos\theta, \\
\Sigma\mathcal{E}
 &=& -A\sin^2\theta-e^2\left(a^2+r^2\cos^2\theta\right)
 \nonumber \\
 & & + 2{\Im}a\left[\Sigma\left(3-\cos^2\theta\right)+a^2\sin^4\theta-
 re^2\sin^2\theta\right]\cos\theta.
\end{eqnarray*}


The electromagnetic field can be written in terms of orthonormal LNRF (locally non-rotating frame) components,
\begin{eqnarray*}
H_{(r)}+{\ri}E_{(r)} &=& A^{-1/2}\sin^{-1}\!\theta\,\Phi^{\prime}_{,\theta},\\
H_{(\theta)}+{\ri}E_{(\theta)} &=&-\left(\Delta/A\right)^{1/2}\sin^{-1}\!\theta\,\Phi^{\prime}_{,r},
\end{eqnarray*}
where $\Phi^{\prime}(r,\theta)=\Lambda^{-1}\left(\Phi-\frac{1}{2}\beta\mathcal{E}\right)$.
The horizon is positioned at $r=r_+=1+\sqrt(1-a^2-e^2)$, i.e., independent of $\beta$. As in the non-magnetized case, the horizon exists only for $a^2+e^2\leq1$.


There is an issue with this solution. Namely, by applying the above-mentioned solution generating technique a conical singularity is produced.
The problem arises from the fact that the mathematical prescription guarantees that Einstein--Maxwell equations are satisfied {\em locally}; however, one still needs to check the global properties of the solution. And indeed, the conical singularity can be removed by rescaling 
the range of azimuthal angle to an enlarged interval $0\leq\phi<2\pi|\Lambda_0|^2$ \citep{1981JMP....22.1828H,1988BAICz..39...30K}, where 
\begin{equation}
|\Lambda_0|^2\equiv|\Lambda(\sin\theta=0)|^2= 1+\textstyle{\frac{3}{2}}\beta^2e^2+2\beta^3ae+\beta^4\left(\textstyle{\frac{1}{16}}e^4+a^2\right).
\end{equation}
The effect of the rescaling operation can be revealed by calculating the (scalar) magnetic flux threading the horizon. 
The total electric charge $Q_{\rm{H}}$ and the magnetic flux $\Phi_{\rm{m}}(\theta)$ across a cap in axially symmetric position on the horizon 
(with the rim of the cap defined by $\theta={\rm{const}}$),
\begin{eqnarray}
Q_{\rm{H}} &=& -|\Lambda_0|^2\,\Im{\rm{m}\,}\Phi^{\prime}\left(r_+,0\right),
\\ 
\Phi_{\rm{m}} &=& 2\pi|\Lambda_0|^2\,\Re{\rm{e}\,}\Phi^{\prime}
 \left(r_+,\bar{\theta}\right)\Bigr|\strut^{\theta}_{\bar{\theta}=0}.
\end{eqnarray}

\newpage\noindent

\clearpage
\phantomsection
\section*{Abbreviations and symbols}
	\noindent
	Here we provide a glossary of constants, symbols and abbreviations that have occurred in the Lecture Notes. Because our text covers a variety of topics seen from different perspective, it would be rather unnatural or even misleading to attempt a fully uniform notation over all chapters. Instead, the partially overlapping definitions are exposed in the following list. 
	
	\subsubsection*{Constants}
	 \begin{enumerate}
	 	\item \gls{speedlight}
	 	\item \gls{gravconst}
	 	\item \gls{me}
	 	\item \gls{mp}
	 	\item \gls{hbar}
	 	\item \gls{kkbb}
	 	\item \gls{Lsun}
	 	\item \gls{masssun}
	 \end{enumerate}

\noindent
The actual meaning of each of the above-listed constants can be searched by a clickable hyperlink in the electronic pdf version of the text.
	\glsaddall
	
\glssetwidest{ZZZZZZ}
	\printglossary[type=functions,nonumberlist]

\glssetwidest{FR type II}
	\printglossary[type=\acronymtype,nonumberlist]

~\\[10pt]~\noindent
Finally, let us note that classical three-vectors are denoted by bold letters, i.e. for example $\BoldM{X}$. We employ Cartesian $(x,y,z)$ and spherical $(r,\theta,\phi)$ coordinates. Within the often used cylindrical system, spatial coordinates are denoted as $(R,Z,\phi)$.
\tiny
\newpage


\phantomsection
\renewcommand\bibname{References} 

\markboth{\bibname}{\bibname}

\bibliographystyle{apalike}
\bibliography{articles}

\vspace*{6em}\normalsize

\newpage
\markboth{About authors}{About authors}

\phantomsection
\textbf{\textsc{Vladim\'{\i}r Karas}} (\orcidicon{0000-0002-5760-0459} \textsc{orcid 0000-0002-5760-0459}) is Professor of Astrophysics at the Astronomical Institute of the Czech Academy of Sciences, where he obtained his second doctorate in 2001. Before coming to the Prague section of the Institute, Vladim\'{\i}r was a Lecturer in astrophysics at Charles University (Faculty of Mathematics and Physics) and Visiting Scientist at several places: Trieste (International School for Advanced Studies), Rome (Universit\`a Roma Tre), Paris (Observatoire de Paris-Meudon), G\"{o}teborg (Chalmers University), Baltimore (Johns Hopkins University), and Cambridge, MA (Massachusetts Institute of Technology). Vladim\'{\i}r Karas has a fruitful collaboration with the Institute of Physics of the Silesian University in Opava, where he has served on Science Council and co-authored several volumes of the {\em RAGtime Proceedings\/} from ``Workshops on Black Holes and Neutron Stars''. Vladim\'{\i}r has developed fast numerical approaches to study the effects of strong gravity. His more recent work deals with signatures of frame-dragging: an interplay between Mach and Meissner effects near compact stars and black holes. Over past decades he initiated numerous research programs, including the Center for Theoretical Astrophysics (CTA) in Prague. As a Director, Vladim\'{\i}r lead the Astronomical Institute which includes Ond\v{r}ejov observatory in 2012--2022. 

\textbf{\textsc{Ji\v{r}\'{\i} Svoboda}} (\orcidicon{0000-0003-2931-0742} \textsc{orcid 0000-0003-2931-0742}) is Research Scientist and the head of {\em Space for the Mankind}, AV\,21 Strategy Program at the Astronomical Institute of the Czech Academy of Sciences. He received Ph.D. in Astronomy, Astrophysics and Theoretical Physics from Charles University in Prague in 2010. He was Research Fellow at the European Space Astronomy Centre of the European Space Agency in 2011-13 in Spain. After returning back to Prague he succeeded in expanding the Institute space activities by joining two upcoming large missions devoted to explore the Universe via X-rays and gravitational waves. In his own research Ji\v{r}\'{\i} focuses on Active Galactic Nuclei and X-ray Binaries. He studies radiation coming from where the space-time is curved by extremely strong gravity. Besides the research, Ji\v{r}\'{\i} is also active in teaching and public outreach.

\textbf{\textsc{Michal Zaja\v{c}ek}} (\orcidicon{0000-0001-6450-1187} \textsc{orcid 0000-0001-6450-1187}) is Research Scientist in the Department of Theoretical Physics and Astrophysics at Faculty of Science, Masaryk University in Brno, which he has joined in summer 2021. Michal completed his Masters in astronomy and astrophysics by defending the thesis titled ``Neutron stars near a galactic centre'' (Charles university, Faculty of Mathematics and Physics in Prague, 2014). He was selected for Ph.D. program at the International Max Planck Research School (IMPRS) in Cologne and Bonn (Germany), where he focused on the dynamics in galactic nuclei, especially near the Galactic Center. By combining models with observations he constrained the properties of near-infrared excess sources and the flaring activity of Sagittarius A*. For one year, he was a postdoctoral researcher in the VLBI group of the Max Planck Institute for Radioastronomy in Bonn. Then, for two and half years (2019--2021), Michal was a postdoctoral fellow in the Center for Theoretical Physics of the Polish Academy of Sciences in Warsaw, where he contributed to the research on quasars in the cosmological context.

\phantomsection
\section*{Acknowledgments}
The lectures were presented at {\em RAGtime Workshops on Black Holes and Neutrons Stars} (Institute of Physics, Silesian University in Opava, 2021). In the original form at {\em Summer School on Relativistic Accretion: Theoretical Models and Their Application to Observations} at the Center of Applied Space Technology and Microgravity (ZARM, University of Bremen, 2016).

Over several year of preparation of the Lecture Notes, work of the authors was partially supported within the Czech Science Foundation projects titled ``Largest Black Holes in the Sky: Origin and Evolution of Horizon-Scale Structures'' (ref.\ 19-01137J), and ``Mass and Charge Currents in General Relativity and Astrophysics (ref.\ 21-11268S). Preliminary versions of different chapters were ``tested'' at various lectures and seminars with students of astronomy, astrophysics and theoretical physics at the University of Bremen (ZARM), Silesian University in Opava (Institute of Physics), and Charles University in Prague (Astronomical Institute). We also acknowledge the Czech PRODEX project ``Hardware contribution to the Chinese X-ray mission eXTP'' (ref.\ 4000132152) to support the exploration of accreting black holes at the Astronomical Institute of the Czech Academt of Sciences in Prague and the Institute of Physics in Opava. The contribution of Michal Zaja\v{c}ek to this work was supported partially by Deutsche Forschungsgemeinschaft (DFG) SFB956 -- \textit{Conditions and Impact of star formation}, then by the Polish Funding Agency National Science Centre, project 2017/26/A/ST9/00756 (MAESTRO 9), and currently by the Czech Science Foundation EXPRO project ``Exploring the Hot Universe and Understanding Cosmic Feedback'' (No. 21-13491X) and MUNI Award for Science and Humanities 3. The Czech Ministry of Education, Youth and Sports project No.\ CZ.02.2.69/0.0/0.0/18\_058/0010238 supported printing of this volume.

We thank V\'aclav Pavl\'{\i}k for designing the cover page. Last but not the least, the authors would like to thank Debora Lan\v{c}ov\'a and all our colleagues at the Institute of Physics of the Silesian University for their efficient and welcoming assistance. We thank Kris Schroven for very useful comments to the manuscript and for helping us with the preparation of the glossary that is included at the end of this volume.

\vspace*{2em}
\hrule
\newpage

\end{document}